\documentclass{aa}  % for twocol version
\usepackage{graphicx}
\usepackage{txfonts}
\usepackage{subfigure}
\usepackage{dblfloatfix}
\usepackage{pdfpages}
\usepackage{hyperref}
\begin{document}

   \title{The population of SNe/SNRs in the starburst galaxy \object{Arp 220}
   \thanks{Data, images and analysis scripts presented in this paper are
available in electronic form via the CDS as described in \ref{app:cds}.}
}

    \subtitle{A self-consistent analysis of 20 years of VLBI monitoring}

      \author{E.~Varenius \inst{\ref{inst:chalmers},\ref{inst:JBCA}} \and
J.~E.~Conway \inst{\ref{inst:chalmers}} \and 
F.~Batejat \inst{\ref{inst:chalmers}} \and 
I.~Mart\'i-Vidal \inst{\ref{inst:chalmers}} \and 
M.~A.~P\'erez-Torres \inst{\ref{inst:iaa},\ref{inst:unizar}} \and \\
S.~Aalto \inst{\ref{inst:chalmers}} \and
A.~Alberdi \inst{\ref{inst:iaa}} \and
C.~J.~Lonsdale \inst{\ref{inst:MIT}} \and
P.~Diamond \inst{\ref{inst:SKA}}
}

   \institute{
              Department of Earth and Space Sciences,
              Chalmers University of Technology
              Onsala Space Observatory,
              439 92 Onsala, 
              Sweden \\
              \email{varenius@chalmers.se}
              \label{inst:chalmers}
              \and
              Jodrell Bank Centre for Astrophysics, The University of Manchester, Oxford Rd, Manchester M13 9PL, UK.
              \label{inst:JBCA}
              \and
			  Instituto de Astrof\'isica de Andaluc\'ia (IAA, CSIC), Glorieta de las Astronom\'ia, s/n, E-18008 Granada, Spain.
              \label{inst:iaa}
              \and
              Departamento de F\'isica Teorica, Facultad de Ciencias, Universidad de Zaragoza, Spain.
              \label{inst:unizar}
              \and
			  Massachusetts Institute of Technology, Haystack Observatory, Westford, MA 01886, USA.
              \label{inst:MIT}
              \and
              SKA Organisation, Jodrell Bank Observatory, Lower Withington, Macclesfield, Cheshire SK11 9DL, UK.
              \label{inst:SKA}
          }

\date{Received February 15, 2017; accepted February 10, 2019}

% \abstract{}{}{}{}{} 
% 5 {} token are mandatory
 
  \abstract
  % context heading (optional)
  % {} leave it empty if necessary  
   { The nearby ultra-luminous infrared galaxy (ULIRG) Arp\,220 is an excellent
   laboratory for studies of extreme astrophysical environments. For 20 years,
   Very Long Baseline Interferometry (VLBI) has been used to monitor a
   population of compact sources thought to be supernovae (SNe), supernova
   remnants (SNRs), and possibly active galactic nuclei (AGNs).  SNe and SNRs
   are thought to be the sites of relativistic particle acceleration powering
   star formation induced radio emission in galaxies, and are hence
   important for studies of for example the origin of the FIR-radio correlation.
   }
  % aims heading (mandatory)
   {
In this work we aim for a self-consistent analysis of a large collection of
Arp\,220 continuum VLBI data sets.  With more data and improved consistency in
calibration and imaging, we aim to detect more sources and improve source
classifications with respect to previous studies. Furthermore, we aim to
increase the number of sources with robust size estimates, to analyse the
compact source luminosity function (LF), and to search for a
luminosity-diameter (LD) relation within Arp\,220.
   }
  % methods heading (mandatory)
   {
Using new and archival VLBI data spanning 20 years, we obtained 23
high-resolution radio images of Arp\,220 at wavelengths from 18\,cm to 2\,cm.
From model-fitting to the images we obtained estimates of flux densities and
sizes of detected sources. The sources were classified in groups according
to their observed lightcurves, spectra and sizes. We fitted a multi-frequency
supernova light-curve model to the object brightest at 6\,cm to estimate
explosion properties for this object.
   }
  % results heading (mandatory)
   {
       We detect radio continuum emission from 97 compact sources and present
       flux densities and sizes for all analysed observation epochs. The
       positions of the sources trace the star forming disks of the two nuclei
       known from lower-resolution studies.  We find evidence for a LD-relation
       within Arp\,220, with larger sources being less luminous.  We find a
       compact source LF $n(L)\propto L^\beta$ with $\beta=-2.19\pm0.15$,
       similar to SNRs in normal galaxies, and we argue that there are many
       relatively large and weak sources below our detection threshold.  The
       brightest (at 6\,cm) object 0.2195+0.492 is modelled as a radio SN with
       an unusually long 6\,cm rise time of 17 years. 
   }
  % conclusions heading (optional), leave it empty if necessary 
   {
       The observations can be explained by a mixed population of SNe and SNRs,
where the former expand in a dense circumstellar medium (CSM) and the latter
interact with the surrounding interstellar medium (ISM).   Nine sources are
likely luminous SNe, for example type IIn, and correspond to few percent of the total
number of SNe in Arp\,220. Assuming all IIns reach these luminosities, and
no confusion with other SNe types, our data are consistent with a total
SN-rate of 4\,yr$^{-1}$ as inferred from the total radio emission given a
normal stellar initial mass function (IMF). Based on the fitted luminosity
function, we argue that emission from all compact sources, also below our
detection threshold, make up at most 20\% of the total radio emission at GHz
frequencies.  However, colliding SN shocks and the production of secondary
electrons through cosmic ray (CR) protons colliding with the dense ISM may
cause weak sources to radiate much longer than assumed in this work. This could
potentially explain the remaining fraction of the smooth synchrotron component.
Future, deeper observations of Arp\,220 will probe the sources with lower
luminosities and larger sizes. This will further constrain the evolution of
SNe/SNRs in extreme environments and the presence of AGN activity.  }

	   \keywords{Galaxies: star formation, starburst, individual: Arp\,220 --
	   ISM: supernova remnants -- Stars: supernovae: general}

   \maketitle
%
%________________________________________________________________

\section{Introduction}
The ultra luminous infrared galaxy (ULIRG) Arp\,220 \citep{arp1966} is a
merging system with two nuclei (east and west) about 1$''$ (370\,pc) apart
\citep{norris1988}. The nuclei are hosts of intense star formation (for example
\citealt{parra2007}) as well as possible active galactic nuclei (AGN) activity
\citep{downes2007}.  Arp\,220 has been subject of global very long baseline
interferometry (VLBI) monitoring for 20 years, starting with the discovery of
multiple compact (<1pc) sources by \cite{smith1998} at 18\,cm.  Subsequent
observations revealed approximately 50 sources thought to be a radio supernovae
(SNe) and/or supernova remnants (SNRs), with many sources seen at multiple
wavelengths \citep{rovilos2005, lonsdale2006,
parra2007,batejat2011,batejat2012}. 

Multiple sources in Arp\,220 were resolved by \cite{batejat2011} and shown to
be consistent with a luminosity-diameter (LD) relation of $L\propto D^{-9/4}$
when extrapolating from the lower density galaxies LMC and M82.  The data
presented by \cite{batejat2011} were however not sensitive enough to determine
an LD-relation within Arp 220 itself.  Evidence for possible variability in
three sources was presented by \cite{batejat2012}, but these sources are very
weak, and multiple sensitive observing epochs are required to determine the
nature of the apparent variability.

With its large population of SNe/SNRs in a very dense ISM
($\sim10^5$\,cm$^{-3}$; \cite{scoville2015}) with strong (mGauss) magnetic
fields \citep{lacki+beck2013,yoast-hull2016}, Arp\,220 is an excellent
laboratory to study the physics of star formation in extreme environments.
SNe and SNRs are also the sources of relativistic particles which are responsible
for the star formation induced synchrotron radio emission in galaxies.
Observational constraints on how SNe interact with a dense ISM may provide
useful input to theories seeking to explain the well known FIR-radio
correlation.  Furthermore, the SNe blast waves are thought to be important for
stellar feedback in galaxies, since they provide the driving force behind the
large scale winds observed from dense starburst regions such as M\,82. 

The focus of this paper is to present a self-consistent set of data for the
compact objects in Arp\,220. We present both new and previously published data
at multiple frequencies spanning 20 years.  The new observations were designed
to
(1) improve our understanding of the evolutionary status of the SNe/SNRs,
(2) investigate the LD-relation within Arp\,220, 
(3) determine the nature of the variable sources.
In addition to new data, archival data was re-analysed using similar
calibration and imaging strategies to allow intercomparison.
This paper presents the full data set and discusses the general properties of
the population of compact objects. In future papers we intend to model in
detail individual objects and the overall population of compact sources to
constrain the evolutionary history of the compact radio objects.

This paper is organised as follows. In Sect. \ref{sect:obs} we describe the
observations as well as the method used to extract flux densities and sizes
from the observed data.  In Sect. \ref{sect:results} we present the results of
our analysis. These results are discussed in Sect.  \ref{sect:discussion}
focusing on the statistical properties of the source population rather than
individual objects.  Finally, we summarise our conclusions and future work in
Sect.  \ref{sect:summary}.  

Throughout this paper we have assumed an angular size distance to Arp\,220 of
77\,Mpc, that is  1\,mas=0.37\,pc, and luminosity distance of 80\,Mpc, as
obtained from \cite{wright2006} (with $H_0=69.9$, $\Omega_M$=0.286 and
$\Omega_\mathrm{vac}=0.714$) using $cz=5469$\,km/s
\citep{devancouleurs1991}\footnote{Via 
http://vizier.u-strasbg.fr/viz-bin/VizieR?-source=VII\%2F155.}.  All spectral
indices $\alpha$ assume the flux density follows $S_\nu\propto\nu^{+\alpha}$.

\section{Observations and data reduction}
\label{sect:obs}
This paper presents data from 23 VLBI epochs at wavelengths from 18 to 2\,cm,
see Table \ref{table:obslist}, spanning 20 years. Nine of these data sets have
not been published before. All data were processed into final images starting
from the archival raw data. Most observations were
performed with global VLBI using typically 15-20 antennas all over the world,
including the VLBA and the EVN.  The experiments GC031, BB297 and BB335
included multiple observations spanning a few months, the purpose being to look
for short term variability.  We note, however, that the epochs BB297B and
BB297C were severely limited in sensitivity due to lack of fringes to multiple
antennas, and therefore their respective 18\,cm and 3.6\,cm epochs (which are
the least sensitive) have been excluded from the analysis in this paper.

All calibration and imaging was performed using the 31DEC16 release of the
Astronomical Image Processing System (AIPS) \citep{greisen2003} from the
National Radio Astronomy Observatory (NRAO).  The full reduction process, from
loading of data to final images, was scripted for all data sets using the
Python-based interface to AIPS, ParselTongue \citep{kettenis2006} 2.3. All the
scripts are available via the CDS as described in appendix \ref{app:cds}.  A
general description of the calibration strategy employed can be found in
Appendix \ref{sect:calibration}.  The interested reader may find all details,
such as task parameters, in the ParselTongue scripts. Furthermore, all the FITS
images obtained when running the calibration and imaging scripts, that is two
images (east and west nuclei) of each of the 23 experiments listed in Table
\ref{table:obslist}, which are the basis of the analysis presented in this
paper, are also available via the CDS as described in appendix \ref{app:cds}.

\begin{table*}
\caption{List of the 23 observations processed and analysed in this work. The
RMS noise was measured in source free regions of the map, as noted in Sect.
\ref{sect:fittinguncert}.}
\label{table:obslist}
\centering
\begin{tabular}{l r r r r r}
\hline\hline
Exp. & Obs. date & $\lambda$ & Beam & Beam position & Image RMS \\ 
     & YYYY-MM-DD & (cm)      & (mas)&  angle [deg] & ($\mu$Jy/beam) \\ 
\hline
GL015 & 1994-11-13 & 18.1 & 8.2x2.5 & -9.2 & 36\\ 
GL021 & 1997-09-15 & 18.2 & 8.4x2.8 & -8.7 & 40\\ 
GL026 & 2002-11-16 & 18.2 & 6.4x2.9 & -16.5 & 12\\ 
GD017A & 2003-11-09 & 18.2 & 6.7x2.8 & -18.3 & 11\\ 
GD017B & 2005-03-06 & 18.2 & 6.2x3.1 & -15.7 & 15\\ 
BP129 & 2006-01-09 & 13.3 & 6.4x3.5 & -3.7 & 173\\ 
BP129 & 2006-01-09 & 6.0 & 2.7x1.4 & 0.1 & 100\\ 
BP129 & 2006-01-09 & 3.6 & 1.8x0.9 & 0.1 & 91\\ 
GD021A & 2006-06-06 & 18.2 & 6.6x2.9 & -21.9 & 16\\ 
GC028 & 2006-11-28 & 3.6 & 1.4x0.5 & -10.4 & 39\\ 
GC028 & 2006-12-28 & 2.0 & 1.4x0.4 & -13.0 & 25\\ 
GC031A & 2008-06-10 & 6.0 & 2.2x0.8 & -13.2 & 17\\ 
GC031B & 2008-10-24 & 6.0 & 1.9x0.8 & -17.6 & 13\\ 
GC031C & 2009-02-27 & 6.0 & 2.0x0.8 & -17.6 & 13\\ 
BB297A & 2011-05-16 & 18.2 & 10.8x4.0 & -16.6 & 40\\ 
BB297A & 2011-05-16 & 6.0 & 2.1x1.0 & -13.4 & 16\\ 
BB297A & 2011-05-16 & 3.6 & 1.6x0.5 & -6.5 & 30\\ 
BB297B & 2011-05-17 & 6.0 & 3.0x0.8 & -16.1 & 32\\ 
BB297C & 2011-06-11 & 6.0 & 2.7x1.0 & 29.2 & 33\\ 
BB335A & 2014-08-01 & 6.0 & 3.0x0.9 & -9.8 & 9\\ 
BB335A & 2014-08-01 & 3.6 & 1.9x0.6 & -5.0 & 17\\ 
BB335B & 2014-10-13 & 6.0 & 3.2x1.0 & -11.0 & 10\\ 
BB335B & 2014-10-13 & 3.6 & 1.8x0.6 & -5.5 & 19\\ 
\hline
\end{tabular}
\end{table*}

We note that in this paper we sometimes use the letters L, S, C, X and U to
designate frequency bands, similar to the IEEE standard letter designations for
radar frequency bands (IEEE Std 521-2002). Table \ref{table:bands} provides a quick
reference for translation between the letters used in this paper and the
approximate central wavelength of observations.  We note that the central
observing wavelength is listed for each observation in Table
\ref{table:obslist}.

\begin{table}
\caption{Frequency band letter designations used in this paper. 
}
\label{table:bands}
\centering
\begin{tabular}{r l}
\hline\hline
     Band & Obs. wavelength\\ 
     \hline
     L &  18~cm\\ 
     S &  13~cm\\ 
     C &  6~cm\\ 
     X &  3.6~cm\\ 
     U &  2~cm\\ 
     \hline
     \end{tabular}
\end{table}

Previous studies of Arp\,220 have used different (although similar) methods to
analyse the data. For example, different criteria have been used to determine
the catalogue of sources to analyse. In the following subsections we describe how
we form a source catalogue, how we fit sizes and flux densities and estimate the
corresponding uncertainties, and how we obtain more robust size-estimates by
averaging measurements close in time.

\subsection{Building a source catalogue}
\label{sect:sourcefinding}
\begin{figure*}[tbp]
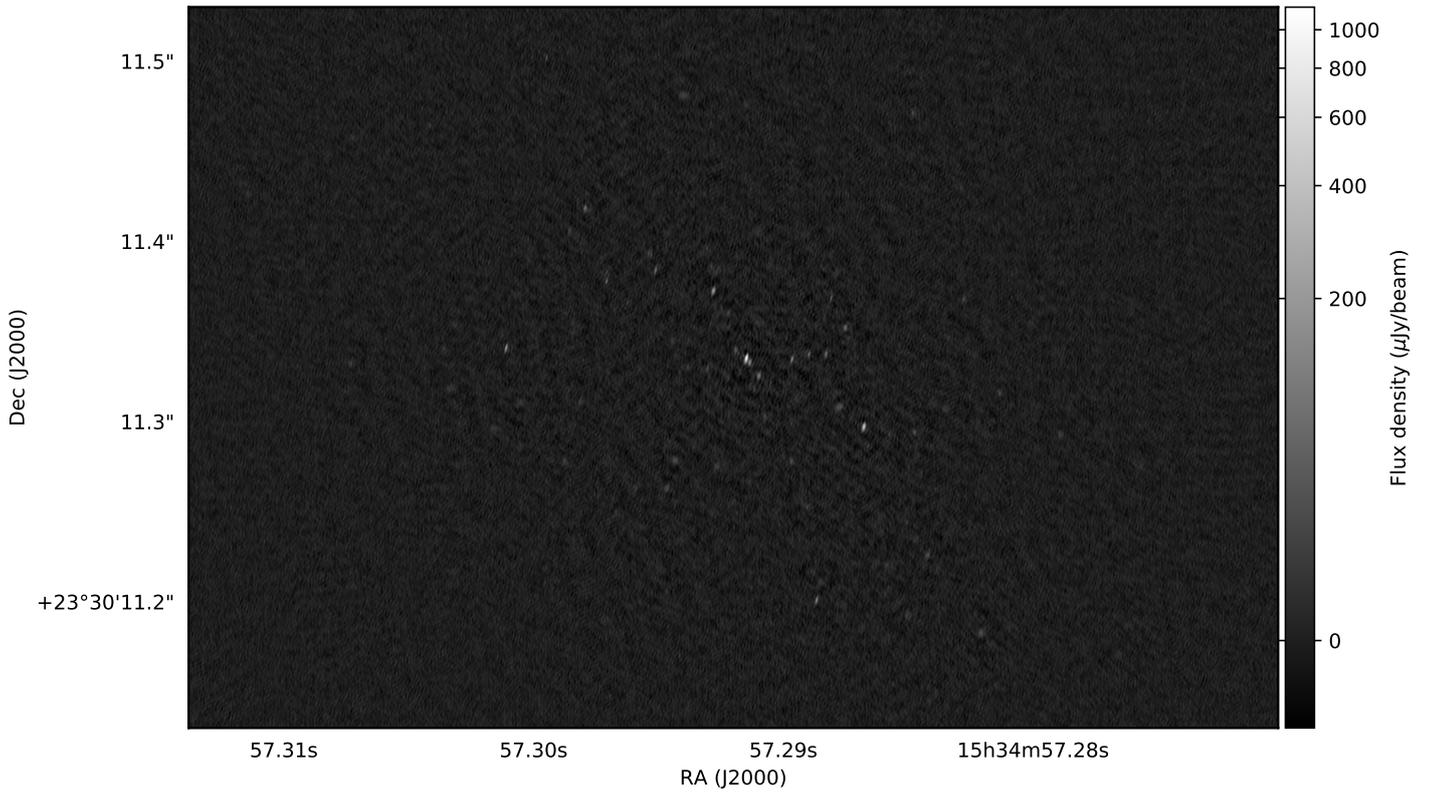
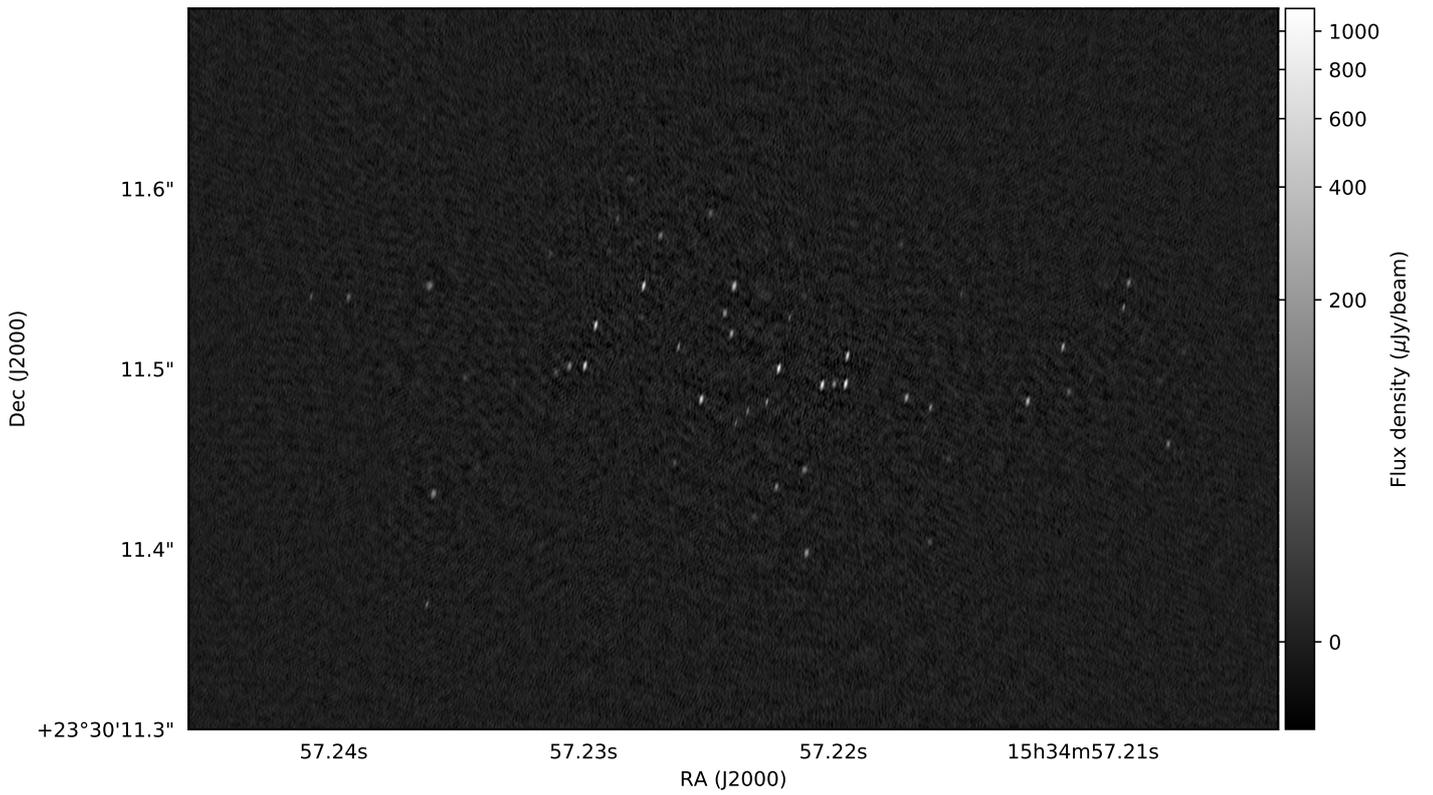

\centering
\subfigure[The eastern nucleus of Arp 220]{
        \includegraphics[height = 0.42\textheight]{figures/{{EAST_C_BB297AREF_STACKED-crop}}}
        \label{fig:stackedeast}
}
\subfigure[The western nucleus of Arp 220]{
        \includegraphics[height = 0.42\textheight]{figures/{{WEST_C_BB297AREF_STACKED-crop}}}
        \label{fig:stackedwest}
}
\caption{Cutouts of the central parts of the stacked 6\,cm images towards the
	east and west nuclei of Arp\,220, with off-source RMS noise
	$\sigma=4\mu$Jy\,beam$^{-1}$. The two panels are displayed using the same
	\emph{arcsinh} grey-scale from $-20\mu$Jy\,beam$^{-1}$ ($-5\sigma$) to
	$1140\mu$Jy\,beam$^{-1}$ (the maximum brightness value). The full
	$0.8192''\times0.8192''$ images of both nuclei were used for the source
	finding, as described in Sect.  \ref{sect:sourcefinding}. The positions of
	the sources studied in this work is shown in Fig. \ref{fig:positions}.}
\label{fig:stacked}
\end{figure*}

To increase our sensitivity for source detection we average all 6\,cm images
together, as a simple weighted average where each image was weighted by its
RMS noise given in Table \ref{table:obslist}, to the power of $-2$.
\footnote{We note that the images were averaged together without accounting
	for differences in respective synthesised beams. This should however only
	have a minor effect on finding the source positions to within $\pm$1\,mas.
	We note that all sizes and flux densities in this work are measured from the
single-epoch images.}  By such averaging, we obtain the deepest maps yet
(off-source RMS noise $\sigma=4.3\,\mu$Jy/beam) of the two nuclei. The central
parts of these two stacked images are presented in Fig. \ref{fig:stacked}.  

To produce a source catalogue we apply the source
finding software PyBDSM version 1.8.6 \citep{mohan2015} on the stacked image,
with a threshold of six times the (stacked) image RMS noise, resulting in a
catalogue of 75 sources at 6\,cm.  Similarly, we stacked the 18\,cm and 3.6\,cm
images (stacked sensitivities of 5 and 10$\mu$Jy/beam respectively) and
detected 22 additional sources seen only at 18\,cm (no new additions were found
in the 3.6\,cm images). \footnote{We note that 15 sources were only detected at
	6\,cm, but care should be taken when comparing frequencies given the
	different sensitivities.  In particular, the spectral indices of the
sources (affected by for example no, little or severe free-free absorption) combined
with differences in luminosity may explain the differences in numbers at
different frequencies.}  The total catalogue hence contains 97 sources and is
presented in Tables \ref{table:east} and \ref{table:west}.  We note that three
additional sources were detected in OH-maser emission only, that is no
continuum, as briefly discussed in Appendix \ref{sect:ohmasers}. The maser
channels in the data were however excluded when making the FITS images analysed
in this paper.

To estimate the expected number of false positives  we use a beam of major
and minor FWHM 3\,mas and 1\,mas respectively (similar to the most sensitive
epochs which have the largest impact on the weighted average).  Given the
number of independent beam elements searched across the two nuclei, a threshold
of $6\sigma$ implies an expected number of false positives of $6\cdot10^{-4}$,
that is well below one. We note that lowering the threshold to $5\sigma$ would
give one expected false detection.

\subsection{Fitting source sizes and flux densities}
\label{sec:fitting}

In principle, it is best to fit models directly to the calibrated visibilities,
instead of the cleaned images, to avoid any effects introduced by 
deconvolution (for example\citealt{marti-vidal2014}). However, the number of sources
in Arp\,220, which need to be fitted simultaneously in the Fourier domain,
make visibility fitting impractical. We therefore
decided to fit models to the cleaned images. All sources were assumed to be
optically thin shells with a fractional shell width of 30\%, similar to
the value observed for SN1993J (see for example \citealt{martividal2011SN1993J}).
\footnote{We note that there may be ejecta opacity effects present in some sources,
as well as asymmetrical structure. Such details are however beyond the scope of this
work, and are the subject of a future, more detailed analysis.} 
The fitting
was done using the bounded least-squares algorithm implemented in the Python
package SciPy \footnote{Using the function scipy.optimize.least\_squares which
	offers bounded fitting since early 2016.}. More details and examples of fitting
results are presented in Appendix \ref{sect:fitting}. A comparison of the
results obtained in this work with previous published values using the same
data is presented in Appendix \ref{sect:prevstudycomp}.
We note that although we fit sizes at all wavelengths, the angular resolution
at 18\,cm and 13\,cm is not sufficient to obtain meaningful sizes at these
wavelengths.

\subsection{Uncertainties}
\label{sect:fittinguncert}
Based on the observed scatter of the fitted positions (see Appendix
\ref{sect:posacc}), we estimate that the catalogue positions given in Tables
\ref{table:east} and \ref{table:west} are accurate to within $\pm1$\,mas.
For flux densities we estimate the uncertainty as
\begin{equation}
E_f = \sqrt{(3\sigma)^2+(0.1F)^2}
\label{eqn:ef}
\end{equation}
where $\sigma$ is the map RMS and $F$ is the measured flux density. This
includes both a conservative estimate of the image noise effects, as well a
10\% uncertainty on the absolute flux density calibration.
Finally, for source sizes, we adopt an uncertainty based on Eq.7 by
\cite{martividal2012} of 
\begin{equation}
E_s=3b_\text{maj}/\sqrt{\text{STN}}
\label{eqn:es}
\end{equation}
where $b_\text{maj}$ is the major axis of the fitted CLEAN beam, and STN is the
signal-to-noise ratio defined as $F$ divided by the respective map RMS noise.
We find the uncertainties calculated above to reflect well the observed scatter
in simulated data, see Appendix \ref{sect:fitting}.

We note that the RMS values given in Table \ref{table:obslist} are measured in
source-free regions of the map. Specifically, the RMS is measured from the
leftmost 128x8192 pixels of each 8192x8192 pixel image of the western nucleus.
We quote these RMS values because the source-free regions measure the performance
of the instrument, that is telescope sensitivity, rather than imaging limitations.
Hence, the off-source RMS is a good way to compare the different observing epochs.
We acknowledge that the RMS in the central regions of the nuclei are generally
higher, due to residual sidelobes from imperfect calibration and/or imaging as
well as sources just below our detection threshold. This is the reason we use 
$3\sigma$ in for example  Eq.  \ref{eqn:ef} rather than just $1\sigma$ 
(as used by multiple previous studies analysing parts of these data).

\subsection{Averaging of multiple epochs}
\label{sect:avgsize}
Although we fit source flux densities and sizes at each epoch separately, the
uncertainties are sometimes very large, in particular on the sizes of the
weakest sources.  We therefore, in some sections of this paper, calculate and
discuss average sizes and flux densities using measurements from multiple
epochs.  An average size is calculated as the weighted average 
\begin{equation}
	\label{eqn:avgs}
	\bar{s} = \frac{\sum s_i/E_{s_i}^2}{\sum 1/E_{s_i}^2}
\end{equation}
where the uncertainty of the weighted mean is calculated as
\begin{equation}
	\label{eqn:avge}
	\bar{E}_s=\frac{1}{\sqrt{\sum 1/E_{s_i}^2}}
\end{equation}
where $s_i$ is a size fitted from an image and $E_{s_i}$ is the corresponding
uncertainty according to Eq. \ref{eqn:es}. An average flux density is
calculated in the same way, using the measured flux densities and their respective
uncertainties instead of the sizes in Eqns. \ref{eqn:avgs} and \ref{eqn:avge}.

\subsection{Source classification}
\label{sect:classification}
To facilitate a discussion of the data, we classified all sources in three
groups based on their 6\,cm evolution between the experiments GC031 and BB335.
The sources were labelled either \emph{Rise}, \emph{Fall}, or \emph{Slowly
varying} (abbreviated S-var in many places in this paper) according to the
following criteria.  First, average flux densities and flux density
uncertainties were calculated (as described in Sect. \ref{sect:avgsize}) for
the GC031 and BB335 experiments, so that we obtained two flux
density values per source. Let now $F_\mathrm{GC}$ and $E_\mathrm{GC}$ be the
average flux density and flux density uncertainty for a given source obtained
for the GC031 experiment, and $F_\mathrm{BB}$ and $E_\mathrm{BB}$ be the
corresponding values for the BB335 experiment. A source was classified as
\emph{Rise} if $(F_\mathrm{GC} + E_\mathrm{GC}) < 0.9\times(F_\mathrm{BB} -
E_\mathrm{BB})$ and as \emph{Fall} if $(F_\mathrm{GC} - E_\mathrm{GC}) >
1.1\times(F_\mathrm{BB} + E_\mathrm{BB})$.  If none of these conditions were
satisfied, that is if there was no strong 6\,cm evidence that a source
was increasing or decreasing, the source was classified as \emph{Slowly varying}
with respect to the noise. We note that weak sources in the S-var category are
not ruled out from increasing or decreasing by more than 10\%, but there is no
positive evidence for such variations given the measured uncertainties. A few 
sources were not detected in either GC031 or BB335 and are hence labelled
\emph{N/A} in Tables \ref{table:east} and \ref{table:west}.

\subsubsection{Motivation}
\label{sect:motivation}
Note the classification presented in Sect. \ref{sect:classification} is purely
observational: it does not imply any particular source nature (such as
SNe/SNRs).  A detailed interpretation of the nature of the sources should of
course use all the available information. One may therefore ask why we decided
to base the classification only on the 6\,cm lightcurves?  There are multiple
reasons for this choice. 

The purpose of this classification is to facilitate the discussion.  However,
we find it instructive to divide the sources in groups already for the
presentation of the results in Sect. \ref{sect:results}.  Given the number of
sources and the high level of detail in these data, a physical classification
(in contrast to this observational one) requires an extensive discussion to
motivate the different possible classes. Such a discussion would be out of
place before all the results are presented.

One may argue that if one devises classifications solely based on one waveband,
one should use 18\,cm since it spans the longest period in time. However, the
18\,cm behaviour varies considerably between sources which have similar trends
at 6\,cm. Indeed, multiple studies (for example
\citealt{smith1998b,lonsdale2006,varenius2016}) argue that
free-free absorption may significantly impact the flux densities measured at
1.4\,GHz, which indicate that a classification based on 18\,cm lightcurves may
not reflect intrinsic source properties.  Finally, the 18\,cm observations do
not have enough angular resolution to properly disentangle the emission of all
sources, as a few sources are very close, thus mixing the lightcurves of 
blended sources.

In principle, also the 6\,cm measurements may be affected by free-free
absorption, and therefore one should use the 3.6\,cm measurements to classify
sources based on intrinsic properties.  However, given the available data, we
find the 6\,cm observations to provide more robust lightcurves. The reason for
this is twofold: first, the 6\,cm observations are, in general, more sensitive
than the 3.6\,cm observations. Second, although the recent 3.6\,cm observations
in experiment BB335 reach a relatively high sensitivity, we lack an experiment
with comparable sensitivity far back in time.  At 6\,cm we have the experiments
GC031 and BB335, both with high sensitivity, separated by about 6 years in
time. Given the above arguments, we decided to use the 6\,cm lightcurves as the
basis of our classification.  We find that these simple classes provide an
excellent foundation for the presentation of results and subsequent discussion
of the data. We note that the physical nature of the sources is discussed in
Sect. \ref{sect:discussion}.

\section{Results}
\label{sect:results}
We detect 97 compact continuum sources in Arp\,220 with positions given in
Tables \ref{table:east} and \ref{table:west}.  The sources show a variety of
lightcurves, spectra and sizes.  A comprehensive summary of all data available
for each source is presented in Appendix \ref{app:book}. In this appendix, all
measured flux densities and sizes are plotted as function of time for all
sources. An approximate spectrum is also shown for each source. Note that all
measured flux densities and sizes are also available in machine-readable format
via CDS as described in appendix \ref{app:cds}.  The diameters and
corresponding uncertainties listed in Tables \ref{table:east} and
\ref{table:west} are calculated using Eqs.  \ref{eqn:avgs} and \ref{eqn:avge}
from sizes fitted in the four images obtained at 6\,cm and 3.6\,cm in
experiments BB335A and BB335B. 

\begin{figure*}[htbp]
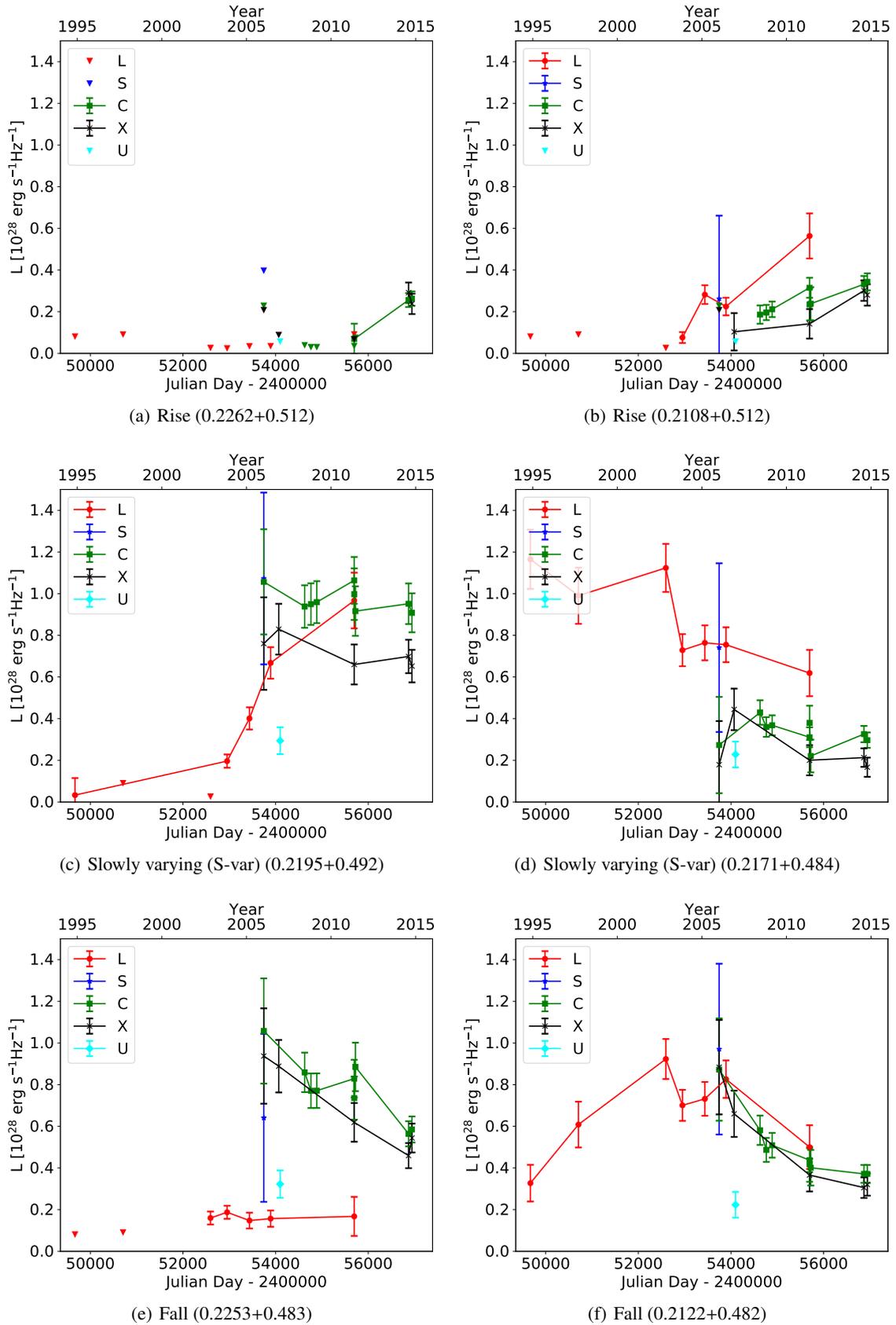

\centering
\subfigure[Rise (0.2262+0.512)]{
        \includegraphics[width=0.40\textwidth]{figures/{{Rise_0.2262+0.512}}}
        \label{fig:riselowL}
}
\subfigure[Rise (0.2108+0.512)]{
        \includegraphics[width=0.40\textwidth]{figures/{{Rise_0.2108+0.512}}}
        \label{fig:rise}
}
\subfigure[Slowly varying (S-var) (0.2195+0.492)]{
        \includegraphics[width=0.40\textwidth]{figures/{{S-var_0.2195+0.492}}}
        \label{fig:svarlowL}
}
\subfigure[Slowly varying (S-var) (0.2171+0.484)]{
        \includegraphics[width=0.40\textwidth]{figures/{{S-var_0.2171+0.484}}}
        \label{fig:svar}
}
\subfigure[Fall (0.2253+0.483)]{
        \includegraphics[width=0.40\textwidth]{figures/{{Fall_0.2253+0.483}}}
        \label{fig:falllowL}
}
\subfigure[Fall (0.2122+0.482)]{
        \includegraphics[width=0.40\textwidth]{figures/{{Fall_0.2122+0.482}}}
        \label{fig:fall}
}
\caption{Lightcurves to illustrate the source classifications
\emph{Rise}, \emph{Slowly varying} (or S-var) and \emph{Fall} used in this
paper. For the purpose of the discussion, two sources are shown for each class,
to show the differences in multi-frequency behaviour within the classes. All
panels have the same scale for easy comparison. The horizontal axis below the
panels show time in Julian Days, while the horizontal axis on top of the panels
show the time in decimal years.  
In Sect. \ref{sect:nature} we argue that these lightcurves are consistent with
SNe/SNRs in different evolutionary stages, where the approximate time order
corresponds to panels a, c, f, b, d. Panel e may represent a stage similar to f
but with significant free-free absorption. Note that the label L on the vertical
axis here denotes spectral luminosity.
\label{fig:classex}} 
\end{figure*}

The sources were classified as described in Sect. \ref{sect:classification}.
Examples of sources in the three categories \emph{Rise}, \emph{Fall}, and
\emph{Slowly varying} (or S-var) are shown in Fig.  \ref{fig:classex}. For the
purpose of the discussion we show two sources for each class to illustrate
the variations in multi-frequency lightcurves within the classes.

\begin{table*}
\caption{\label{table:east}List of the 45 sources analysed in this work in the eastern nucleus.
$\Delta$R.A. and $\Delta$Dec. are given with respect to the position R.A.
15$^h$34$^m$57.0000$^s$ and Dec. 23$^\circ$30$'$11.000$''$. Flux densities
F$_\mathrm{6\,cm}$ are measured by fitting a spherical shell to BB335B. Average
diameters calculated according to Sect. \ref{sect:avgsize} from the 3.6\,cm and
6\,cm measurements in experiments BB335A and BB335B.}
\centering
\begin{tabular}{r r l r r r }
\hline\hline
$\Delta$R.A. [s] & $\Delta$Dec. [$''$] & Legacy name & Class & F$_\mathrm{6\,cm}$ [$\mu$Jy] & Average diameter [mas]\\
\hline
0.2758 & 0.276 & E22 & Rise & $128\pm30$& $5.06\pm1.31$\\ 
0.2789 & 0.293 & E21 & S-var & $109\pm29$& $3.21\pm2.17$\\ 
0.2801 & 0.353 &  & S-var & $49\pm28$& $3.84\pm4.18$\\ 
0.2814 & 0.316 &  & S-var & $68\pm28$& $2.00\pm2.41$\\ 
0.2821 & 0.183 & E20 & S-var & $181\pm33$& $3.03\pm1.21$\\ 
0.2835 & 0.308 &  & Rise & $116\pm30$& $3.72\pm2.04$\\ 
0.2840 & 0.151 & E19 & S-var & $<27$& $2.62\pm3.39$\\ 
0.2842 & 0.226 & E18 & S-var & $104\pm29$& $2.00\pm1.91$\\ 
0.2848 & 0.294 &  & Fall & $77\pm28$& $1.29\pm2.74$\\ 
0.2848 & 0.471 & E16 & Fall & $94\pm29$& $2.75\pm1.36$\\ 
0.2850 & 0.193 & E17 & Fall & $107\pm29$& $2.70\pm2.37$\\ 
0.2858 & 0.013 &  & S-var & $<27$& N/A\\ 
0.2868 & 0.297 & E14 & S-var & $553\pm61$& $0.87\pm0.53$\\ 
0.2875 & 0.352 &  & Fall & $121\pm30$& $2.57\pm0.80$\\ 
0.2878 & 0.308 & E13 & Fall & $184\pm33$& $3.03\pm1.12$\\ 
0.2881 & 0.369 &  & Rise & $156\pm31$& $0.78\pm1.06$\\ 
0.2883 & 0.338 &  & S-var & $110\pm29$& $1.83\pm0.94$\\ 
0.2887 & 0.201 &  & S-var & $166\pm32$& $1.01\pm0.98$\\ 
0.2890 & 0.337 &  & S-var & $115\pm30$& $0.97\pm1.12$\\ 
0.2891 & 0.253 &  & S-var & $123\pm30$& $3.13\pm1.27$\\ 
0.2897 & 0.278 &  & S-var & $97\pm29$& $2.16\pm1.95$\\ 
0.2897 & 0.335 &  & S-var & $172\pm32$& $1.48\pm0.80$\\ 
0.2910 & 0.325 & E11 & S-var & $180\pm33$& $1.77\pm0.86$\\ 
0.2914 & 0.333 &  & Fall & $216\pm35$& $0.98\pm0.80$\\ 
0.2915 & 0.335 & E10 & Fall & $853\pm89$& $0.56\pm0.40$\\ 
0.2915 & 0.476 & E9 & S-var & $43\pm28$& $0.97\pm4.50$\\ 
0.2919 & 0.340 &  & S-var & $93\pm29$& $1.63\pm1.30$\\ 
0.2928 & 0.373 & E24 & S-var & $286\pm39$& $1.34\pm0.68$\\ 
0.2931 & 0.330 & E8 & S-var & $73\pm28$& $0.44\pm2.40$\\ 
0.2938 & 0.344 &  & Fall & $<27$& $2.74\pm1.25$\\ 
0.2940 & 0.481 &  & S-var & $125\pm30$& $3.86\pm1.88$\\ 
0.2943 & 0.279 & E7 & S-var & $127\pm30$& $2.22\pm1.48$\\ 
0.2947 & 0.263 & E6 & S-var & $74\pm28$& $1.85\pm2.18$\\ 
0.2951 & 0.384 &  & Rise & $215\pm35$& $0.86\pm0.94$\\ 
0.2954 & 0.393 &  & S-var & $122\pm30$& $3.07\pm1.91$\\ 
0.2959 & 0.263 &  & S-var & $89\pm29$& $2.84\pm2.38$\\ 
0.2971 & 0.378 &  & S-var & $63\pm28$& $3.14\pm0.95$\\ 
0.2979 & 0.419 & E5 & S-var & $137\pm30$& $1.89\pm0.87$\\ 
0.2995 & 0.502 &  & S-var & $54\pm28$& $0.67\pm1.78$\\ 
0.3011 & 0.341 &  & Rise & $312\pm41$& $0.51\pm0.68$\\ 
0.3073 & 0.332 & E3 & Rise & $80\pm28$& $1.80\pm2.49$\\ 
0.3073 & 0.456 &  & N/A & $<27$& N/A\\ 
0.3090 & 0.618 &  & N/A & $<27$& N/A\\ 
0.3103 & 0.575 &  & N/A & $<27$& N/A\\ 
0.3125 & 0.494 &  & N/A & $<27$& N/A\\ 
\hline
\end{tabular}
\end{table*}

\begin{table*}
\caption{\label{table:west}List of the 52 sources analysed in this work in the western nucleus.
$\Delta$R.A. and $\Delta$Dec. are given with respect to the position R.A.
15$^h$34$^m$57.0000$^s$ and Dec. 23$^\circ$30$'$11.000$''$. Flux densities
F$_\mathrm{6\,cm}$ are measured by fitting a spherical shell to BB335B. Average
diameters calculated according to Sect. \ref{sect:avgsize} from the 3.6\,cm and
6\,cm measurements in experiments BB335A and BB335B.}
\centering
\begin{tabular}{r r l r r r }
\hline\hline
$\Delta$R.A. [s] & $\Delta$Dec. [$''$] & Legacy name & Class & F$_\mathrm{6\,cm}$ [$\mu$Jy] & Average diameter [mas]\\
\hline
0.2066 & 0.458 & W46 & S-var & $235\pm37$& $2.00\pm0.86$\\ 
0.2082 & 0.548 & W44 & S-var & $267\pm39$& $1.60\pm0.82$\\ 
0.2084 & 0.534 &  & Rise & $197\pm34$& $0.61\pm0.92$\\ 
0.2106 & 0.487 &  & S-var & $167\pm33$& $2.08\pm0.95$\\ 
0.2108 & 0.512 &  & Rise & $448\pm53$& $0.88\pm0.64$\\ 
0.2122 & 0.482 & W42 & Fall & $485\pm56$& $1.20\pm0.61$\\ 
0.2149 & 0.542 &  & S-var & $83\pm29$& $0.79\pm1.66$\\ 
0.2154 & 0.450 &  & S-var & $77\pm29$& $2.33\pm2.32$\\ 
0.2161 & 0.479 & W40 & S-var & $225\pm36$& $1.15\pm0.86$\\ 
0.2162 & 0.404 & W41 & S-var & $112\pm30$& $1.86\pm1.46$\\ 
0.2171 & 0.484 & W39 & S-var & $387\pm48$& $1.44\pm0.73$\\ 
0.2173 & 0.569 &  & Fall & $89\pm30$& $2.90\pm1.16$\\ 
0.2194 & 0.507 & W58 & Rise & $820\pm86$& $0.58\pm0.51$\\ 
0.2195 & 0.492 & W34 & S-var & $1185\pm122$& $0.51\pm0.40$\\ 
0.2200 & 0.492 & W33 & Fall & $173\pm33$& $2.37\pm1.00$\\ 
0.2205 & 0.491 & W56 & S-var & $1073\pm111$& $0.72\pm0.40$\\ 
0.2211 & 0.398 & W30 & S-var & $383\pm47$& $1.85\pm0.69$\\ 
0.2212 & 0.444 & W29 & S-var & $317\pm42$& $2.38\pm0.73$\\ 
0.2212 & 0.540 & W28 & Rise & $92\pm30$& $3.20\pm2.02$\\ 
0.2218 & 0.528 &  & Rise & $116\pm31$& $0.52\pm0.86$\\ 
0.2222 & 0.500 & W25 & Fall & $950\pm99$& $0.59\pm0.41$\\ 
0.2223 & 0.435 & W26 & Fall & $227\pm36$& $1.48\pm0.83$\\ 
0.2227 & 0.482 & W55 & Fall & $127\pm31$& $0.86\pm0.92$\\ 
0.2234 & 0.477 &  & Rise & $230\pm36$& $0.45\pm0.61$\\ 
0.2239 & 0.470 &  & S-var & $67\pm29$& $1.37\pm0.94$\\ 
0.2240 & 0.546 & W18 & S-var & $737\pm79$& $1.44\pm0.51$\\ 
0.2241 & 0.520 & W17 & S-var & $353\pm45$& $1.16\pm0.66$\\ 
0.2244 & 0.531 & W16 & S-var & $293\pm41$& $1.60\pm0.76$\\ 
0.2249 & 0.586 &  & S-var & $178\pm33$& $1.89\pm1.15$\\ 
0.2253 & 0.483 & W15 & Fall & $763\pm81$& $1.00\pm0.49$\\ 
0.2262 & 0.512 &  & Rise & $343\pm44$& $0.59\pm0.69$\\ 
0.2264 & 0.448 &  & S-var & $96\pm30$& $2.46\pm0.96$\\ 
0.2269 & 0.574 & W14 & S-var & $244\pm37$& $2.30\pm0.75$\\ 
0.2276 & 0.546 & W60 & S-var & $997\pm103$& $0.45\pm0.43$\\ 
0.2277 & 0.528 &  & S-var & $67\pm29$& $2.43\pm1.47$\\ 
0.2281 & 0.605 &  & Fall & $81\pm29$& $3.12\pm2.23$\\ 
0.2286 & 0.583 &  & S-var & $134\pm31$& $1.15\pm1.74$\\ 
0.2290 & 0.564 &  & S-var & $71\pm29$& $2.94\pm1.30$\\ 
0.2295 & 0.524 & W12 & S-var & $904\pm94$& $0.61\pm0.48$\\ 
0.2299 & 0.502 & W11 & Rise & $872\pm91$& $0.90\pm0.45$\\ 
0.2306 & 0.502 & W10 & S-var & $338\pm44$& $2.02\pm0.70$\\ 
0.2310 & 0.496 &  & S-var & $<28$& $2.45\pm3.89$\\ 
0.2311 & 0.498 &  & S-var & $115\pm30$& $2.68\pm1.18$\\ 
0.2317 & 0.402 & W9 & S-var & $65\pm29$& $1.58\pm3.71$\\ 
0.2347 & 0.495 &  & S-var & $106\pm30$& $2.79\pm1.35$\\ 
0.2360 & 0.431 & W8 & S-var & $381\pm47$& $2.49\pm0.68$\\ 
0.2362 & 0.546 & W7 & Fall & $329\pm43$& $2.94\pm0.70$\\ 
0.2363 & 0.369 &  & S-var & $120\pm31$& $0.62\pm1.17$\\ 
0.2364 & 0.639 &  & S-var & $39\pm29$& $0.66\pm3.62$\\ 
0.2394 & 0.540 & W2 & S-var & $196\pm34$& $1.77\pm0.97$\\ 
0.2395 & 0.637 &  & N/A & $<28$& N/A\\ 
0.2409 & 0.540 &  & Rise & $105\pm30$& $0.22\pm1.65$\\ 
\hline
\end{tabular}
\end{table*}

\begin{figure*}
\centering
\includegraphics[width = \textwidth]{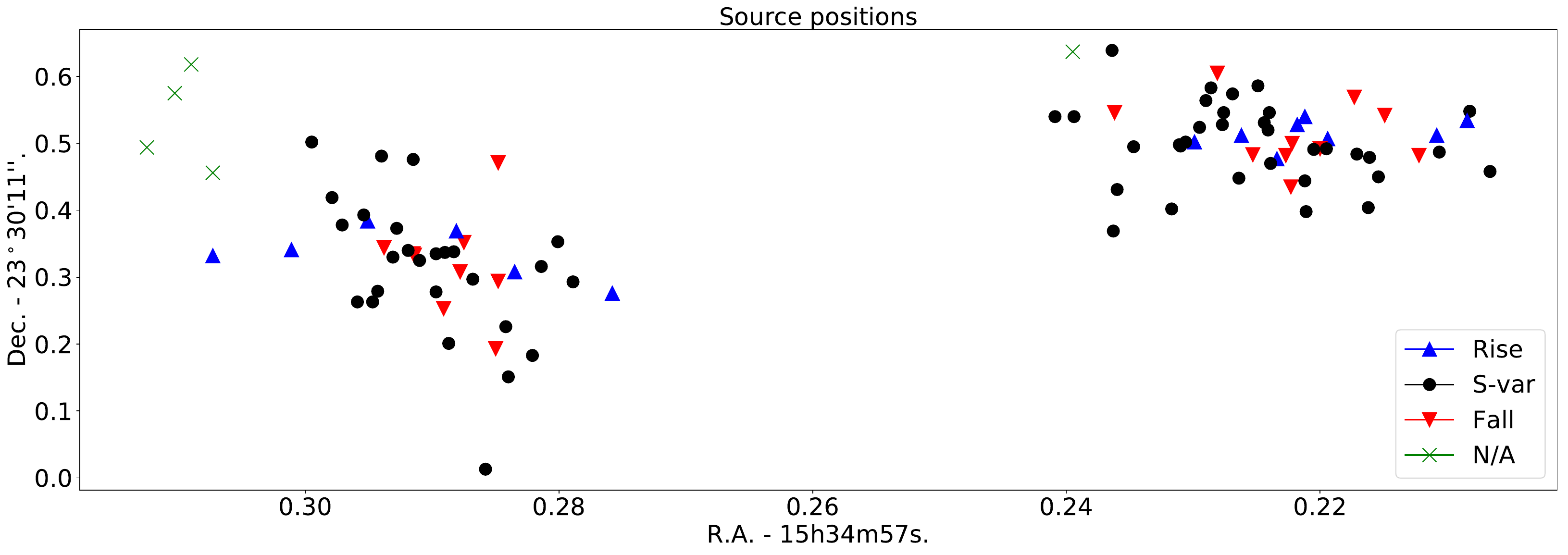} 
\caption{Spatial distribution of all sources given in Tables
\ref{table:east} and \ref{table:west} with the eastern nucleus to the left and the western
nucleus to the right. Each source is coloured by its observational
classification (including not classified, or N/A), as described in Sect.
\ref{sect:classification}. We note that a few sources are detected in the eastern
nucleus outside the central cutout region shown in Fig.  \ref{fig:stackedeast}.
\label{fig:positions} }
\end{figure*}
In Fig. \ref{fig:positions}, we show the positions of the sources, coloured by
classification. The sources trace a region extended from east to west in the
western nucleus, and from south-west to north-east in the eastern nucleus. This
is consistent with the orientation of the disks seen at lower spatial
resolution at 5\,GHz by \cite{barcosmunoz2015}.  

\subsection{The luminosity-diameter relation}
\label{sect:LD}
\begin{figure*}
\centering
        \includegraphics[width=0.85\textwidth]{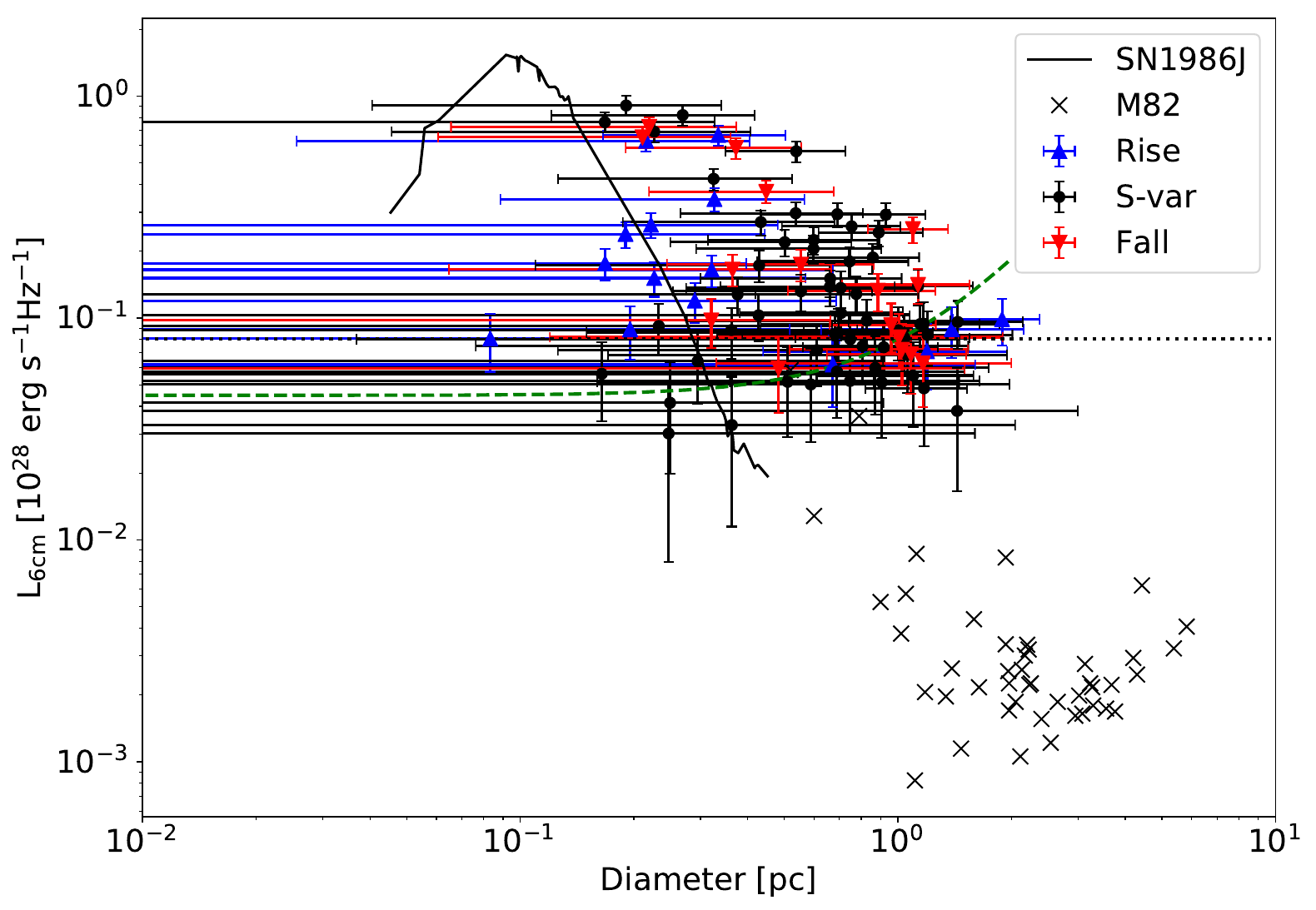}
\caption{The data in Tables \ref{table:east} and \ref{table:west} plotted as
spectral luminosity vs diameter in log-log scale.  The surface brightness
detection limit of BB335B is shown as a dashed line, and the fitted luminosity
completeness limit $L_c$, as derived in Sect.  \ref{sect:lumhist}, is shown as
a dotted line.  The lower panel show 45 SNRs in M\,82 plotted as black crosses
(data from \citealt{huang1994}, their Table 2, scaled to 6\,cm assuming
$\alpha=-0.5$). The evolution of SN1986J during its first 30 years is plotted
as a solid curve, from the model described in Sect.  \ref{sect:LD}.
\label{fig:LD}} 
\end{figure*}

Fig. \ref{fig:LD} shows the 6\,cm spectral luminosity versus source diameter
for the sources detected in Arp\,220, that is the values listed in Tables
\ref{table:east} and \ref{table:west}.
The luminosities were calculated from the flux densities measured in BB335B
while the sizes were averaged over the BB335 6\,cm and 3.6\,cm observations as
described in Sect. \ref{sect:avgsize}.
In Fig. \ref{fig:LD} the data are shown in log-log scale together with 45
SNRs observed in the nearby galaxy M82, using data from \cite{huang1994}.  We
note that the two brightest SNRs in M82 overlap the weakest objects detected in
Arp\,220.  For discussion purposes, we have also overlayed the evolution of the
bright radio supernova SN1986J\footnote{SN1986J was chosen because it is the
highest luminosity radio SNe
($L_\text{6cm}^\text{peak}=1.97\times10^{28}$\,erg\,s$^{-1}$\,Hz$^{-1}$;
\cite{weiler2002}) to have good VLBI size measurements.  Sources with higher
peak luminosity have been observed (for example SN1988Z; \cite{weiler2002}) but these
are significantly more distant and hence both weaker and smaller in angular
size.} during its first 30 years, using the 5\,GHz flux density measurements
presented by \cite{weiler1990, bietenholz2002, bietenholz2010} and
\cite{bietenholz2017}, together with the size evolution after $t$ years of
$D[$mas$]=0.86\cdot t^{0.69}$ at distance of 10\,Mpc obtained by combining the
1-year radius of 0.43~mas from \cite{bietenholz2002} with the expansion
$\propto t^{0.69}$ from \cite{bietenholz2010}. 

The sources with BB336B 6\,cm luminosities above
$0.5\times10^{28}$\,erg\,s$^{-1}$\,Hz$^{-1}$ are all consistent with reaching
6\,cm peak luminosities of about $10^{28}$\,erg\,s$^{-1}$\,Hz$^{-1}$ during
their evolution, see the lightcurves in appendix \ref{app:book}. 
From Fig. \ref{fig:LD} we note that most rising sources are relatively small
($D<0.4$\,pc). Given their current lightcurves, some of the weaker rising
sources are unlikely to reach 6\,cm peak luminosities of
$\sim10^{28}$\,erg\,s$^{-1}$\,Hz$^{-1}$, for example 0.2084+0.534, while others for example
0.3011+0.341 may reach these luminosities in a few years time.

The slowly varying sources occupy all regions of Fig. \ref{fig:LD}. They are
all consistent with either being near their 6\,cm peak luminosity, or in a state of
slow decline. Some of these sources have slowly varying 18\,cm lightcurves,
for example Fig. \ref{fig:svarlowL}, while others show rapidly changing 18\,cm
lightcurves, for example Fig. \ref{fig:svar}. Some are optically thin (for example
$L_\text{18\,cm}>L_\text{6\,cm}>L_\text{3.6\,cm}$) while others show almost no 18\,cm emission while
still clearly detected in multiple epochs at shorter wavelengths.

The falling sources also occupy all regions of Fig. \ref{fig:LD}. Some of these,
for example 0.2227+0.482, show relatively rapid (50\% in 10 years) optically thick
decline from a peak luminosity of $>10^{28}$\,erg\,s$^{-1}$\,Hz$^{-1}$ without
any clear 18\,cm detection. Others, for example 0.2122+0.482, show an optically thin
decline with prominent 18\,cm emission since 1994.

\subsection{Source expansion from 2008 to 2014?}
\label{sect:sizevstime}
Given that we have data spanning many years one could try to measure if sources
are expanding between the observing epochs.  However, an expansion speed of
$<10000$\,km/s, as expected for the SNe in Arp\,220 \citep{batejat2011},
translates to an increase in diameter of $<0.02$\,pc in one year.  Even using
the most sensitive 6\,cm data, such a small increase is very challenging to
detect between two single epochs given the uncertainties on our size
measurements.  Also, older sources are expected to decelerate as well as fade,
making any expansion harder to detect.

To reduce uncertainties we average the sizes (as described in Sect.
\ref{sect:avgsize}) for the most sensitive series of 6\,cm epochs close in
time, that is the epochs within experiments GC031 and BB335, to obtain more robust
size measurements separated by about 6 years in time. However, even with these
average sizes the uncertainties as too large to probe any expansion.  New
sensitive high-resolution observations cold possibly help to measure expansion
velocities.

\begin{figure}
\centering
 \includegraphics[width=0.48\textwidth]{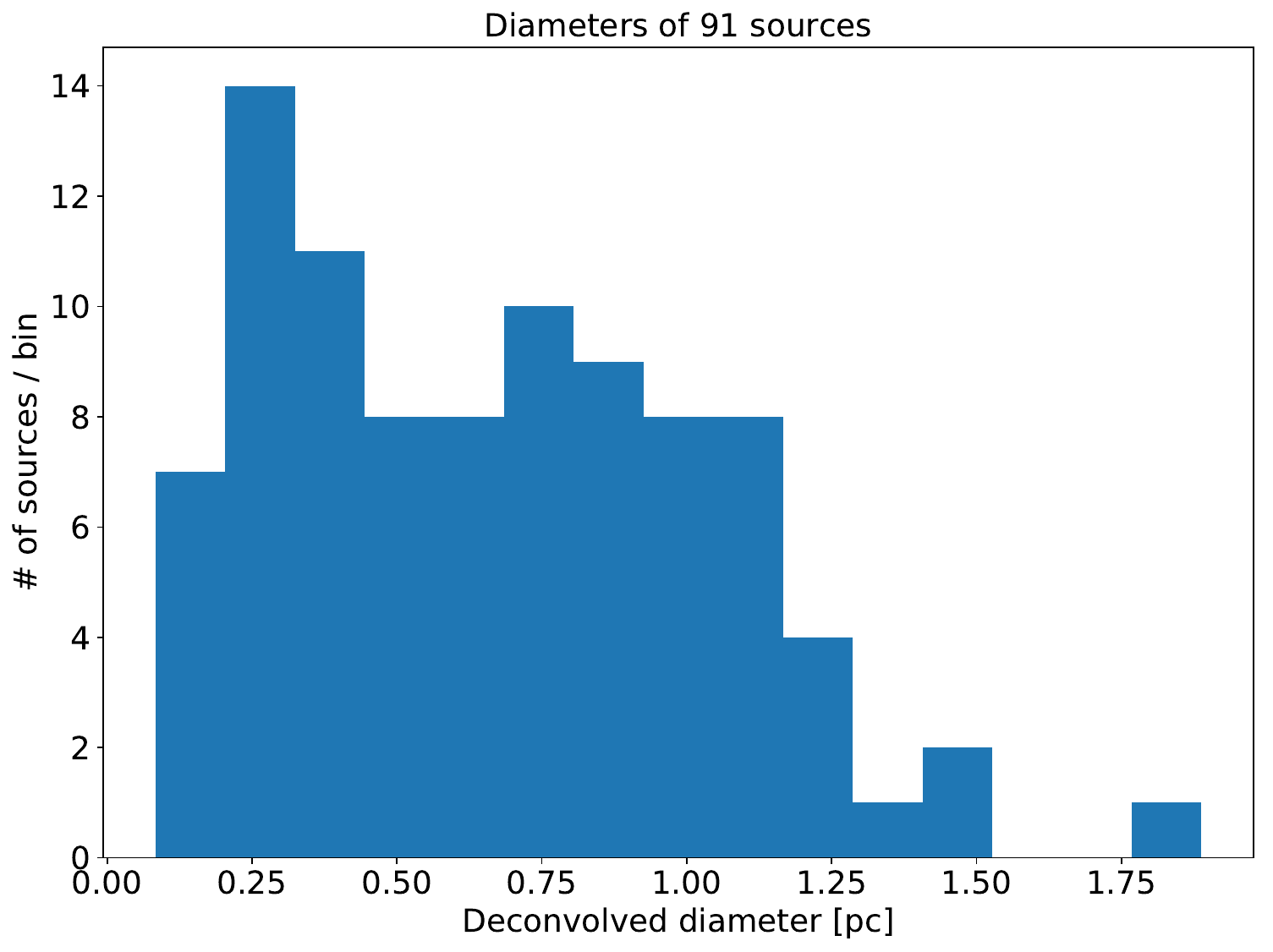}
 \caption{The distribution of measured source diameters appears bi-modal,
	 with peaks around 0.25\,pc and 0.8\,pc. We note that three of the 97
	 sources have no size estimates in either BB335A or BB335B and are
	 therefore not included in this figure.
 \label{fig:siz}}
 \end{figure}

\subsection{The measured distribution of source diameters}
\label{sect:folding}
The measured source sizes are shown as a histogram in Fig.  \ref{fig:siz}, that is
a projection of Fig. \ref{fig:LD} onto the horizontal axis. The size distribution
appears double humped, with peaks around 0.25\,pc and 0.8\,pc, and falls off
around a diameter of 1\,pc. We note that the measured sizes have
significant uncertainties and stress that Fig. \ref{fig:siz} alone is 
not proof of a bimodal distribution.
In general, source sizes may be in error where
the actual source structure deviate significantly from our assumed spherical
shell model, such as for examplethe case of SN1986J where there is a compact central
component present \citep{bietenholz2017}.
The fitted sizes for the smaller sources may also be affected by a Ricean bias,
where the fit favours slightly larger sources because negative sizes are
not allowed. This effect is tentatively supported by the simulations
presented in Table \ref{table:simvals}, as eight of nine cases with simulated
diameter Sim.D $\leq0.56$\,mas were fitted with a larger least-squares
diameter. Therefore, the peak around $0.25$\,pc may in reality be shifted
towards zero. It is hard to estimate how much this would affect the apparent
bimodality. If all of the $0.25$\,pc sources were in fact point sources, the
peak would be shifted but the distribution still bimodal. If, on the other
hand, nine sources around 0.3\,pc are biased upwards and are in fact
unresolved, with no other changes, then the observed distribution would
essentially appear uniform.  While uncertain, the shape of the distribution is,
however, consistent with two underlying distributions truncated by the
observational limits in surface brightness. This idea is discussed further in
Sect.  \ref{sect:discussion}.

\subsection{The source luminosity function}
\label{sect:lumhist}
\begin{figure}
\centering
 \includegraphics[width=0.48\textwidth]{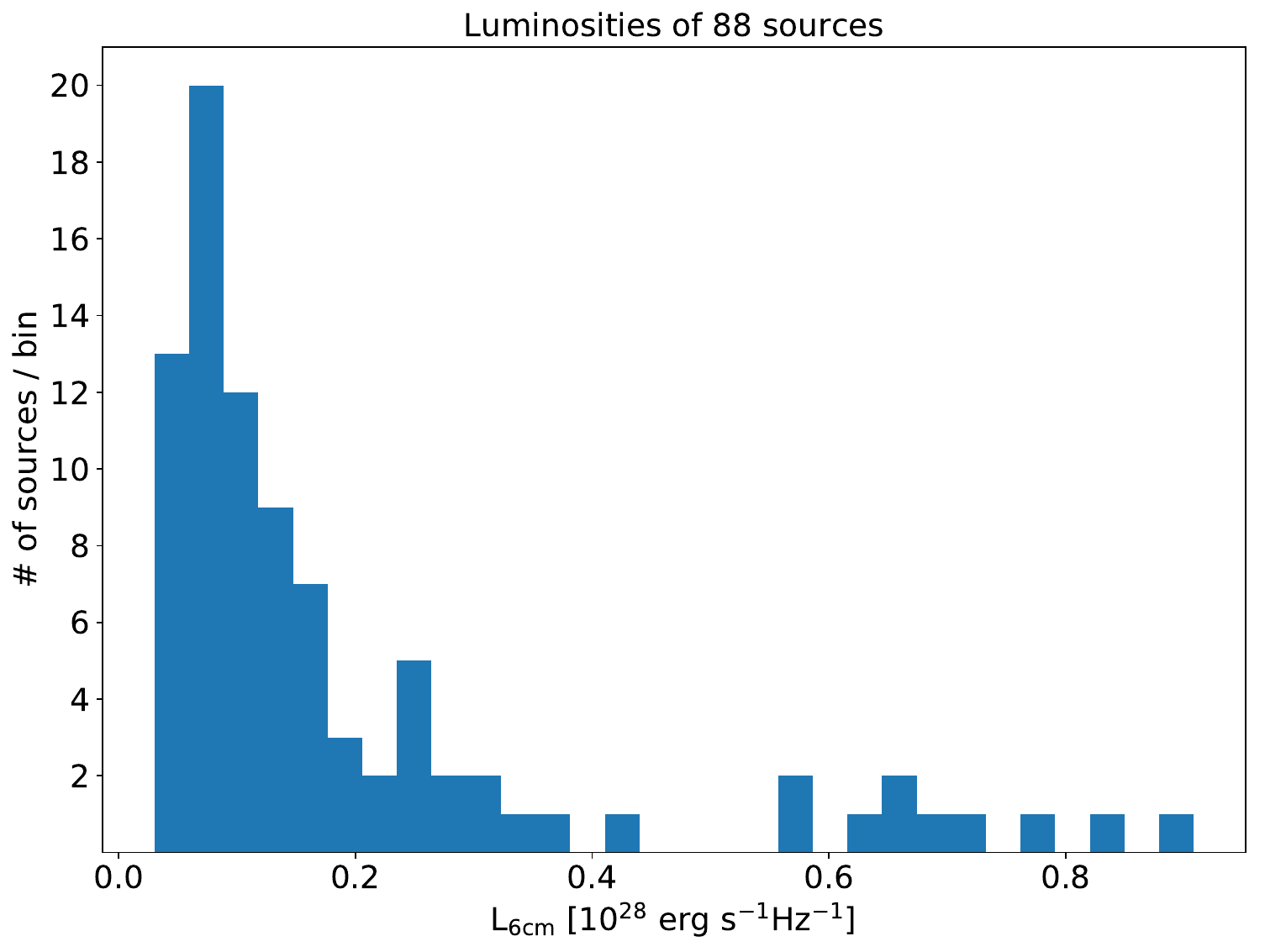}
	\caption{The distribution of measured spectral source luminosities, as measured in the epoch
		BB335B\_C. Nine of the 97 sources were not detected in this epoch, and have therefore
		been excluded from this figure.
 \label{fig:lumhist}}
 \end{figure}

\begin{figure}
\centering
 \includegraphics[width=0.48\textwidth]{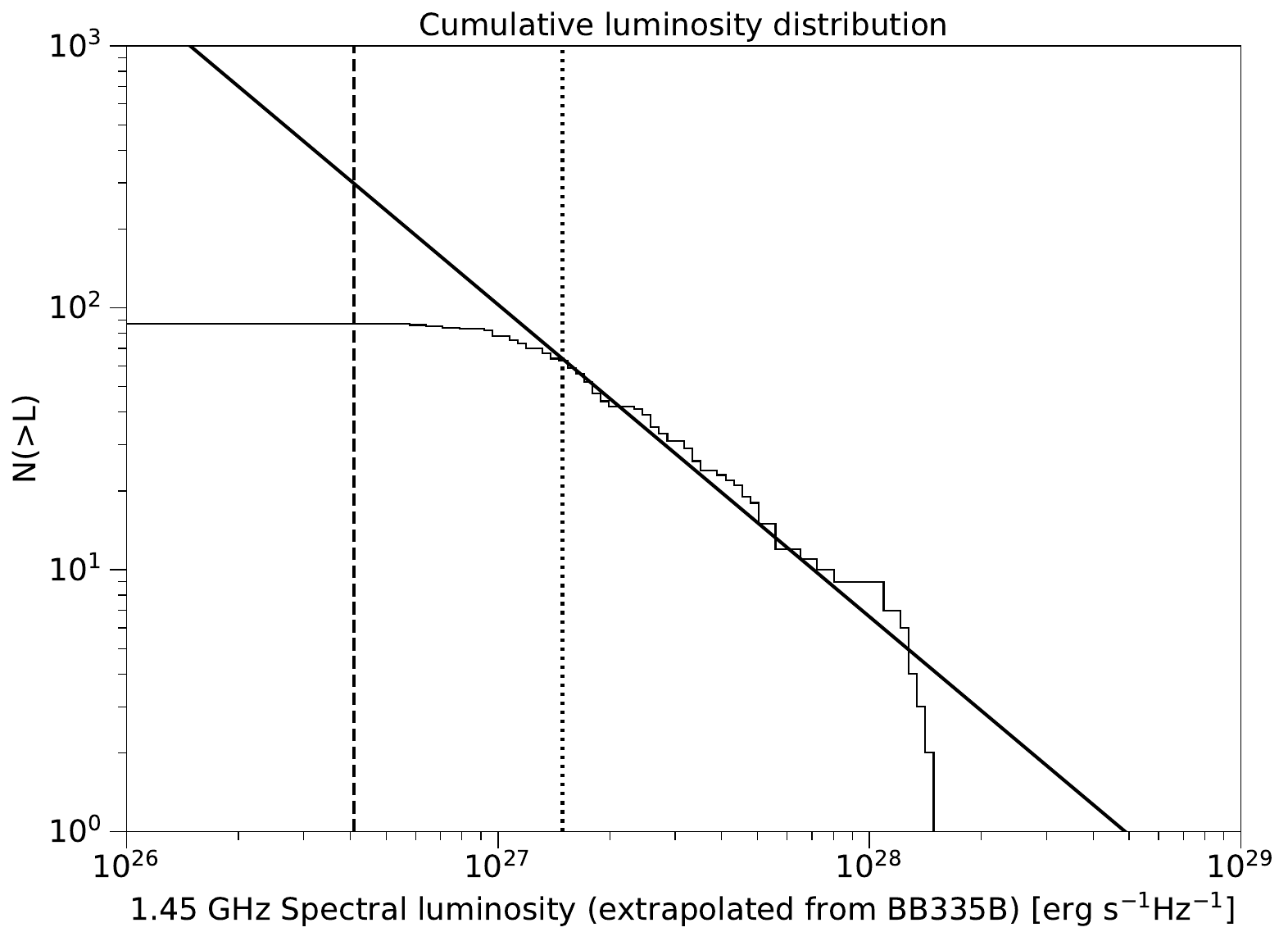}
 \caption{The cumulative luminosity function for the 88 sources detected in
	 BB335B\_C (of the total 97 sources detected in all epochs).  To be directly
	 comparable with Fig. 1 of \citealt{chomiuk2009}, the luminosities
	 have been extrapolated from Tables \ref{table:east} and \ref{table:west} to  1.45\,GHz,
	 assuming a spectral index of $-0.5$. A power-law fit it shown as a solid
	 line.  The point source detection limit is shown as a vertical black
	 dashed line to the left, and fitted completeness limit (where the power
	 law turns over) is shown as a dotted line.}
            \label{fig:lum}
 \end{figure}

The source luminosity function is shown as a histogram in Fig.
\ref{fig:lumhist} and as a cumulative function in Fig.  \ref{fig:lum}.
\cite{chomiuk2009} investigate the SNR LF in 18 `normal' galaxies at
1.45\,GHz by fitting the power law
\begin{equation}
	\label{eqn:lf}
	n(L)=AL_{24}^\beta
\end{equation}
where $A$ is a scaling factor (accounting for example SFR), $\beta$ is a negative
number (predicted by \cite{chomiuk2009} to be -2.1) and $L_{24}$ is the
1.45\,GHz luminosity given in units of $10^{24}$\,erg\,s$^{-1}$\,Hz$^{-1}$.
Using maximum likelihood estimation methods (MLE) \cite{chomiuk2009} find a
slope $\beta=-2.07\pm0.07$ from combining 258 SNRs in all 18 galaxies, in very
good agreement with the predicted value. 

In addition to the 18 galaxies, \cite{chomiuk2009} investigate the LF in
Arp\,220 using eight SNRs listed by \cite{parra2007}. They find
$\beta=-3.00\pm1.89$ which, although consistent with the predicted value,
is uncertain because of the small sample size.  

We analyse the Arp\,220 LF as sampled by the 88 sources detected at 5\,GHz in
experiment BB335B. Assuming an average spectral index of $\alpha=-0.5$ (as
modelled by \cite{chomiuk2009} assuming the CR electron energy spectrum can be
described as a power law $E^{-2}$), we extrapolate the flux densities measured
at 5\,GHz (given in Tables \ref{table:east} and \ref{table:west})
to 1.45\,GHz for direct comparison with \cite{chomiuk2009}.  We note that this
extrapolation likely gives a better estimate of the intrinsic 1.45\,GHz
emission than directly using values measured at closer frequencies because of
the significant free-free absorption affecting the radio emission from Arp\,220
below 2\,GHz \citep{smith1998b,lonsdale2006,varenius2016}. In principle,
also the 5\,GHz emission may be (slightly) affected by absorption. If so, 
fitting the luminosity function using 5\,GHz data may result in a (slight) underestimate 
of the intrinsic spectral luminosity function. However, as noted in Sect.
\ref{sect:motivation}, available observations at higher frequencies have lower
sensitivity and hence detect fewer sources which we estimate to introduce more significant
uncertainties.

To determine $\beta$ in Eq.\ref{eqn:lf} we use the MLE methods of
\cite{clauset2009}, that is the same as used by \cite{chomiuk2009}, as implemented
in the Python package \emph{powerlaw} by \cite{alstott2014}. This enables
determination of a completeness limit, that is  the luminosity where the
powerlaw turns over due to for example incomplete sampling of faint sources.  The limit is
found by creating a power-law starting from each unique value in the data set,
then selecting the one that results in the minimal Kolmogorov-Smirnov distance
between the data and the fit (see also the Appendix by \citealt{chomiuk2009}).
We note that this limit is significantly higher than the point source detection
threshold, see Fig.  \ref{fig:lum}.  

Using the 64 sources above our completeness limit (at 1.45\,GHz) of
$L_c=1.49\cdot10^{27}$\,erg\,s$^{-1}$\,Hz$^{-1}$ we fit a slope
$\beta=-2.19\pm$0.15 for Arp\,220, in good agreement with the $-2.1$ predicted
by \cite{chomiuk2009}.  We note that some sources in our sample (in particular
those labelled \emph{rising}) may not yet have reached the SNR stage. However,
these are relatively few and excluding those 14 sources from the fit only
changes the slope very marginally well within the given uncertainty (to
$\beta=-2.21\pm$0.17).

The observed cumulative luminosity function as well as the best fit of Eq.
\ref{eqn:lf} is shown in Fig. \ref{fig:lum} along with the $3\sigma$ point
source detection limit (in the BB335B epoch, that is a level similar to the
$6\sigma$ limit used in the stacked map) and the fitted completeness limit
$L_c$. The fact that the completeness limit is significantly above the point
source detection limit is not unexpected. Sources at, or below, the limit have
resolved sizes so their luminosities are spread over multiple beam areas.
Hence, in each beam area they would fall below the point source detection
limit.  We note that the fitted completeness limit could be a real lower limit
to the SNR LF in Arp\,220 and not just an observational effect, although this
seems unlikely given that the turnover occurs close to where it is expected
given that most weak sources are resolved. 

By integration of Eq. \ref{eqn:lf} we obtain an expression for the number
of sources above a certain luminosity as
\begin{equation}
	\label{eqn:NLmin}
	N(L>L_\text{min})=\int_{L_\text{min}}^{\infty}n(L)\,dL = -A\frac{L_\text{min}^{\beta+1}}{\beta+1}.
\end{equation}
Knowing that $N(L>L_c)=64$, this expression can be used to estimate a value for
$A$. Even though the parameter $\beta$ is assumed to have Gaussian probability
distribution (using MLE methods), the distribution of $A$ is clearly asymmetric
because of the exponential dependence on $\beta$. We determine $A$ using a
Monte-Carlo approach where we sample its distribution using $100,000$ random
draws of $\beta$ from a Gaussian distribution with mean $-2.19$ and standard
deviation $0.15$.  From the resulting distribution of A-values we fit the
cumulative distribution function using the interpolation approach implemented
in the package \emph{SciPy}\footnote{Using the function
scipy.stats.mstats.mquantiles.} to obtain three empirical quantiles
corresponding to the $-1\sigma$, mean, and $+1\sigma$ values for a Gaussian
distribution. From this we obtain a robust estimate of
$A=461000^{+1068000}_{-325000}$ where the uncertainty reflects the uncertainty
in $\beta$.  We note that this numeric value of $A$ assumes a luminosity given
in units of $10^{24}$\,erg\,s$^{-1}$\,Hz$^{-1}$ in Eq.  \ref{eqn:lf}.  We also
note that if we instead sample $\beta$ from a Gaussian distribution with mean
$-2.07$ and standard deviation $0.07$, that is the average value found by
\cite{chomiuk2009}, we obtain $A=170000^{+131000}_{-74000}$. 
Adding the uncertainties in quadrature, we find our estimate to formally differ
from the value of $A=57000^{+39000}_{-19000}$ obtained by \cite{chomiuk2009}
for Arp 220. The difference may be due to various factors not explicitly taken
into account in the formal uncertainty estimates, such as differences in flux
density measurements for the BP129 observations (which provide the spectral
index estimates used by \cite{chomiuk2009} for their Arp 220 sources), as
discussed in Appendix \ref{sect:prevstudycomp}.  Finally we note that if we fix
$\beta=-2.1$ we obtain $A=218000$.

\section{Discussion}
\label{sect:discussion}
As argued by previous studies, for example  \cite{parra2007}, we expect
a large ($\sim4$\,yr$^{-1}$) rate of supernovae in Arp\,220, of which only
the most radio luminous objects are bright enough to be detected.  Detailed studies of
the radio emission from these radio luminous SNe in Arp\,220 may allow us to
constrain the properties of their CSM, and thus for example the mass loss histories of
the progenitor stars.  However, a detailed discussion of the evolution of each
individual object is currently very challenging, given the complexity of SN
evolution and the limited time range sampled by the observations at wavelengths
shorter than 18\,cm, and is therefore deferred to a future paper.  In this
section we discuss the radio emission following SN explosions, focusing on
general properties of the population of compact sources.

\subsection{The radio SN/SNR dichotomy}
The evolution of the radio emission arising after supernova explosions is
often split in a radio SN-stage, where the blast wave is thought to interact
primarily with the CSM, and a SNR-stage, where the blast wave interacts with
the surrounding ISM.  The passage of SN to SNR phase occurs when the supernova
blast wave reaches the radius where the pressure of the ISM is approximately
equal to the ram pressure of the wind. This is expected to happen earlier for
SNe exploding in the high-pressure environments of starbursts than in normal
galaxies.

We note that although this dichotomy is useful in many cases, it is an
oversimplification of a broad range of possible evolutionary paths, depending
on for example the progenitor mass loss history (that is CSM structure), the explosion
geometry, the progenitor nature (single/binary), the structure and density of
the surrounding ISM, and the magnetic fields in the CSM and ISM (see for example
\citealt{weiler2002,vink2012,dubner2015} and references therein).  However, as
a detailed discussion of the nature of individual objects is planned for a
later paper, for the purpose of this discussion we adopt these simple
labels of SNe for objects interacting primarily with their CSM, and SNRs for
objects interacting primarily with the surrounding ISM. 

\subsection{Radio emission from SNe and SNRs}
\label{sect:radiofromSN}
The intensity of radio emission from core-collapse SNe versus time is closely related to the
density profile versus radius in the CSM. This relationship is complicated
because of the presence of HII-regions and wind-blown bubbles which may have
formed around the progenitor stars before the explosion. The structure of the
CSM may be complex with for example layers having different densities and temperatures
depending on the wind history of the progenitor (for example
\cite{weiler2002,vink2012}).  

Core-collapse SNe may show prominent emission within a few days after the
explosion (for example SN1993J; see \citealt{martividal2011SN1993J} and references
theirin). In other cases it can take years for the radio emission to appear.

The observed radio emission is thought to come from a region right behind the
SN blast wave, where charged particles are accelerated to relativistic energies
and trapped in strong magnetic fields \citep{weiler2002}.  If the SN explodes
in a low density CSM, very little radio emission is expected until the blast
wave reaches the surrounding ISM.
If, on the other hand, the SN explodes in a relatively high density CSM, the SN
is expected to show a characteristic radio lightcurve, with the peak appearing
later at longer wavelengths due to the free-free absorption by the ionised CSM
\citep{weiler2002}.

In a simple model, the radio emission is expected to increase (again) when the
shock reaches the dense ISM (i.e the SNR phase) and particle acceleration
becomes efficient. The maximum luminosity during this phase is expected to
occur at the \emph{Sedov} radius, where the mass swept up by the blast wave is
equal to the mass of the SN ejecta and the energy in relativistic electron
reaches a maximum.  

In the case of a very dense progenitor wind, that is a dense CSM, it is possible
that the Sedov radius is reached already before the shock enters the ISM. For
such objects there may not be a clear observational distinction between
the SNe and SNR phases.  Progenitors with such dense winds are however
expected to be rare, and the majority of SNe explosions in Arp\,220 will likely
produce weak and undetectable radio emission in the initial CSM phase and the
sources will rise to a peak radio emission only when they begin interacting
with the ISM. 

\subsection{The nature of the compact objects in Arp\,220}
\label{sect:nature}
Although most radio SNe may be relatively weak and hence below our detection
limit, some may brighten enough to be detected when they reach the ISM.  In
addition to these SNRs (that is shocks interacting with the ISM), we expect to see
the most radio luminous SNe which interact strongly with their dense CSM.
According to Fig. 1 by \citep{chevalier2006}, the most radio luminous SNe
are of types 1b/c or IIn. Type 1/bc SNe have a relatively short rise time
before they start to decline. Since the data presented in this paper sample
the evolution sparsely in time, for example every few years or months, we are unlikely to
see type Ib/c SNe near these high peak luminosities.  However, SNe of type IIn
also reach very high luminosities and appear to have significantly longer rise
times (few years). These would be seen in our observations, and should be present
with similar luminosities in adjacent observing epochs with similar frequency.
While type IIn SNe may be a likely explanation for bright sources in our data,
these particular type of SNe are, however, thought to be rare and likely make
up only a few percent of the total SN population
\citep{smith2011,eldridge2013}. It is therefore of interest to study the
population of the brightest sources in our data, and to estimate the
rate of these luminous events in Arp 220.

\subsubsection{The nature of the brightest objects}
The brightest objects in Fig.  \ref{fig:LD} all show lightcurves (see Appendix
\ref{app:book}) consistent with expansion in a dense, ionised CSM.  Indeed,
given expansion velocities of a few thousand km/s and high luminosities,
multiple objects have significantly longer rise times than
expected if synchrotron self-absorption is the dominant absorption mechanism
(see Fig. 2 of \citealt{chevalier2006}). It is thus likely that free-free
absorption from the ionised CSM is significant in these objects.

Given the background presented in Sect. \ref{sect:radiofromSN} we expect to see
a few objects in the early radio SN-stage, rising at 3.6 and 6\,cm but with
weak or undetected 18\,cm emission, such as 0.2262+0.512 (see  Fig.
\ref{fig:riselowL}).  Objects with high explosion energies evolving in a dense,
ionised CSM are then expected to reach their respective peak luminosities later
at longer wavelengths, such as 0.2195+0.492 (see Fig.
\ref{fig:svarlowL}), followed by an optically thin decline such as 0.2122+0.482
(see Fig. \ref{fig:fall}).  As the shock wave reaches the ISM, the
lightcurves may, in case of a sharp boundary to a higher-density ISM, rise at
all frequencies, such as 0.2108+0.512 (see Fig. \ref{fig:rise}) or,
in case of no sharp boundary, flatten (as seems to be the case for 0.2122+0.482
in 2014) due to the constant density.  The subsequent SNR phase, when the blast
wave expands in the constant density ISM, is expected to show a slow optically
thin decline, such as observed in 0.2171+0.484 (see Fig.
\ref{fig:svar}).  Because multiple studies (for example
\citealt{smith1998b,lonsdale2006,varenius2016}) argue that (foreground)
free-free absorption may significantly reduce the measured 18\,cm flux
densities, we expect to see a few sources, in particular towards the centres of
the nuclei, with relatively weak 18\,cm emission. This could explain the spectrum 
of 0.2122+0.482. Also 0.2253+0.483 (see Fig.
\ref{fig:falllowL}) could be an example of a blast wave hidden behind
significant foreground absorption.

\subsubsection{Modelling the lightcurves of the brightest object}
\label{sect:weilerfit}
\begin{figure}
\centering
\includegraphics[width = 0.48\textwidth]{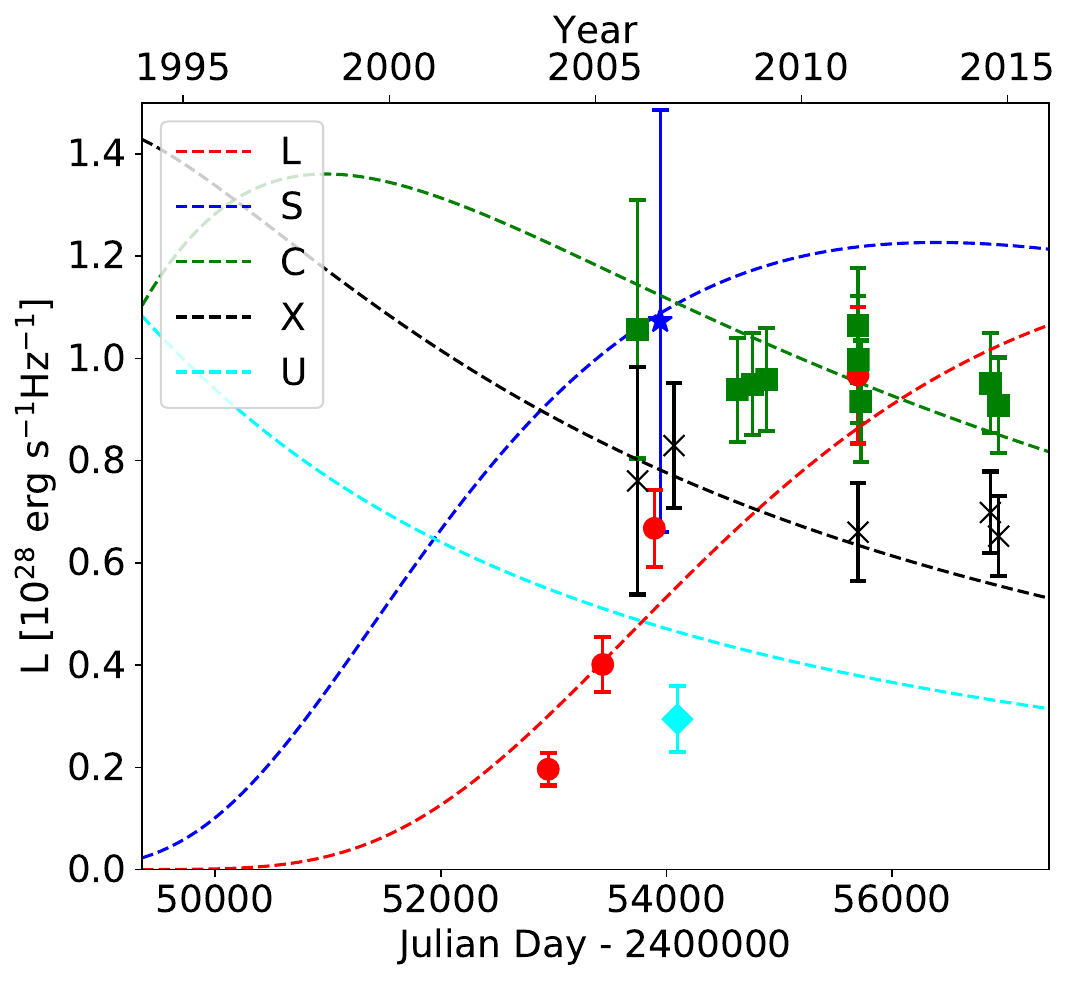}
\caption{The best-fit lightcurve model described in Sec. \ref{sect:weilerfit}
overplotted on the observed detections of 0.2195+0.492. L on the vertical axis
denotes spectral luminosity. The S-band data point has been displaced 200 days
to the right to make it clearly visible in this plot.}
\label{fig:weilerfit} 
\end{figure}

It is challenging to derive accurate age estimates from the observed
lightcurves, mainly because of the limited time coverage at 5\,GHz and higher
frequencies. Based on the lightcurves the brightest sources are, in general,
likely a few decades old with 6\,cm rise times of a few years.  One way to
estimate the age of the brightest sources is to use model-fitting of the
lightcurve for an example object. As an example, we choose the brightest object
0.2195+0.492, given that this object has the highest signal-to-noise and
therefore should be the first candidate for model fitting from a data quality
point of view.  We used a simplified version of the model presented by
\cite{weiler2002}, as described by Eqns. 1 and 2 of \cite{marchili2010}. 
This model follows the one presented by \cite{chevalier1982a,chevalier1982b}
where the radial density profile of the CSM is assumed to be $\propto r^{-2}$,
The best-fit parameters are $\alpha=-0.93$, K$_1$=319\,Jy,
K$_2=4.0\times10^{9}$, $\beta=-1.33$, and an explosion date of 1981-05-23 that is
an observed age estimate (in BB335) of about 33 years, and a corresponding
6\,cm peak time of about 17 years.  This is significantly longer than the 1210 days listed as the longest
rise-time in \cite{weiler2002} (for SN1986J). The model is shown together with the data
in Fig.  \ref{fig:weilerfit}. The fitted values imply a deceleration parameter
$m=0.87$ (from Eq. 7 by \citealt{weiler2002}) and a pre-supernova mass-loss
rate of $3.9\times10^{-4}$\,M$_\odot$yr$^{-1}$, assuming a pre-supernova wind speed of
10\,km/s for a type II SN (\citealt{weiler2002}; Eq. 18). We note that the
estimated mass-loss rate is similar to known type II SNe such as SN1988Z
as presented in Table 3 by \cite{weiler2002}. Given the age (and relatively modest
deceleration) obtained and the measured diameter of 0.2195+0.492 of about
0.2\,pc, a linear estimate implies an expansion velocity of $\sim3500$\,km/s,
that is consistent with the few thousand km/s expected for these objects.

\subsubsection{The observed number of luminous SNe}
In Sect. \ref{sect:weilerfit} the brightest object is estimated to be about 30 years
old.  The lightcurves of other objects with similar current luminosity and size
indicate a similar evolutionary history.  If we, to increase sample size, select
as luminous SNe the nine objects with
L$_\mathrm{6\,cm}>0.5\times10^{28}$\,erg\,s$^{-1}$\,Hz$^{-1}$ and diameter
$<0.4$\,pc, and assume these are at most 50\,years old, this would correspond
to an observed rate of the most luminous SNe of $\sim0.2$\,yr$^{-1}$.

Given the total star formation rate of Arp\,220 of 230\,M$_\odot$yr$^{-1}$
\citep{varenius2016} and using Eq. 2 in \cite{smith1998}, assuming the same
$m_l=1\text{M}_\odot$, $m_u=45\text{M}_\odot$ and
$m_\text{sn}=8\text{M}_\odot$, we estimate a total SN-rate of about
4\,yr$^{-1}$. If 5.6\% (average of \cite{smith2011,eldridge2013}) of these end up
as type IIn during 50 years of continuous star formation, we expect to see 10
such objects, or a rate of luminous SNe of $0.2$\,yr$^{-1}$, that is in excellent
agreement with the observed number. This suggests that a standard IMF in
Arp\,220 can explain the number of bright SNe observed in Arp\,220. However,
this assumes not only that all the brightest SNe are type IIns, but also that
all type IIn SNe in Arp 220 reach these high luminosities. 

If some IIn peak below our detection threshold, our measured rate would be too
low. Furthermore, even though unlikely given the sparse sampling in
time, we may still detect some very rapidly evolving SN of types Ib/c and misinterpret
these as a IIn.  If this is the case, our measured rate could be too high.
We stress that both the measured and expected values likely suffer from small
sample numbers, where for example a few of the small rising sources may rise above
$5\times10^{28}$\,erg\,s$^{-1}$\,Hz$^{-1}$ within a few years. 
Finally, our SN-rate estimate are subject to uncertainties related to for example the
age estimates of the observed source population.

A more thorough analysis of the data to constrain the IMF in Arp\,220 should
therefore include careful, detailed modeling of multiple compact sources to
better constrain for example source ages. This is however beyond the scope of this work,
and will be the subject of a future paper. In addition to the data presented
here, this future work will include new global VLBI observations.  These data
will help to further constrain the evolution of these objects and hence the
rate of very luminous SNe in Arp\,220. 

\subsubsection{The nature of the weaker objects}
\label{sect:faintest}
The majority of objects in Fig. \ref{fig:LD} have luminosities below
$5\times10^{27}$\,erg\,s$^{-1}$\,Hz$^{-1}$, and these sources are also (based
on visual inspection of the lightcurves in appendix \ref{app:book}) unlikely to
have reached much higher luminosities during their evolution.  The position of
this population in Fig. \ref{fig:LD} is consistent with these being smaller
diameter versions of the ISM interacting SNRs in M82.  \cite{lacki+beck2013}
argue that particles accelerated in the SNe in such environments as the nuclei
of Arp\,220 would have a maximum radiative lifetime of about a thousand years
(see their Fig.  1).  These calculations  assume the electrons are embedded in
an ISM  magnetic field of 2 mG in Arp\,220. In fact, fields in the SNR shells
may be a factor of 10 larger \citep{batejat2011} and hence lifetimes shorter,
but in this paper we assume a conservative figure of 1000 yrs.  Given the
SN-rate of 4 year$^{-1}$, we would hence expect at most 4000 radiating
SNe/SNRs present in the galaxy.  The fact that we observe only $\sim 2.5\%$ of
that number could be explained by a majority of the SNRs having significantly
shorter lifetimes. However, given that many sources detected in this work are
close to our detection limit, there may very well be a significant number of
weaker sources present but not yet detected. 

\cite{BV04} show that less energetic SNRs expanding in lower densities reach
lower peak luminosities at larger sizes (see their Fig. 4). This means such
SNRs may only peak when larger than 0.5\,pc and be missed by our surface
brightness limited observations. In addition, the dependence of peak radio
luminosity on explosion energy $E_{\text{sn}}$ is very strong: in the Sedov
phase the luminosity is expected to scale with $E_{\text{sn}}^{7/4}$, that is a
factor of four less energy implies 10 times weaker peak luminosity.  Since many
sources are observed just above our detection limit, it is reasonable to assume
that these are the brightest objects of the radiating population.

We conclude that the fainter objects may very well be the tip of a distribution
of supernova remnants, where the majority have luminosities below our detection
limit. Indeed, the brightest of these fainter objects sources, with diameter
$>0.4$\,pc but $L_\mathrm{6\,cm}<3\cdot10^{28}$\,erg\,s$^{-1}$Hz$^{-1}$ (see
Fig.  \ref{fig:LD}), for example 0.2171+0.484, 0.2211+0.398, and 0.2360+0.431 are all
consistent with being observed close to their 6\,cm peak (see for example panel d in
Fig.  \ref{fig:classex} and appendix \ref{app:book}).  Furthermore, these three
sources are all long lived with stable lightcurves, suggesting they are more
likely in the more-slowly-evolving SNR phase, rather than in SN phase like the
most luminous sources discussed above.  Also, these sources show prominent
18\,cm emission, indicating the blast wave has reached outside the ionised CSM.

\subsection{The distribution of source sizes}
We argue above that some small ($D<0.4$\,pc) and luminous
$L_\mathrm{6\,cm}>5\cdot10^{28}$\,erg\,s$^{-1}$Hz$^{-1}$ objects are examples
of the relatively rare radio luminous SNe, which interact strongly with a dense
and ionised CSM. We also argue that the larger (and weaker) sources are SNRs,
that is where the SN blast wave is interacting with the dense ISM. Two source
populations are also consistent with the tentative bimodal size distribution,
see Fig. \ref{fig:siz}.  We stress, however, that most size estimates have
significant uncertainties.  

\subsection{Is the smooth GHz-emission of Arp\,220 a collection of compact objects?}
\cite{barcosmunoz2015} image Arp\,220 at 4.7\,GHz with lower angular resolution
and measure a total flux density of 222\,mJy.  Their measurement includes both
the emission from compact objects, detected in VLBI observations, as well as
smooth emission on scales resolved out by the VLBI observations presented in
this paper.  The total flux density measured in the 88 continuum sources from
Tables \ref{table:east} and \ref{table:west} is 22.5\,mJy, that is 10\% of the
total flux density measured by \cite{barcosmunoz2015}. The fitted completeness
level of the observed luminosity function in Sect.  \ref{sect:lumhist} and the
detection limit shown in Fig. \ref{fig:LD} strongly suggests that there are
many more radiating sources in Arp\,220 than we observe in this work. Could the
sources below our VLBI surface brightness detection limit be the source of the
smooth synchrotron emission detected by for example \cite{barcosmunoz2015}?

As a simple empirical estimate of the relative contribution of compact
sources to the integrated flux density, we take the observed luminosity function
and extrapolate this to luminosities below our detection threshold.
If we assume all SNe give rise to an SNR, and an upper radiative lifetime of
1000 years, we expect, given the above SN-rate of about 4\,yr$^{-1}$, at most
4000 objects in Arp\,220, of which we observe the brightest $\sim2$ percent.
Given a total number of 4000 SNRs, we can use Eq.\ref{eqn:NLmin} to obtain a
distribution of
$L_\text{min}=4.6_{-1.8}^{+2.2}\times10^{25}$\,erg\,s$^{-1}$\,Hz$^{-1}$ (at 1.45\,GHz) given a
distribution of $\beta$. The distribution of $A$ is determined from Eq.
\ref{eqn:NLmin} assuming 64 sources brighter than our completeness limit as
described above. 

If we integrate $n(L)L$ from $L_\text{min}$ to the observed
$L_\text{max}\approx2\times10^{28}$ (at 1.45\,GHz) we obtain the distribution of
the total luminosity from compact sources as
\begin{equation}
	L_\text{tot}=\int_{L_\text{min}}^{L_\text{max}}AL^{\beta+1}\,dL =\begin{cases}
		\frac{A}{\beta+2}\left(L_\text{max}^{\beta +2}-L_\text{min}^{\beta +2}\right), & \text{if $\beta\neq-2$}.\\
		A\ln(L_\text{max}/L_\text{min}), &\text{if $\beta=-2$}.
	\end{cases}
\end{equation}
Using 100,000 random draws of $\beta$ (with mean
$-2.19$ and $\sigma=0.15$) we sample the distribution of $L_\text{tot}$ to
estimate uncertainties. In this calculation, the parameter A is again a
distribution determined from Eq. \ref{eqn:NLmin} assuming 64 sources brighter
than our completeness limit as described above.
Assuming $\alpha=-0.5$ we obtain a total 5\,GHz flux
density from 4000 compact sources of $L_\text{tot}=56^{+9}_{-8}$\,mJy. As
done above, the uncertainties correspond to $\pm1\sigma$ for a Gaussian
distribution, calculated using fitting to the sampled cumulative distribution
function. Given that 4000 radiating sources is an upper limit, it follows
that a source population described by the powerlaw distribution in luminosity
defined by Eq.  \ref{eqn:lf} can account for at most 25\% of the total
flux density observed from Arp\,220 at GHz frequencies. So what is the source
of the remaining radio emission measured by \cite{barcosmunoz2015}?

The above argument assumes a maximum lifetime of CRs in the SN shocks
of a thousand years. While the actual lifetime of accelerated CRs is likely
shorter given the strong magnetic fields in the shocked regions (as noted
in Sect. \ref{sect:faintest}), the dense ISM in Arp\,220 means that the SNR
shocks will encounter significant particle densities also after leaving the
CSM. Although each accelerated CR cools relatively quickly, the blast wave may
posses significant kinetic energy for hundreds of years. During this time,
particles in the ISM may be (re-)accelerated to emitting energies, long after
the initial CRs accelerated in the ISM when entering the Sedov phase have
faded. In addition, the massive progenitor stars likely form in dense star
clusters of sizes of a few parsec \citep{wilson2006}.  In this case, SN shock
waves with diameters larger than 1\,pc (which may be too faint to be detected
in Fig. \ref{fig:LD}) may collide with each other. The combined mechanical
energy may be enough to compress the ISM to even higher densities and
accelerate CRs to cause significant radio emission \citep{bykov2014}. In
addition, given the high densities in Arp\,220, also protons accelerated in the
SN shocks may cause significant synchrotron radiation via collisions with ISM
which produce secondary electrons and positrons, thereby more efficiently
radiating the energy in the SN shocks \citep{lacki2010one}. This would imply
that the fitted luminosity function has a significant tail of a large number of
weak sources, compared to when extrapolated from the powerlaw in Fig. \ref{fig:lum}.
This could potentially explain the origin of the smooth radio emission.

Another explanation could be an AGN contribution to the radio emission.
However, a significant AGN contribution would likely manifest itself in a
radio-bright compact core, and/or radio jets.  While we do not find any clear
support in terms of for example  radio jets to support a recent significant AGN
contribution to the radio emission, it is still
possible that an AGN could play an important role in this galaxy, since its
activity could be episodic and any jet like structure dissolved into smoother
structure by merger forces, and the emitting structure thereby hidden from VLBI
observations due to lack of spatial scales sampled.

Future observations with very high sensitivity, possibly combined with
stacking of the current available data as well as tapering and statistical
analyses of possible non-Gaussian contributions to the image noise, may reach
sufficient sensitivities to detect or exclude such a population of weak
sources, and/or further constrain any AGN contribution.

\section{Summary and outlook}
\label{sect:summary}
In this paper we have presented data from 20 years of VLBI monitoring of Arp\,220. 
We detect radio continuum emission from 97 compact sources, and find they
follow a luminosity function 
$n(L)\propto L^\beta$ with $\beta=-2.19\pm0.15$, similar to SNRs
in normal galaxies.
The spatial distribution of sources trace the star forming disks of the
two nuclei seen at lower resolution.

We find evidence for a Luminosity-Diameter relation within Arp\,220,
where larger sources are less luminous. The exact form of the
relation is however hard to quantify because of a range of
selection effects.  The observed distributions of source luminosities and sizes
are consistent with two underlying populations. One group consists of very
radio luminous SNe where the emitting blast wave is still inside the dense, ionised
CSM. The other group consists of less luminous and larger sources which are
thought to be SNRs interacting with the surrounding ISM. 

Assuming all SNE type IIn reach the highest luminosities, and assuming
the brightest sources we detect are IIns, we find that the observed number of very
luminous SNe is consistent with expectations given a standard initial mass
function and the total integrated star formation rate of the galaxy. This
result should however be taken with care, as more detailed modeling of multiple
sources is needed to better constrain the evolution, and in particular the ages,
of the most luminous SNe.
	   
When extrapolating the observed luminosity function below our detection
threshold we find that the population make up at most 20\% of the total radio
emission from Arp\,220 at GHz frequencies.  However, secondary CR produced when
protons accelerated in the SNRs interact with the dense ISM and/or
re-acceleration of cooled CRs by overlapping SNR shocks may increase radio
emission from the sources below our detection threshold, compared to the
extrapolated value. This mechanism may provide enough emission to 
explain the remaining fraction of the total radio flux density, and could
potentially be constrained by future high-sensitivity observations. 
	   
Continued high-sensitivity VLBI monitoring of Arp\,220 will likely detect many
more fainter sources and therefore probe the distribution of lower luminosities
and larger sizes which may further constrain the evolution of SN/SNRs in
extreme environments. Results from similar ongoing monitoring of other galaxies,
such as the closer LIRG Arp\,299 (Perez-Torres et al. in prep.) will be
interesting for comparison to the results presented in this work.

\begin{acknowledgements}
EV, JC, SA, and IM-V. all acknowledge support from the Swedish research
council.  MAP-T and AA acknowledge support from the Spanish MINECO through
grant AYA2015-63939-C2-1-P, partially supported by FEDER funds.  The European
VLBI Network is a joint facility of independent European, African, Asian, and
North American radio astronomy institutes. Scientific results from data
presented in this publication are derived from the project codes listed in
Table \ref{table:obslist}.  The National Radio Astronomy Observatory is a
facility of the National Science Foundation operated under cooperative
agreement by Associated Universities, Inc.  This work made use of the Swinburne
University of Technology software correlator, see \cite{deller2011}, developed
as part of the Australian Major National Research Facilities Programme and
operated under licence.  The Arecibo Observatory is operated by SRI
International under a cooperative agreement with the National Science
Foundation (AST-1100968), and in alliance with Ana G. Méndez-Universidad
Metropolitana, and the Universities Space Research Association.  This research
made use of APLpy 1.0, an open-source plotting package for Python
\cite{aplpy2012}. Finally, we thank the anonymous referee for their careful
reading of our manuscript which improved the quality and clarity of this
paper.

\end{acknowledgements}

\bibliographystyle{aa}
\bibliography{allrefs}

\appendix

\section{Data and scripts in the CDS}
\label{app:cds}

Data, images and analysis scripts presented in this paper are available in
electronic form at the CDS via anonymous ftp to cdsarc.u-strasbg.fr
(130.79.128.5) or via http://cdsweb.u-strasbg.fr/cgi-bin/qcat?J/A+A/.  The
material is split in five directories. The \emph{calibration}
directory contains one subdirectory per experiment. Each subdirectory contains
a ParselTongue script, and in some cases also some additional files, used to
calibrate archival UVFITS data to obtain FITS images. The \emph{fitsimages}
directory contains the FITS images obtained for all observations analysed in
the paper. There are two images for each experiment, one for each nuclei in Arp
220. The \emph{stacking} directory contains the stacked images obtained by
stacking together all images at each band, as well as the script used to
perform the stacking. The stacked 6~cm images are also displayed in Fig.
\ref{fig:stacked}. The \emph{fitting} directory contains the Python scripts
used to fit spherical shell models to sources in the FITS images for all
epochs. The output of the fitting is stored in the directory \emph{fitresults}
as ASCII files. These contain the fitted values for position, flux density and
diameter for all sources in all experiments.

\section{Calibration and imaging}
\label{sect:calibration}

\subsection{Phase calibration and positional uncertainty}
For most epochs phase calibration started by removing bulk residual delays and
rates, using one of the three bright sources J1516+1932 (ICRF
J151656.7+193212), J1613+3412 (ICRF J161341.0+341247) or OQ208 (ICRF
J140700.3+282714) by running the AIPS task \verb!FRING!.  This should also
remove relative phase differences between the spectral windows (AIPS IFs).
However, multiple epochs showed time variable phase differences between the
IFs.  This was corrected for by running \verb!FRING! also on the
phase-reference source J1532+2344 (ICRF J153246.3+234405) for epochs where this
source was included in the observations. 

Experiments BP129, GC028, GC031, BB297 and BB335 were all phase-referenced to
the nearby (0.55$^\circ$) compact calibrator J1532+2344, assumed to be located
at R.A. $15^h32^m46^s.3452$, Dec. $23^\circ44'05''.268$, and Arp 220 was
correlated at position R.A. $15^h34^m57^s.250$, Dec. $23^\circ30'11''.33$, that is
between the two nuclei.  After transferring the phase-reference corrections,
an initial image was made of Arp\,220. In many cases, the resulting image
showed beam like artefacts around bright sources consistent with residual phase
 errors. To remove the residual phase-errors, phase
self-calibration (see for example  \citealt{cornwell1981}) was used with the initial
phase-referenced image as calibration model.

The self-calibration step removed obvious phase-errors, and generally improved
the images.  However, because Arp\,220 is weak at mas-scales (5-35\,mJy in
these data), a low baseline SNR threshold of 2 was required. Such a low
threshold may introduce spurious sources or in other ways impact the data in a
negative way \citep{marti-vidal2008}.  To assure the validity of the
self-calibrated results, we checked
\begin{enumerate}
\item The general image quality: Did the self-calibration cycle remove obvious
phase-errors, such as convolution like artefacts on all sources?
\item The RMS noise in central regions with many sources: Did it decrease
	after self-calibration?
\item The source flux densities: Did the source flux densities increase as
	expected after self-calibration?
\item The phase-solutions: Although noisy, were the solutions tracing slowly
varying atmospheric errors and/or clear antenna offsets?
\item The consistency with all available data: Were the flux densities and
sizes, measured for a particular epoch, consistent with the information from all
other epochs?
\end{enumerate}
In summary, we found clear signs of improvements in image quality for most
epochs, and no clear sign of corruption due to self-calibration.  We note that
we only performed one cycle of self-calibration, which leaves little room for
spurious sources to grow bright, as can sometimes happen after many cycles of
poorly constrained self-calibration. We note that only phase-corrections, that is no
amplitude corrections, were derived using self-calibration of Arp\,220.

In contrast to the phase-referencing above, four 18\,cm experiments,
GL021, GL026, GD017, and GD021, were phase-referenced to compact OH-maser
emission within Arp\,220 itself.  Phase corrections were derived from
self-calibration using the strongest maser channel.  Initially, a point source
model was used, but in all cases the elongated structure of W1 (see
Fig. 1a of \cite{lonsdale1998}) was recovered after a single iteration. One
or two additional self-calibration iterations were carried out to ensure the
structure of the maser was correctly taken into account.  The corrections
derived for this single channel were then applied to all channels of all
spectral windows. No further self-calibration of Arp 220 was performed for
these epochs. Note that no high-resolution spectral data were used (although in
some cases available in the archive). Instead, we used a single broad frequency
channel in the continuum data, containing the maser emission, to obtain
corrections for residual phase errors.

The three 18\,cm experiments GL021, GL026, and GD017 used a correlation
position of R.A.  15$^h$34$^m$57$^s$.2247, Dec. 23$^\circ$30$'$11$''$.564 for
Arp 220, while GD021 used the same as BP129.  By phase self-calibration of the
brightest maser channel (in the continuum data), the position of all compact
sources were anchored to the peak of the compact maser.  In this work, we
assume a peak position of the maser W1 of R.A. $15^h34^m57^s.22435$, Dec.
$23^\circ30'11''.6644$. This position was obtained from imaging the maser
emission in the phase-referenced 18\,cm data taken in experiment BB297A.  Since
BB297A was phase-referenced to J1532+2344 (and did not use the maser for phase
calibration), we thereby align the maser calibrated epochs to the common
reference position of J1532+2344 used for the other epochs. We assume that the
maser position does not vary between the observations.  Cross-correlation of
the resulting continuum images among the different epochs supports this
assumption.

Given that many sources are close to our surface brightness detection limit,
and the fact that many are resolved which may impact the peak position,  we
assume a conservative uncertainty of 1\,mas for all source positions within Arp
220. Notes on position offsets for particular sources can be found in Appendix
\ref{sect:posacc}.

\subsection{Amplitude calibration}
\label{sect:ampcal}
The amplitude calibration of all epochs was anchored to the measurements of
system temperatures and gains of VLBA antennas. The  procedure applied to
ensure the best possible amplitude calibration was as follows:
\begin{enumerate}
    \item Apply á priori amplitude calibration to all antennas, using measured
        system temperatures and antenna gain curves.
	\item Determine a set of good, self-consistent, antennas (usually the
VLBA), excluding antennas with obvious amplitude offsets seen in amp vs.
UV-distance plots.
    \item Perform amplitude self-calibration of the good antennas to correct minor
        errors, using imaging of a bright observed source, that is J1516+1932,
        J1613+3412 or OQ208, depending on data set.
    \item Amplitude self-calibration of previously excluded antennas, fixing
        the good antennas, using the same bright source. We note that if
        sensitive antennas with a major error are included from the start, the
        whole flux scale may be shifted due to their large weight.
    \item
        Either, for phase-referenced epochs: derive amplitude corrections for
the target by imaging and (self-)calibrating the phase-reference source, or,
    \item For some maser-calibrated epochs: derive amplitude corrections for the 
target by imaging and (self)-calibrating the maser emission.
    \item Apply cumulative antenna based corrections to the Arp\,220 observations.
    \end{enumerate}
We adopt an absolute flux calibration uncertainty of 10\% for all data.  

We
note that the noise we obtained in some re-reduced observations is larger than
in some previous publications (for example \citealt{lonsdale2006}).  This may be
explained by several factors.  First, multiple non-VLBA antennas had to be
excluded from our processing, since no system temperatures or gain measurements
could be found (these seem to have been lost as they are not available via for example
the NRAO or JIVE online archives). Second, because of the large amount of data
processed in this work and our efforts to use a similar calibration and imaging
strategy for all data, we may have missed opportunities to improve specific
epochs by for example elaborate weighting of very sensitive antennas, or extra careful
editing of bad data. 

For the 6\,cm data sets GC031B and GC031C, a careful analysis of the final
images, made after using the calibration strategy outlined above, revealed both
these epochs to be systematically too bright for all sources, although their
relative flux densities were reasonable. These offsets were also found when
comparing the amplitudes of the uncalibrated visibilities on multiple VLBA
baselines to the corresponding baselines in BP126 and BB297A. We also found
consistent offsets in the recovered flux densities of the bright calibrator
J1516+1932 in these epochs. We believe that these epochs were affected by an
error, scaling the visibilities to higher values before storing them in the
archive.  Further investigations of this error is however beyond the scope of
this work. For the purpose of this work, we derive a scaling factor for each of
the two epochs, based on a requirement of continuity for the light-curves of
the brightest sources, as well as the observed differences in the raw
uncalibrated visibilities. The correction factor derived for both epochs was
0.70, and it was applied to the images before any analysis of these epochs.

\subsection{Bandpass calibration}
For all epochs, bandpass corrections were derived using a bright observed
source, that is J1516+1932, J1613+3412 or OQ208, depending on data set, assuming
these sources to have a flat spectrum across the observing bandwidth.

\subsection{Imaging}
All epochs were imaged using the CLEAN deconvolution algorithm
as implemented in the task IMAGR in AIPS, using two fields to
simultaneously clean the two nuclei.  All epochs were imaged in a two-step
auto-boxing procedure, as implemented in IMAGR.  First, boxes were
automatically placed by IMAGR on the brightest sources, defined as having peak
signal-to-noise > 10 and source island level > 5. This minimised the risk of
cleaning strong side lobes caused by the synthesised beam. Then, the boxes were
removed and a few hundred clean iterations ran without any box-restrictions to
find weaker sources. The imaging was stopped when the CLEAN algorithm converged,
that is when no significant amount of flux was recovered in a few hundred iterations.
The final images were exported to FITS for further analysis outside AIPS.  The
number of pixels used were 8192 in R.A. and Dec. for all images.  For the 2\,cm
data, a pixel size of 0.05\,mas was used. All other epochs used a pixel size of
0.1\,mas. Robust weighting \citep{briggs1995} of 0.5 was used for all images.

\section{Model fitting}
\label{sect:fitting}
In this work we fit models of sources to CLEAN images.  In principle, it is
best to fit models directly to the calibrated visibilities to avoid any effects
introduced by the deconvolution (for example\citealt{marti-vidal2014}).  However,
visibility fitting of multiple nearby sources involve simultaneous fitting of
all parameters for all sources. Because of the large number of sources in Arp
220, and the large number of datasets analysed in this work, simultaneous
visibility fitting is not feasable.  We therefore decided to fit models to the
CLEAN images, which allows us to work on a small subimage encompassing each
source.  We decided to use least-squares (LS) optimisation to find the optimal model
for each source. We verified the validity of this approach by comparing it with
visibility fitting of simulated data, see appendix \ref{sect:fitsim}.  Below we
describe in detail how the fitting was done for each source. We note that the
Python code used to do the fitting is available via the CDS as described in \ref{app:cds}.

\subsection{Preparing a stamp image}
First a small subimage, hereafter called a stamp, was extracted from the
cleaned image, centred on the source catalogue position in Tables
\ref{table:east} and \ref{table:west}. The stamp was selected as $128\times128$
pixels at bands C and X, and $256\times256$ pixels at bands L, S and U. The
number of pixels are chosen to fully cover the source being fitted, and any
nearby sources, with enough source-free pixels around to avoid edge-effects,
and then rounded upwards to the next power of 2 to enable FFT optimizations.  A
threshold was imposed to only fit stamps with clear emission present in the
centre. If the central $2\times2$\,mas did not contain a peak value of at least
three times the experiment RMS noise (see Table \ref{table:obslist}), no fit
was attempted.

\subsection{Defining the model}
We assume each stamp to be an estimate of the true source convolved with a beam
and sampled on a discrete grid of pixels.  Each source is assumed to be a
projection of an optically thin spherically symmetric shell with 30\%
fractional shell width. The convolving beam was taken to be the Gaussian CLEAN
restoring beam for each stamp.  We chose to discretise the expressions for
shells and beams in Fourier space, replacing the computationally intensive
convolution with a multiplication. In other words, in each fitting-iteration,
we define and discretise our shell and beam in Fourier-space. The convolved
shell-model, constructed in Fourier space, is then transform to an image for
calculation of LS residuals, that is comparison with the CLEAN stamp.  We
stress that this approach gives results with similar accuracy to visibility
fitting on simulated data, see Appendix \ref{sect:fitcomp}.

\subsubsection{Constructing a shell}
We construct a model of a shell by subtraction of two spheres. The spheres are
defined in Fourier space using the analytical expression of a projected sphere
of unit flux density, that is Eq.  16-28  of \cite{pearson1999}:
\begin{equation}
\mathcal{F}(f(x,y,d)) = \frac{3}{(\pi d \rho)^3}\left[\sin(\pi d \rho) - \pi d \rho \cos(\pi d \rho)\right]
\label{eqn:1628}
\end{equation}
where $d$ is the sphere diameter, $\rho=\sqrt{u^2+v^2}$ and $u,v$ are the
spatial frequencies along the $x,y$ directions respectively. 

A thin shell $f_s$ with flux density $I$ is formed as the difference of two spheres
with 30\% fractional width, scaled in flux density by their respective volumes,
as
\begin{equation}
f_s(x,y,d,I) = I\frac{4\pi}{3}\frac{\left(f(x,y,d)\cdot d^3-f(x,y,0.7d)\cdot(0.7d)^3\right)}{(4\pi d^3/3-4\pi (0.7d)^3/3)}
\label{eqn:imshell}
\end{equation}
where $I$ is the integrated flux density of the shell with outer diameter $d$
and inner diameter $0.7d$. This expression is deliberately not maximally
simplified, to better illustrate the constituents.  For computational
efficiency, we discretise also the Gaussian beam directly in Fourier space.
This allows us to calculate the elementwise product of the Fourier transforms
of the shell and beam.  We then used an inverse Fast Fourier transform (IFFT),
as implemented in the python module \verb!numpy.fft.ifft2!, to obtain the
pixelated convolved shell $m_{i,j}$ in the image domain for comparison with the
stamp.  According to the well known convolution theorem, this is equivalent to
convolution in the image domain, that is  $m(x,y,d,I) = \mathcal{F}^{-1}(F\cdot
H)$ where $m$ denote the concolved shell in the image domain, and $F$ and $H$
denote the shell and beam in Fourier space. 

The residual vector used by the LS minimisation routine was calculated
by forming the weighted pixelwise difference as 
\begin{equation}
\mathrm{res}_{ij} = \frac{(m_{ij}-d_{ij})^2}{E_{f_{ij}}^2}
\end{equation}
where $m_{ij}$ is the convolved model and $d_{ij}$ is the image data. $E_{f}$ is
defined in Eq. \ref{eqn:ef} with subscripts $ij$ referring to the flux density
in pixel number $i$ in R.A. and $j$ in Dec. The residual was minimised using
the Python library \verb!scipy.optimize.least_squares!  which enables bounded
LS fitting. The minimisation was allowed to simultaneously vary the
flux density, size, and position of the shell. The flux density was required to
be $>0$, the size $\leq8$\,mas, and the position in R.A. and Dec. was
restricted to be within $\pm1$\,mas of the stacked catalogue position listed in
Tables \ref{table:east} and \ref{table:west}. The initial guess was the same
for all fits: flux density 0.5\,mJy, size 0.1\,mas (1 pixel), and position as
in Tables \ref{table:east} and \ref{table:west}.

\begin{figure*}[htbp]
\centering
        \includegraphics[width=\textwidth]{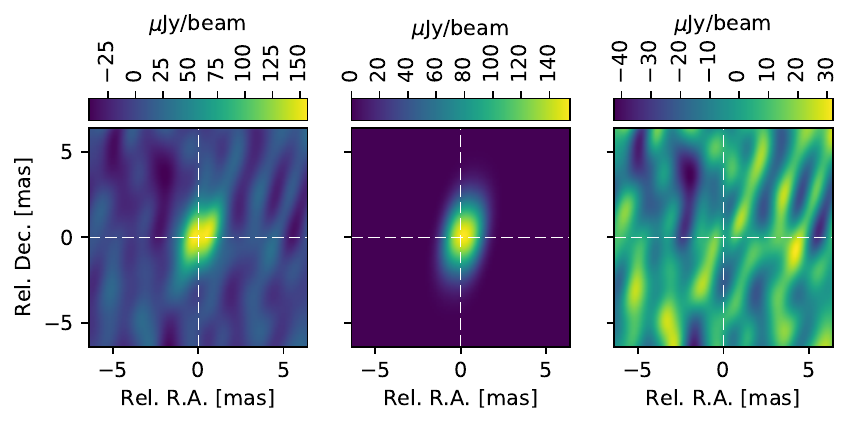}
\caption{Example fitting of a clearly resolved source 0.2212+0.444 in the 6\,cm
image of BB335B. The left panel shows the CLEANED image (the observation), the
mid panel the best fit model convolved with the CLEAN beam, and the right panel
the residual that is image$-$model.  The white cross marks the centre of each
panel, that is the position guess obtained from PyBDSM.
	This particular fit was chosen because it shows that the position found by
	PyBDSM may be off when PyBDSM's Gaussian fitting finds the peak of one of
	the two beam-features of a weak resolved shell.  However, since the
	position is allowed to vary, this taken into account in the fitting, as
	seen here where the fitted position is a little to the right of the white
cross.\label{fig:fittingexample}}
\end{figure*}

Most sources were very well centred, but a few clearly resolved shells had
minor offsets since the PyBDSM algorithm had found the peak at one side of the
shell. An example of this can be seen in Fig. \ref{fig:fittingexample}.
When the fit converged, the best-fit model was transformed to the image plane
and subtracted from the working-copy of the cleaned image to simplify fitting
of other nearby sources. The best-fit parameters were saved to disk and values
for all attempted fits are available as electronic tables through the CDS as
described in \ref{app:cds}.

\subsection{A self-consistent check of the catalogue positions}
\begin{figure*}[htbp]
\centering
\subfigure[Right Ascension]{
        \includegraphics[width=0.48\textwidth]{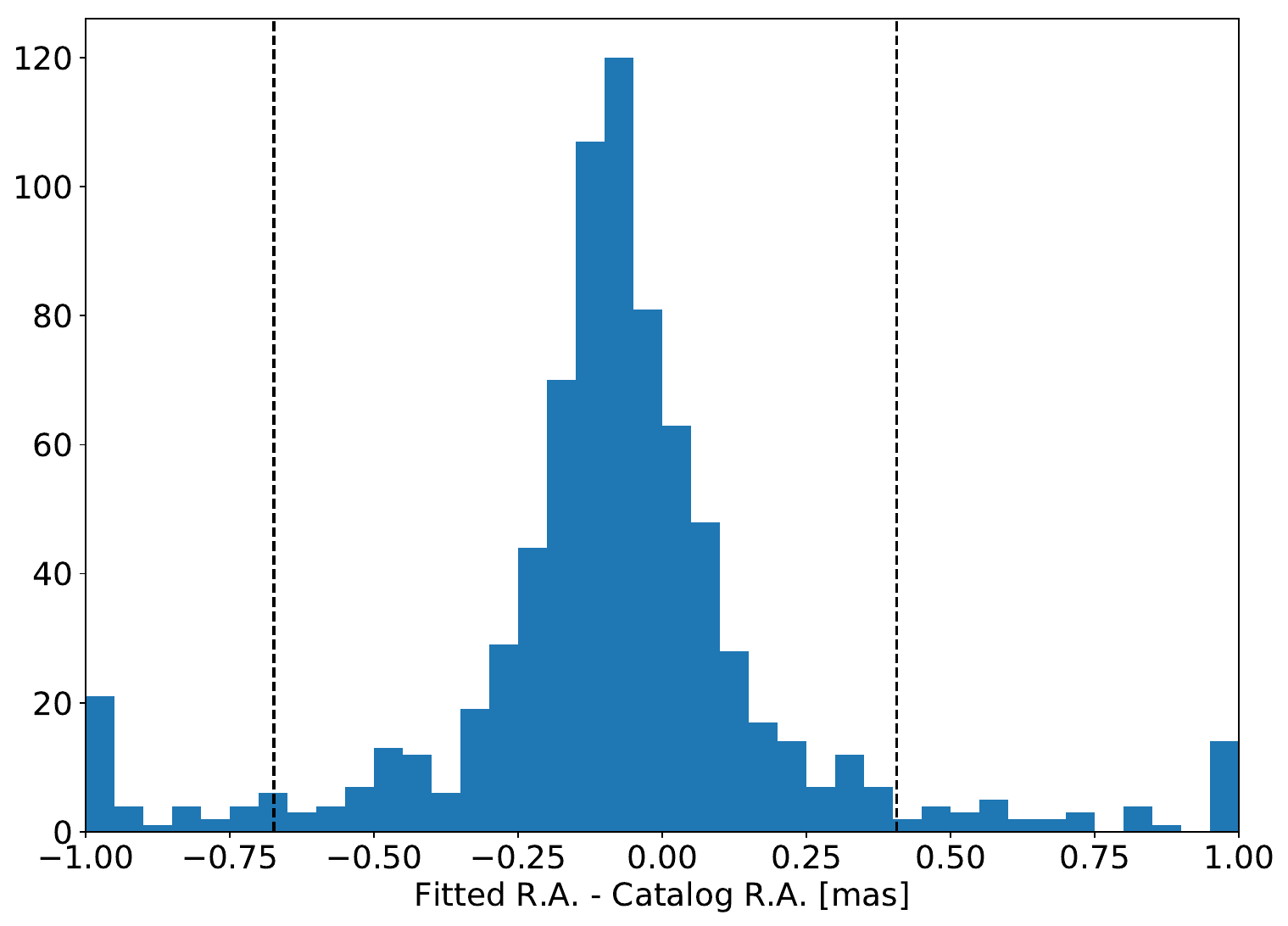}
        \label{fig:radiff}
}
\subfigure[Declination]{
        \includegraphics[width=0.48\textwidth]{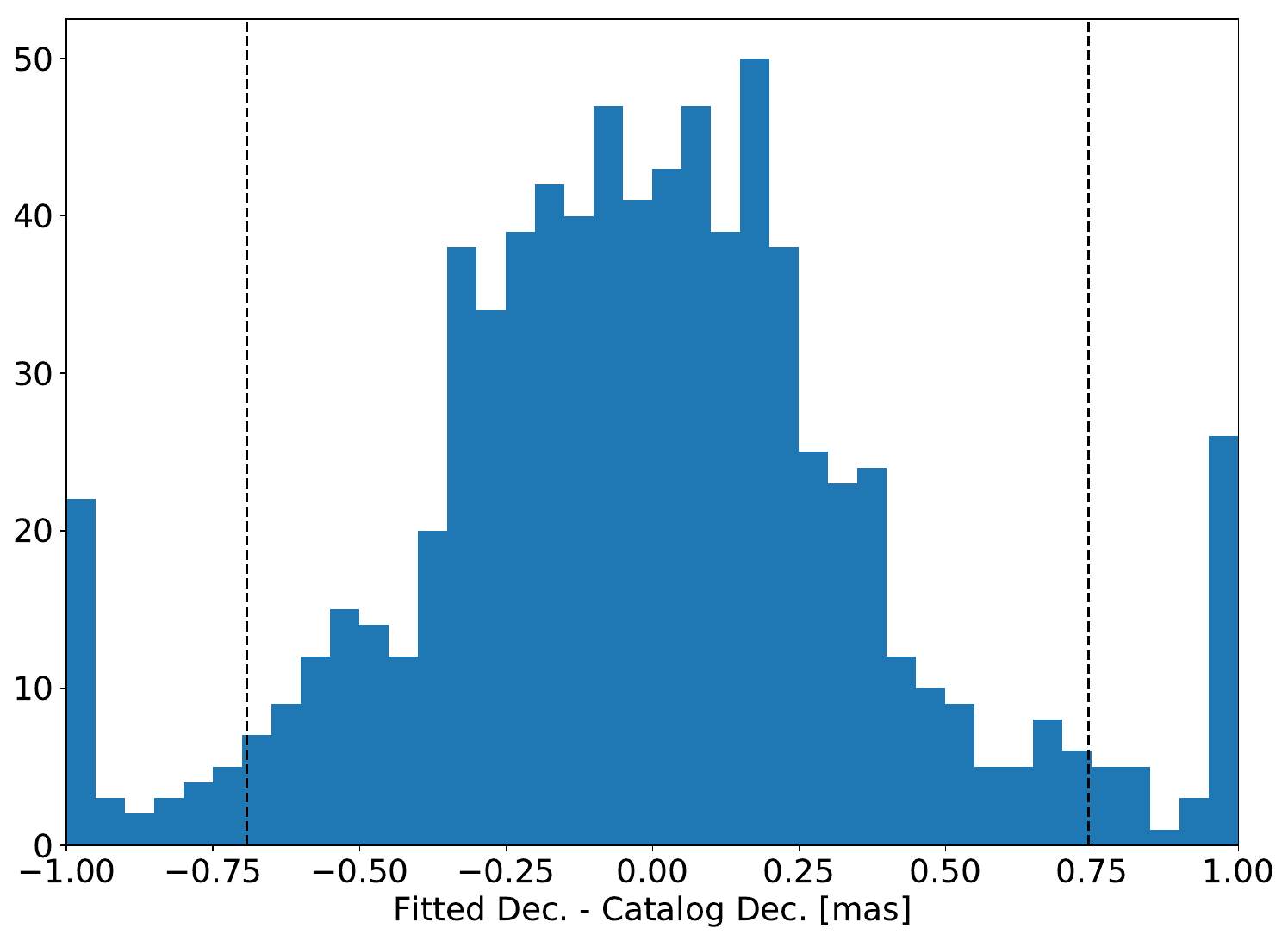}
        \label{fig:decdiff}
}
\caption{Comparison of fitted positions with catalogue positions given in
Tables \ref{table:east} and \ref{table:west}.  The dashed lines mark the 5\%
and 95\% percentiles. The majority of fitting results are well within our
target accuracy of $\pm$1\,mas. The larger dispersion in Dec. compared to R.A.
is explained by the synthesised beam being elongated approximately north-south
in most epochs. Note that 18\,cm fits are not included in this figure; the
spread in R.A.  is very similar for 18\,cm, but the fitting in Dec. is much
less well constrained because of the relatively large major axis of the
synthesised beam.
\label{fig:posdiffs}}
\end{figure*}

The source positions in Tables \ref{table:east} and \ref{table:west} were
obtained by PyBDSM as described in Sect. \ref{sect:fitting}. However, one may
ask, do the stacked images really provide a good reference for the catalogue,
or for initial guesses for the fitting routine? To check this we compiled two
figures of the difference between the catalogue R.A. and Dec. and the fitted
R.A.  and Dec for all modelfits done in this work, see Fig. \ref{fig:posdiffs}.

The majority of fitting results are well within our target accuracy of
$\pm$1\,mas. Note that 18\,cm fits are not included in this figure; the spread
in R.A.  is very similar for 18\,cm, but the fitting in Dec. is much less well
constrained because of the relatively large major axis of the synthesised beam
which in most epochs extends towards north-south. 

\section{Comparison of fitting methods}
\label{sect:fitcomp}
In this work we have used LS optimisation to fit source flux
densities and sizes to deconvolved images.  LS-fitting is widely used and is
very quick. It is, however, challenging to obtain meaningful uncertainty
estimates from the fitting procedure itself (see for example
\cite{2010arXiv1008.4686H}). Therefore, we instead calculate our uncertainties
as explained in Sect.  \ref{sect:fittinguncert}. This way of calculating
uncertainties should however be verified by simulations, in particular to
ensure the numerical factor in Eq.  \ref{eqn:es} is adequately chosen.  To
assess the validity of the LS-method and ensure that our uncertainty estimates
are valid, we compare our method to state-of-the-art visibility fitting on
simulated data.  Visibility fitting, or UV-fitting, is in theory the optimal
way to model-fit interferometric data since it avoids deconvolution biases
introduced by CLEAN.  In this work, we compare our LS-method to visibility
fitting as implemented in the CASA tool UV-multifit \citep{marti-vidal2014}.  

\subsection{Simulations}
\label{sect:fitsim}
We used the CASA simulator tools to construct 50 measurement sets (MSs), where 
each contained a single optically thin shell plus noise.  The
flux density and diameter of the shell was chosen from a 5x10 grid spanning the
ranges 0.1-1.0\,mJy and 0.001-5\,mas in flux density and diameter respectively,
that is similar to our observed values.  We simulated each shell as located in the
centre of Arp\,220 (at R.A.  15$^h$34$^m$57.25$^s$, Dec.
23$^\circ$30$'$11.33$''$) and observed with the VLBA during 12 hours with 8
hours on source. The noise was adjusted to yield a realistic off source RMS
level of 30\,$\mu$Jybeam$^{-1}$ in the imaging domain.

\subsection{Fitting simulated data}
For each source we now ran UV-multifit and fitted a shell model to each source,
obtaining a fitted flux density, diameter, position, and the respective
uncertainties as given by UV-multifit. The initial guess was the same as used
in LS-fitting our real observations, that is a flux density of 0.5\,mJy and size
0.1\,mas, and the same bounds was applied as for the LS-fitting.

To test the LS-method on these data, we converted the simulated MSs to UVFITS
and imaged each simulated source in AIPS using IMAGR with the same parameters
as used to process our real observations, and the images were saved as FITS
files.  The LS-fitting was performed as described in Sect. \ref{sect:fitting},
that is in exactly the same way as for our real observations, and the
uncertainties calculated according to Eqns. \ref{eqn:ef} and \ref{eqn:es}.

For relative comparison we note that on average, on the same computer,
UV-fitting took about 4 minutes per source, while LS-fitting took about 1
second per source.  

\subsection{Results of method comparison}
\begin{figure*}[htbp]
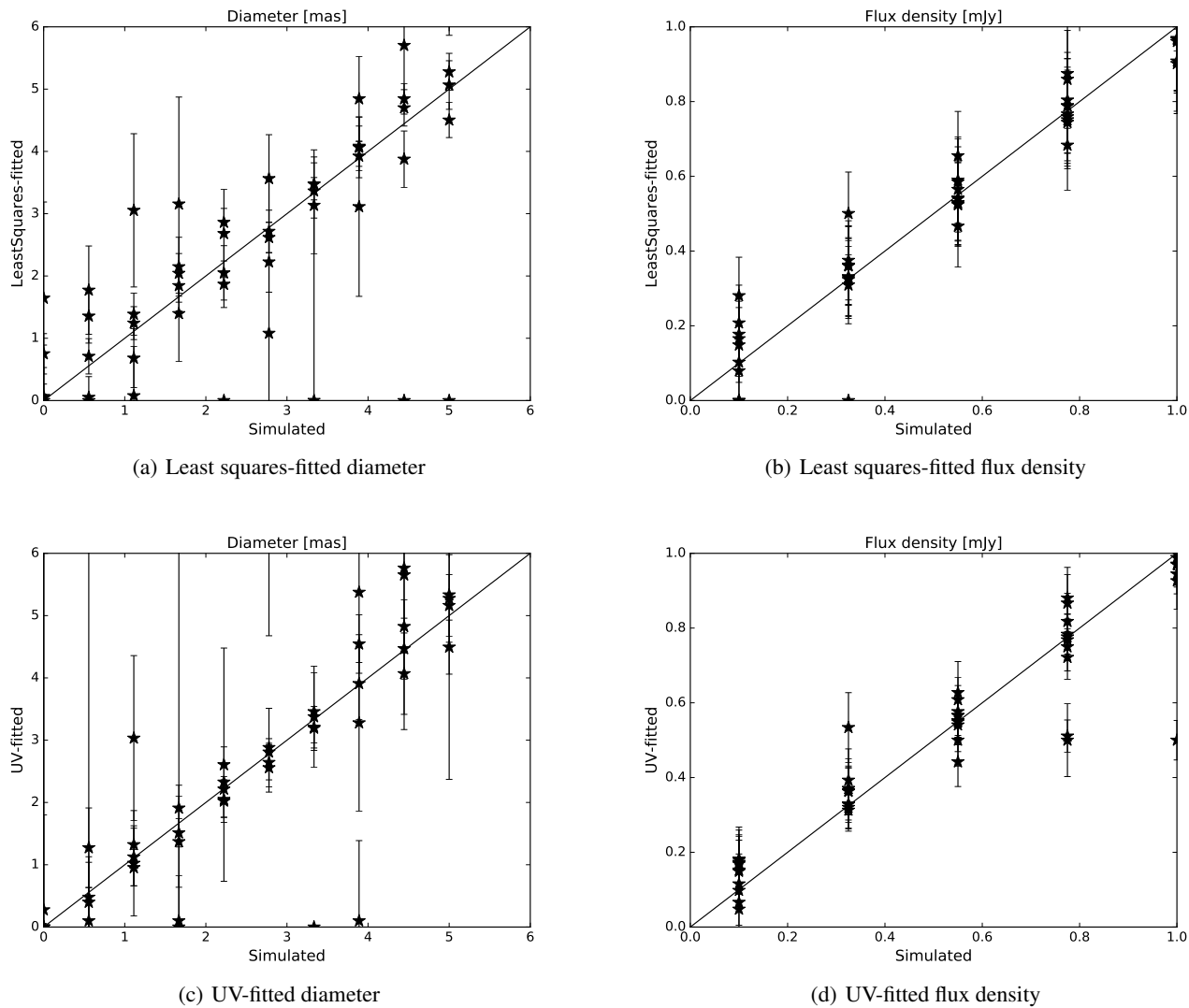

\centering
\subfigure[Least squares-fitted diameter]{
        \includegraphics[width=0.48\textwidth]{figures/{{D_LS_SIM}}}
        \label{fig:dlssim}
}
\subfigure[Least squares-fitted flux density]{
        \includegraphics[width=0.48\textwidth]{figures/{{F_LS_SIM}}}
        \label{fig:flssim}
}
\subfigure[UV-fitted diameter]{
        \includegraphics[width=0.48\textwidth]{figures/{{D_UV_SIM}}}
        \label{fig:duvsim}
}
\subfigure[UV-fitted flux density]{
        \includegraphics[width=0.48\textwidth]{figures/{{F_UV_SIM}}}
        \label{fig:fuvsim}
}
\caption{Recovered diameters and flux densities for simulated data using three
different methods: LS-fitting (first row) and UV-fitting (second row). The
values for these figures are presented in Table \ref{table:simvals}. Cases
where no fit was attempted in (at least one) method are included for
completeness, but displaced as both value 0 and uncertainty 0 on the
representative method axis.
\label{fig:simcomp}}
\end{figure*}

The fitted values and uncertainties for the 50 simulated data sets are
presented in Table \ref{table:simvals} and Fig. \ref{fig:simcomp}.  We find the
two methods give similar results in terms of accuracy and uncertainty for the
simulated sources.  Because of the amount of data and number of sources
analysed in this work, we have decided to use the (much) faster LS-fitting.
Other methods, for example  Markov-Chain Monte Carlo (MCMC) fitting, could also be
used. However, given that our method already obtains results comparable with
the (in theory) best possible method of visibility fitting, comparison with
other methods such as MCMC is beyond scope of this paper.

\begin{table*}
\caption{Simulated and fitted values to assess validity of LS and UV fitting
methods. Each row represent one of 50 simulated test data sets, as described in
Sect. \ref{sect:fitsim}. The first two columns show the flux density and
diameter used to generate a model of an optically thin shell. The remaining
columns show the values recovered by fitting. Dashes (-) indicate no fit was
attempted since source was too weak.}
\label{table:simvals}
\centering
\begin{tabular}{r r r r r r r r}
\hline\hline
Sim. F & Sim. D & UV F & UV D & LS F & LS D \\ 
mJy & mas & mJy & mas & mJy & mas \\ 
\hline
0.10 & 0.00 & 0.15$\pm$0.04 & 0.00$\pm$0.06  & 0.17$\pm$0.10 & 1.64$\pm$1.54\\
0.33 & 0.00 & 0.33$\pm$0.04 & 0.00$\pm$22.13 & 0.31$\pm$0.10 & 0.06$\pm$0.83\\
0.55 & 0.00 & 0.57$\pm$0.05 & 0.28$\pm$1.52  & 0.53$\pm$0.11 & 0.05$\pm$0.48\\
0.78 & 0.00 & 0.51$\pm$0.04 &              - & 0.79$\pm$0.13 & 0.75$\pm$0.32\\
1.00 & 0.00 & 1.00$\pm$0.05 & 0.00$\pm$61.62 & 0.97$\pm$0.14 & 0.00$\pm$0.26\\
0.10 & 0.56 & 0.07$\pm$0.10 & 7.38$\pm$7.77  &             - &             -\\
0.33 & 0.56 & 0.37$\pm$0.06 & 1.27$\pm$0.64  & 0.36$\pm$0.11 & 1.77$\pm$0.71\\
0.55 & 0.56 & 0.50$\pm$0.07 & 0.10$\pm$0.53  & 0.59$\pm$0.12 & 1.36$\pm$0.43\\
0.78 & 0.56 & 0.78$\pm$0.05 & 0.48$\pm$0.65  & 0.77$\pm$0.13 & 0.05$\pm$0.33\\
1.00 & 0.56 & 0.94$\pm$0.05 & 0.40$\pm$0.64  & 0.90$\pm$0.13 & 0.71$\pm$0.28\\
0.10 & 1.11 & 0.18$\pm$0.06 & 3.03$\pm$1.33  & 0.21$\pm$0.10 & 3.05$\pm$1.23\\
0.33 & 1.11 & 0.32$\pm$0.05 & 1.02$\pm$0.85  & 0.32$\pm$0.10 & 0.08$\pm$0.79\\
0.55 & 1.11 & 0.55$\pm$0.06 & 1.13$\pm$0.46  & 0.54$\pm$0.11 & 0.68$\pm$0.47\\
0.78 & 1.11 & 0.77$\pm$0.06 & 1.32$\pm$0.30  & 0.76$\pm$0.12 & 1.39$\pm$0.34\\
1.00 & 1.11 & 0.97$\pm$0.05 & 0.95$\pm$0.29  & 0.97$\pm$0.14 & 1.24$\pm$0.26\\
0.10 & 1.67 & 0.10$\pm$0.04 & 0.00$\pm$42.03 & 0.15$\pm$0.10 & 3.16$\pm$1.72\\
0.33 & 1.67 & 0.31$\pm$0.06 & 1.37$\pm$0.73  & 0.33$\pm$0.10 & 1.40$\pm$0.77\\
0.55 & 1.67 & 0.57$\pm$0.06 & 1.91$\pm$0.37  & 0.54$\pm$0.11 & 2.15$\pm$0.47\\
0.78 & 1.67 & 0.50$\pm$0.10 & 0.10$\pm$0.72  & 0.80$\pm$0.13 & 2.04$\pm$0.32\\
1.00 & 1.67 & 0.97$\pm$0.06 & 1.51$\pm$0.23  & 0.96$\pm$0.14 & 1.85$\pm$0.27\\
0.10 & 2.22 & 0.12$\pm$0.06 & 2.61$\pm$1.87  &             - &             -\\
0.33 & 2.22 & 0.37$\pm$0.06 & 2.33$\pm$0.57  & 0.36$\pm$0.11 & 2.68$\pm$0.71\\
0.55 & 2.22 & 0.61$\pm$0.06 & 2.02$\pm$0.34  & 0.58$\pm$0.11 & 2.05$\pm$0.44\\
0.78 & 2.22 & 0.72$\pm$0.06 & 2.05$\pm$0.29  & 0.68$\pm$0.12 & 1.87$\pm$0.37\\
1.00 & 2.22 & 1.05$\pm$0.06 & 2.21$\pm$0.20  & 1.15$\pm$0.15 & 2.86$\pm$0.22\\
0.10 & 2.78 & 0.17$\pm$0.10 & 8.00$\pm$3.32  & 0.10$\pm$0.10 & 1.08$\pm$2.49\\
0.33 & 2.78 & 0.36$\pm$0.06 & 2.88$\pm$0.63  & 0.36$\pm$0.11 & 3.56$\pm$0.70\\
0.55 & 2.78 & 0.55$\pm$0.06 & 2.56$\pm$0.39  & 0.53$\pm$0.11 & 2.23$\pm$0.49\\
0.78 & 2.78 & 0.78$\pm$0.06 & 2.64$\pm$0.28  & 0.75$\pm$0.12 & 2.72$\pm$0.34\\
1.00 & 2.78 & 1.04$\pm$0.06 & 2.81$\pm$0.22  & 1.04$\pm$0.14 & 2.62$\pm$0.25\\
0.10 & 3.33 & 0.05$\pm$0.04 & 0.00$\pm$0.01  & 0.08$\pm$0.10 & 0.00$\pm$3.22\\
0.33 & 3.33 & 0.33$\pm$0.07 & 3.38$\pm$0.81  & 0.33$\pm$0.10 & 3.13$\pm$0.78\\
0.55 & 3.33 & 0.44$\pm$0.07 & 3.46$\pm$0.62  & 0.47$\pm$0.11 & 3.48$\pm$0.55\\
0.78 & 3.33 & 0.75$\pm$0.06 & 3.21$\pm$0.33  & 0.74$\pm$0.12 & 3.47$\pm$0.34\\
1.00 & 3.33 & 1.06$\pm$0.06 & 3.19$\pm$0.24  & 1.18$\pm$0.15 & 3.36$\pm$0.22\\
0.10 & 3.89 & 0.18$\pm$0.06 & 3.28$\pm$1.42  & 0.18$\pm$0.10 & 3.11$\pm$1.44\\
0.33 & 3.89 & 0.39$\pm$0.08 & 5.38$\pm$1.13  & 0.38$\pm$0.11 & 4.84$\pm$0.68\\
0.55 & 3.89 & 0.58$\pm$0.07 & 3.91$\pm$0.58  & 0.52$\pm$0.11 & 4.06$\pm$0.49\\
0.78 & 3.89 & 0.87$\pm$0.08 & 4.54$\pm$0.47  & 0.79$\pm$0.13 & 4.08$\pm$0.32\\
1.00 & 3.89 & 0.50$\pm$0.05 & 0.10$\pm$1.29  & 1.08$\pm$0.15 & 3.92$\pm$0.24\\
0.10 & 4.44 & 0.17$\pm$0.09 & 5.76$\pm$2.59  & 0.28$\pm$0.10 & 5.70$\pm$0.91\\
0.33 & 4.44 & 0.37$\pm$0.09 & 5.65$\pm$1.23  &             - &             -\\
0.55 & 4.44 & 0.54$\pm$0.07 & 4.07$\pm$0.65  & 0.56$\pm$0.11 & 3.88$\pm$0.45\\
0.78 & 4.44 & 0.82$\pm$0.08 & 4.47$\pm$0.49  & 0.87$\pm$0.13 & 4.70$\pm$0.29\\
1.00 & 4.44 & 0.99$\pm$0.08 & 4.83$\pm$0.43  & 1.04$\pm$0.14 & 4.84$\pm$0.25\\
0.10 & 5.00 & 0.15$\pm$0.08 & 5.33$\pm$2.96  &             - &             -\\
0.33 & 5.00 & 0.53$\pm$0.09 & 7.06$\pm$0.92  & 0.50$\pm$0.11 & 6.37$\pm$0.51\\
0.55 & 5.00 & 0.63$\pm$0.08 & 5.28$\pm$0.70  & 0.65$\pm$0.12 & 5.07$\pm$0.39\\
0.78 & 5.00 & 0.88$\pm$0.08 & 5.16$\pm$0.50  & 0.86$\pm$0.13 & 5.28$\pm$0.30\\
1.00 & 5.00 & 0.93$\pm$0.08 & 4.49$\pm$0.43  & 0.91$\pm$0.13 & 4.50$\pm$0.28\\
\hline
\end{tabular}
\end{table*}

\section{A comparison of source properties with published literature}
\label{sect:prevstudycomp}
In this section we present comparisons done between our fitted values
for the compact sources and values available in the literature.

\subsection{Source catalogues and completeness}
To compare our results with previously published values we tried to match our
detected sources with the positions published by
\cite{lonsdale1998,lonsdale2006, parra2007, batejat2011}.  The matching was
done by comparing coordinates both automatically and manually to account for
typos and blending effects, as mentioned in Sect. \ref{sect:posacc}, and in
general the difference between our positions and the literature positions were
much less than the beam size.  When we found matches we have included the
legacy name (for example W55, E8) in Tables \ref{table:east} and
\ref{table:west}.  Even though we accounted for different reference positions,
as noted in Sect.  \ref{sect:posacc}, we found no counterpart in our images for
the following sources: E4, E12, W3, W19, W24, W27. These may be false positives
in previous studies, but it is also possible that they were weak when detected
and has been declining since. Although we re-analyse the data used by for
example \cite{lonsdale2006}, our images are not as deep as the ones previously
published. Hence we may miss sources both in these old epochs due to the lower
sensitivity, and in our recent 5\,GHz epochs due to declining lightcurves and
steep source spectra.  Further investigation of these sources are beyond the
scope of this work.

\subsection{Positions of particular sources}
\label{sect:posacc}
As noted by \cite{parra2007}, the maser reference position assumed by
\cite{lonsdale2006} and previous studies was off by about $0.1''$ in
declination, with a declination of $11.564''$ being assumed instead of
$11.664''$, as noted above. However, from cross-referencing of the relative
source positions listed in Table 1 of \cite{lonsdale2006} with the
positions obtained in this work, we find the reference position used for the
relative coordinates listed by \cite{lonsdale2006} at R.A.
$15^h34^m57^s.26255$, Dec.  $23^\circ30'11''.352$, that is different from the
maser peak position used for phase referencing. This difference has been taken
into account when comparing positions and flux densities for these sources.

We note that the R.A. position listed by \cite{lonsdale2006} for W11
corresponds to 57$^s$.2300, which is consistent with the position we find as
well as the position listed by \cite{batejat2011}. However, in \cite{parra2007}
this source is listed with R.A. 57$^s$.2230 which seems to be a typographical
error.  Furthermore, the position listed for E11 by \cite{lonsdale2006}
corresponds to our source 0.2910+0.325, which is about 10\,mas from the source
0.2195+0.335 (E10). However, we also find a source 0.2913+0.333 situated
between these two sources, only about 3\,mas from 0.2195+0.335. From Table
\ref{table:obslist}, it is clear that observations at wavelengths longer than
6\,cm do not have sufficient resolution to distinguish the two sources
0.2195+0.335 and 0.2913+0.333, and hence the flux densities measured at
1.4\,GHz for these two sources, also in this work, should be interpreted with
caution.  We note that \cite{batejat2011} associated 0.2913+0.333 with the
previously listed source E11 instead of associating 0.2910+0.325 with E11 as we
have done. Although 0.2195+0.335 and 0.2913+0.333 may be confused at 1.4\,GHz,
the resolution is good enough to separate 0.2913+0.333 and 0.2910+0.325.  We
therefore believe that \cite{batejat2011} misidentified 0.2913+0.333 as E11,
likely because 0.2910+0.325 is weaker at 6\,cm than 0.2913+0.333, and hence
their spectrum for E11 should be interpreted with caution. 

To avoid typographical errors in this paper we have generated all figures and
tables in the paper directly from the source catalogues generated by the
source-finding algorithm as described in Sect. \ref{sect:sourcefinding},
without any manual editing.

\subsection{Flux densities and sizes}
\begin{figure}[h]
\centering
\subfigure[Flux density vs \cite{parra2007}]{
        \includegraphics[width=0.32\textwidth]{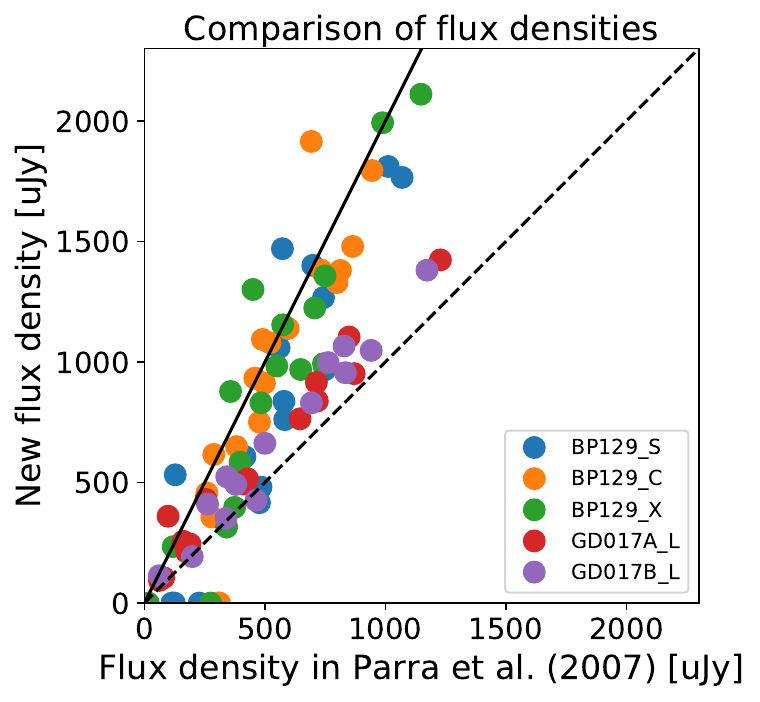}
        \label{fig:parraflux}
}
\subfigure[Flux density vs \cite{batejat2011}]{
        \includegraphics[width=0.325\textwidth]{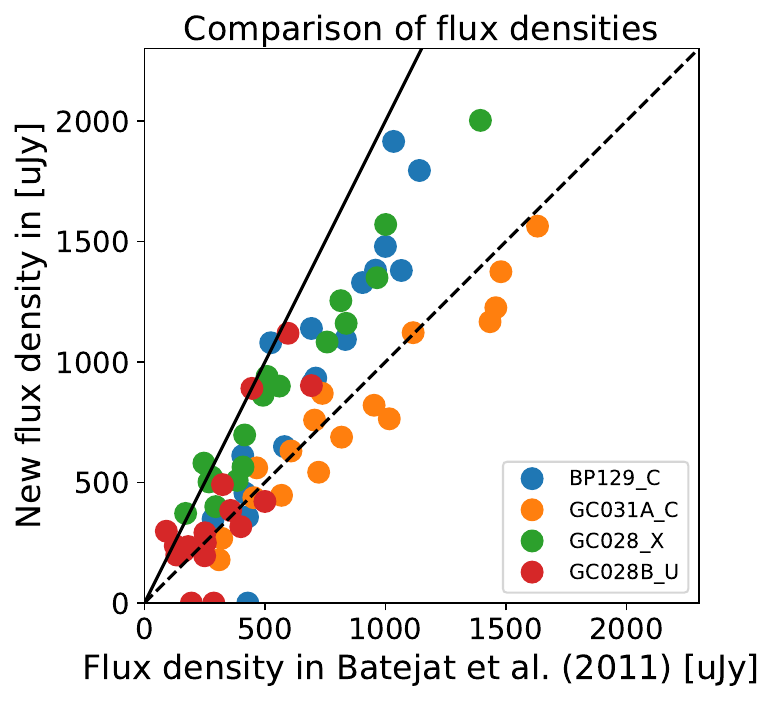}
        \label{fig:batejatflux}
}
\subfigure[Diameter vs \cite{batejat2011}]{
        \includegraphics[width=0.29\textwidth]{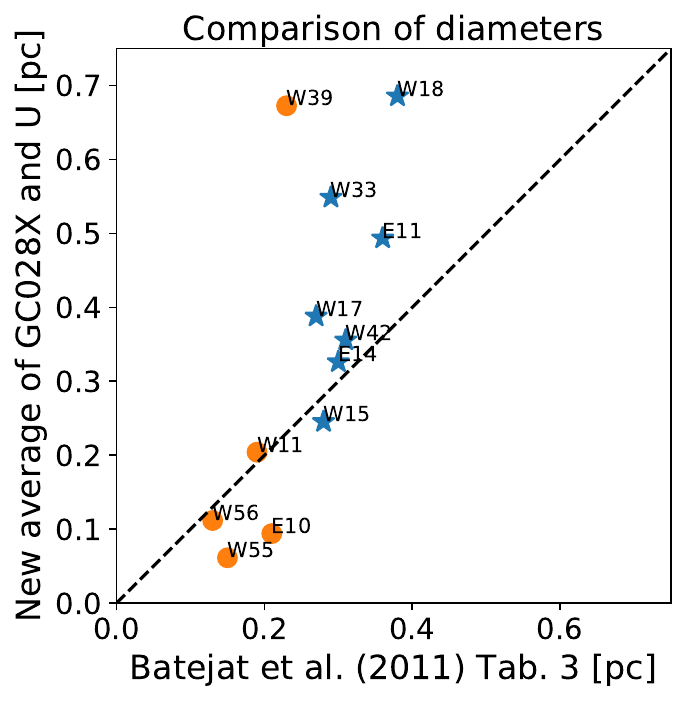}
        \label{fig:batejatsize}
}
\caption{Comparison of flux densities and sizes measured in this work with
previously published values using the same data. The dashed lines indicate a
one-to-one correspondence, and the solid lines indicate the new values to be a
factor of 2 higher than the ones previously published. In Fig.
\ref{fig:batejatsize}, stars mark sources resolved at both bands by
\cite{batejat2011} (bold face in their Table 3) while circles mark the
remaining sources.  See Sect.  \ref{sect:prevstudycomp} for a discussion of
discrepancies.
\label{fig:prevstudycomp}}
\end{figure}

As stated previously, some data included in this work have been published
before.  In Fig. \ref{fig:prevstudycomp}, we compare the flux densities and sizes
measured in this work with previously published values. 

Flux densities for the epochs GD017 and BP129 where reported by
\cite{parra2007}, where the 18\,cm values observed in 2003 (GD017A) were taken
from \cite{lonsdale2006}. In Fig. \ref{fig:prevstudycomp} (panel a), we compare the flux
densities recovered in this work to those previously published by
\cite{parra2007}. We find that we in general find larger values, although the scaling
factor varies between publications and data sets. Notably, we
measure approximately two times higher flux densities than previously reported
for all BP129. Since we find that our fitting method produces good results
compared to for example manual inspection of the images (which was the method used to
extract the flux densities by \cite{parra2007} for BP129) we believe this flux
discrepancy is due to differences in calibration and imaging strategy.
However, a factor of two is much larger than expected given the uncertainties of the 
measurements. We use the same calibration strategy for BP129 as for other epochs. 
We have checked our calibration scripts for BP129 carefully without finding any
reason for too high flux densities. We therefore suspect that values reported for
BP129 by \cite{parra2007} suffer from a systematic error in the calibration 
or imaging strategy, giving too low flux densities. We note that the image noise
levels reported by \cite{parra2007} are very similar to the noise we obtain.
This argues against a simple difference in how the flux scale was set. Instead,
such a flux density reduction suggests incoherent addition of visibilities,
that is  with strong phase errors present in one or more antennas during a
significant period of time.  We note that self-calibration could increase the
flux density of sources in many epochs, which should reduce the RMS noise in
the image. However, we see, for example for the GD017-epochs, an increase in
both RMS noise level and source flux density compared to previous publication.
Such an increase could instead be attributed to differences in amplitude
calibration strategy, where our careful alignment of potentially bad antennas
using bright sources should provide a more accurate amplitude scale.
Further investigation of the details of the calibration performed by
\cite{lonsdale2006} and \cite{parra2007} is beyond the scope of this paper.

Flux densities for epochs GC028 and GC031A are reported in \cite{batejat2011},
together with a revised version of the flux densities for the 6\,cm data of
BP129 (the 13\,cm and 3.6\,cm values for BP129 are taken directly from
\citealt{parra2007}). In Fig. \ref{fig:prevstudycomp} (panel b), we compare the flux
densities recovered in this work to those previously published by
\cite{batejat2011}. We find the flux densities reported for GC031A to be in
excellent agreement with our measurements.  However, we measure flux densities
to be 1.5 times higher than reported for GC028 and BP129-C by \cite{batejat2011}.
Again, using the same argument as in the previous paragraph, we assume that our
new values are correct. We note that \cite{batejat2011} claim an image noise of
41\,$\mu$Jybeam$^{-1}$ for BP129-6\,cm, which is a factor of two lower than what we
obtain. We note that while \cite{batejat2011} left multiple sources unclassified
due to inconsistent lightcurves, we find, given our higher flux density values for
some epochs, the lightcurves for these sources consistent with SNe/SNR
evolution, as noted in Sect. \ref{sect:classerrs}.

Source sizes were reported by \cite{batejat2011} using data at 2\,cm and
3.6\,cm.  In Fig. \ref{fig:prevstudycomp} (panel c), we compare the sizes given in
\cite{batejat2011} with our measurements from the same data.  We only include
the 12 sources where diameters are given in Table 3 of \cite{batejat2011}
that is not the sources with only upper limits. The sources fitted in both epochs
by \cite{batejat2011}, that is bold face in their Table 3, are shown as stars in
our Fig. \ref{fig:prevstudycomp}, while the remaining sources are shown as
circles. As new diameters, we take the average of the fitted values for the
2\,cm and 3.6\,cm images, as done by \cite{batejat2011}. We find the
measurements to be in reasonable agreement.

\subsection{Source classification changes}
\label{sect:classerrs}
The number of sources detected in this work is too large for a detailed
discussion on each object. However, in this section we discuss the
classification of a few particular sources where previous studies have
suggested AGN activity.

The source 0.2171+0.484 (W39) was left unclassified by \cite{batejat2011}
because it showed declining luminosity at long wavelengths and increasing
luminosity at short wavelengths. After re-analysing the data, we find
0.2171+0.484 to have stable or declining lightcurves at all frequencies with a
powerlaw spectrum, consistent with an SNR scenario where the blast wave is
interacting with the ISM.

The source 0.2915+0.335 was also left unclassified by \cite{batejat2011} as it
showed declining luminosity at long wavelengths and increasing luminosity at
short wavelengths. We instead find the lightcurves for 0.2915+0.335 to decline
at both 6\,cm and 3.6\,cm, with approximately the same flux densities in both
bands.  Little or no emission is detected at 18\,cm.  The faint 18\,cm emission
and fact that the 6\,cm emission is not significantly brighter than the 3.6\,cm
emission may indicate significant internal and/or external free-free
absorption.  Although 0.2915+0.335 is today declining at 6\,cm and 3.6\,cm, it
emitted close to $1.5\times10^{28}$\,erg\,s$^{-1}$\,Hz$^{-1}$ at these
wavelengths in 2006.  
Given the relatively small measured size of 0.2\,pc,
the weak 18\,cm emission, this source may be a relatively young SN where the
shock is interacting with a dense ionised CSM.

Finally, the three sources 0.2306+0.502 (W10), 0.2241+0.520 (W17), and
0.2122+0.482 (W42), noted as flat-spectrum AGN candidates by \cite{parra2007},
are likely SNRs.

\subsubsection{OH-maser sources}
\label{sect:ohmasers}
We note that in addition to the continuum sources, we also detect, in our
18\,cm observations, three sources in OH-maser emission, without any continuum
counterpart, at positions 0.2243+0.665, 0.2957+0.341, 0.2912+0.219. We identify
these as the maser objects W1, E1, and E2 discussed by \cite{lonsdale1998}.
These sources are however not included in Tables \ref{table:east} and
\ref{table:west} as a discussion of the OH-maser emission in Arp\,220 is beyond
the scope of this work. We refer the interested reader to for example
\cite{lonsdale1998,rovilos2003}.

We note, however, that at the position of the OH-maser listed as W2 by
\cite{lonsdale1998}, we find a clear (although relatively weak) continuum
counterpart 0.2394+0.540 at multiple epochs and frequencies with stable or
declining lightcurves. Based on the available data we conclude that this source
is likely an SNR.

\section{Source summary slides}
\label{app:book}
The 97 summary pages, one per source, are available in the online version of
this paper. Each page shows the multi-frequency lightcurve, the source spectra
taken from data points close in time, and the size measurements made in all
epochs with sufficient resolution.



\section*{Supplementary material}

# 0.2066+0.458 (W46)

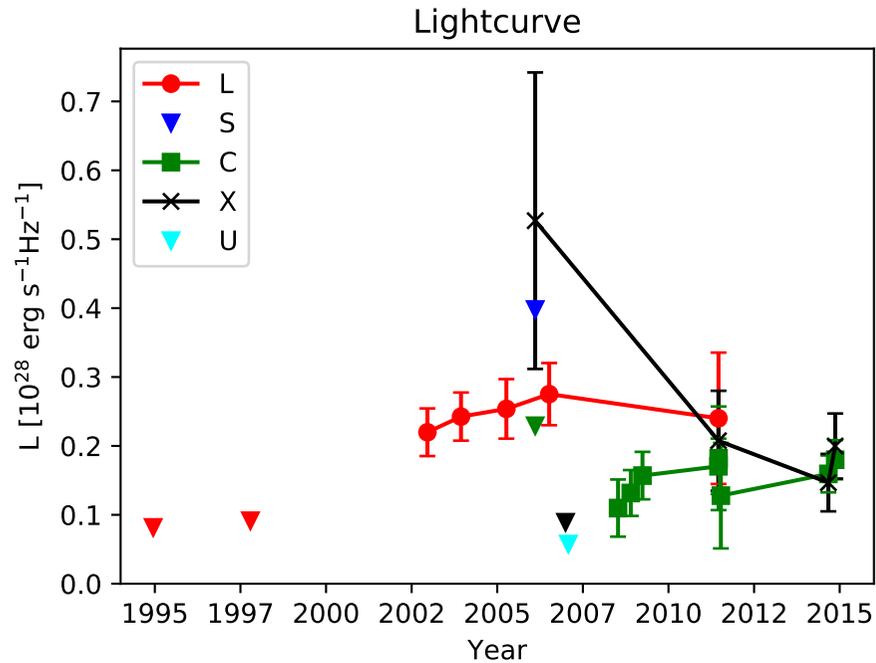

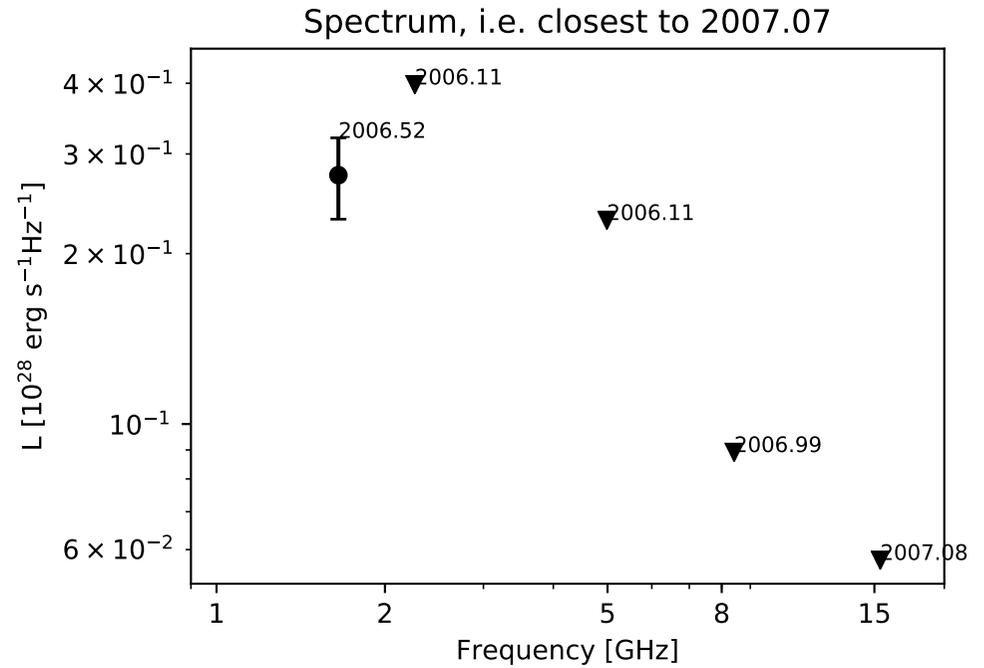

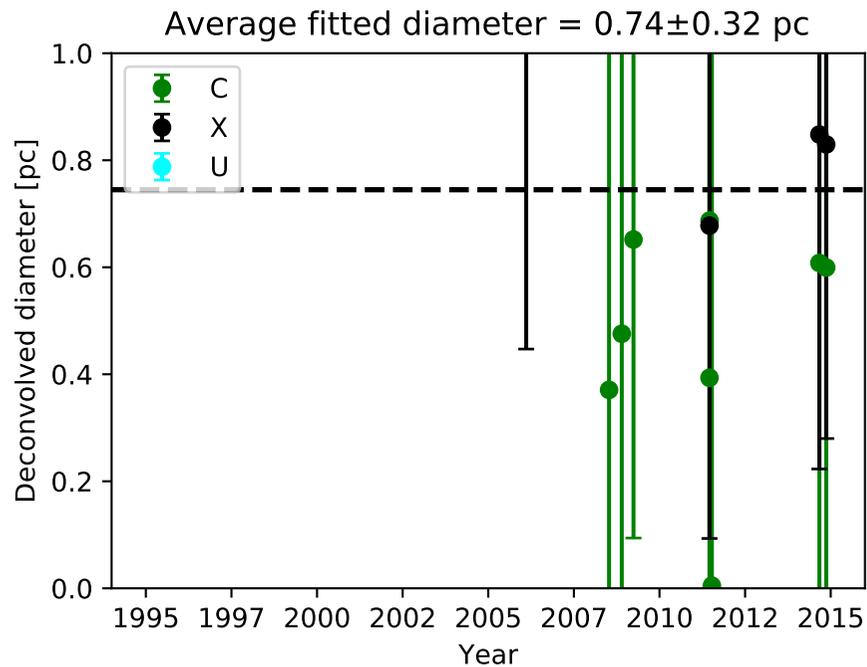

Classification: Slowly varying (S-var)

Figure notes:

Uncertainties calculated as described in Sect. 2.3.

Triangles represent $3\sigma$ upper limits for non-detections.

Sizes shown only for detections in bands C, X, and U.

Average diameters calculated according to Sect. 2.4 from the

3.6 cm and 6 cm measurements in experiments BB335A and BB335B.

# 0.2082+0.548 (W44)

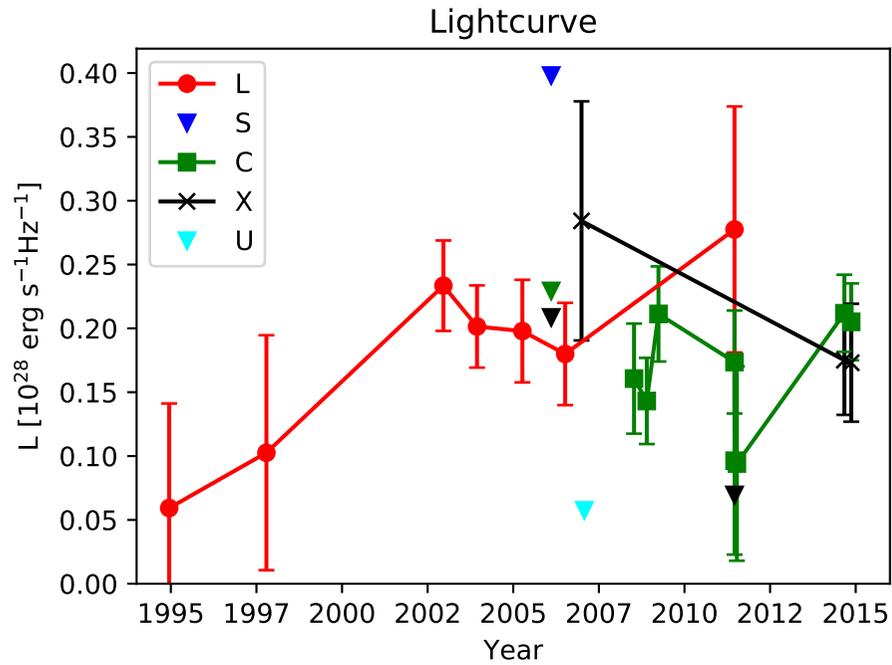

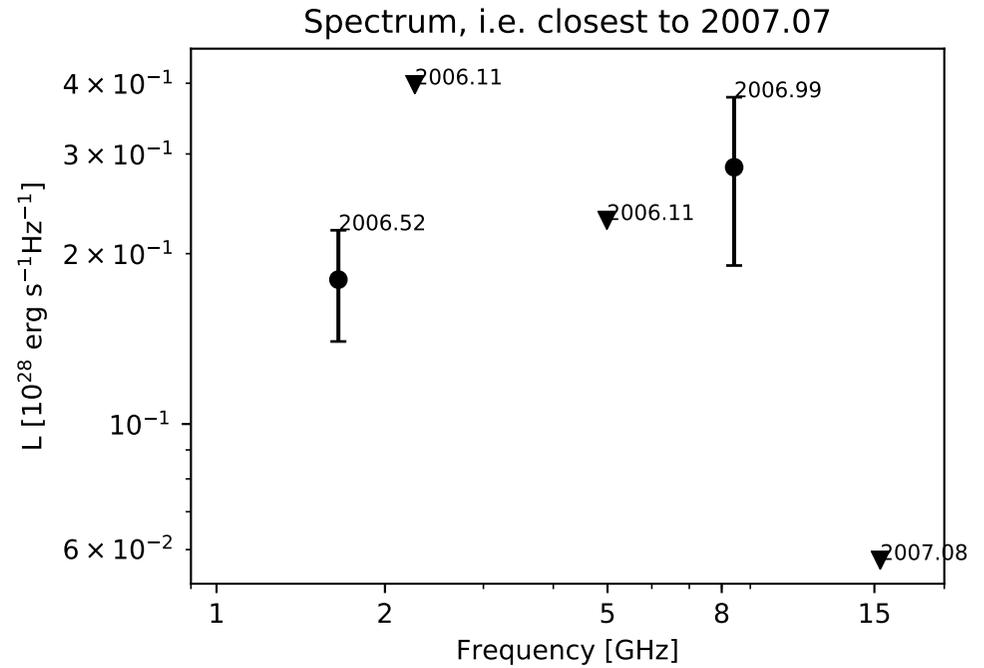

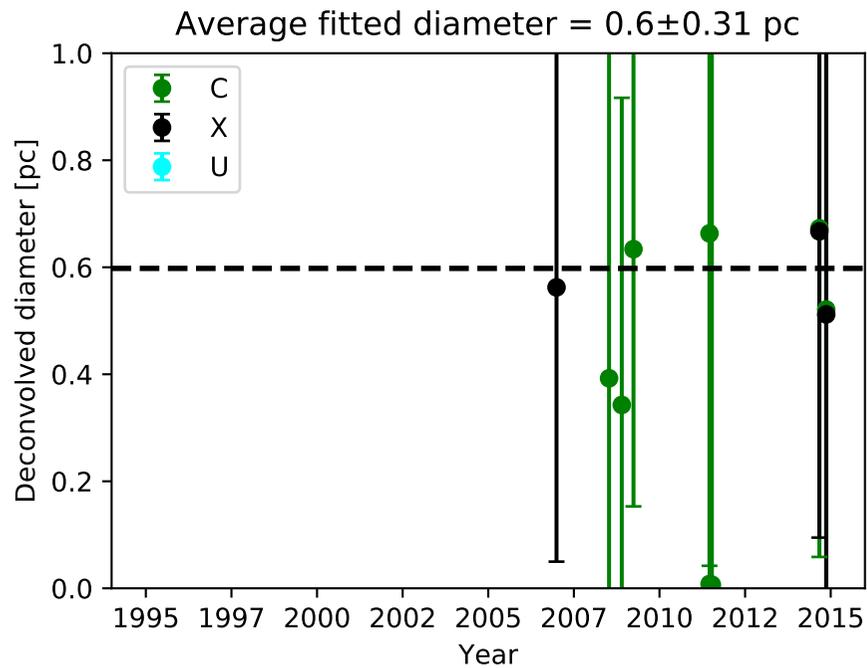

Classification: Slowly varying (S-var)

Figure notes:

Uncertainties calculated as described in Sect. 2.3.

Triangles represent $3\sigma$ upper limits for non-detections.

Sizes shown only for detections in bands C, X, and U.

Average diameters calculated according to Sect. 2.4 from the

3.6 cm and 6 cm measurements in experiments BB335A and BB335B.

# 0.2084+0.534

## Lightcurve

## Spectrum, i.e. closest to 2007.07

## Average fitted diameter = 0.23±0.34 pc

Classification: Rise

Figure notes:

Uncertainties calculated as described in Sect. 2.3.

Triangles represent 3σ upper limits for non-detections.

Sizes shown only for detections in bands C, X, and U.

Average diameters calculated according to Sect. 2.4 from the

3.6 cm and 6 cm measurements in experiments BB335A and BB335B.

# 0.2106+0.487

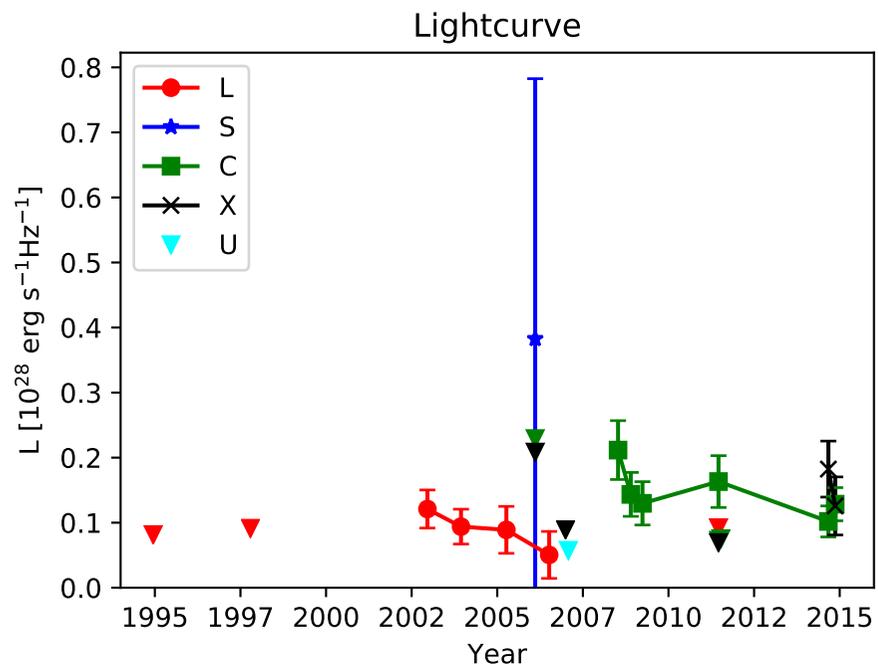

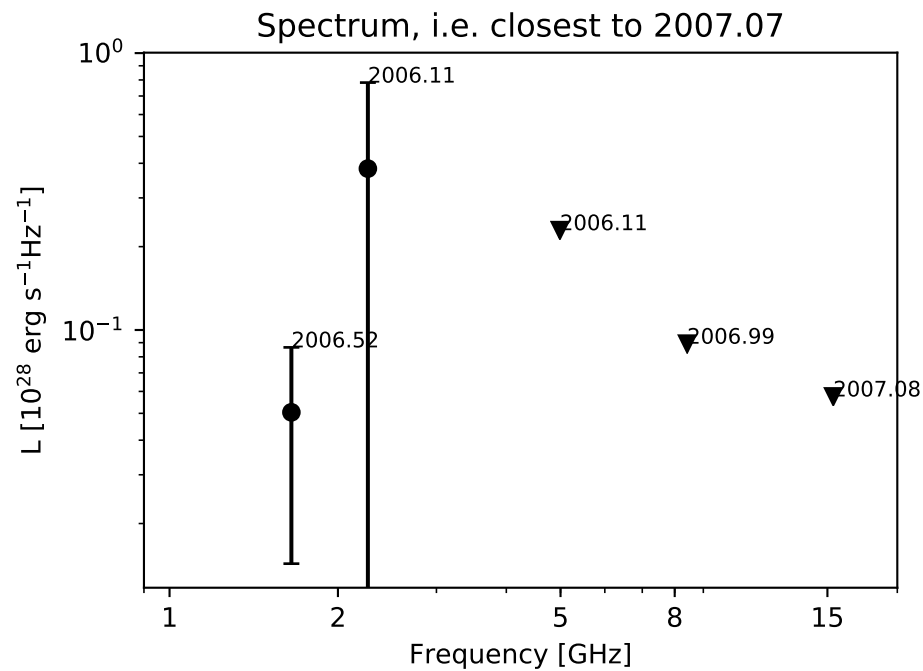

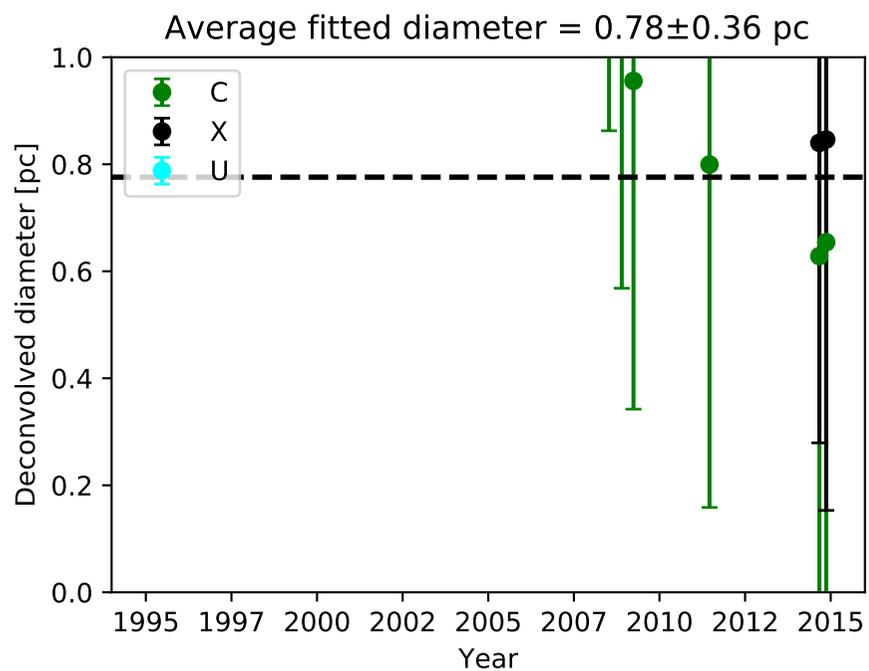

Classification: Slowly varying (S-var)

Figure notes:

Uncertainties calculated as described in Sect. 2.3.

Triangles represent 3σ upper limits for non-detections.

Sizes shown only for detections in bands C, X, and U.

Average diameters calculated according to Sect. 2.4 from the

3.6 cm and 6 cm measurements in experiments BB335A and BB335B.

# 0.2108+0.512

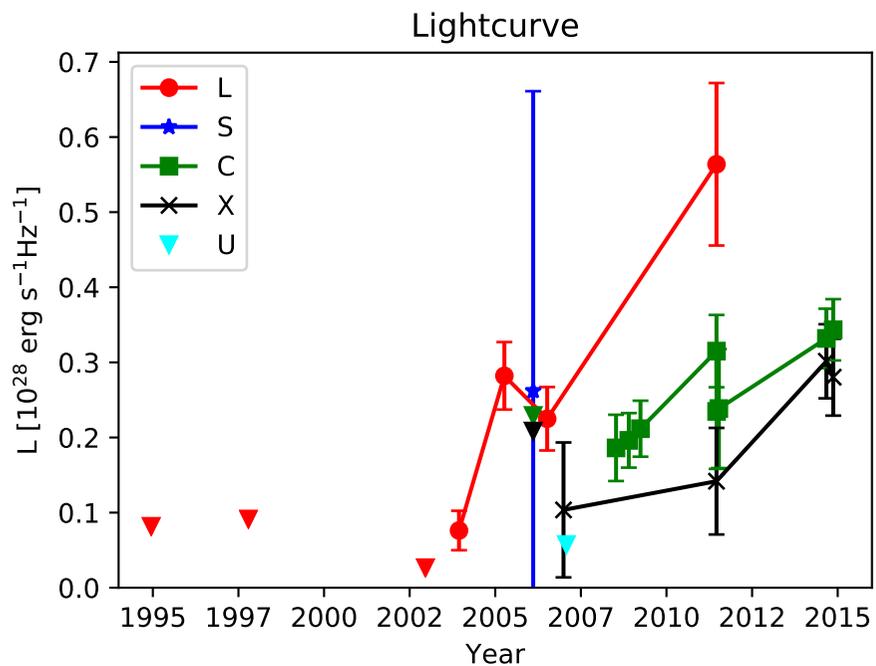
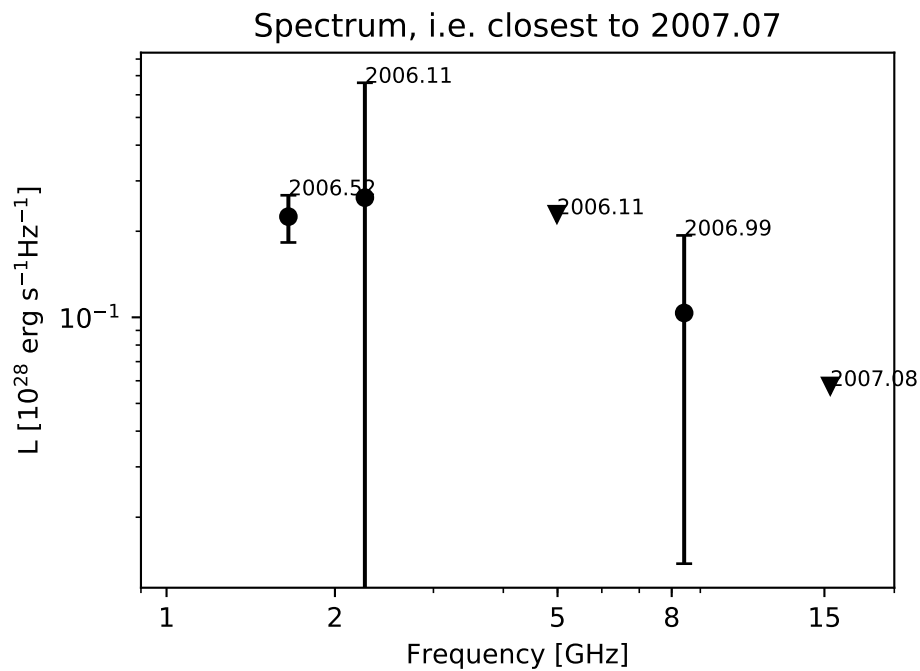
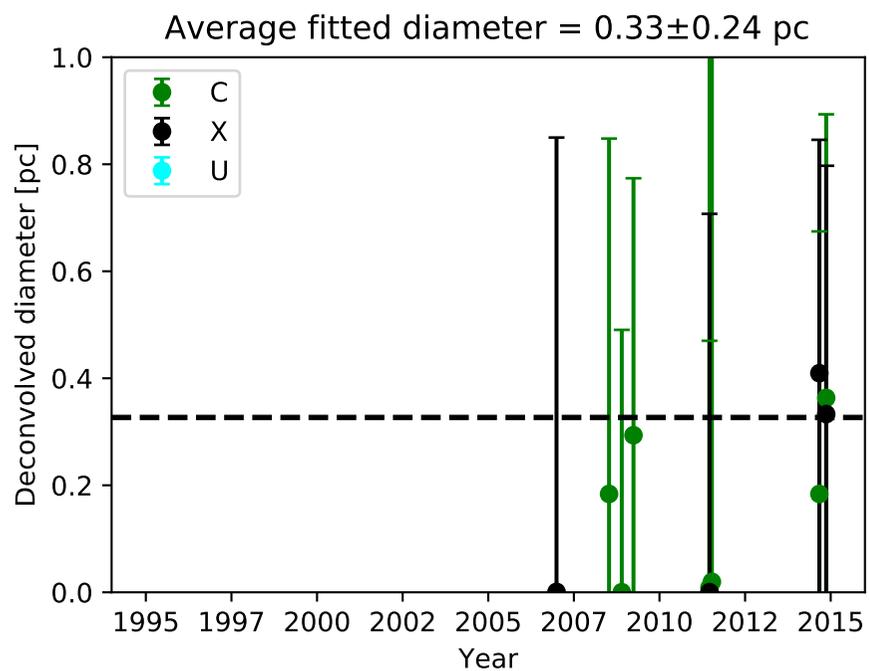

Classification: Rise

Figure notes:

Uncertainties calculated as described in Sect. 2.3.

Triangles represent $3\sigma$ upper limits for non-detections.

Sizes shown only for detections in bands C, X, and U.

Average diameters calculated according to Sect. 2.4 from the

3.6 cm and 6 cm measurements in experiments BB335A and BB335B.

# 0.2122+0.482 (W42)

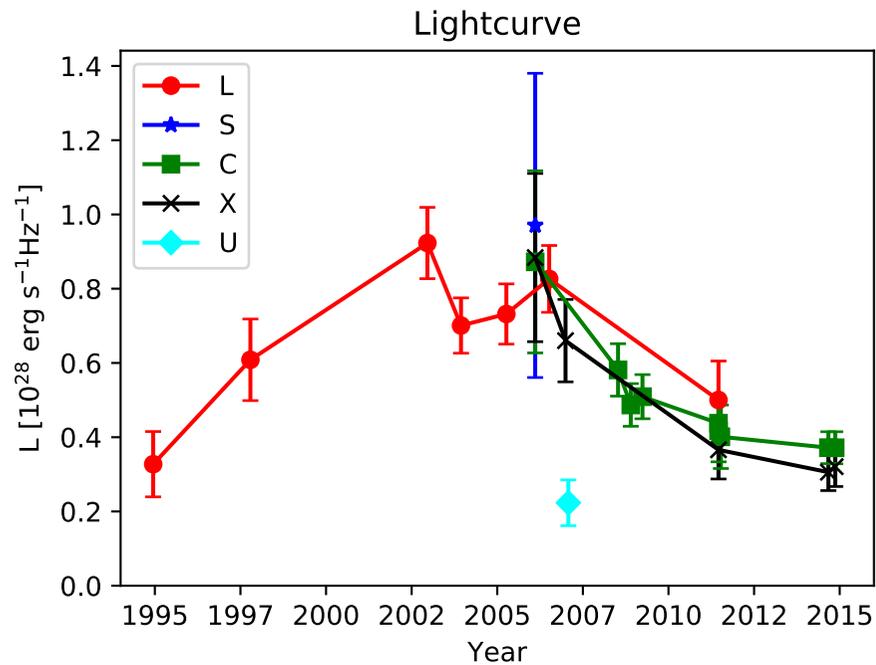
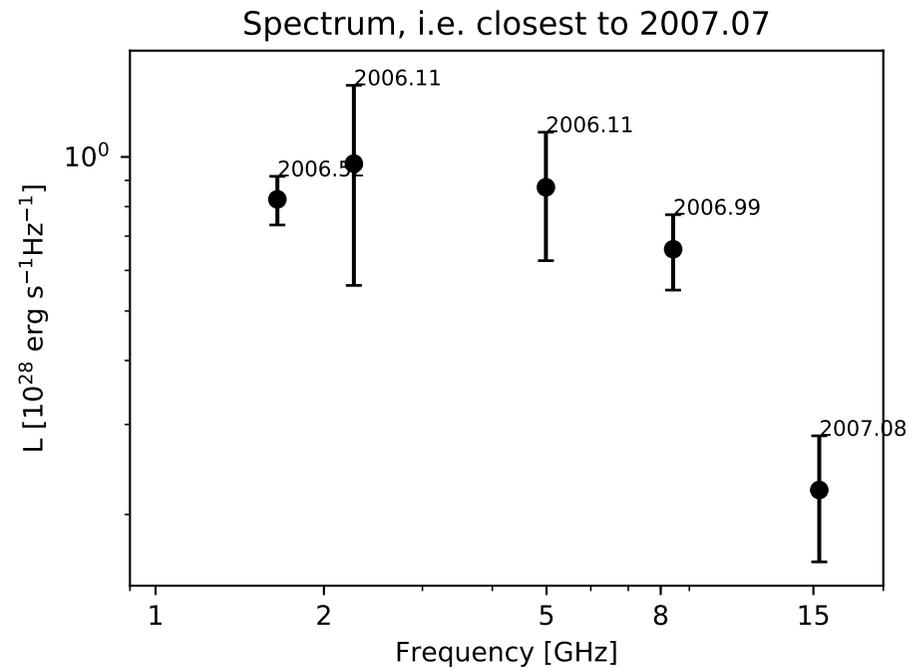
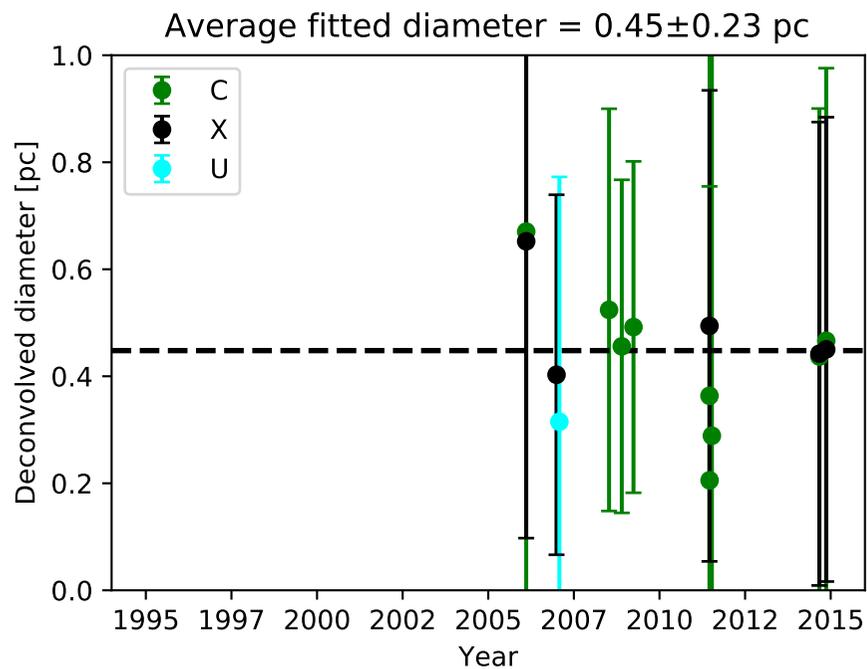

Classification: Fall

Figure notes:

Uncertainties calculated as described in Sect. 2.3.

Triangles represent $3\sigma$ upper limits for non-detections.

Sizes shown only for detections in bands C,X, and U.

Average diameters calculated according to Sect. 2.4 from the

3.6 cm and 6 cm measurements in experiments BB335A and BB335B.

# 0.2149+0.542

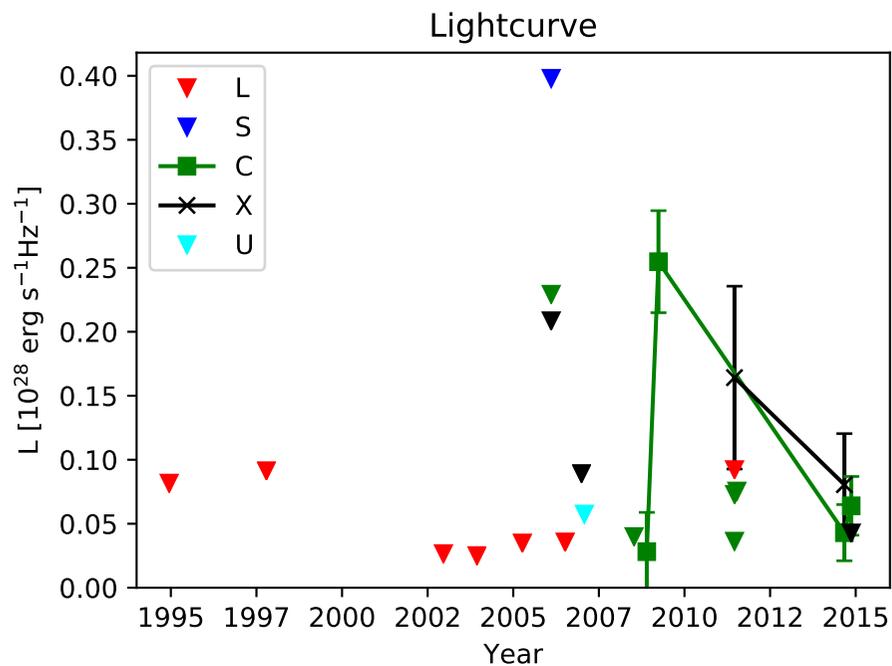
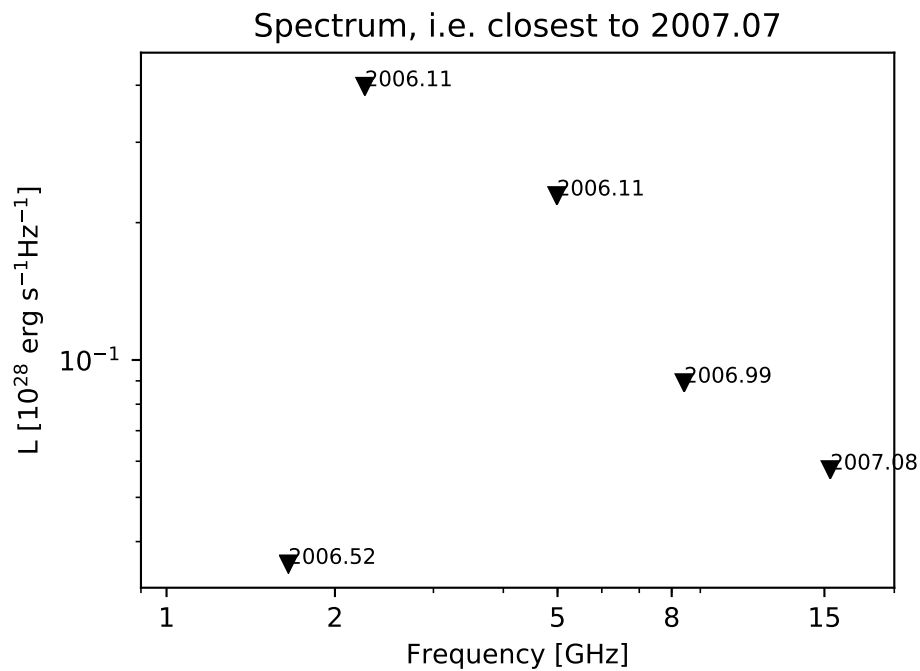
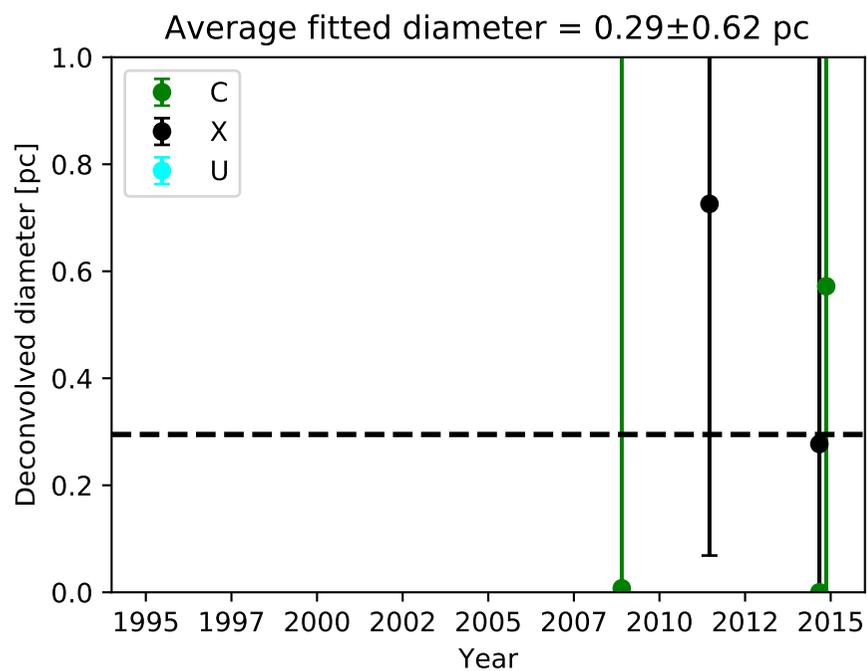

Classification: Slowly varying (S-var)

Figure notes:

Uncertainties calculated as described in Sect. 2.3.

Triangles represent 3σ upper limits for non-detections.

Sizes shown only for detections in bands C, X, and U.

Average diameters calculated according to Sect. 2.4 from the

3.6 cm and 6 cm measurements in experiments BB335A and BB335B.

# 0.2154+0.450

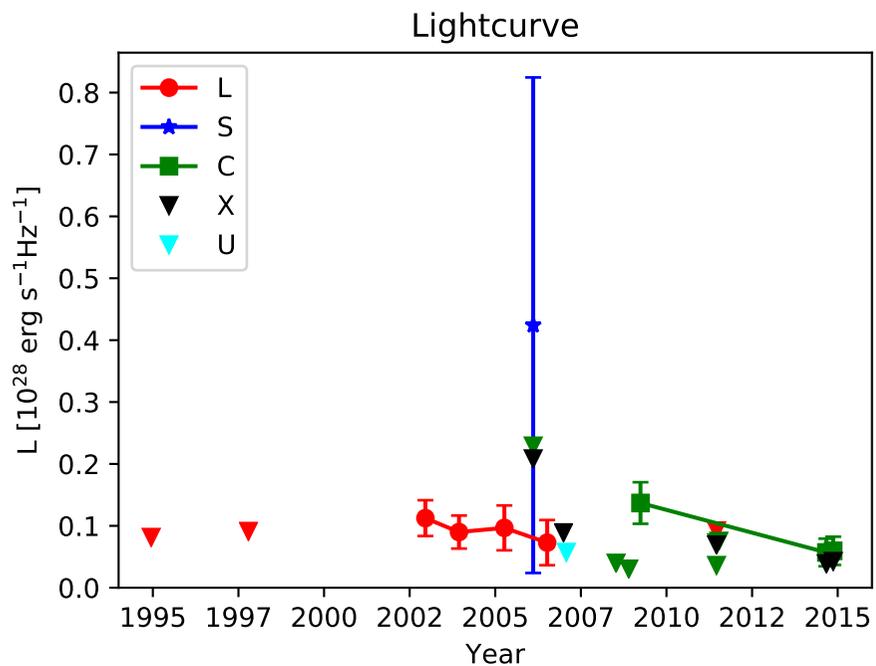
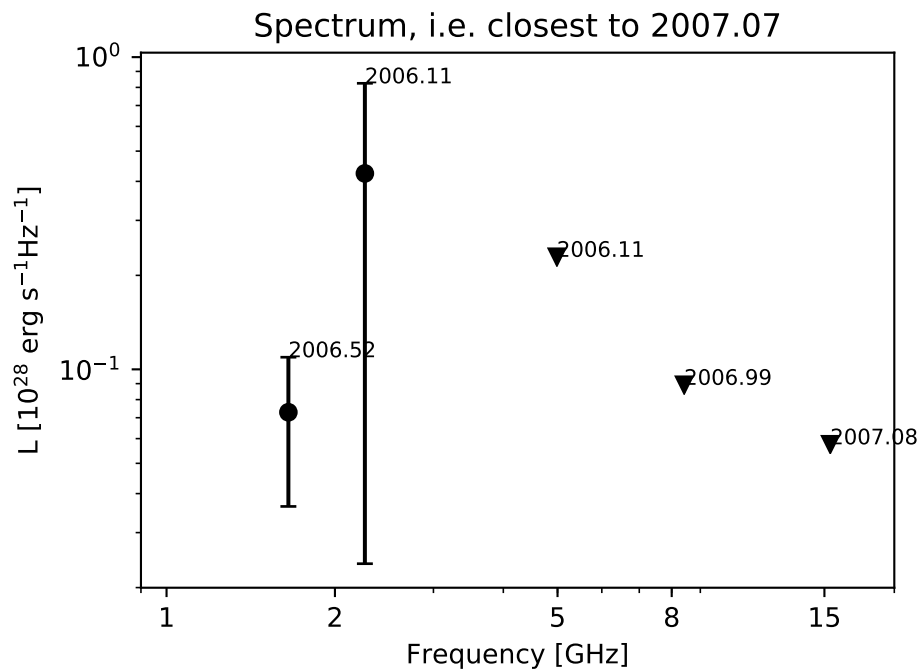
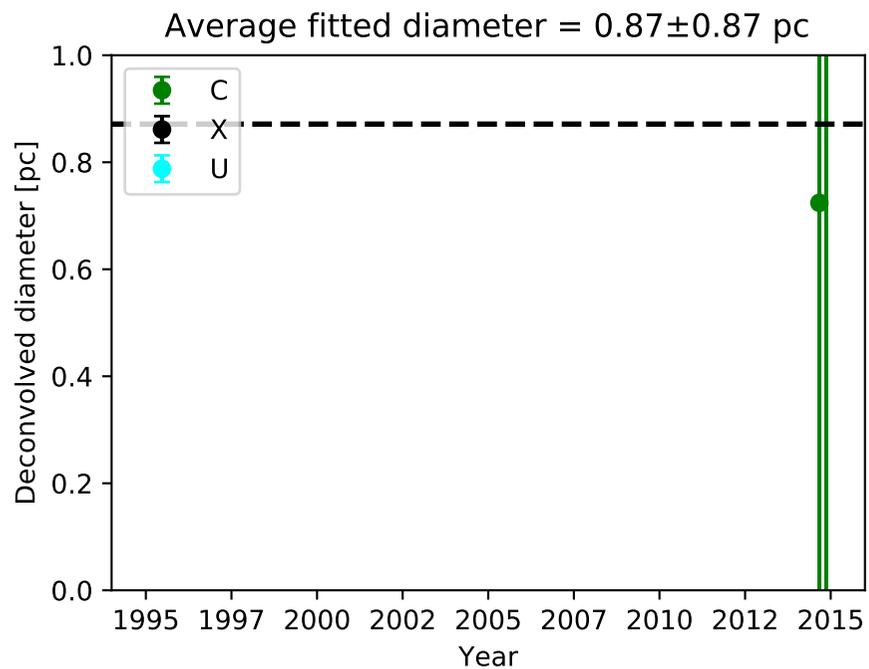

Classification: Slowly varying (S-var)

Figure notes:

Uncertainties calculated as described in Sect. 2.3.

Triangles represent 3σ upper limits for non-detections.

Sizes shown only for detections in bands C, X, and U.

Average diameters calculated according to Sect. 2.4 from the

3.6 cm and 6 cm measurements in experiments BB335A and BB335B.

# 0.2161+0.479 (W40)

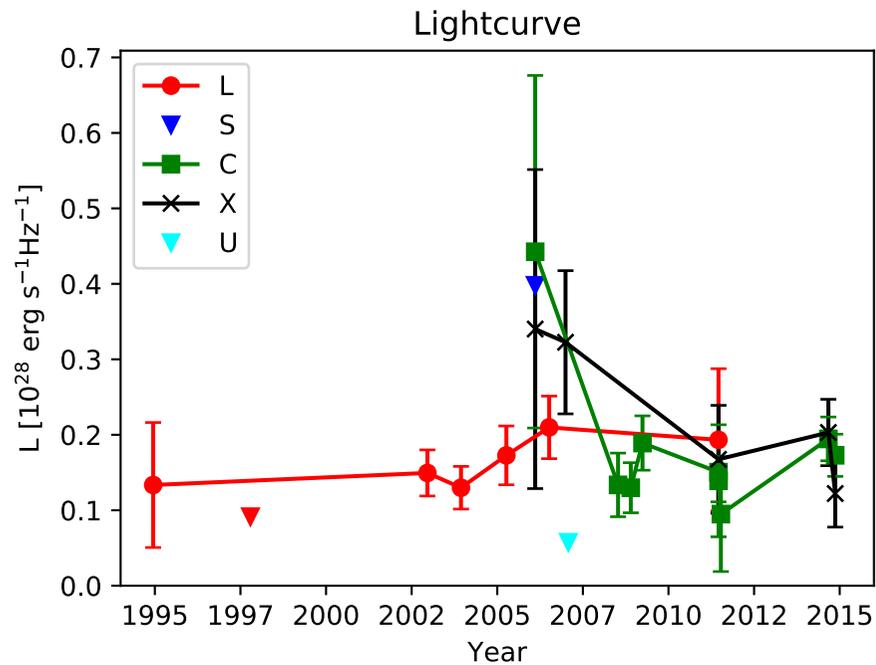
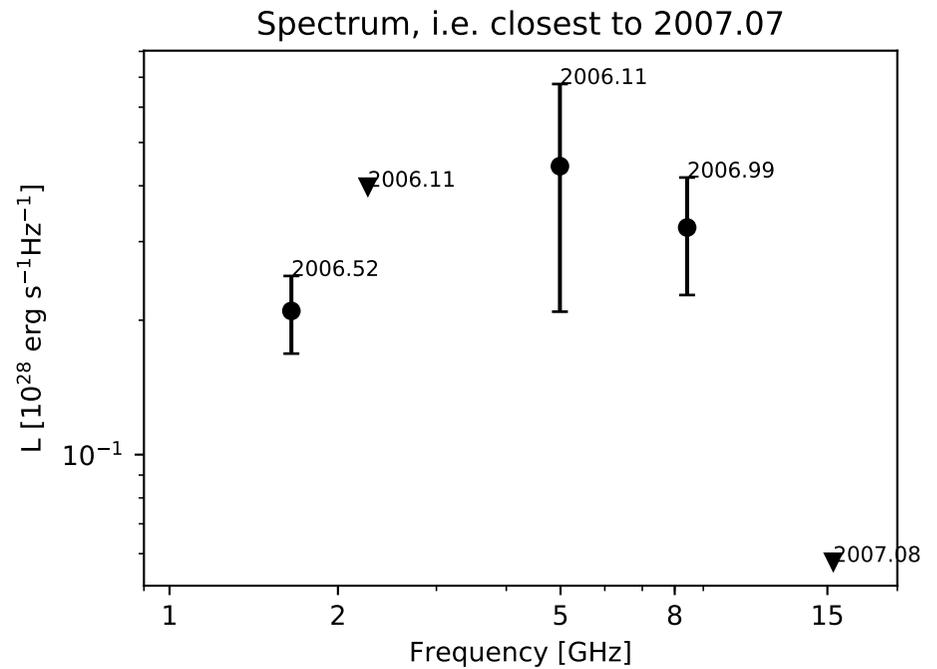
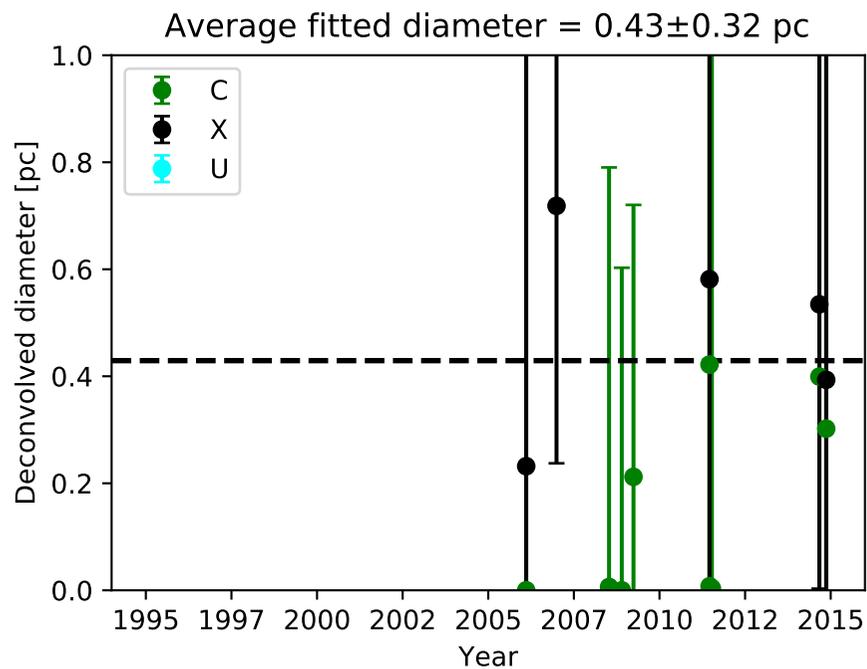

Classification: Slowly varying (S-var)

Figure notes:

Uncertainties calculated as described in Sect. 2.3.

Triangles represent $3\sigma$ upper limits for non-detections.

Sizes shown only for detections in bands C, X, and U.

Average diameters calculated according to Sect. 2.4 from the 3.6 cm and 6 cm measurements in experiments BB335A and BB335B.

# 0.2162+0.404 (W41)

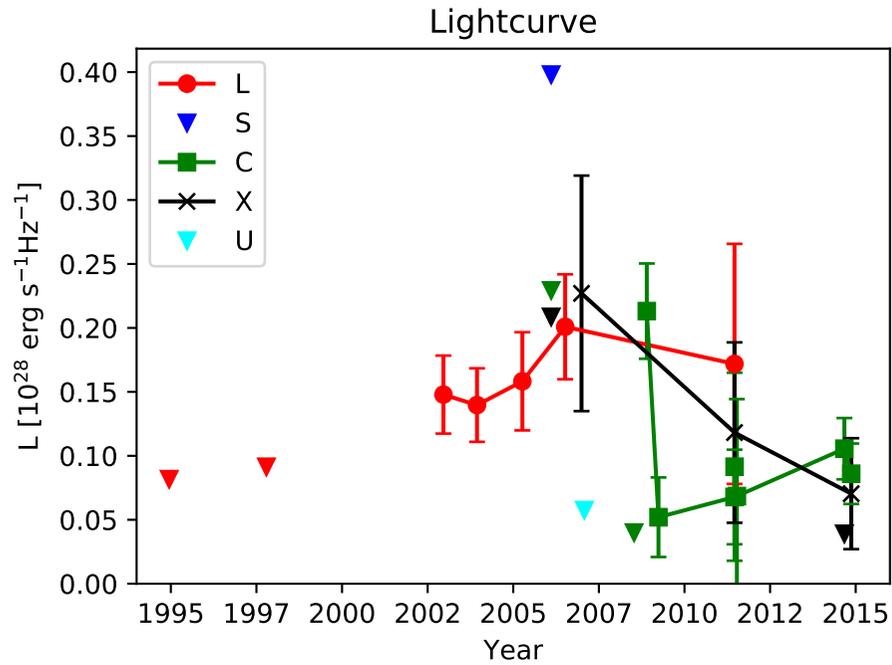
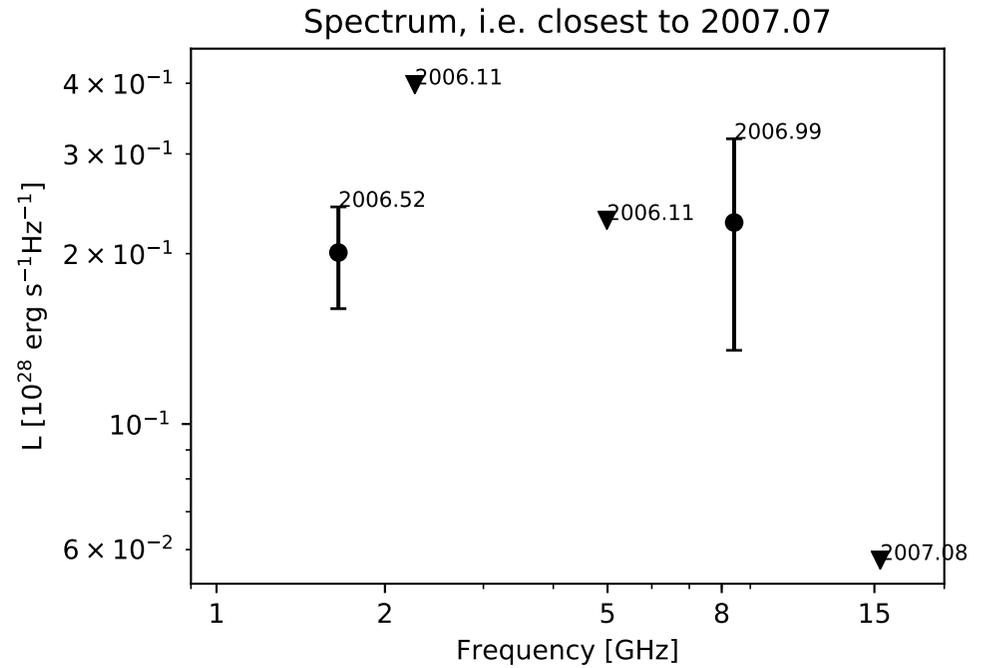
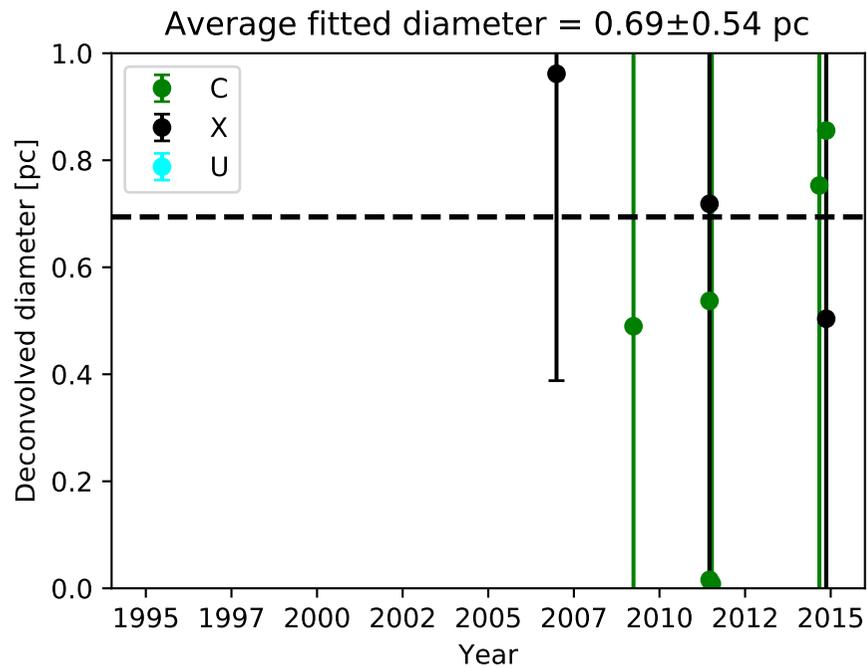

Classification: Slowly varying (S-var)

Figure notes:

Uncertainties calculated as described in Sect. 2.3.

Triangles represent $3\sigma$ upper limits for non-detections.

Sizes shown only for detections in bands C, X, and U.

Average diameters calculated according to Sect. 2.4 from the 3.6 cm and 6 cm measurements in experiments BB335A and BB335B.

# 0.2171+0.484 (W39)

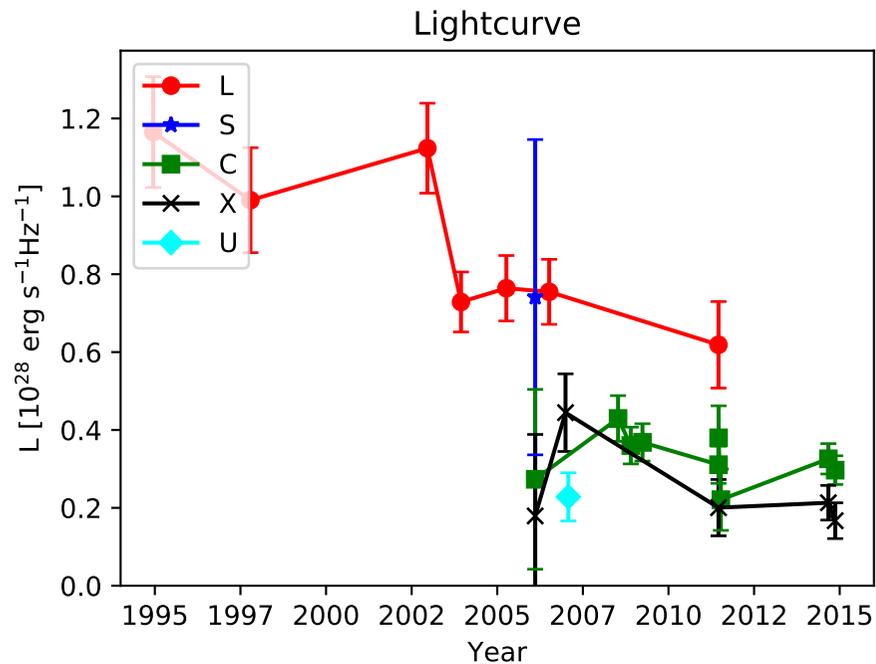

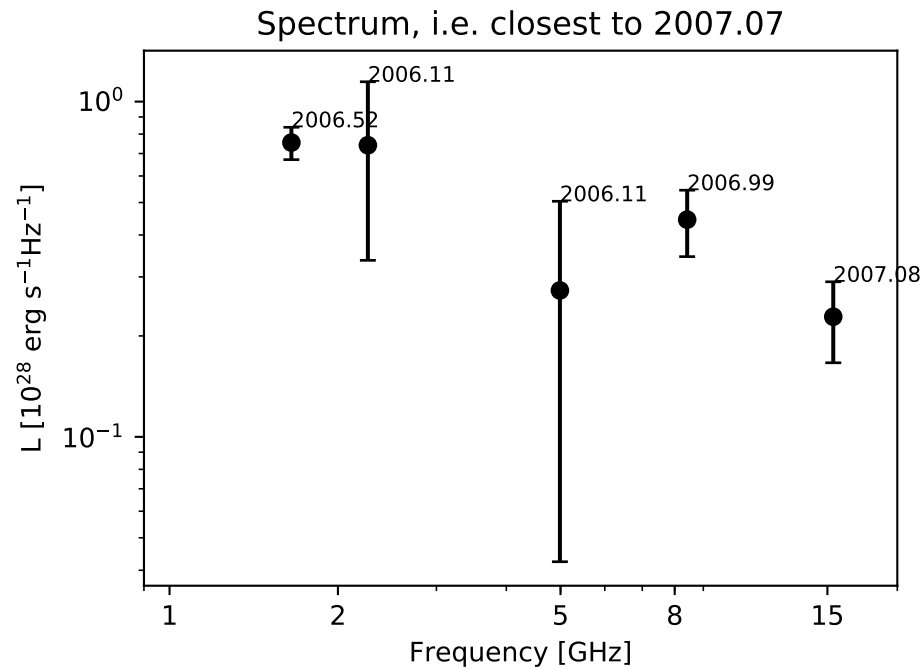

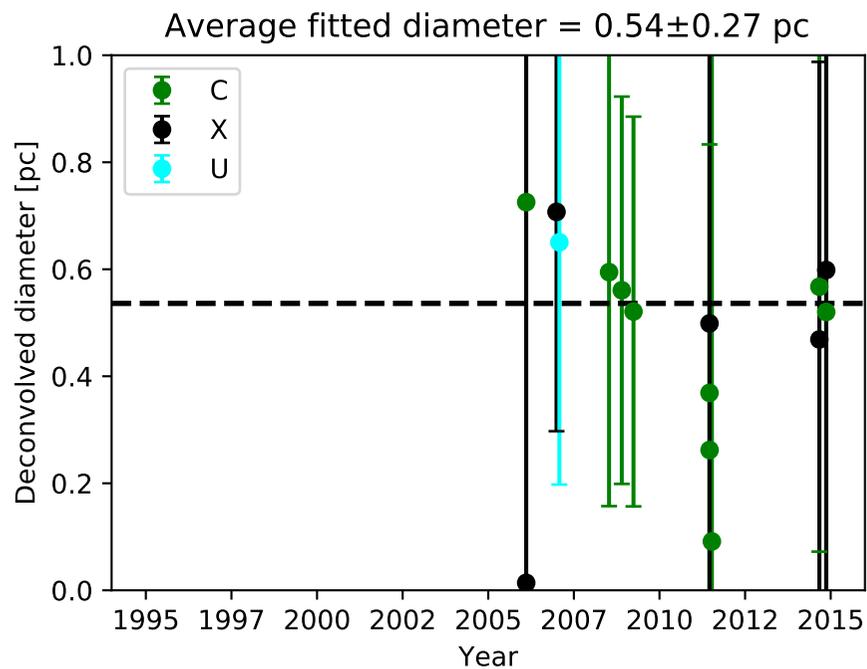

Classification: Slowly varying (S-var)

Figure notes:

Uncertainties calculated as described in Sect. 2.3.

Triangles represent $3\sigma$ upper limits for non-detections.

Sizes shown only for detections in bands C, X, and U.

Average diameters calculated according to Sect. 2.4 from the

3.6 cm and 6 cm measurements in experiments BB335A and BB335B.

# 0.2173+0.569

## Lightcurve

## Spectrum, i.e. closest to 2007.07

## Average fitted diameter = 1.08±0.43 pc

Classification: Fall

Figure notes:

Uncertainties calculated as described in Sect. 2.3.

Triangles represent 3σ upper limits for non-detections.

Sizes shown only for detections in bands C, X, and U.

Average diameters calculated according to Sect. 2.4 from the

3.6 cm and 6 cm measurements in experiments BB335A and BB335B.

# 0.2194+0.507 (W58)

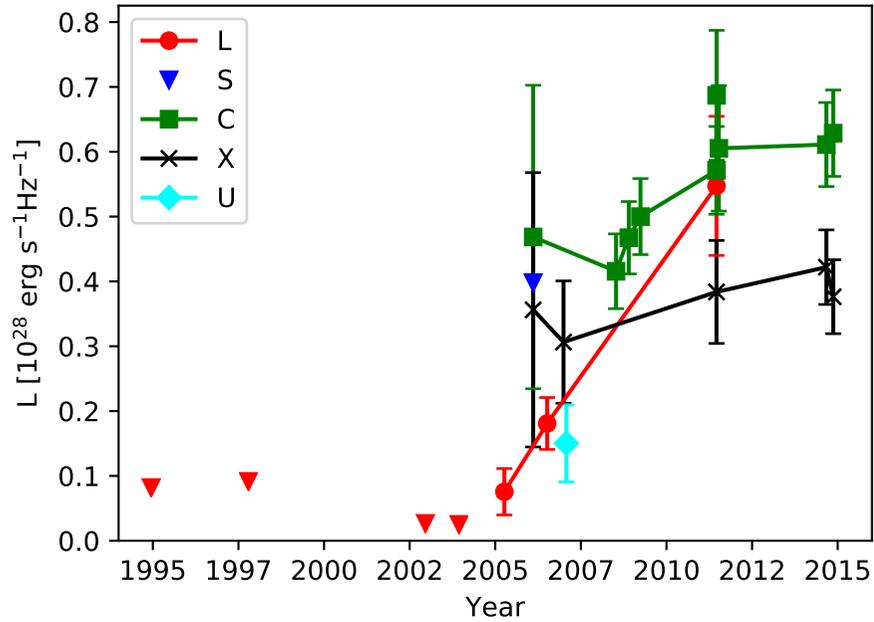

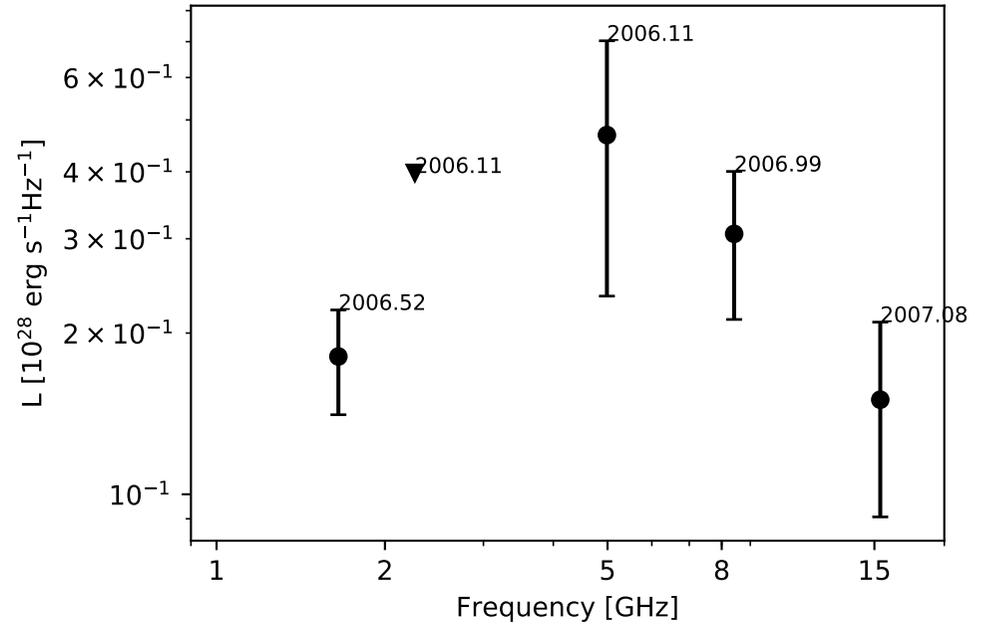

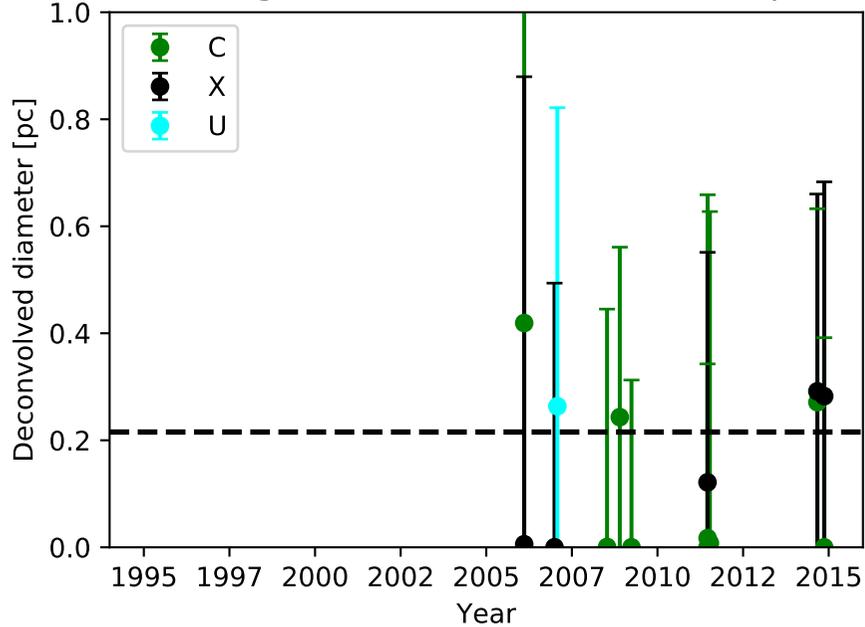

Classification: Rise

Figure notes:

Uncertainties calculated as described in Sect. 2.3.

Triangles represent 3σ upper limits for non-detections.

Sizes shown only for detections in bands C, X, and U.

Average diameters calculated according to Sect. 2.4 from the

3.6 cm and 6 cm measurements in experiments BB335A and BB335B.

# 0.2195+0.492 (W34)

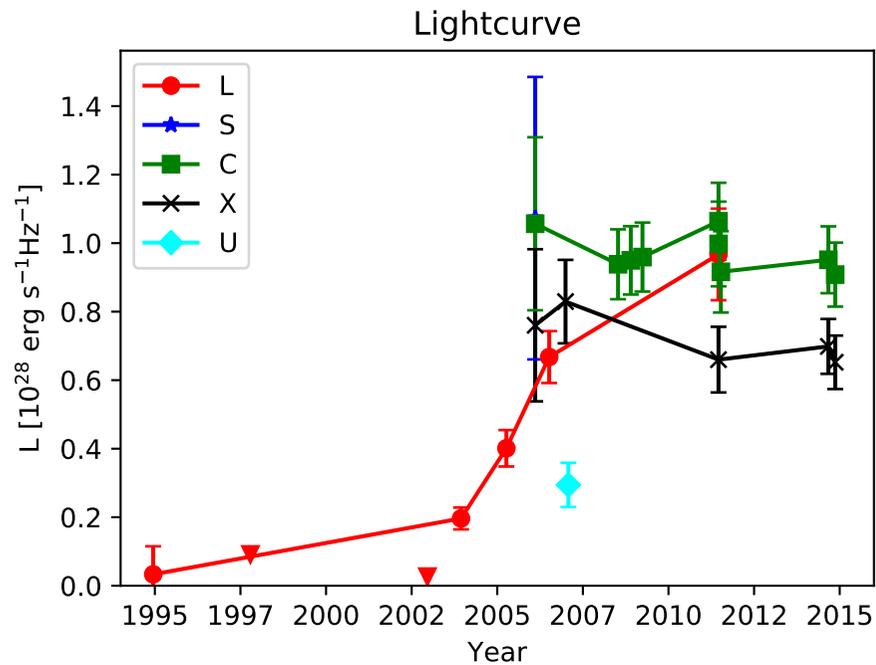

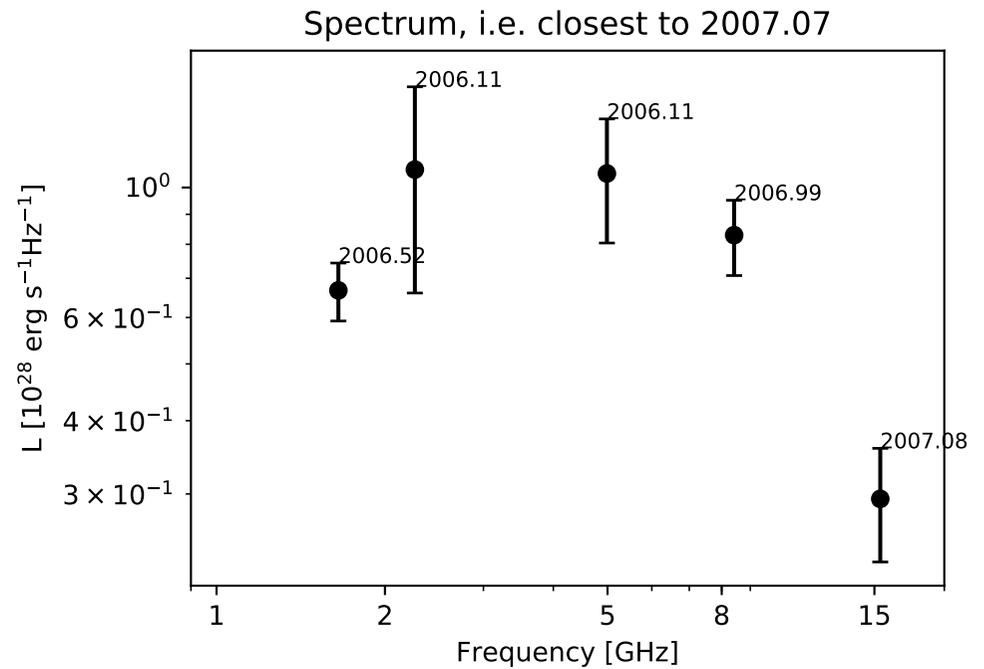

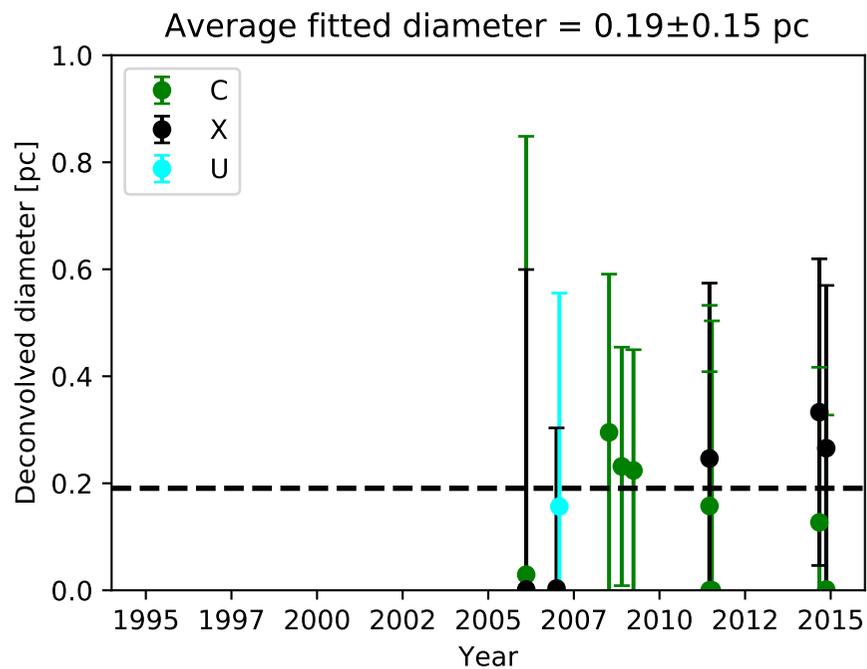

Classification: Slowly varying (S-var)

Figure notes:

Uncertainties calculated as described in Sect. 2.3.

Triangles represent $3\sigma$ upper limits for non-detections.

Sizes shown only for detections in bands C, X, and U.

Average diameters calculated according to Sect. 2.4 from the 3.6 cm and 6 cm measurements in experiments BB335A and BB335B.

# 0.2200+0.492 (W33)

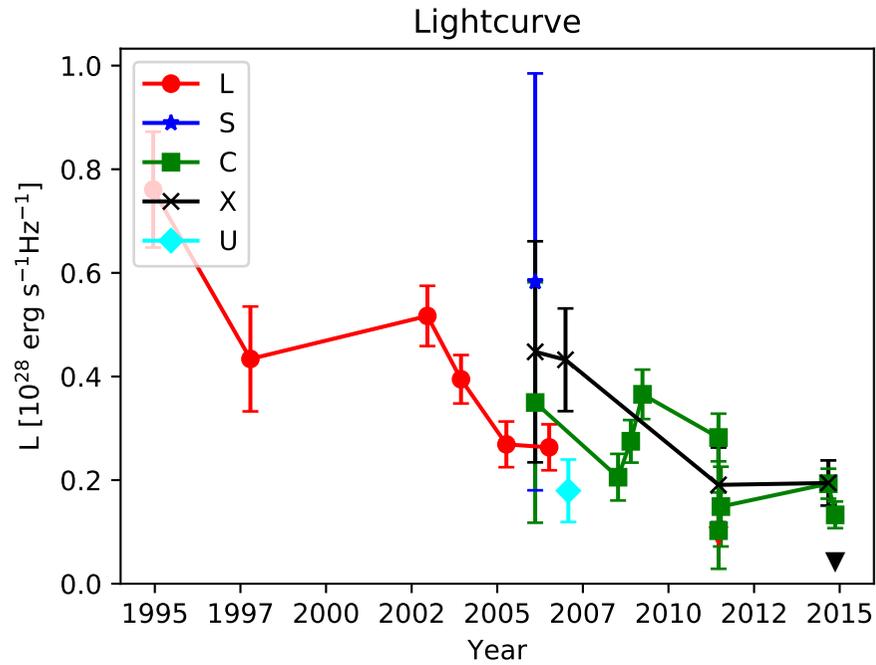
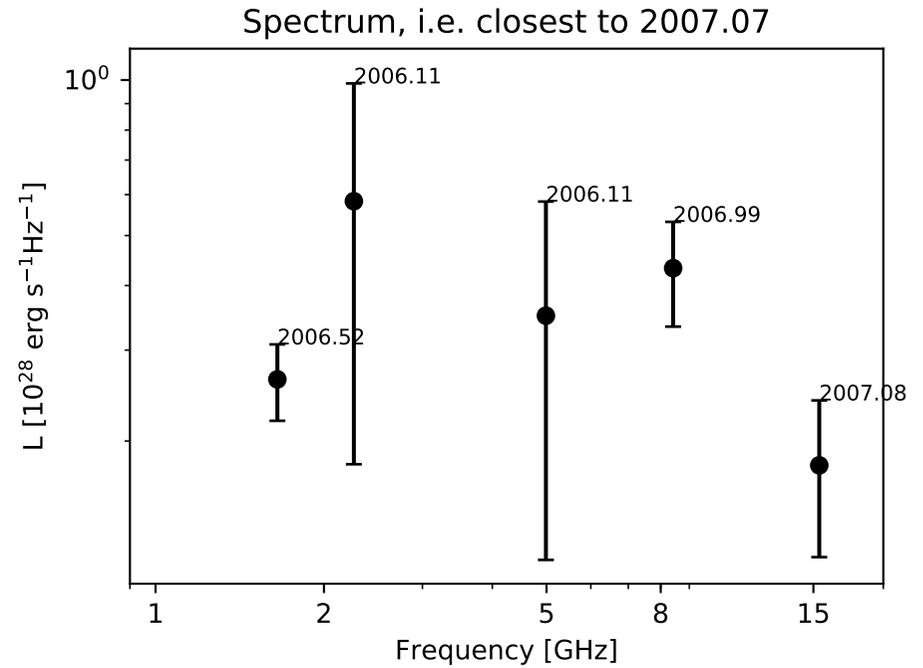
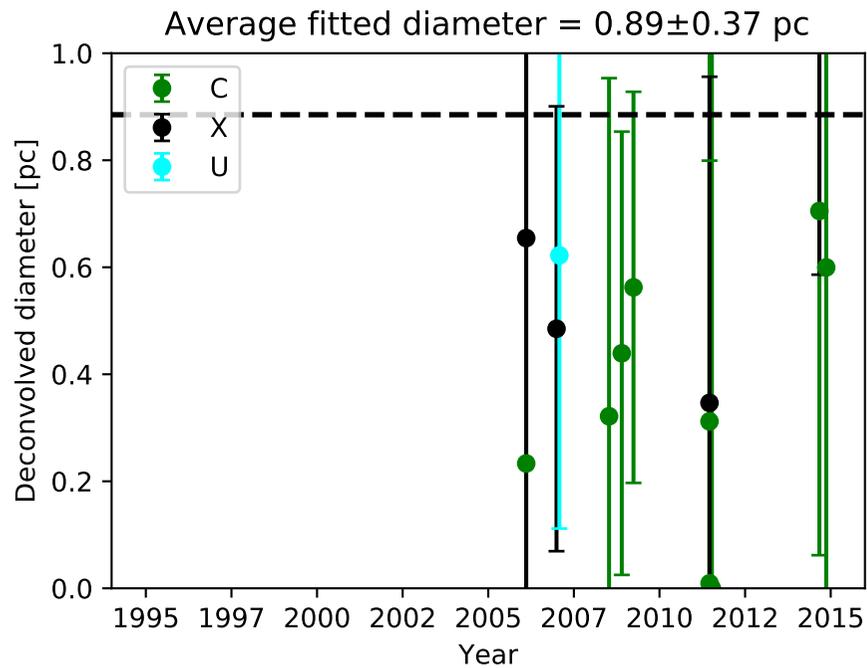

Classification: Fall

Figure notes:

Uncertainties calculated as described in Sect. 2.3.

Triangles represent 3σ upper limits for non-detections.

Sizes shown only for detections in bands C, X, and U.

Average diameters calculated according to Sect. 2.4 from the

3.6 cm and 6 cm measurements in experiments BB335A and BB335B.

# 0.2205+0.491 (W56)

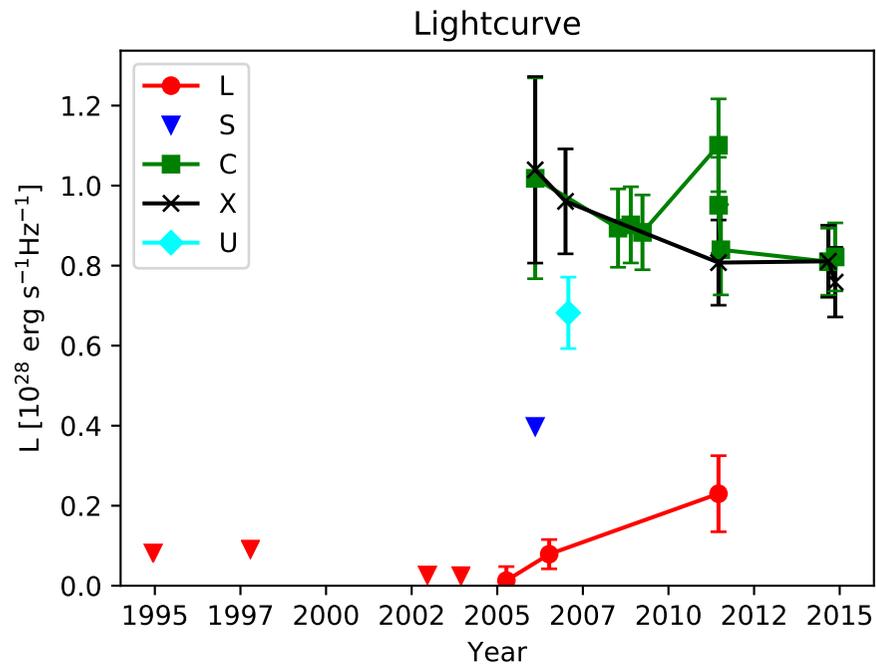

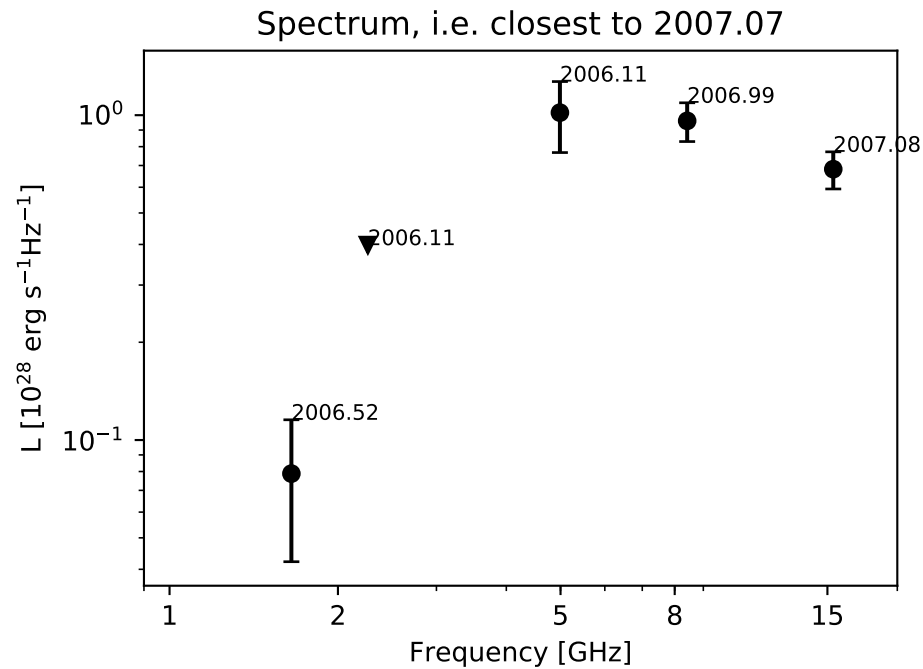

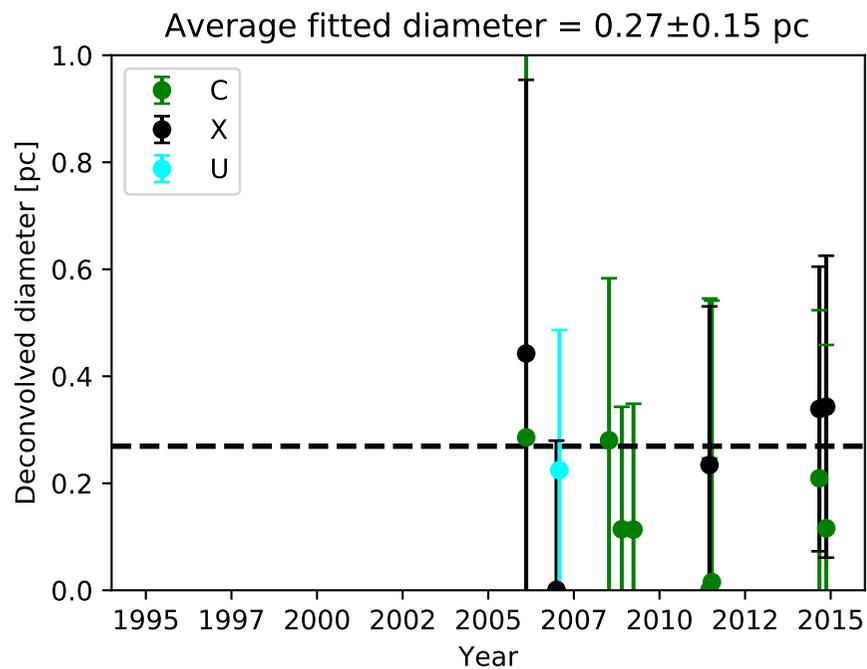

Classification: Slowly varying (S-var)

Figure notes:

Uncertainties calculated as described in Sect. 2.3.

Triangles represent $3\sigma$ upper limits for non-detections.

Sizes shown only for detections in bands C, X, and U.

Average diameters calculated according to Sect. 2.4 from the

3.6 cm and 6 cm measurements in experiments BB335A and BB335B.

# 0.2211+0.398 (W30)

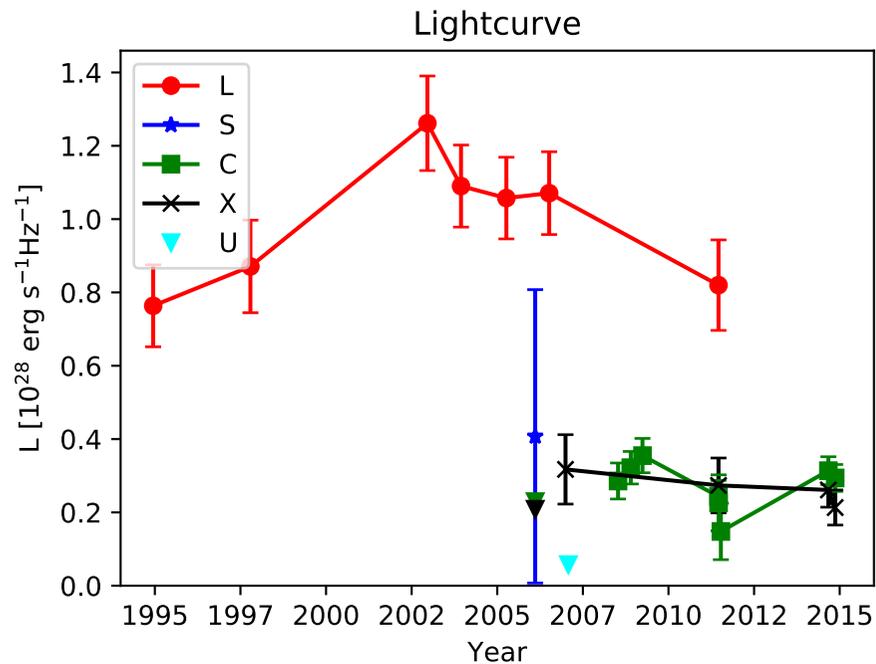
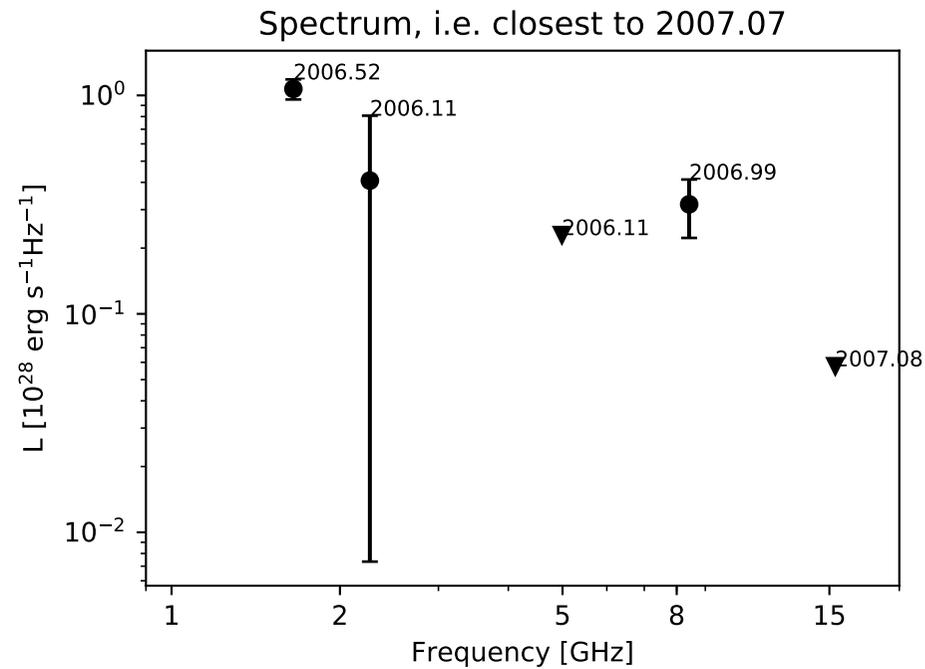
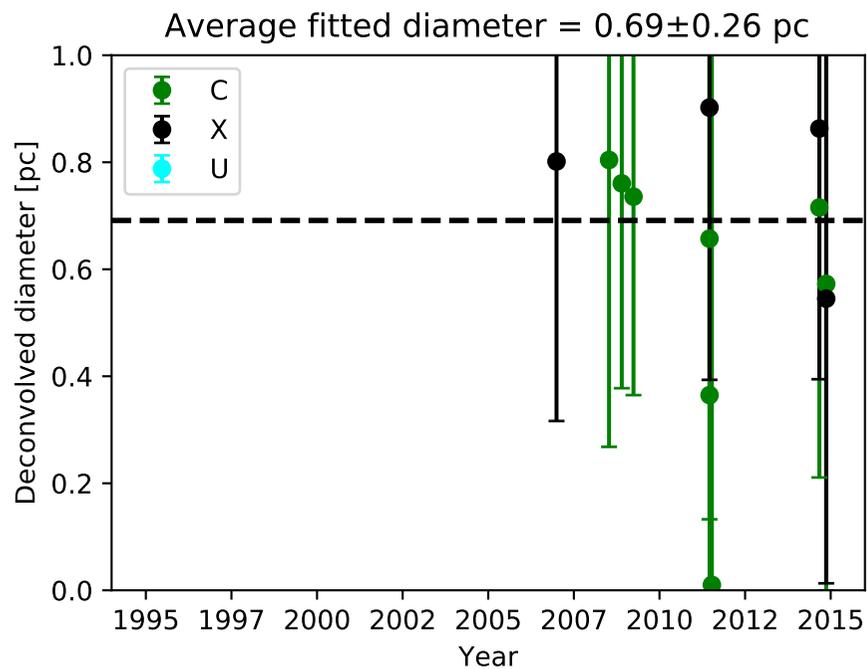

Classification: Slowly varying (S-var)

Figure notes:

Uncertainties calculated as described in Sect. 2.3.

Triangles represent 3σ upper limits for non-detections.

Sizes shown only for detections in bands C,X, and U.

Average diameters calculated according to Sect. 2.4 from the

3.6 cm and 6 cm measurements in experiments BB335A and BB335B.

# 0.2212+0.444 (W29)

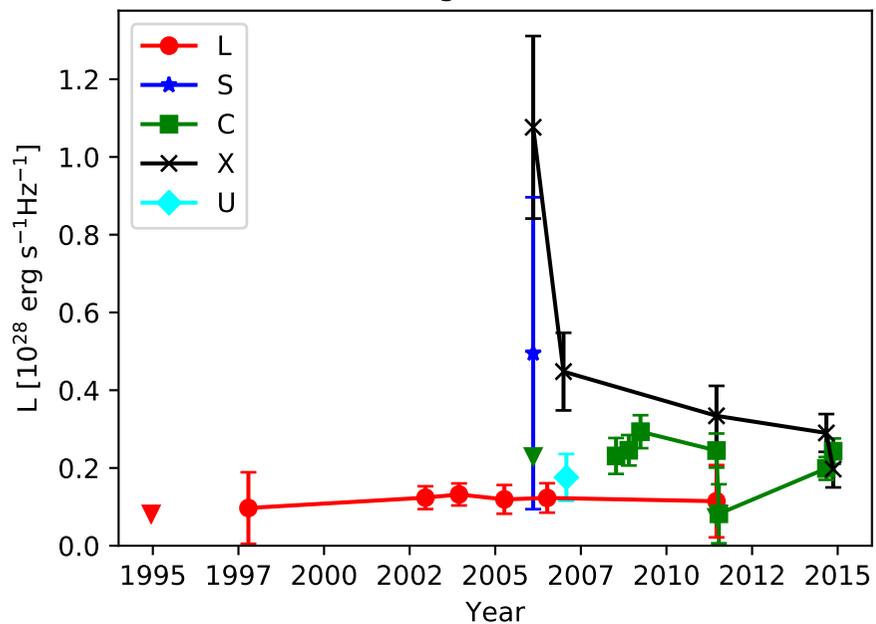
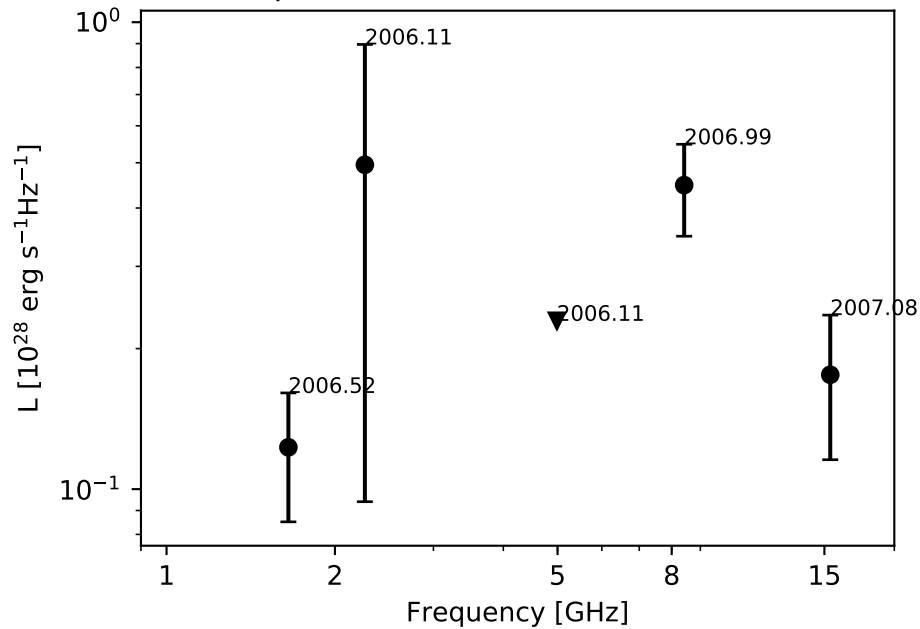
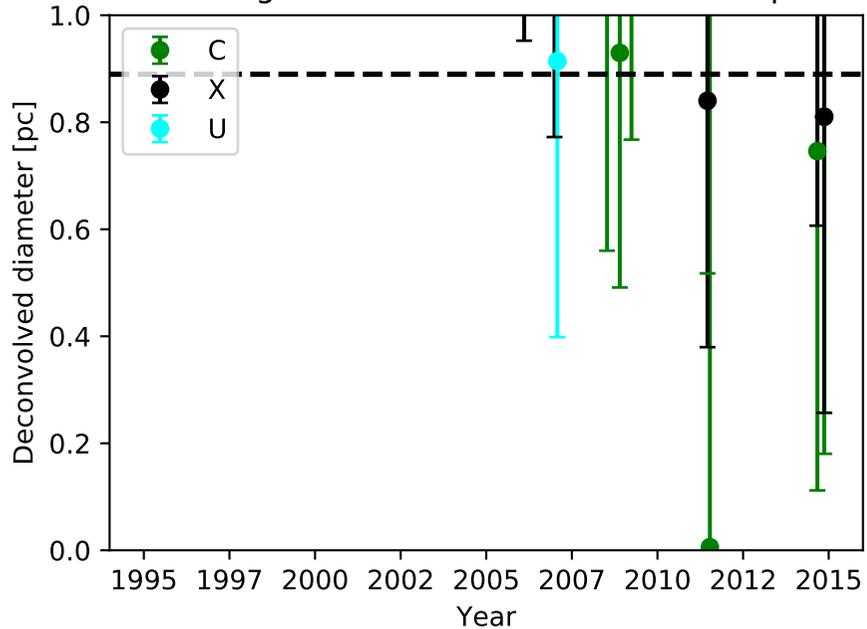

Classification: Slowly varying (S-var)

Figure notes:

Uncertainties calculated as described in Sect. 2.3.

Triangles represent 3σ upper limits for non-detections.

Sizes shown only for detections in bands C, X, and U.

Average diameters calculated according to Sect. 2.4 from the 3.6 cm and 6 cm measurements in experiments BB335A and BB335B.

# 0.2212+0.540 (W28)

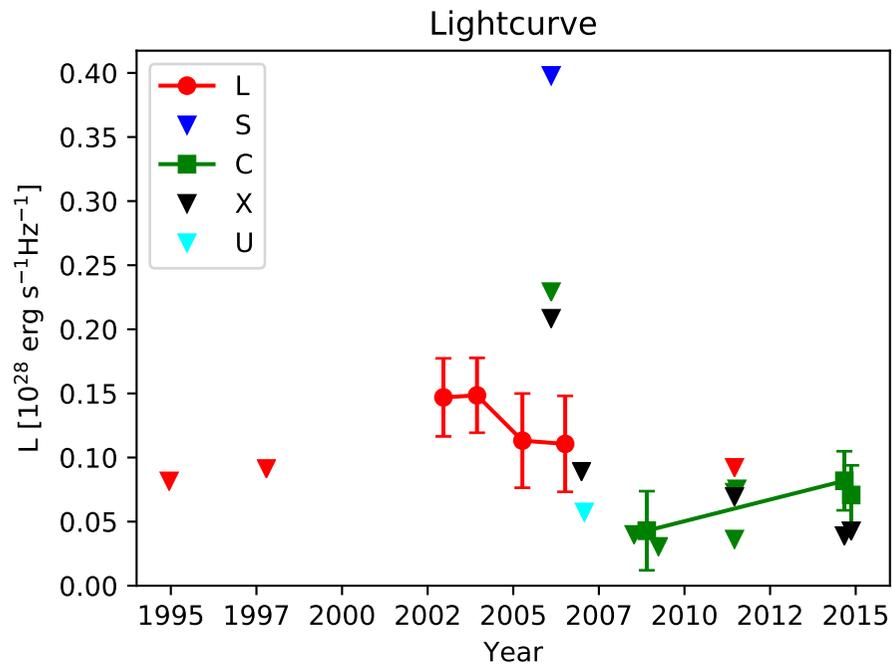

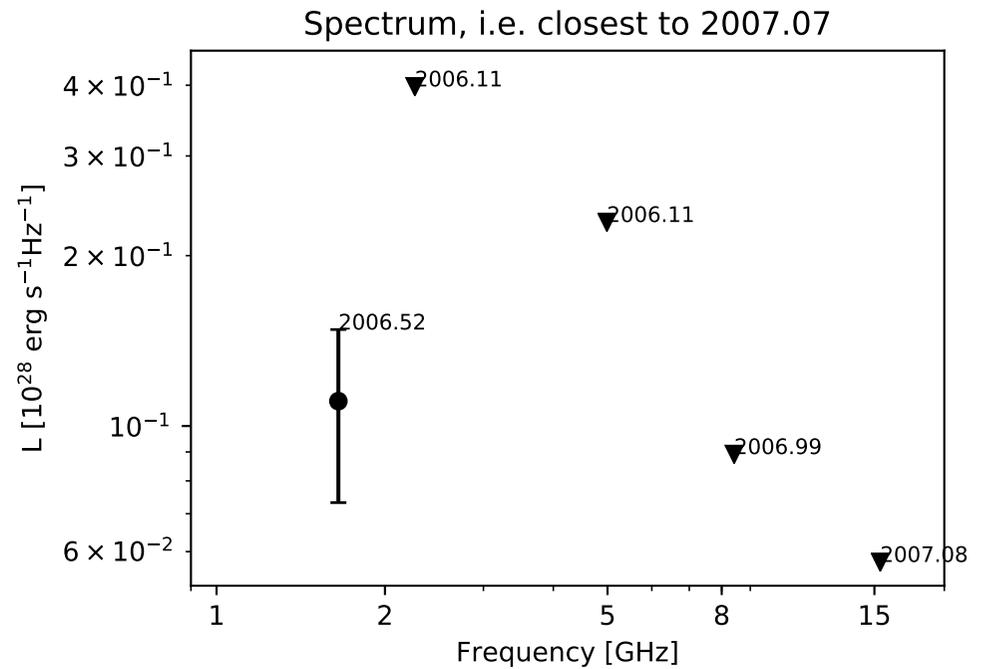

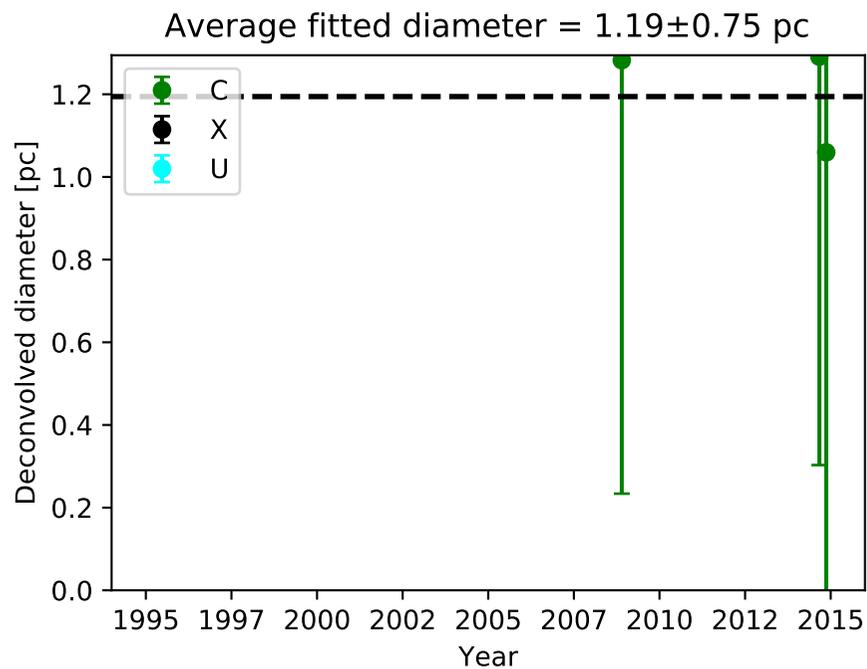

Classification: Rise

Figure notes:

Uncertainties calculated as described in Sect. 2.3.

Triangles represent 3σ upper limits for non-detections.

Sizes shown only for detections in bands C,X, and U.

Average diameters calculated according to Sect. 2.4 from the

3.6 cm and 6 cm measurements in experiments BB335A and BB335B.

# 0.2218+0.528

## Lightcurve

## Spectrum, i.e. closest to 2007.07

## Average fitted diameter = 0.2±0.32 pc

Classification: Rise

Figure notes:

Uncertainties calculated as described in Sect. 2.3.

Triangles represent 3σ upper limits for non-detections.

Sizes shown only for detections in bands C,X, and U.

Average diameters calculated according to Sect. 2.4 from the

3.6 cm and 6 cm measurements in experiments BB335A and BB335B.

# 0.2222+0.500 (W25)

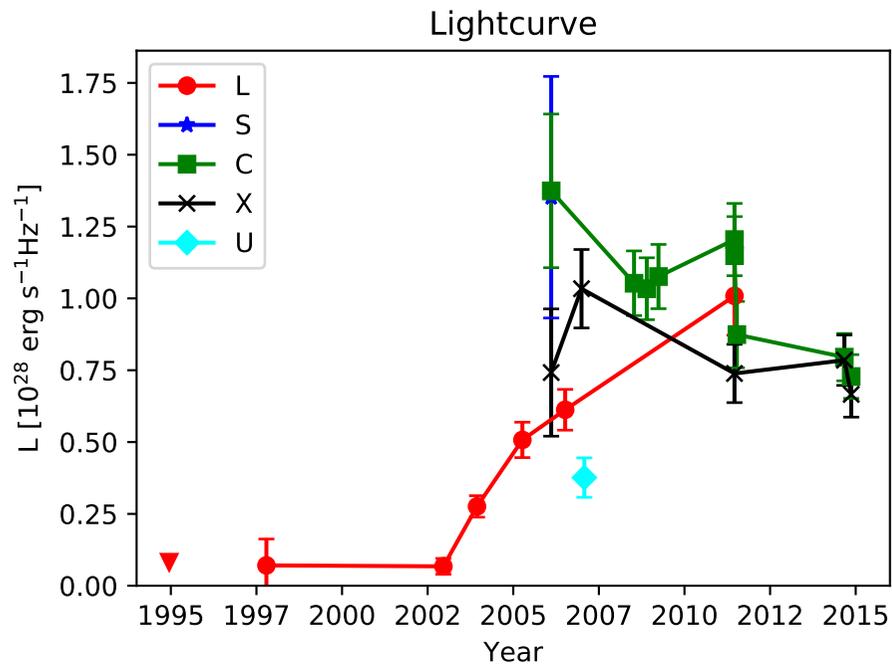
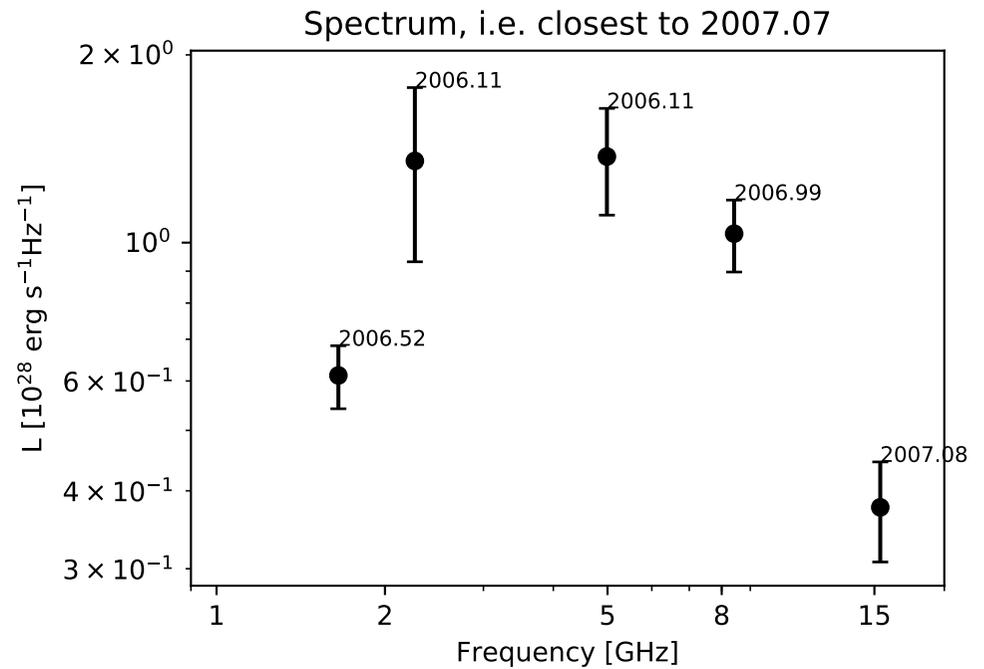
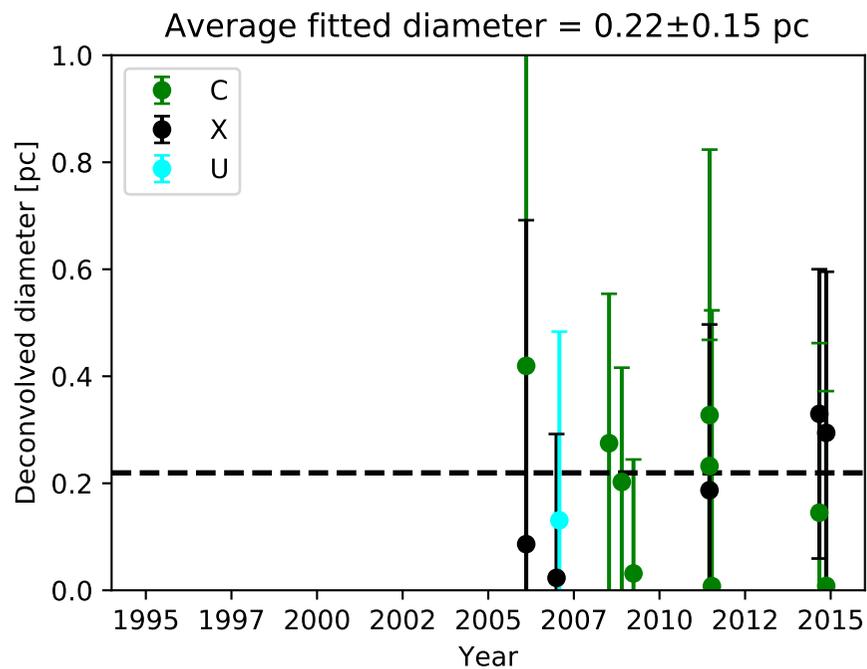

Classification: Fall

Figure notes:

Uncertainties calculated as described in Sect. 2.3.

Triangles represent $3\sigma$ upper limits for non-detections.

Sizes shown only for detections in bands C, X, and U.

Average diameters calculated according to Sect. 2.4 from the 3.6 cm and 6 cm measurements in experiments BB335A and BB335B.

# 0.2223+0.435 (W26)

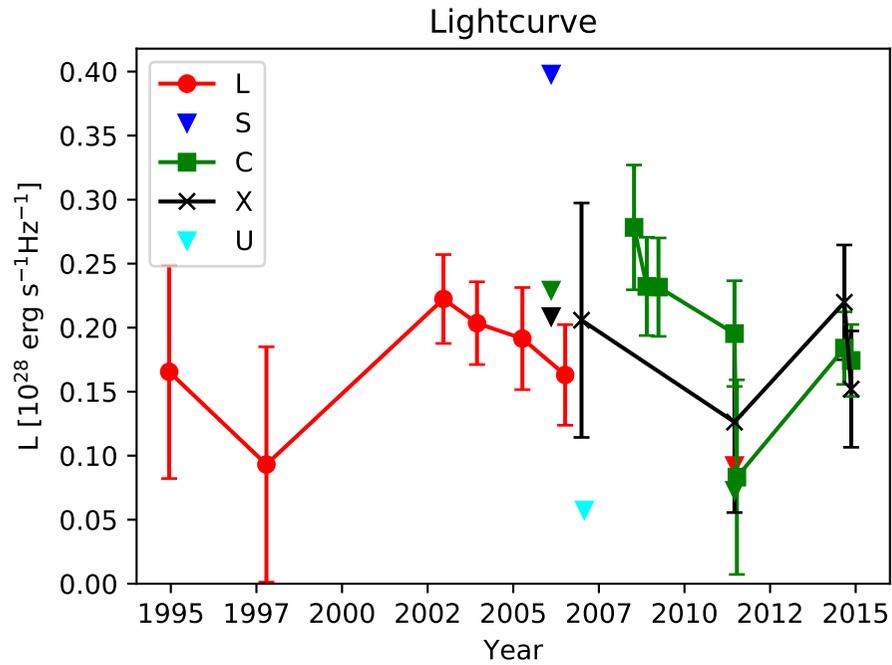
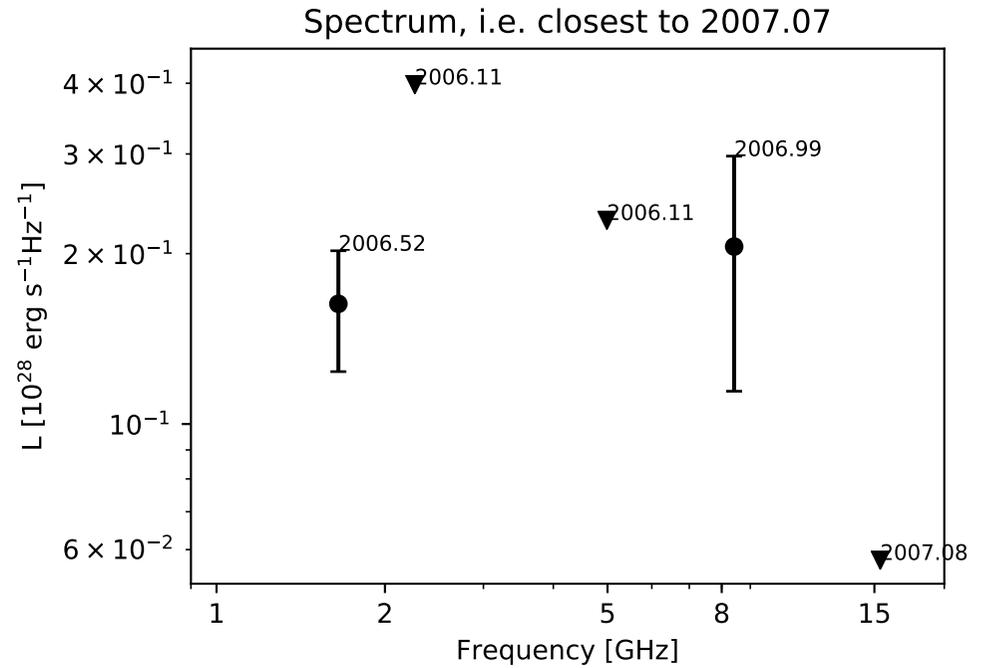
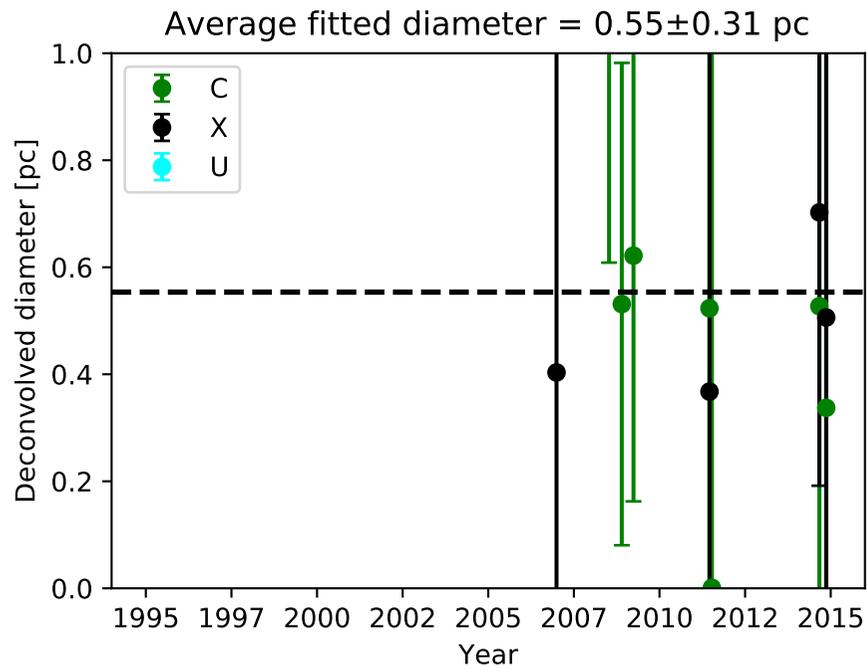

Classification: Fall

Figure notes:

Uncertainties calculated as described in Sect. 2.3.

Triangles represent 3σ upper limits for non-detections.

Sizes shown only for detections in bands C,X, and U.

Average diameters calculated according to Sect. 2.4 from the

3.6 cm and 6 cm measurements in experiments BB335A and BB335B.

# 0.2227+0.482 (W55)

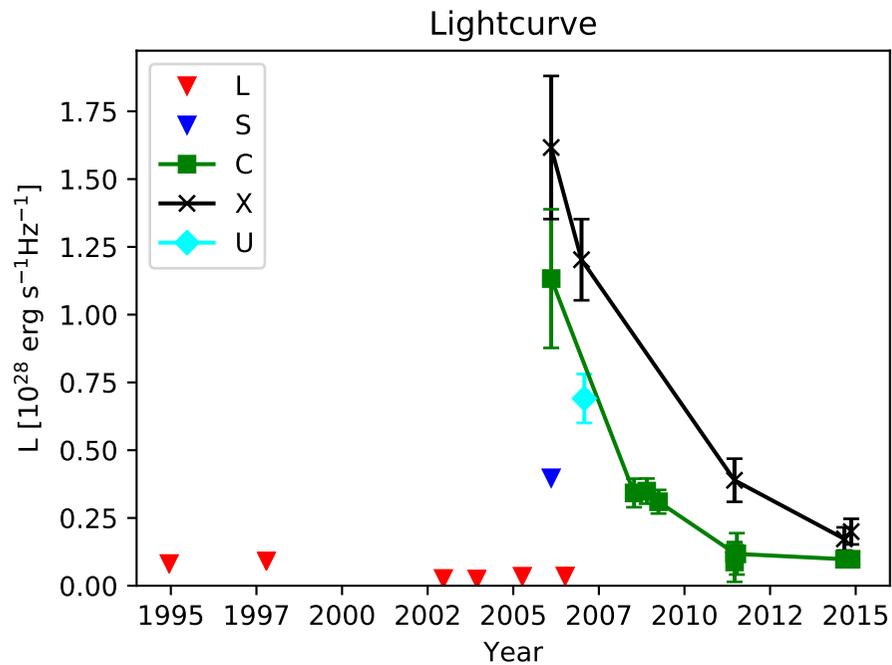
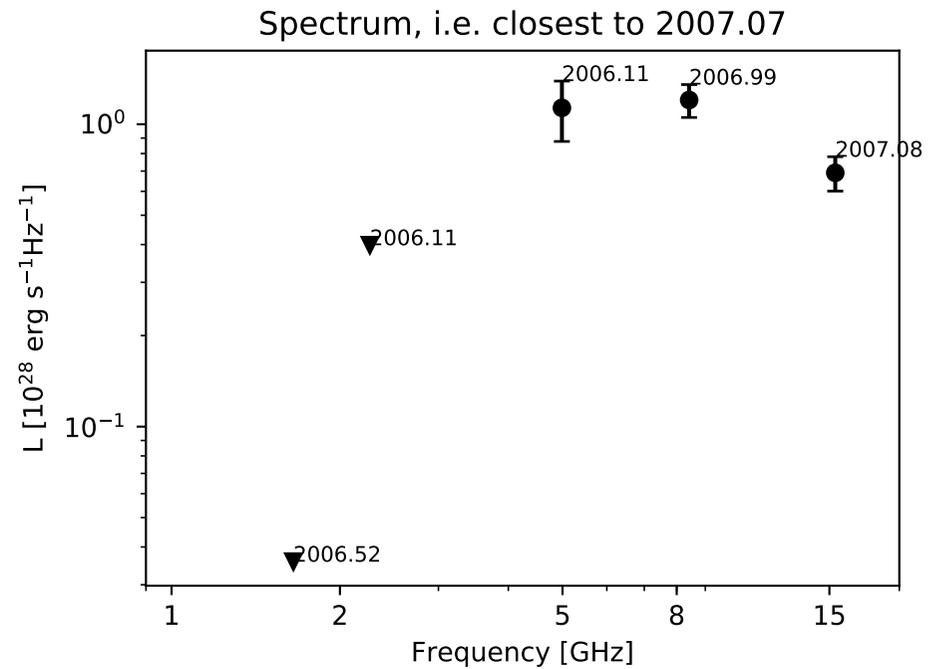
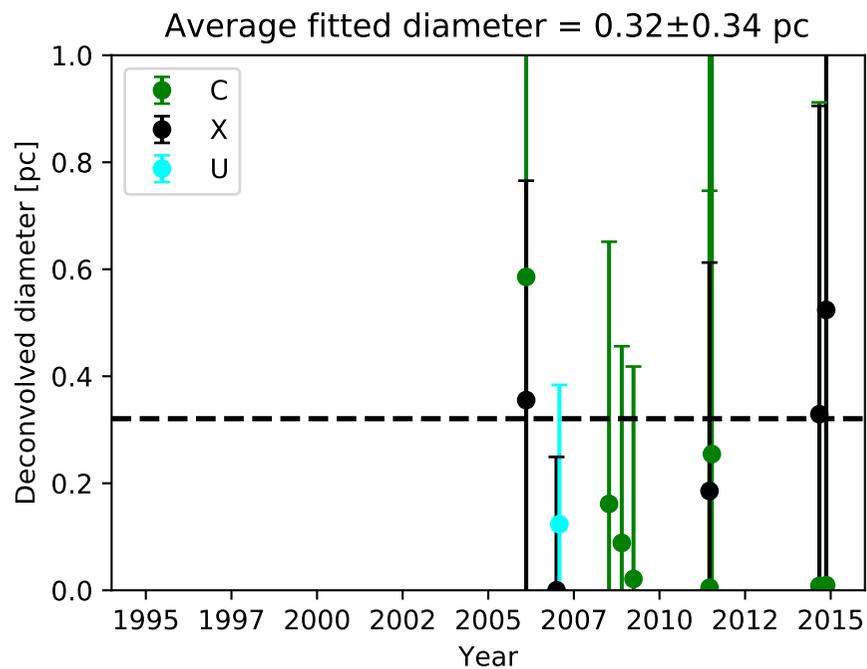

Classification: Fall

Figure notes:

Uncertainties calculated as described in Sect. 2.3.

Triangles represent 3σ upper limits for non-detections.

Sizes shown only for detections in bands C, X, and U.

Average diameters calculated according to Sect. 2.4 from the

3.6 cm and 6 cm measurements in experiments BB335A and BB335B.

# 0.2234+0.477

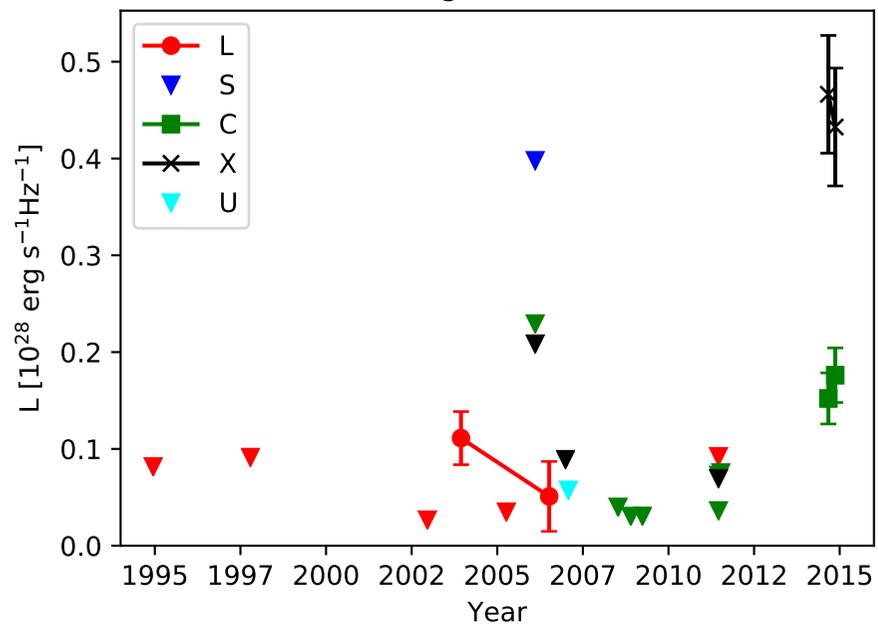
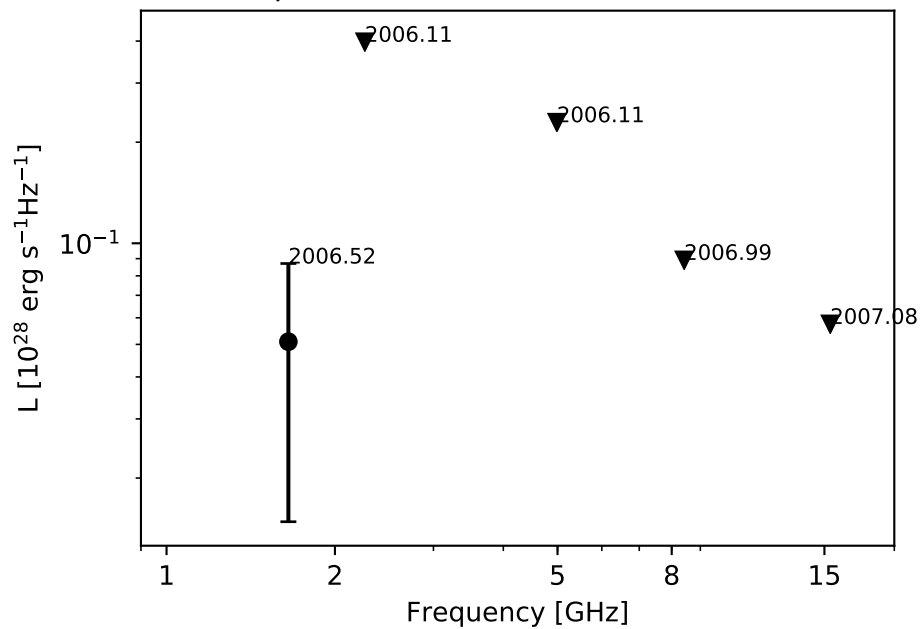
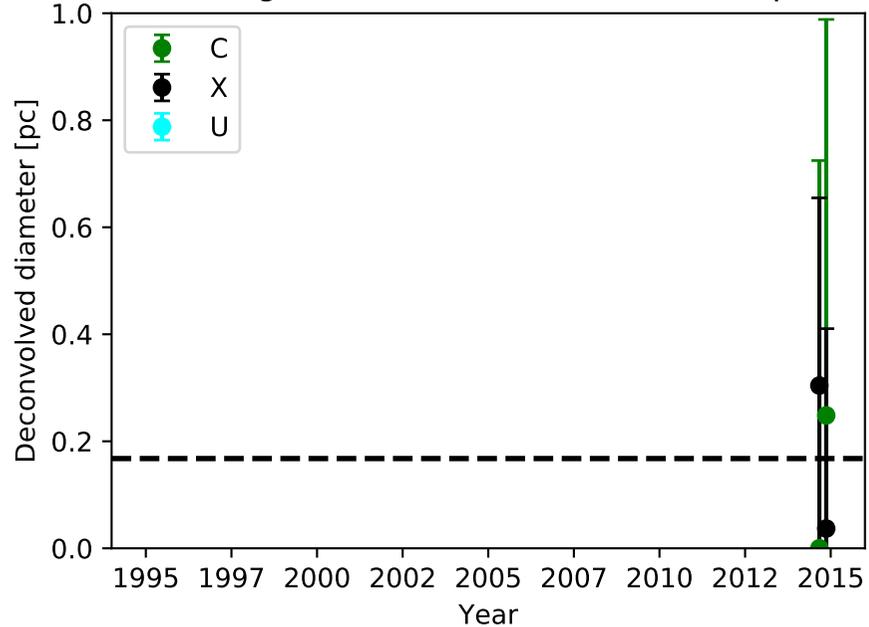

Classification: Rise

Figure notes:

Uncertainties calculated as described in Sect. 2.3.

Triangles represent 3$\sigma$ upper limits for non-detections.

Sizes shown only for detections in bands C,X, and U.

Average diameters calculated according to Sect. 2.4 from the

3.6 cm and 6 cm measurements in experiments BB335A and BB335B.

# 0.2239+0.470

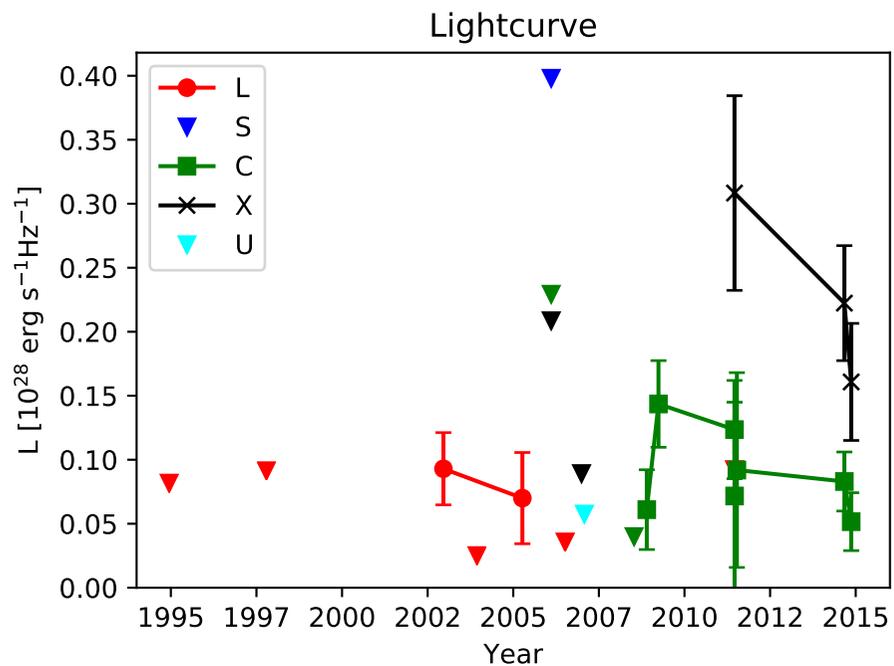
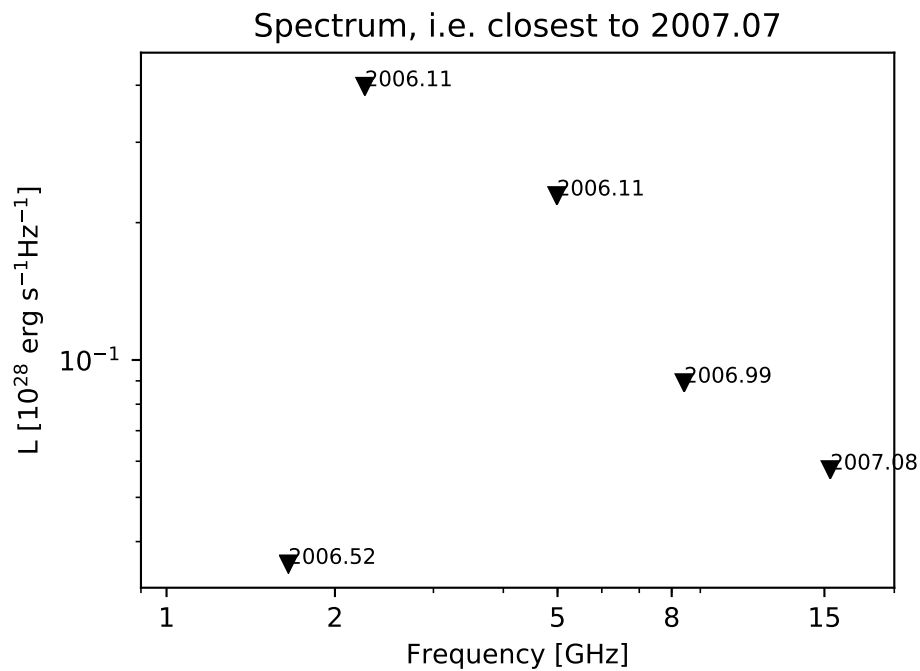
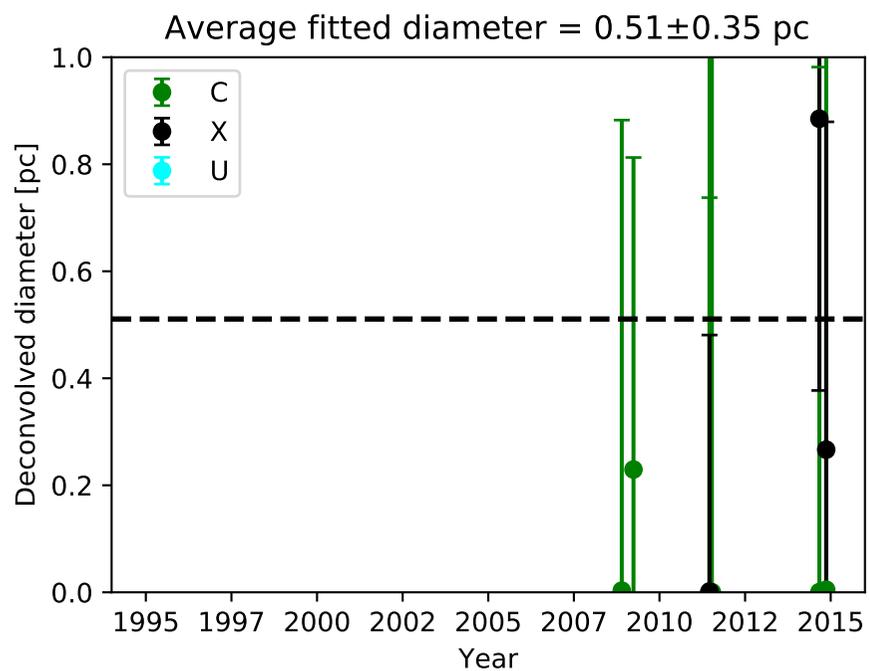

Classification: Slowly varying (S-var)

Figure notes:

Uncertainties calculated as described in Sect. 2.3.

Triangles represent 3σ upper limits for non-detections.

Sizes shown only for detections in bands C, X, and U.

Average diameters calculated according to Sect. 2.4 from the

3.6 cm and 6 cm measurements in experiments BB335A and BB335B.

# 0.2240+0.546 (W18)

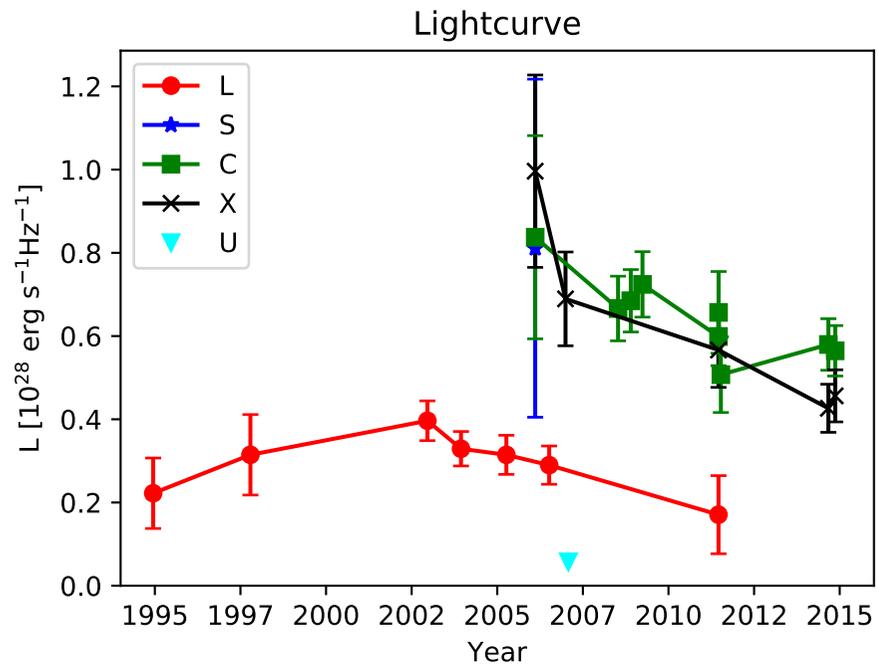
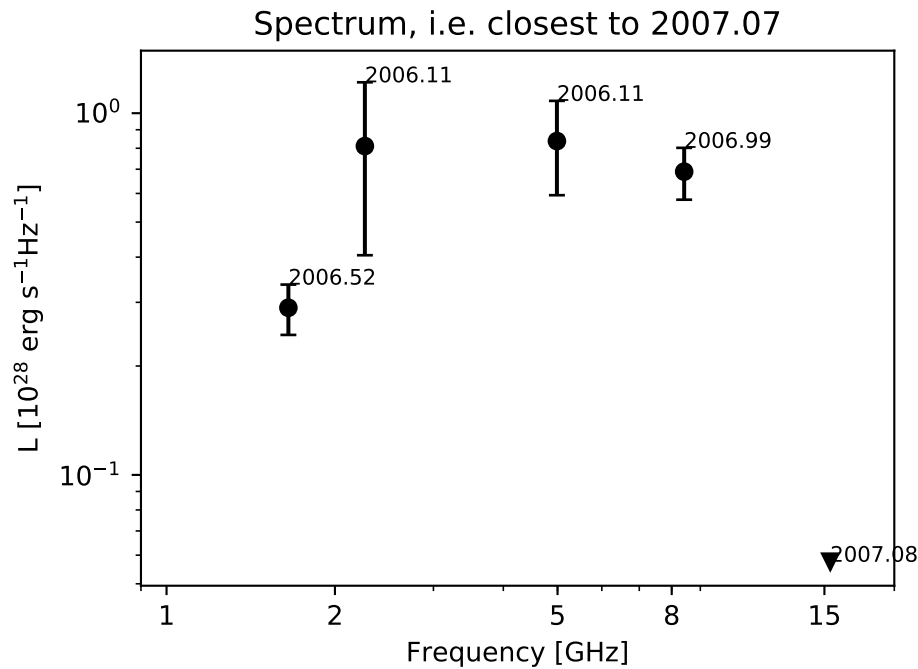
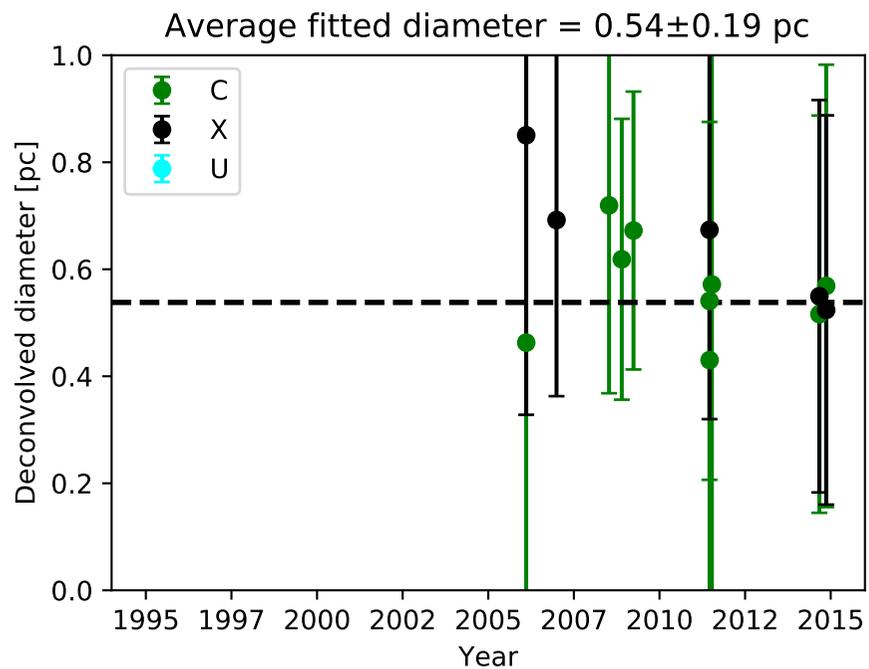

Classification: Slowly varying (S-var)

Figure notes:

Uncertainties calculated as described in Sect. 2.3.

Triangles represent 3σ upper limits for non-detections.

Sizes shown only for detections in bands C, X, and U.

Average diameters calculated according to Sect. 2.4 from the 3.6 cm and 6 cm measurements in experiments BB335A and BB335B.

# 0.2241+0.520 (W17)

## Lightcurve

## Spectrum, i.e. closest to 2007.07

## Average fitted diameter = 0.43±0.25 pc

Classification: Slowly varying (S-var)

Figure notes:

Uncertainties calculated as described in Sect. 2.3.

Triangles represent 3$\sigma$ upper limits for non-detections.

Sizes shown only for detections in bands C, X, and U.

Average diameters calculated according to Sect. 2.4 from the

3.6 cm and 6 cm measurements in experiments BB335A and BB335B.

# 0.2244+0.531 (W16)

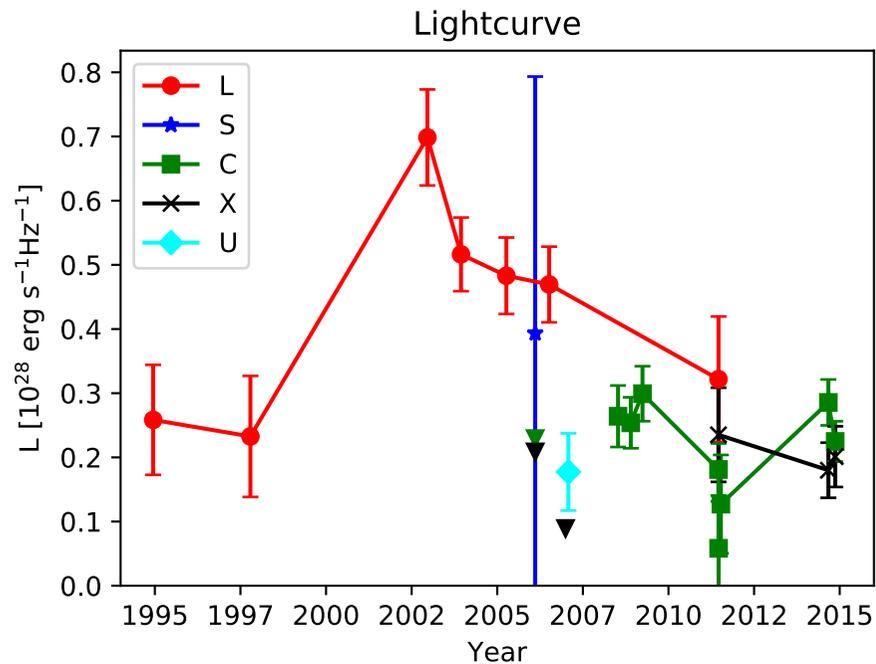
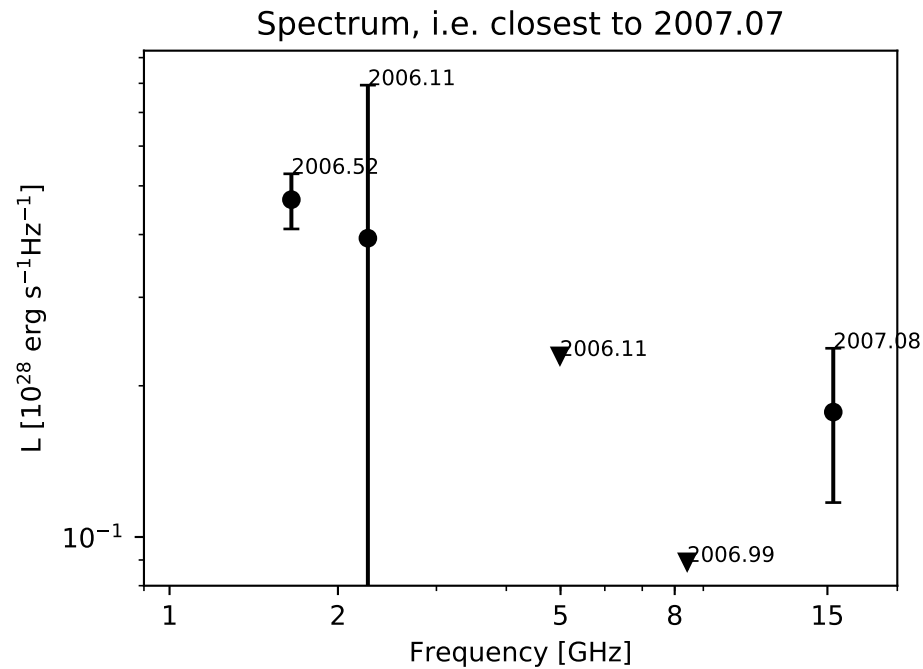
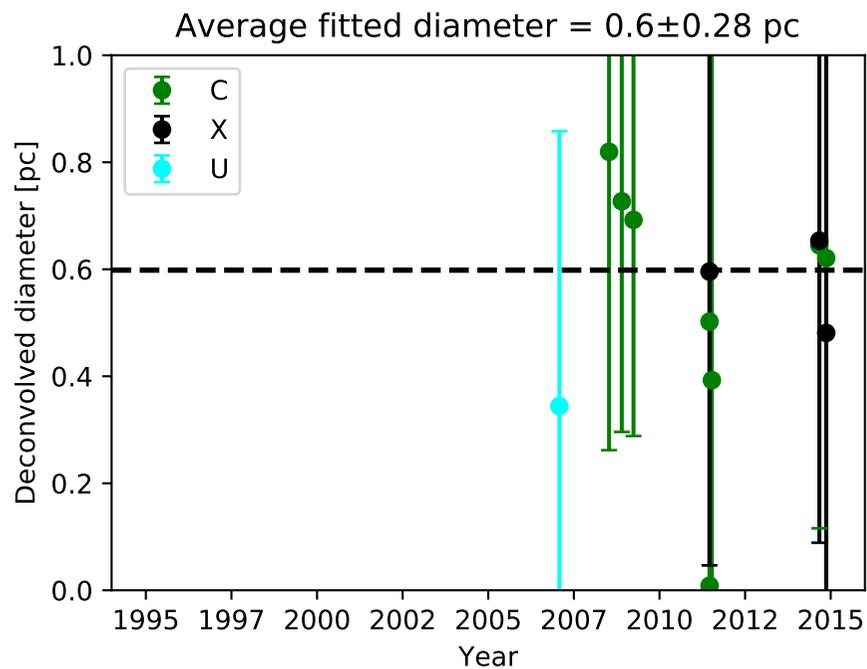

Classification: Slowly varying (S-var)

Figure notes:

Uncertainties calculated as described in Sect. 2.3.

Triangles represent $3\sigma$ upper limits for non-detections.

Sizes shown only for detections in bands C, X, and U.

Average diameters calculated according to Sect. 2.4 from the 3.6 cm and 6 cm measurements in experiments BB335A and BB335B.

# 0.2249+0.586

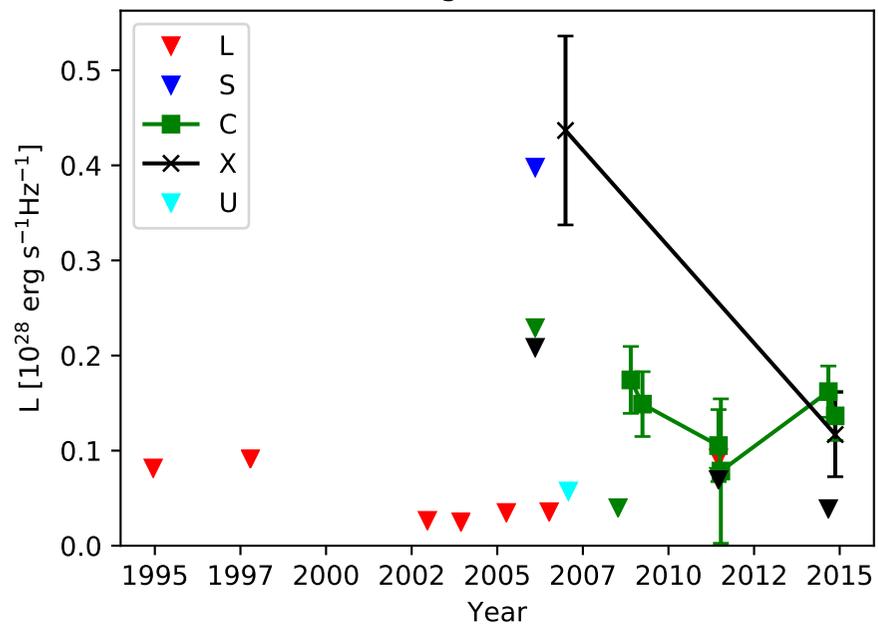

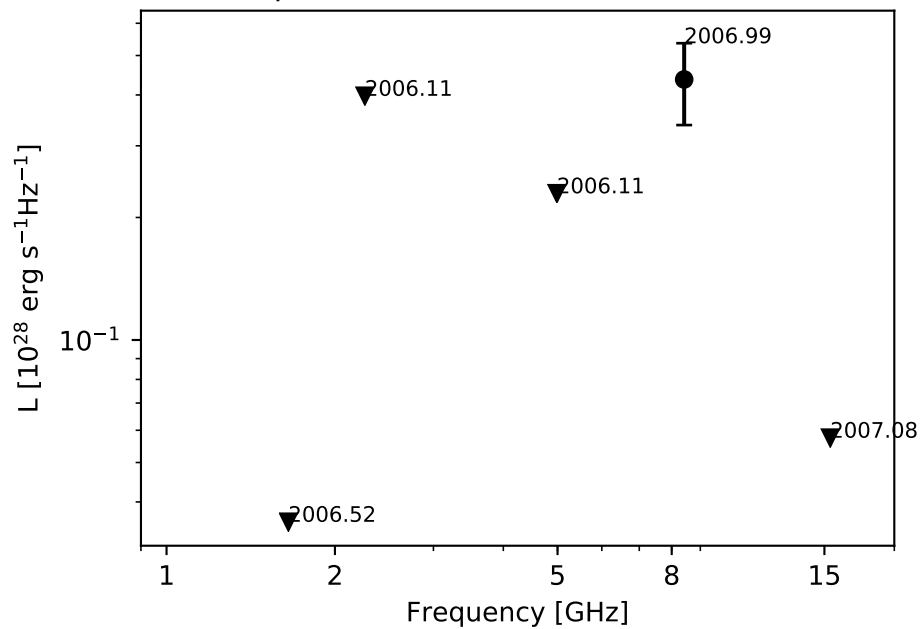

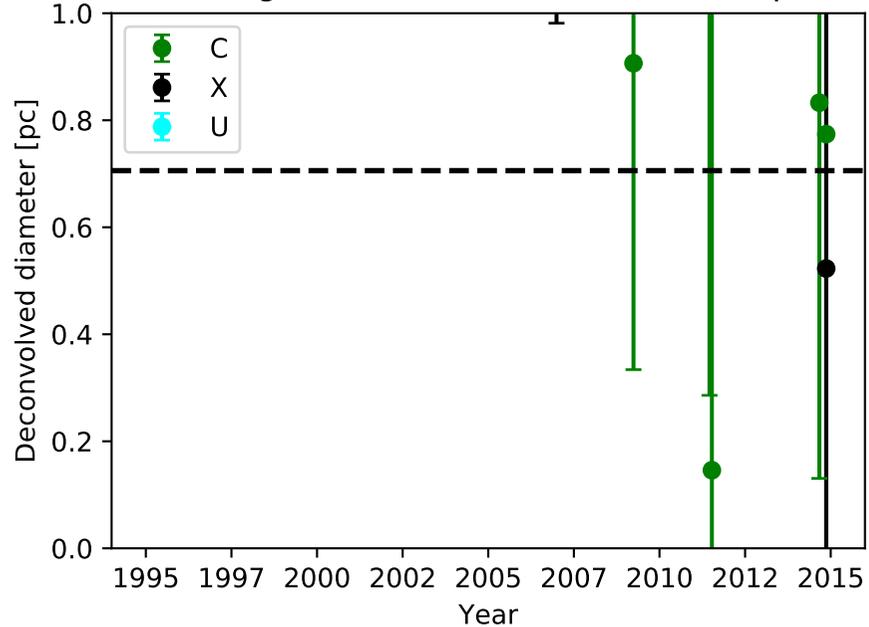

Classification: Slowly varying (S-var)

Figure notes:

Uncertainties calculated as described in Sect. 2.3.

Triangles represent 3σ upper limits for non-detections.

Sizes shown only for detections in bands C, X, and U.

Average diameters calculated according to Sect. 2.4 from the 3.6 cm and 6 cm measurements in experiments BB335A and BB335B.

# 0.2253+0.483 (W15)

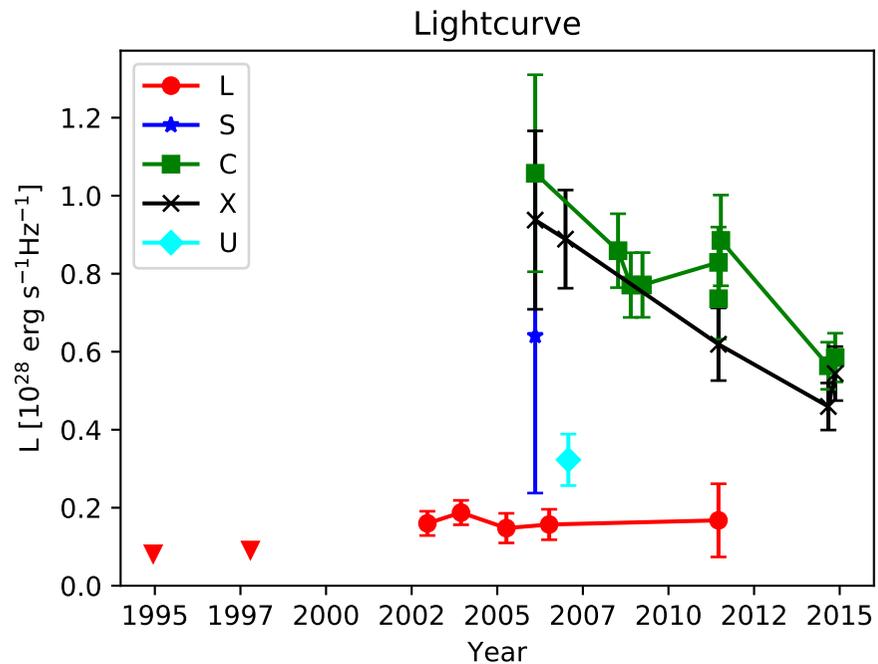

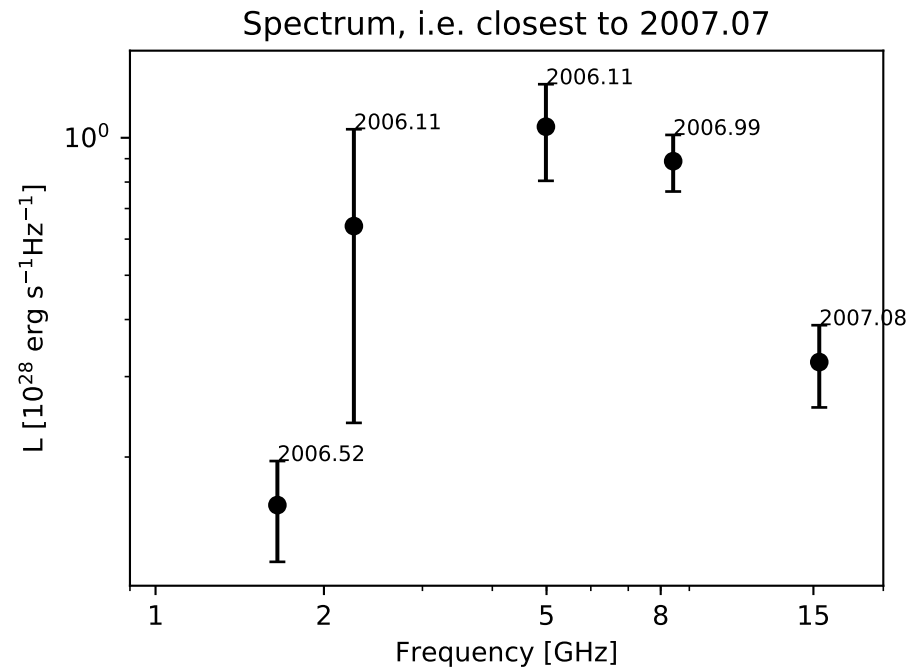

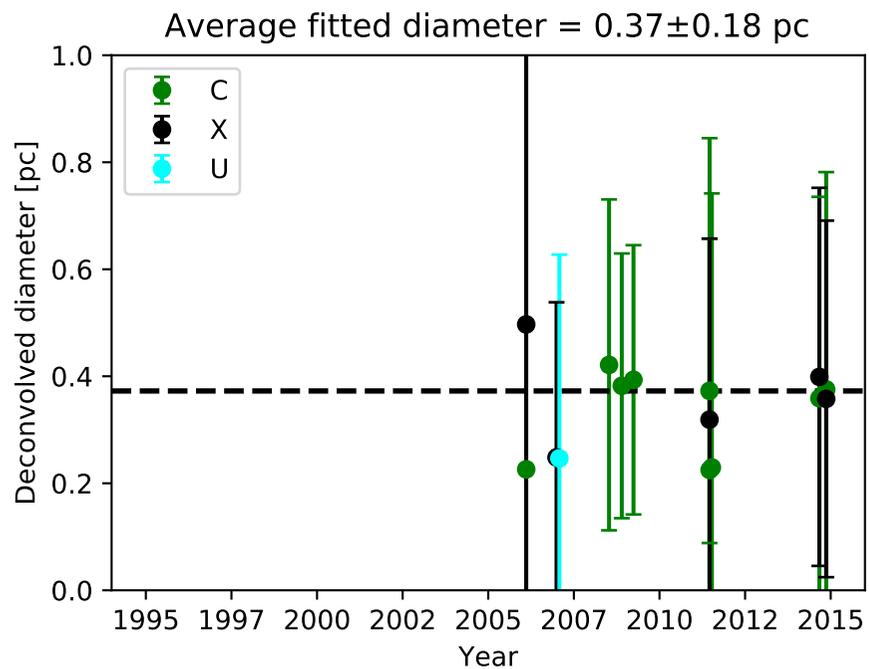

Classification: Fall

Figure notes:

Uncertainties calculated as described in Sect. 2.3.

Triangles represent 3σ upper limits for non-detections.

Sizes shown only for detections in bands C, X, and U.

Average diameters calculated according to Sect. 2.4 from the

3.6 cm and 6 cm measurements in experiments BB335A and BB335B.

# 0.2262+0.512

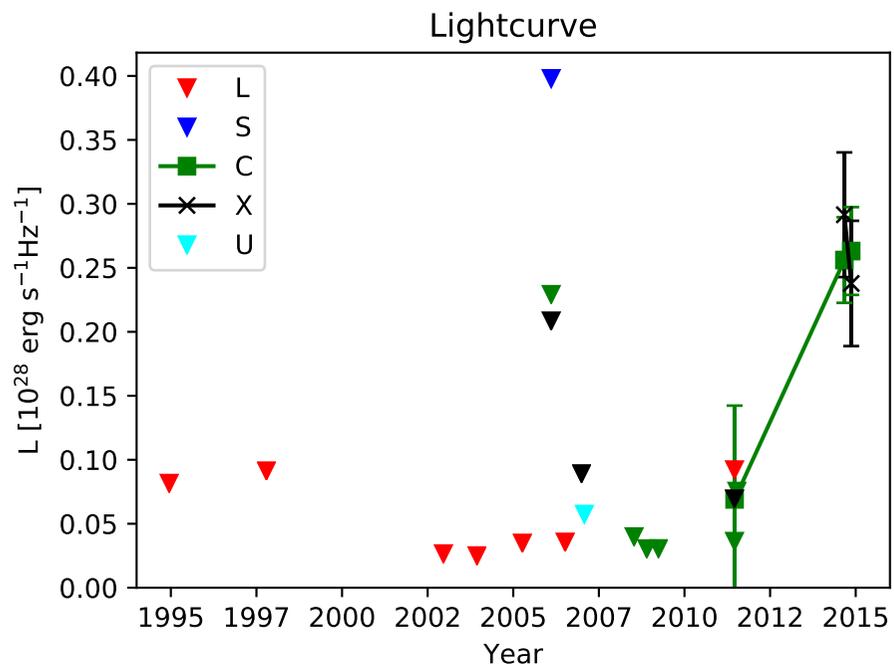

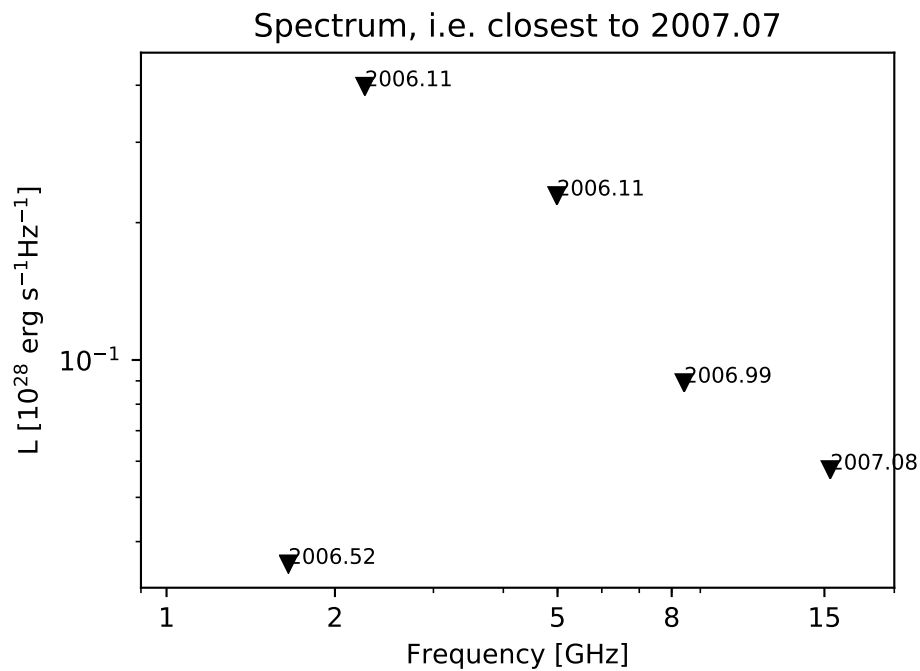

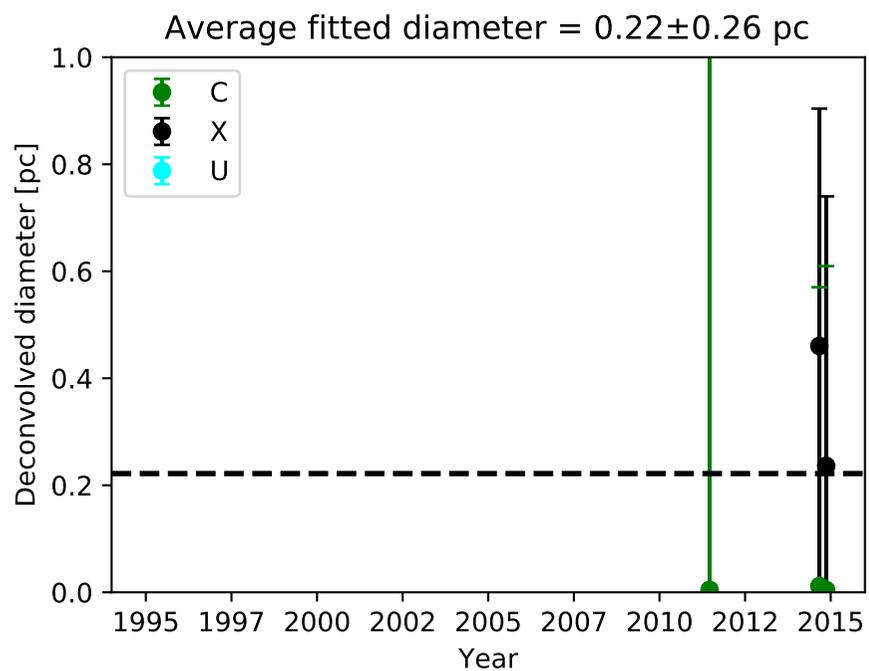

Classification: Rise

Figure notes:

Uncertainties calculated as described in Sect. 2.3.

Triangles represent 3σ upper limits for non-detections.

Sizes shown only for detections in bands C, X, and U.

Average diameters calculated according to Sect. 2.4 from the

3.6 cm and 6 cm measurements in experiments BB335A and BB335B.

# 0.2264+0.448

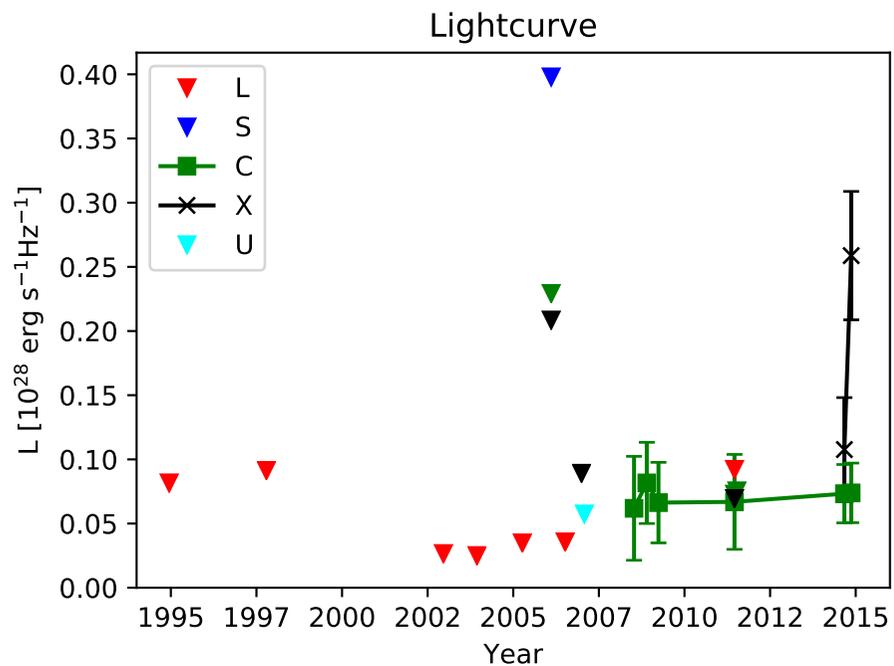
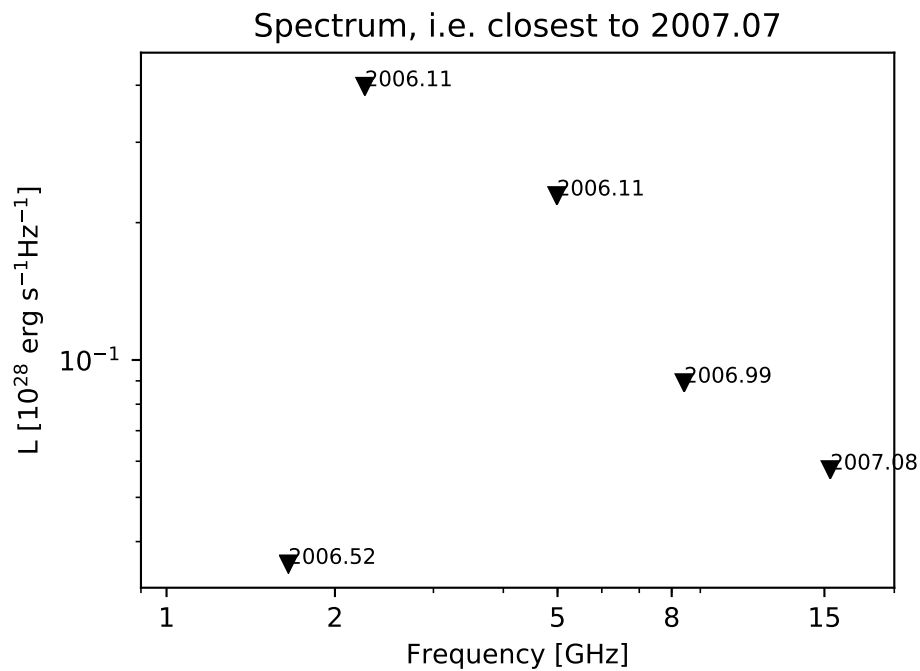
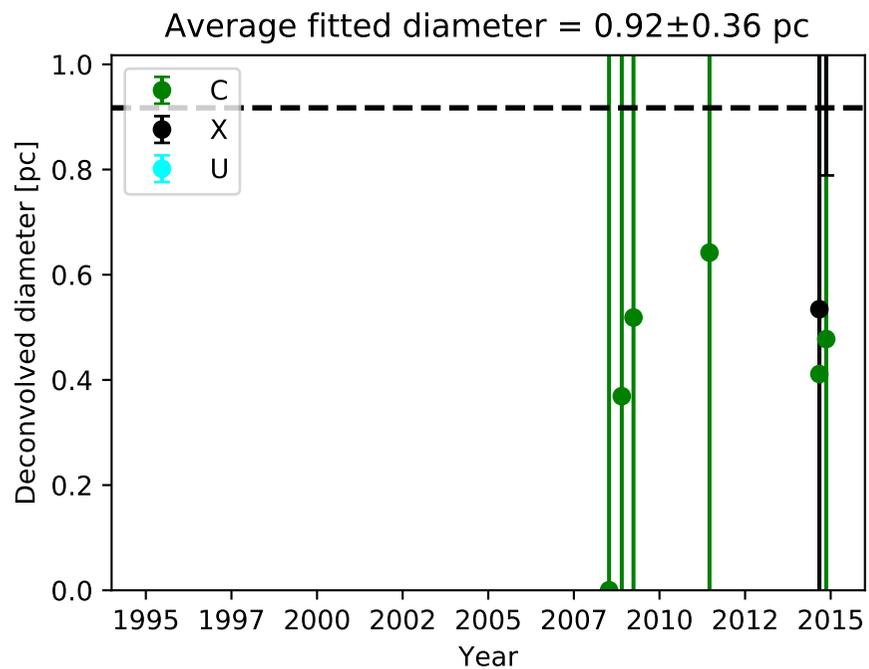

Classification: Slowly varying (S-var)

Figure notes:

Uncertainties calculated as described in Sect. 2.3.

Triangles represent 3$\sigma$ upper limits for non-detections.

Sizes shown only for detections in bands C, X, and U.

Average diameters calculated according to Sect. 2.4 from the

3.6 cm and 6 cm measurements in experiments BB335A and BB335B.

# 0.2269+0.574 (W14)

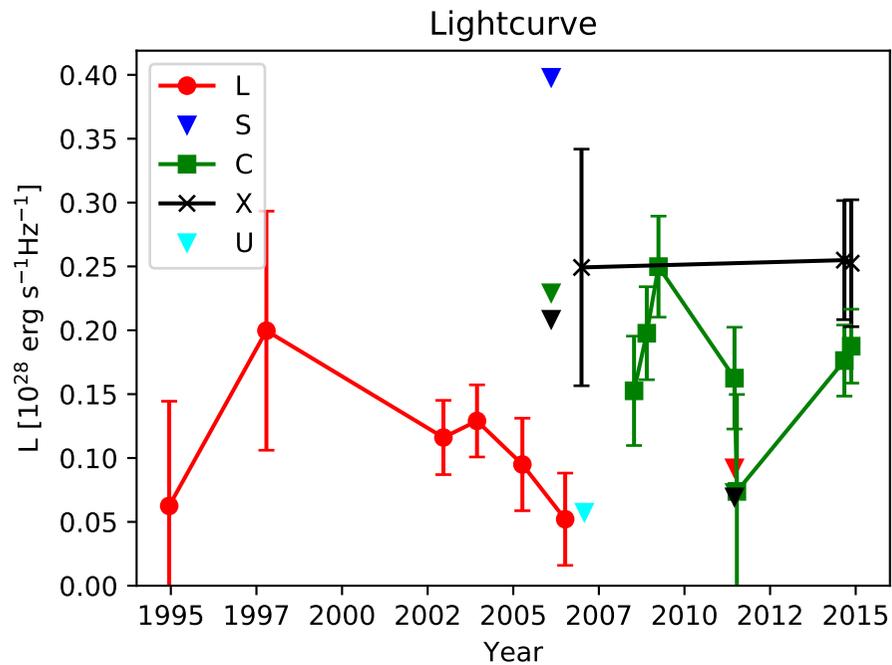
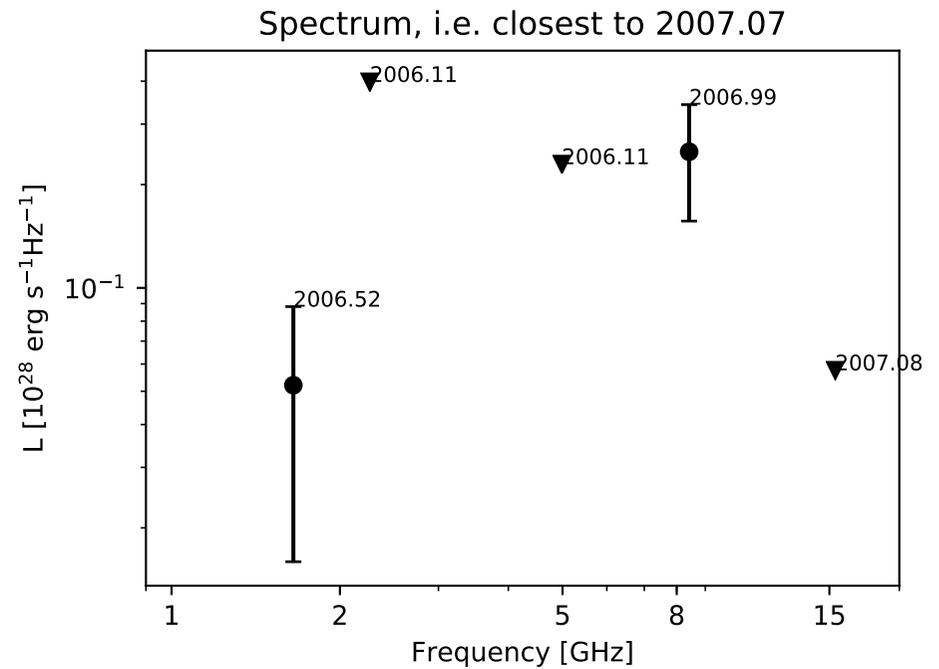
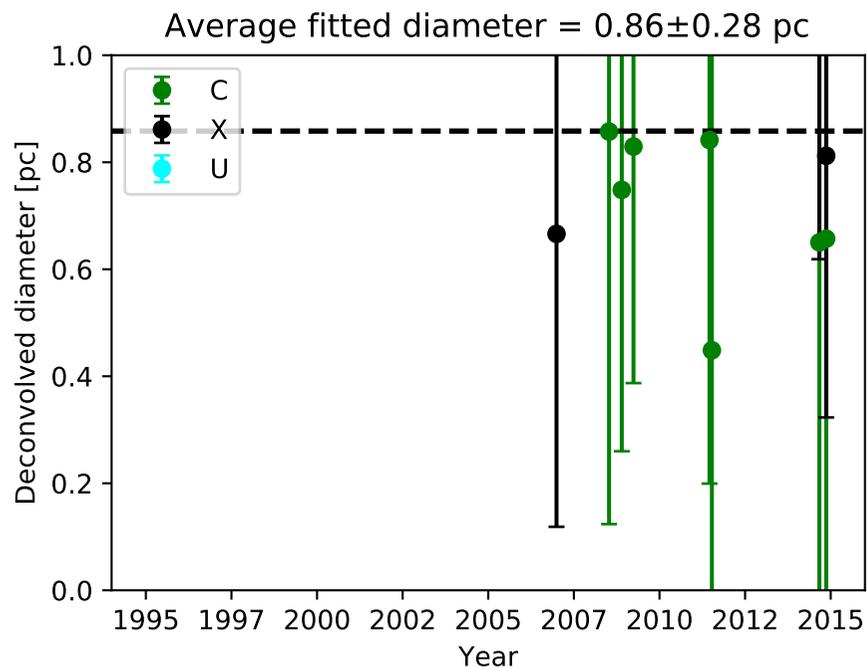

Classification: Slowly varying (S-var)

Figure notes:

Uncertainties calculated as described in Sect. 2.3.

Triangles represent $3\sigma$ upper limits for non-detections.

Sizes shown only for detections in bands C, X, and U.

Average diameters calculated according to Sect. 2.4 from the

3.6 cm and 6 cm measurements in experiments BB335A and BB335B.

# 0.2276+0.546 (W60)

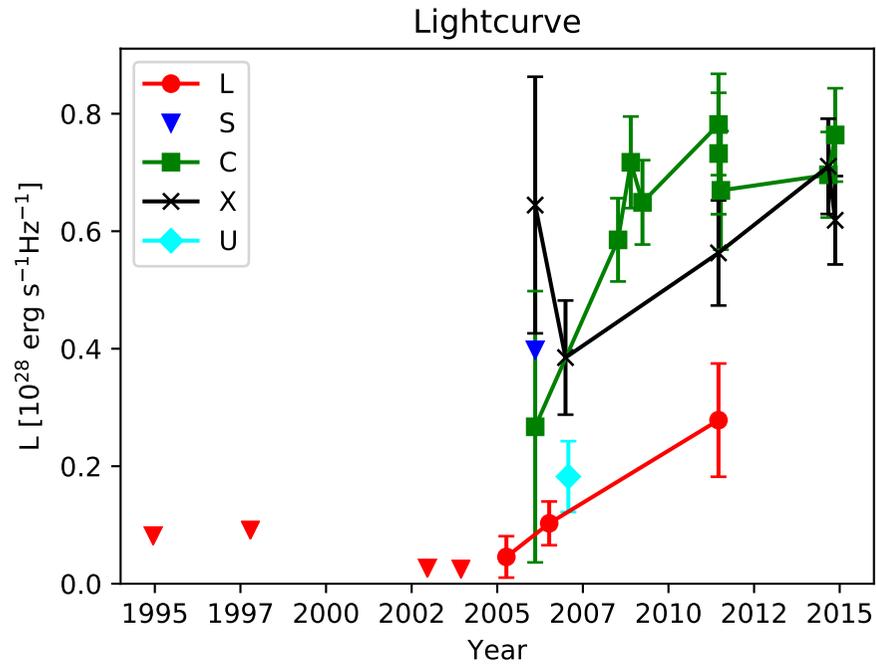
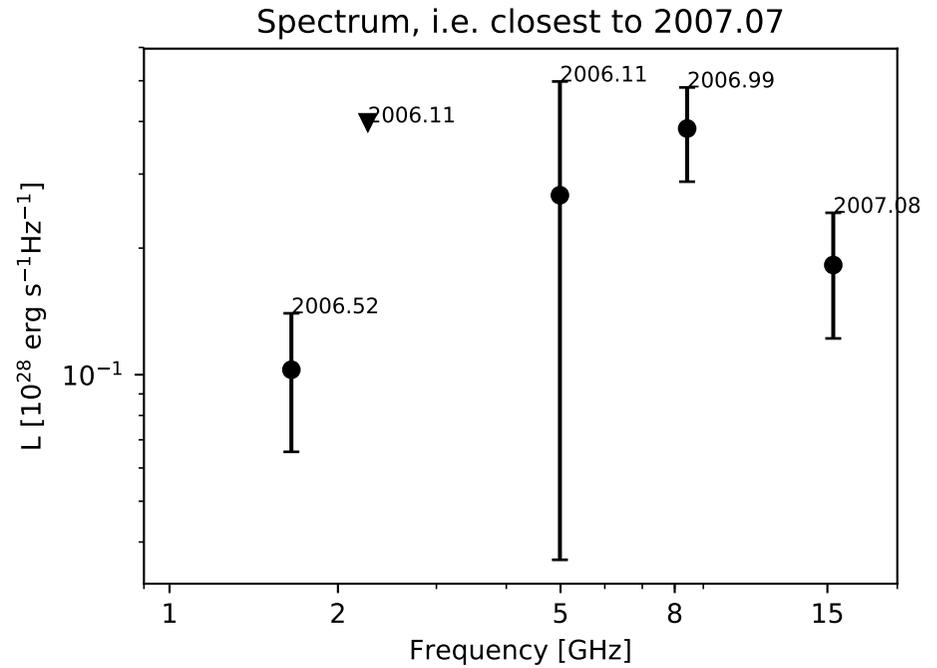
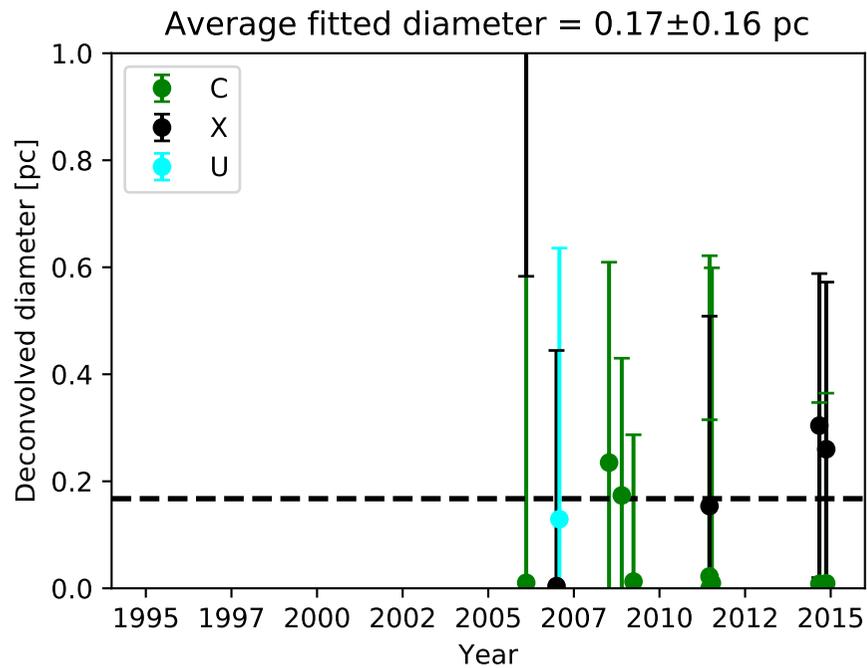

Classification: Slowly varying (S-var)

Figure notes:

Uncertainties calculated as described in Sect. 2.3.

Triangles represent $3\sigma$ upper limits for non-detections.

Sizes shown only for detections in bands C, X, and U.

Average diameters calculated according to Sect. 2.4 from the

3.6 cm and 6 cm measurements in experiments BB335A and BB335B.

# 0.2277+0.528

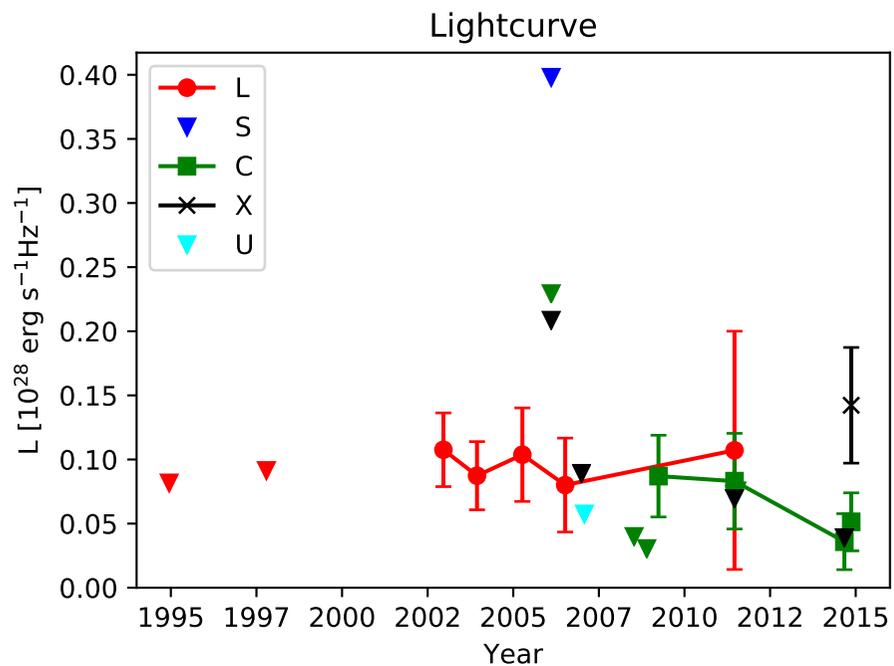
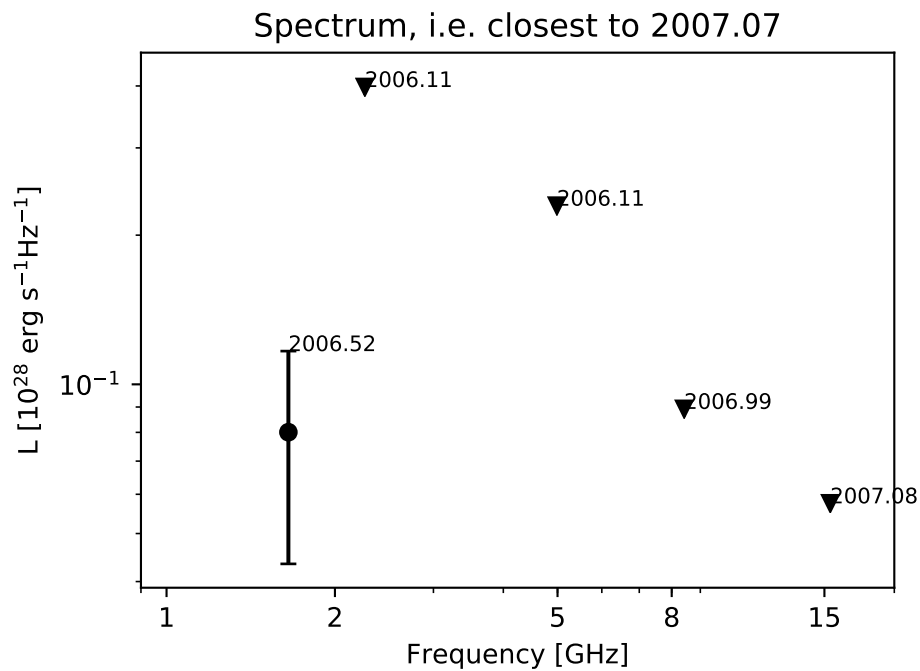
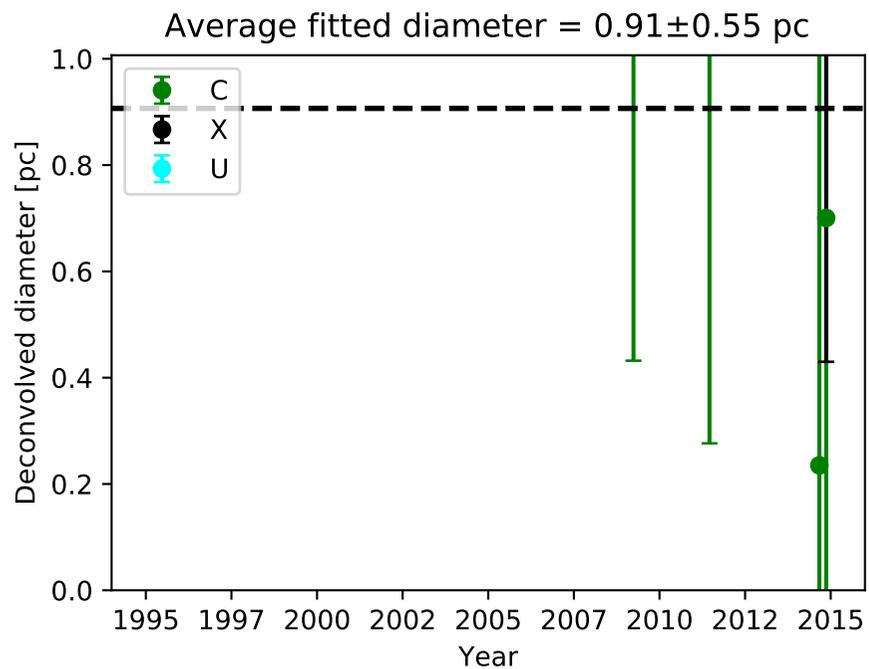

Classification: Slowly varying (S-var)

Figure notes:

Uncertainties calculated as described in Sect. 2.3.

Triangles represent $3\sigma$ upper limits for non-detections.

Sizes shown only for detections in bands C, X, and U.

Average diameters calculated according to Sect. 2.4 from the 3.6 cm and 6 cm measurements in experiments BB335A and BB335B.

# 0.2281+0.605

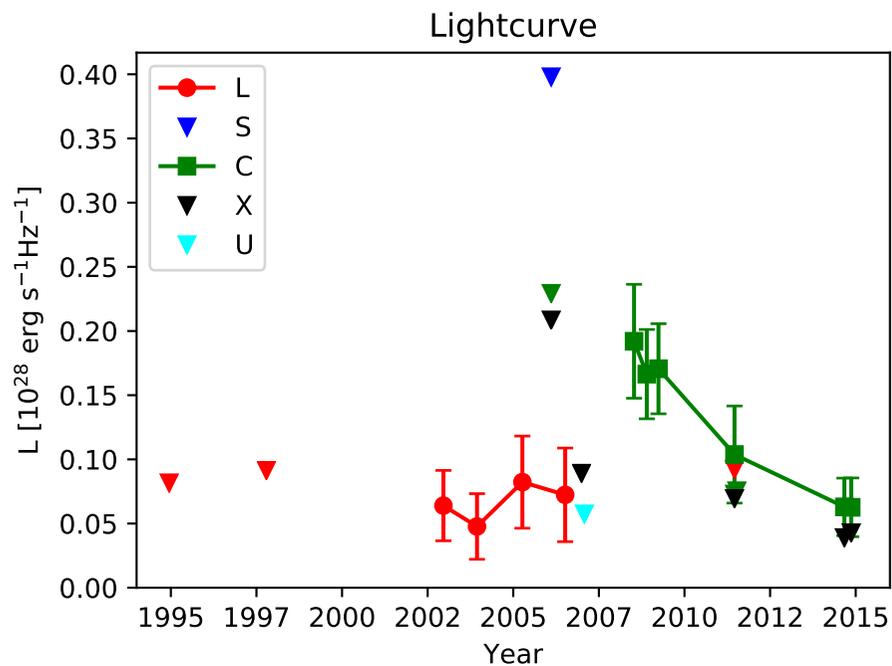

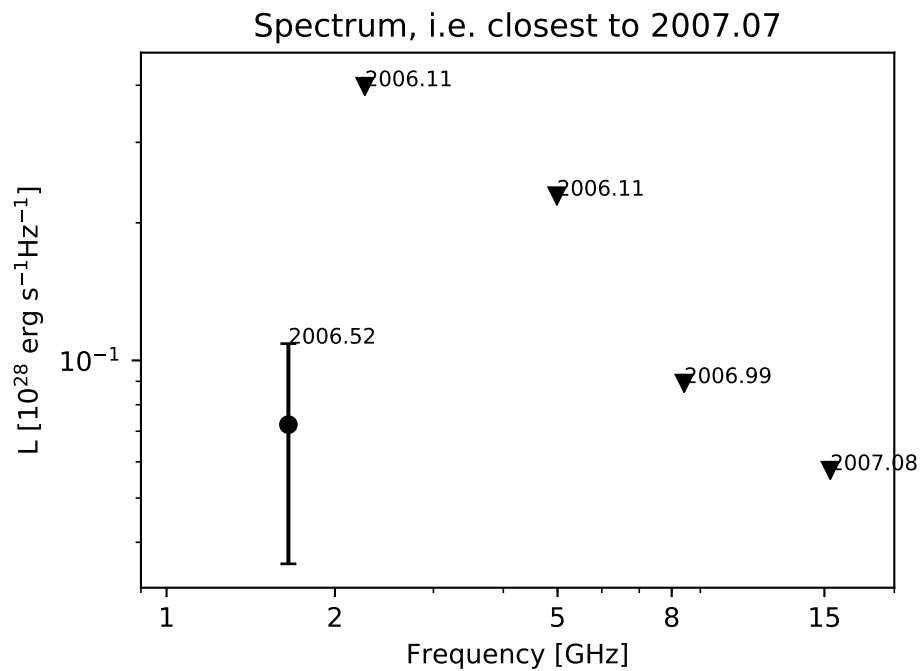

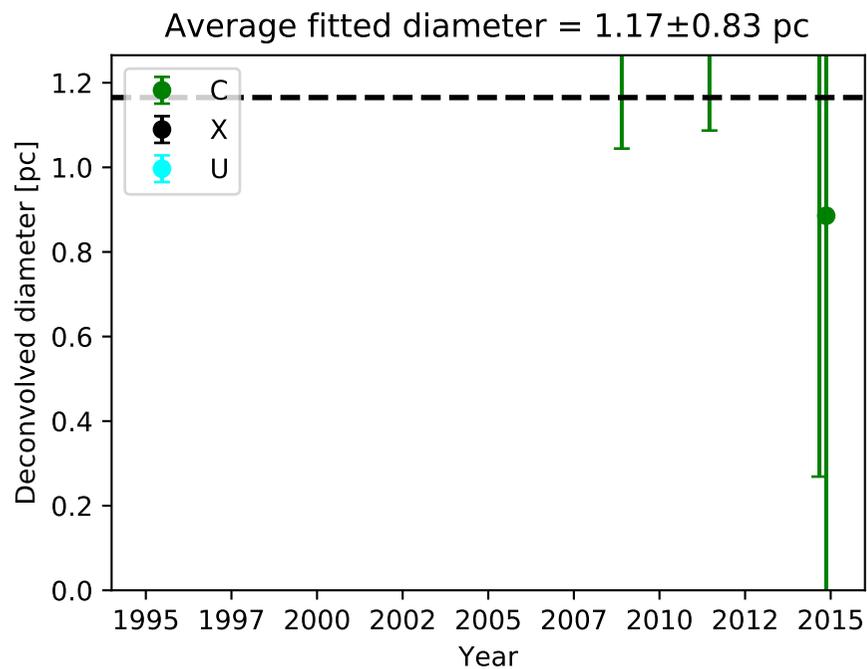

Classification: Fall

Figure notes:

Uncertainties calculated as described in Sect. 2.3.

Triangles represent 3σ upper limits for non-detections.

Sizes shown only for detections in bands C, X, and U.

Average diameters calculated according to Sect. 2.4 from the

3.6 cm and 6 cm measurements in experiments BB335A and BB335B.

# 0.2286+0.583

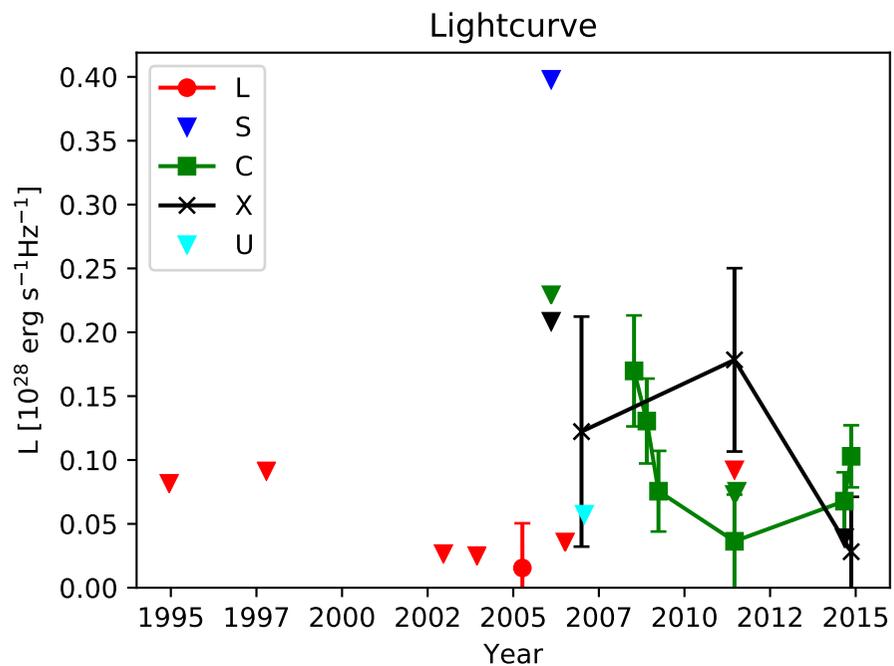

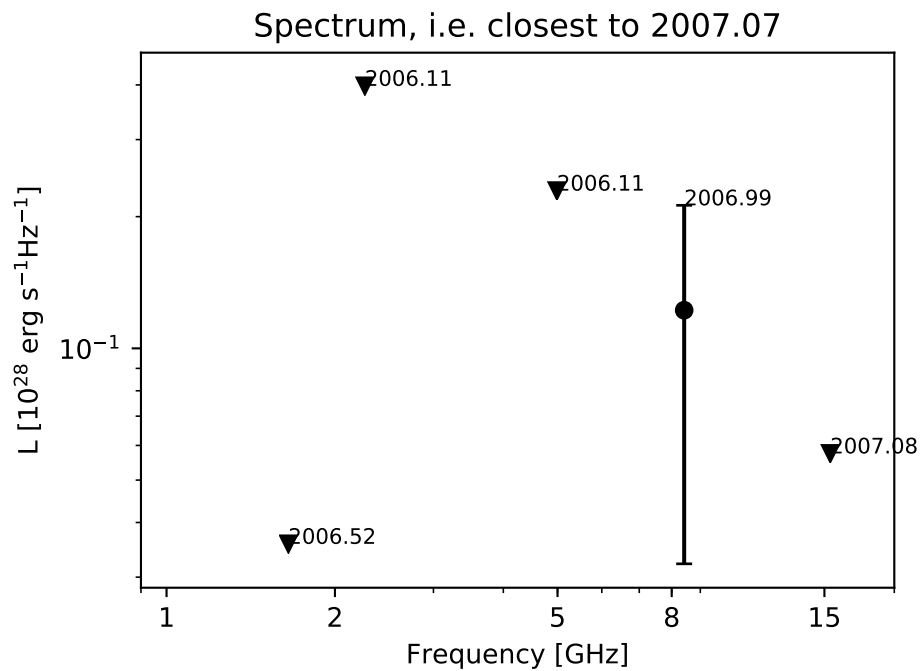

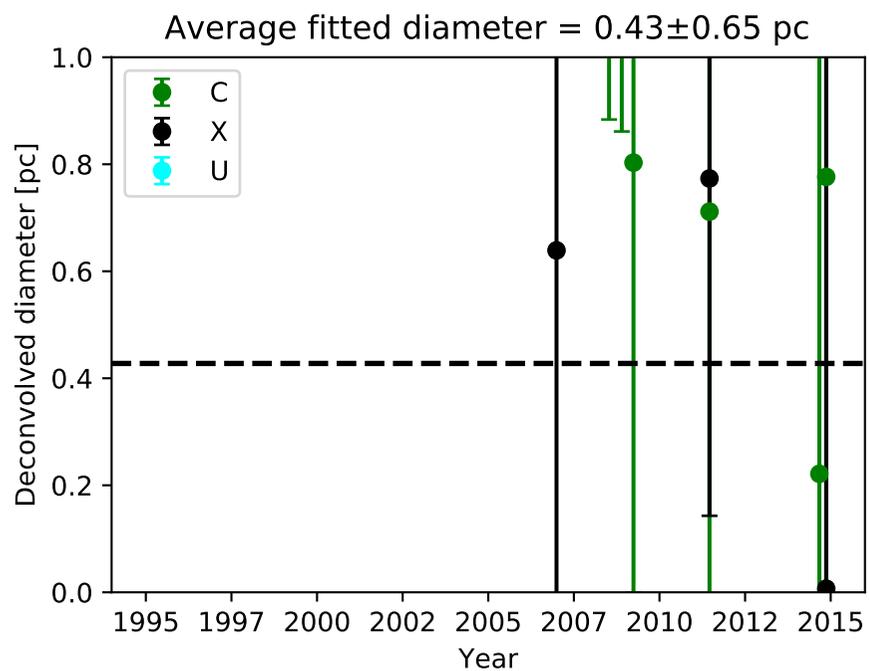

Classification: Slowly varying (S-var)

Figure notes:

Uncertainties calculated as described in Sect. 2.3.

Triangles represent 3σ upper limits for non-detections.

Sizes shown only for detections in bands C,X, and U.

Average diameters calculated according to Sect. 2.4 from the

3.6 cm and 6 cm measurements in experiments BB335A and BB335B.

# 0.2290+0.564

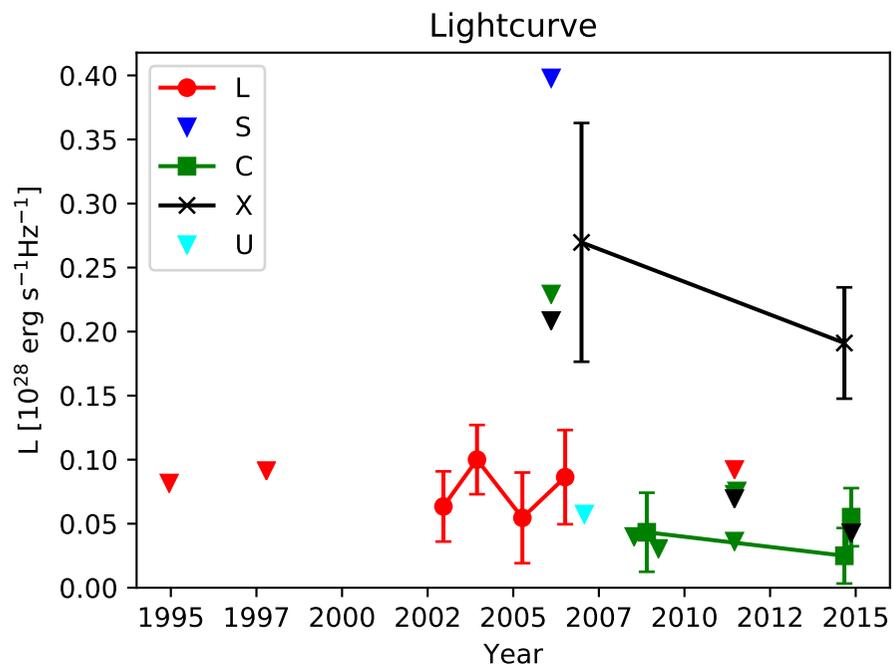
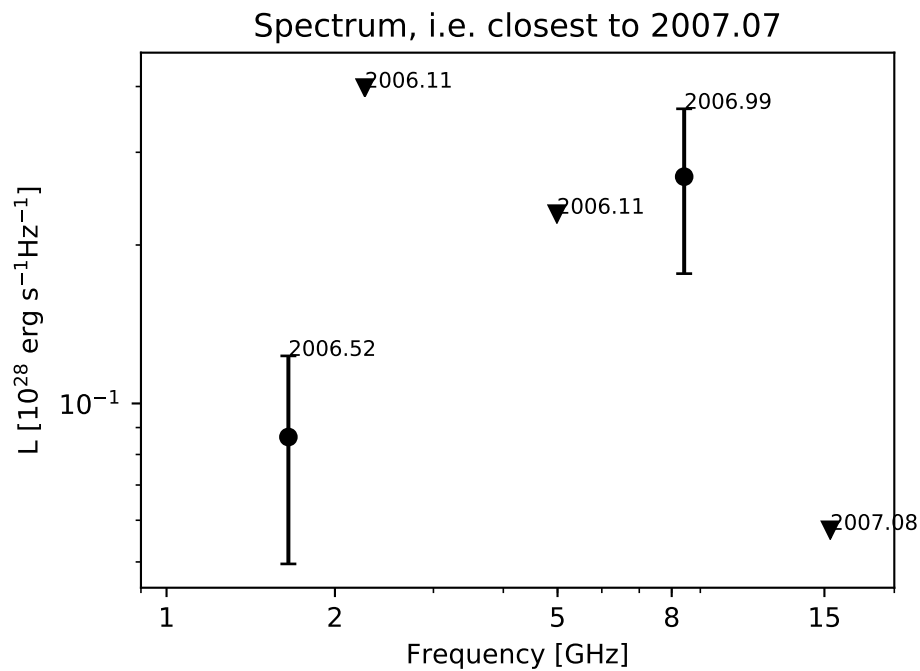
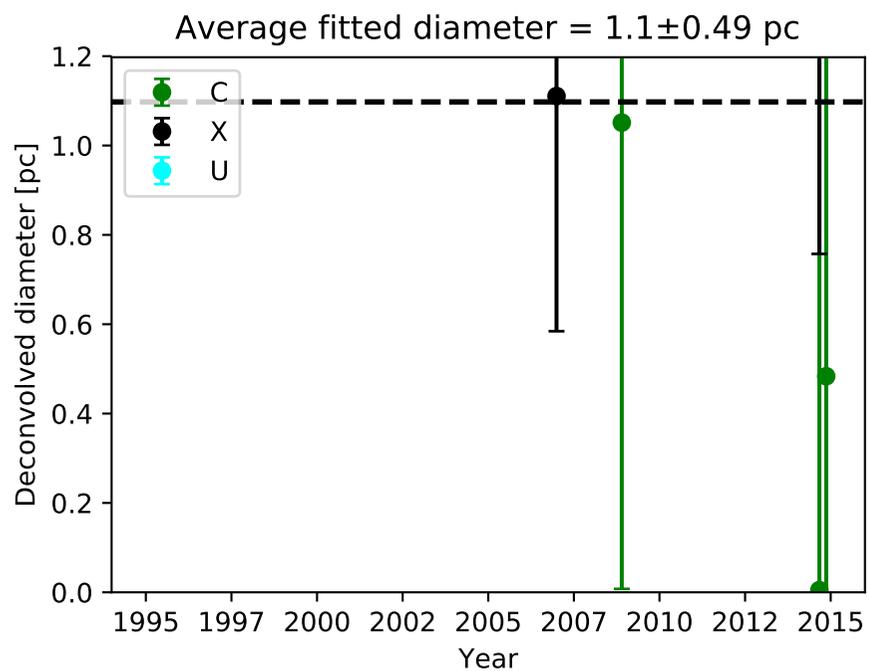

Classification: Slowly varying (S-var)

Figure notes:

Uncertainties calculated as described in Sect. 2.3.

Triangles represent 3σ upper limits for non-detections.

Sizes shown only for detections in bands C, X, and U.

Average diameters calculated according to Sect. 2.4 from the

3.6 cm and 6 cm measurements in experiments BB335A and BB335B.

# 0.2295+0.524 (W12)

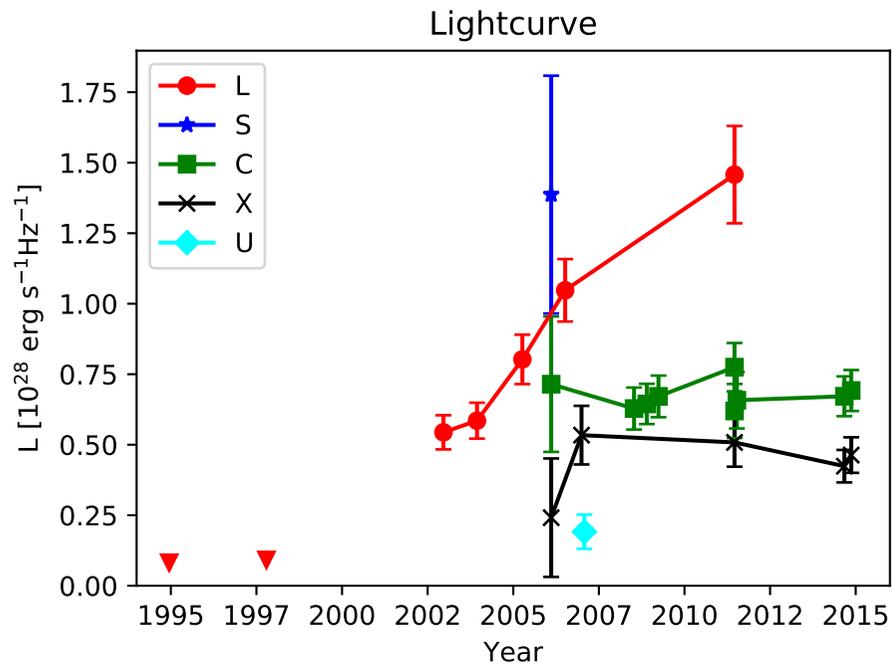
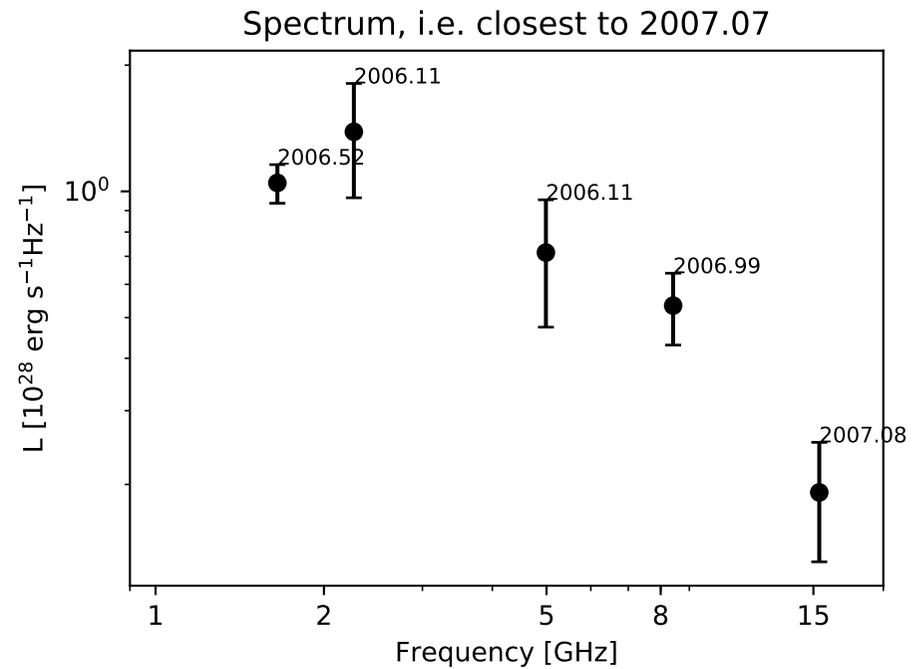
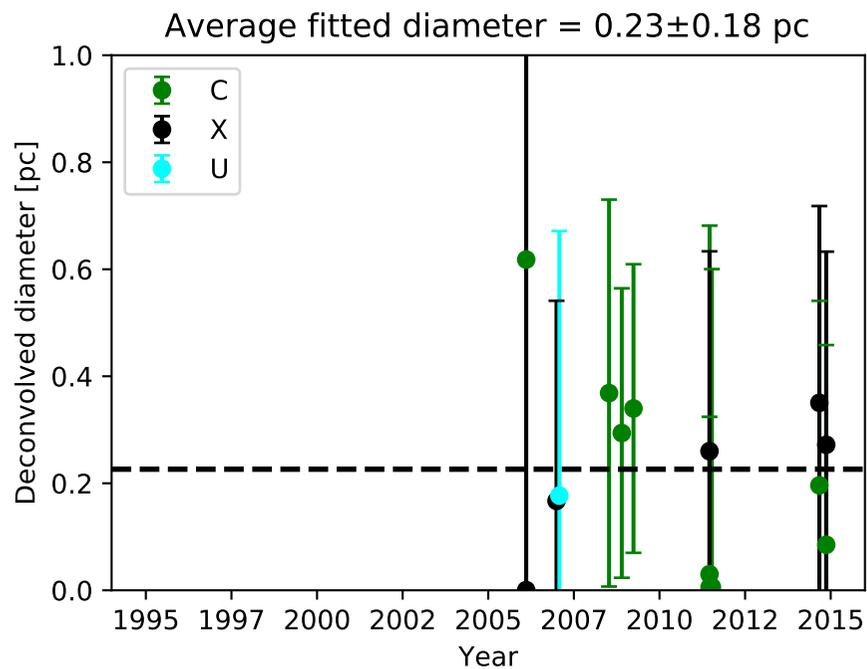

Classification: Slowly varying (S-var)

Figure notes:

Uncertainties calculated as described in Sect. 2.3.

Triangles represent 3σ upper limits for non-detections.

Sizes shown only for detections in bands C, X, and U.

Average diameters calculated according to Sect. 2.4 from the

3.6 cm and 6 cm measurements in experiments BB335A and BB335B.

# 0.2299+0.502 (W11)

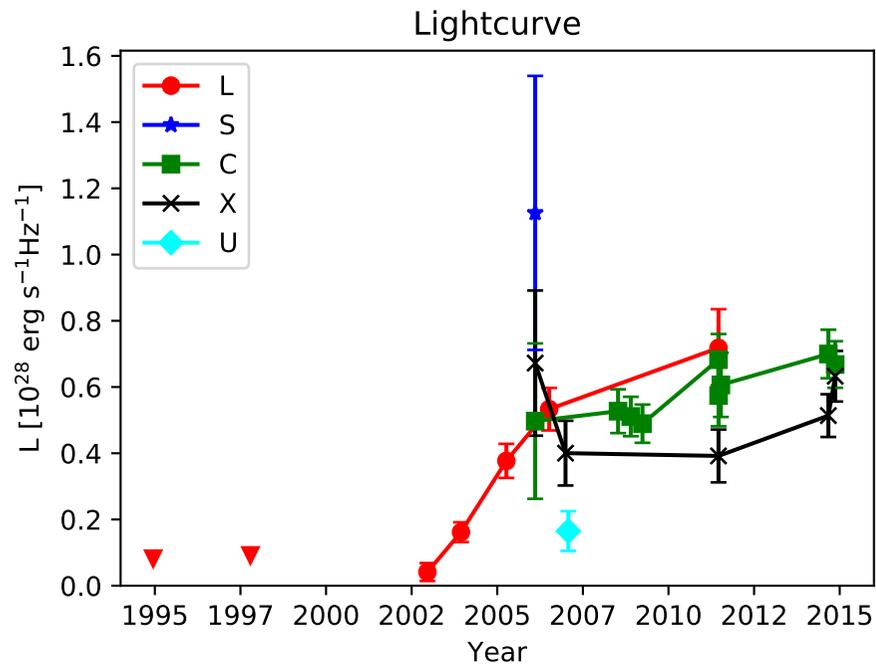
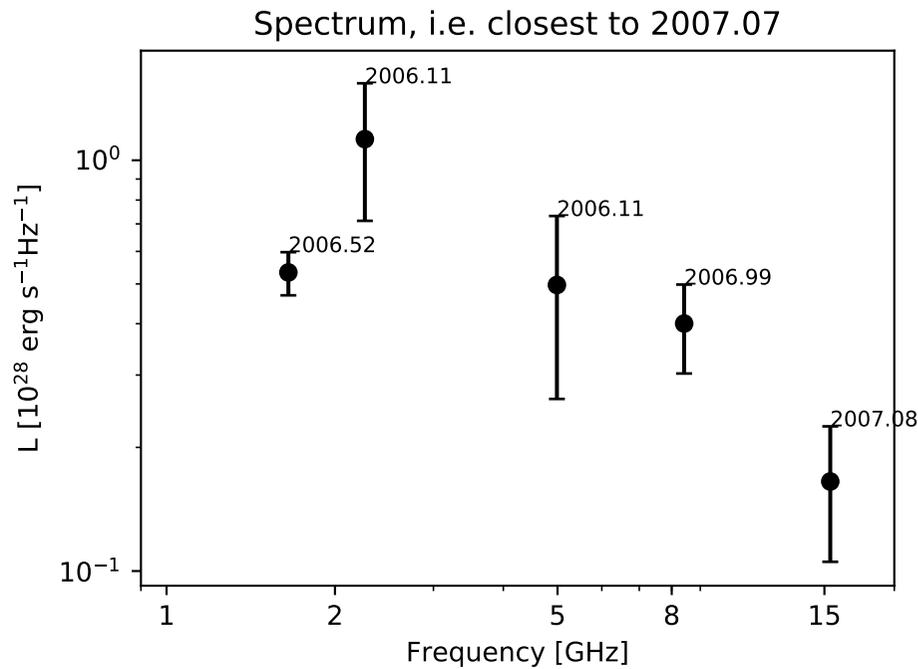
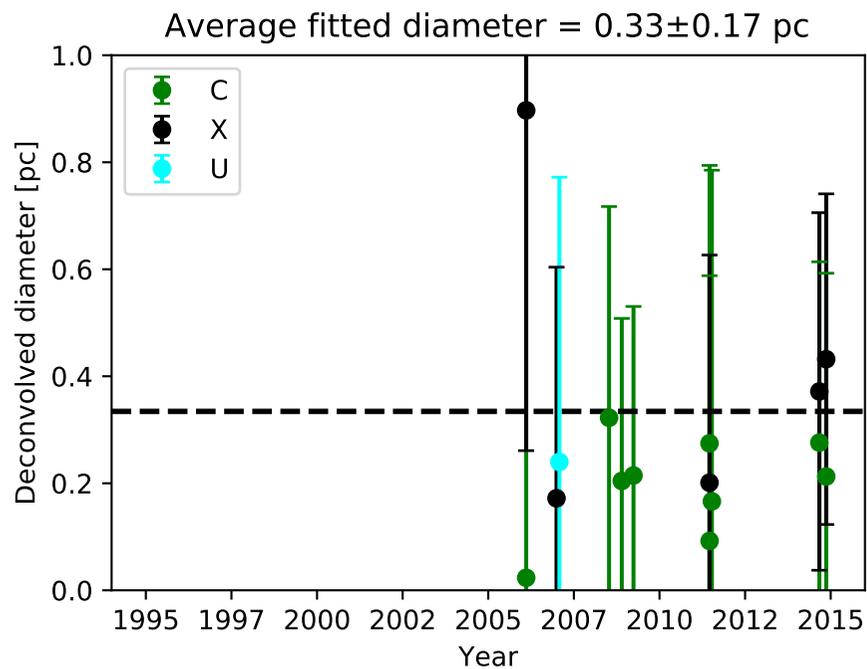

Classification: Rise

Figure notes:

Uncertainties calculated as described in Sect. 2.3.

Triangles represent $3\sigma$ upper limits for non-detections.

Sizes shown only for detections in bands C, X, and U.

Average diameters calculated according to Sect. 2.4 from the

3.6 cm and 6 cm measurements in experiments BB335A and BB335B.

# 0.2306+0.502 (W10)

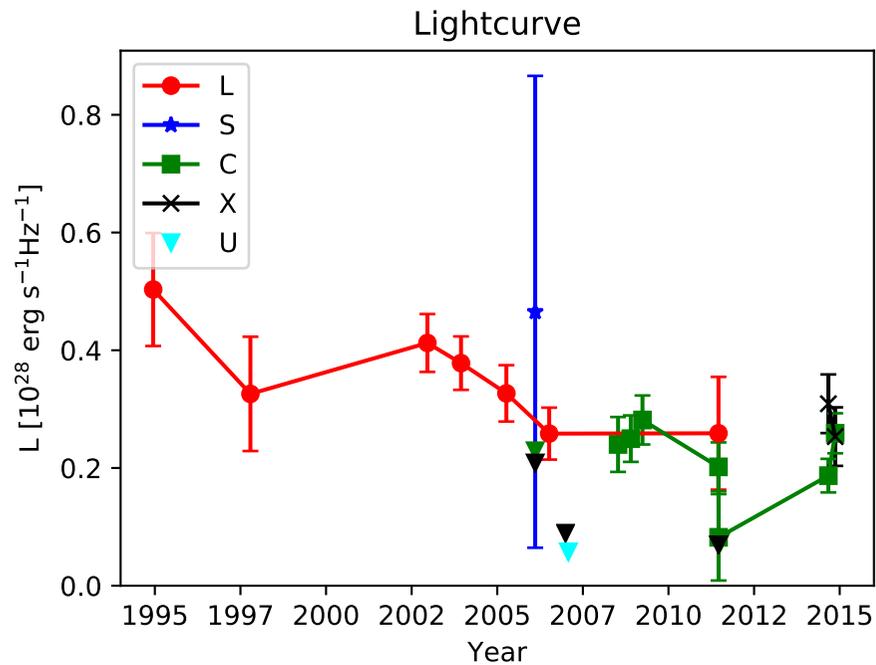
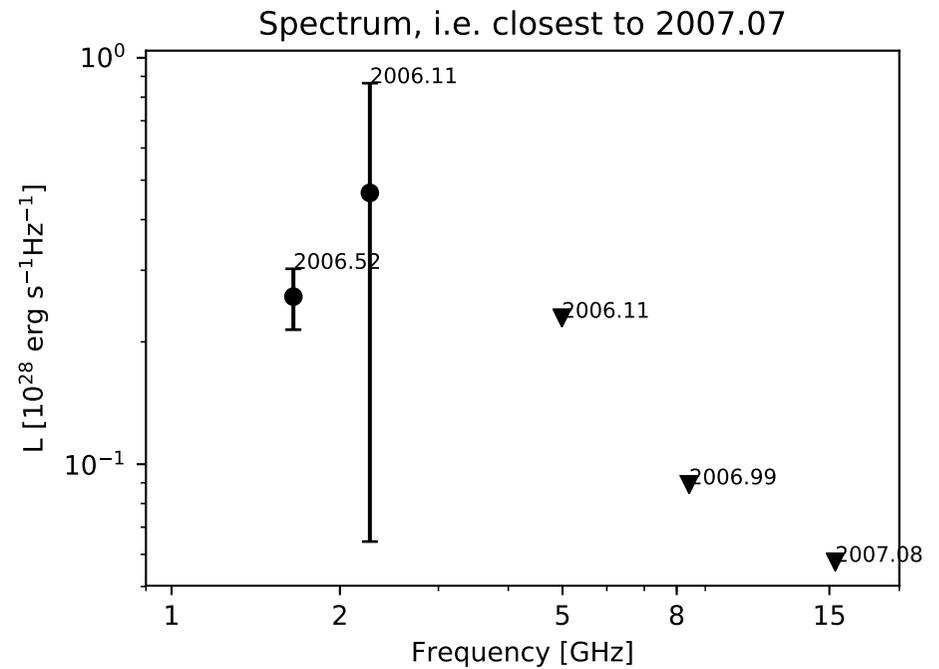
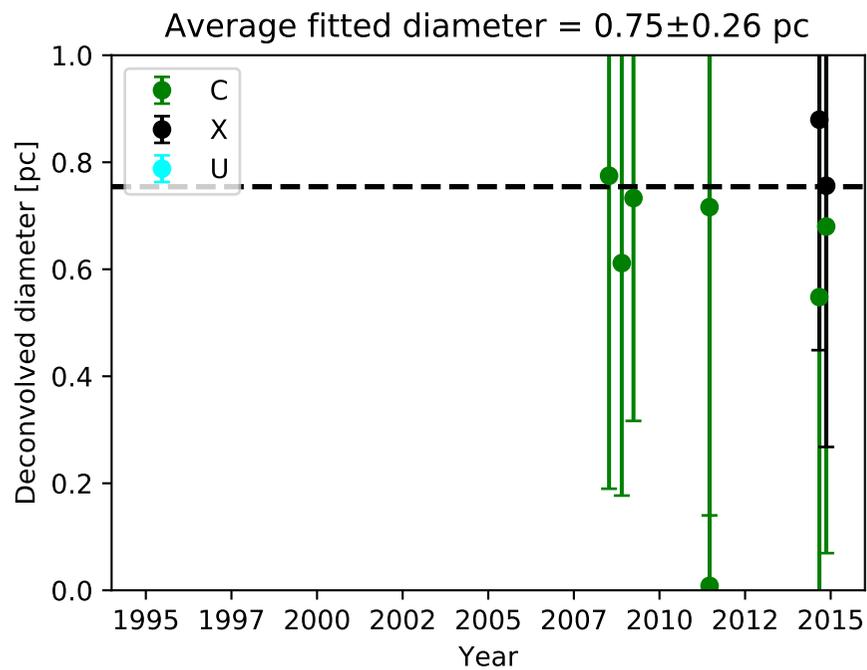

Classification: Slowly varying (S-var)

Figure notes:

Uncertainties calculated as described in Sect. 2.3.

Triangles represent $3\sigma$ upper limits for non-detections.

Sizes shown only for detections in bands C, X, and U.

Average diameters calculated according to Sect. 2.4 from the 3.6 cm and 6 cm measurements in experiments BB335A and BB335B.

# 0.2310+0.496

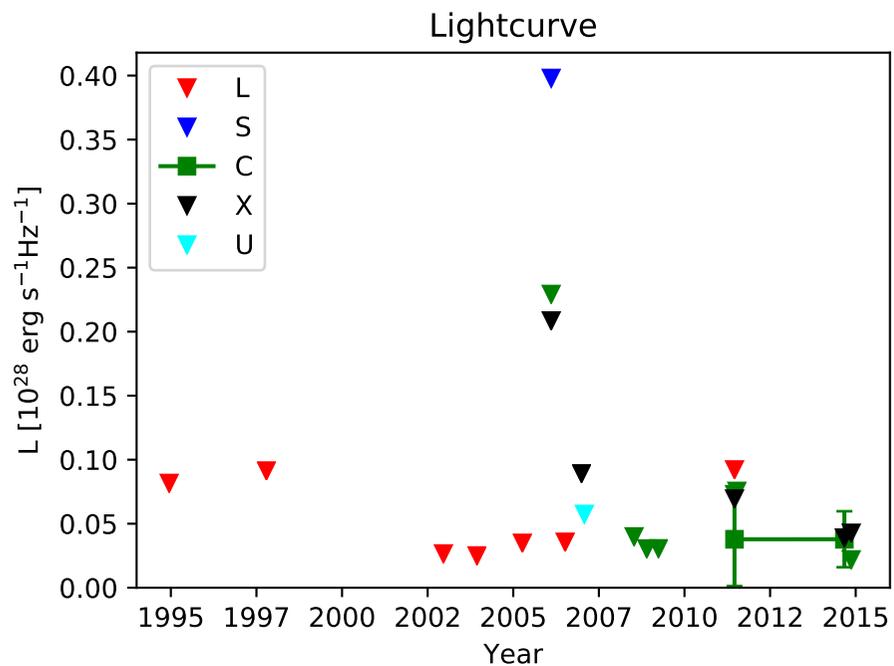

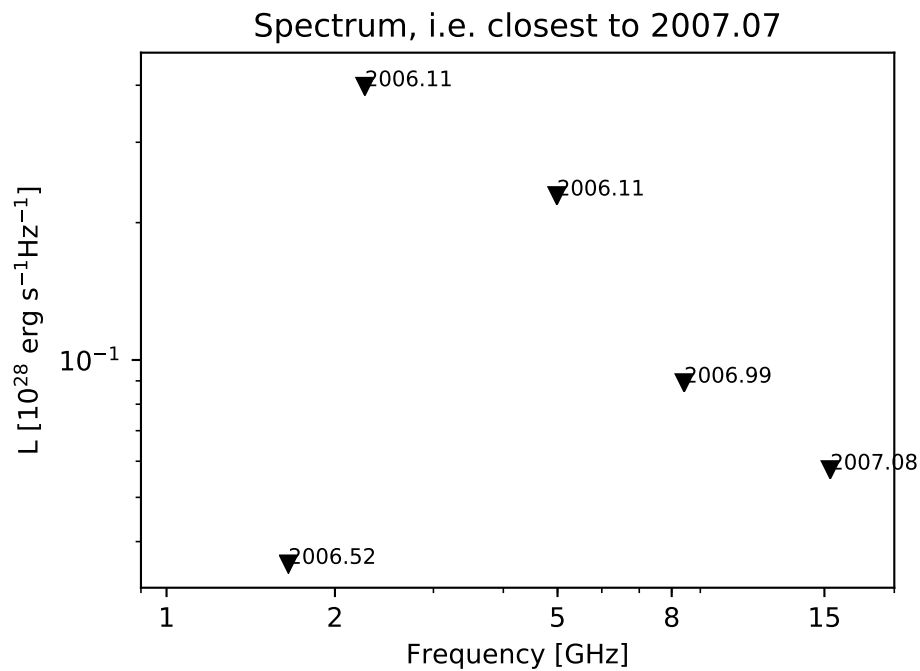

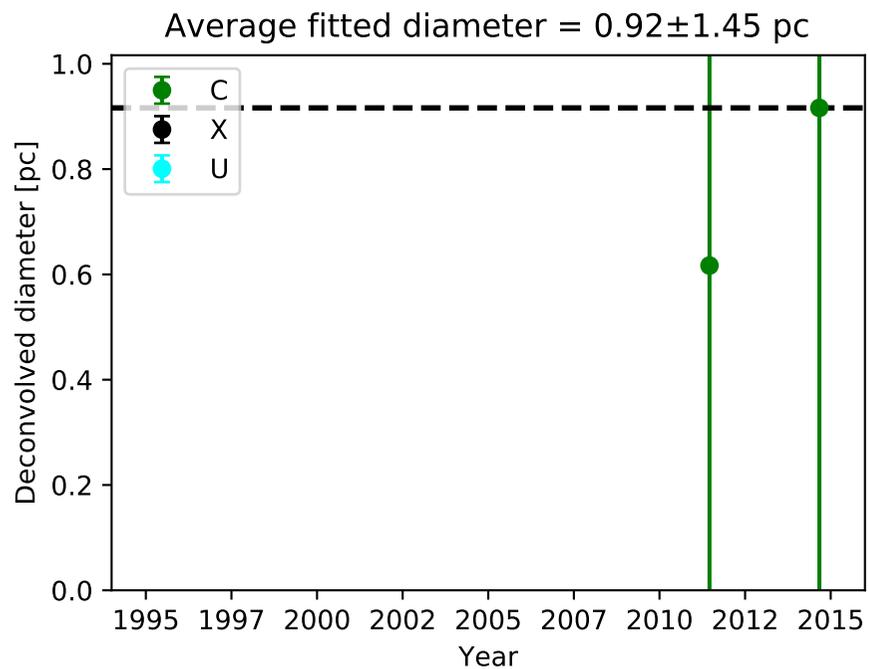

Classification: Slowly varying (S-var)

Figure notes:

Uncertainties calculated as described in Sect. 2.3.

Triangles represent 3σ upper limits for non-detections.

Sizes shown only for detections in bands C, X, and U.

Average diameters calculated according to Sect. 2.4 from the

3.6 cm and 6 cm measurements in experiments BB335A and BB335B.

# 0.2311+0.498

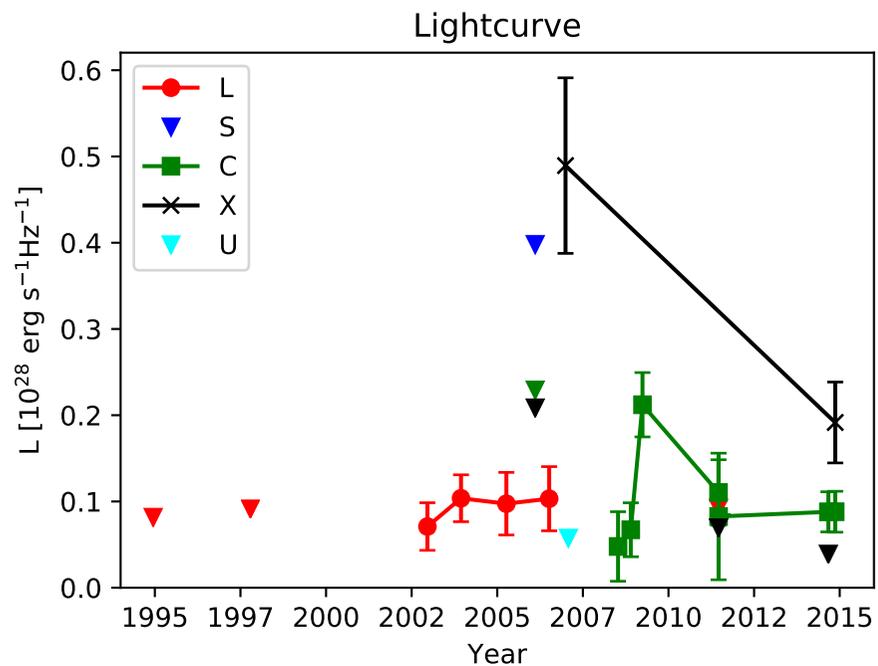
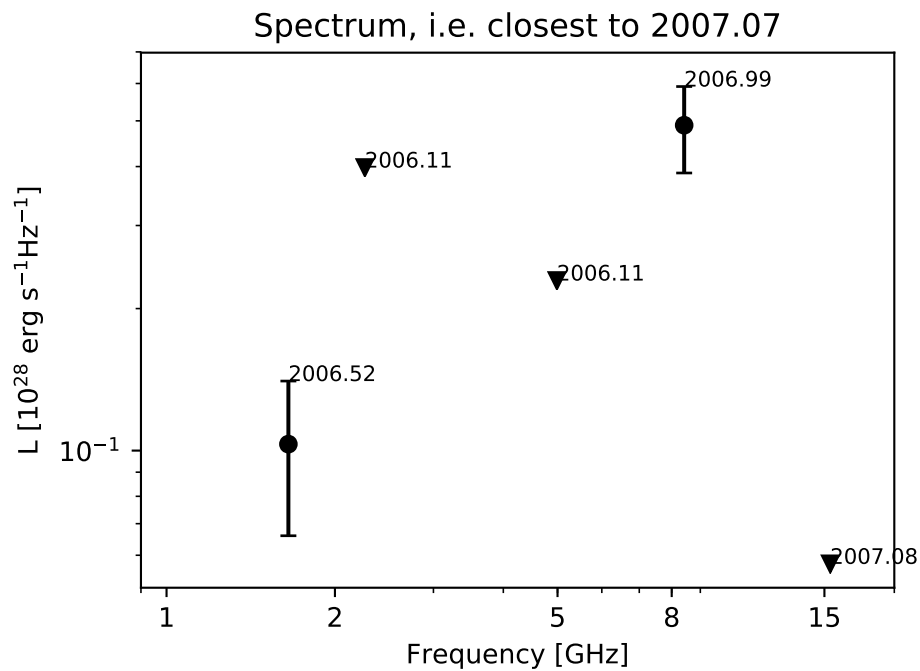
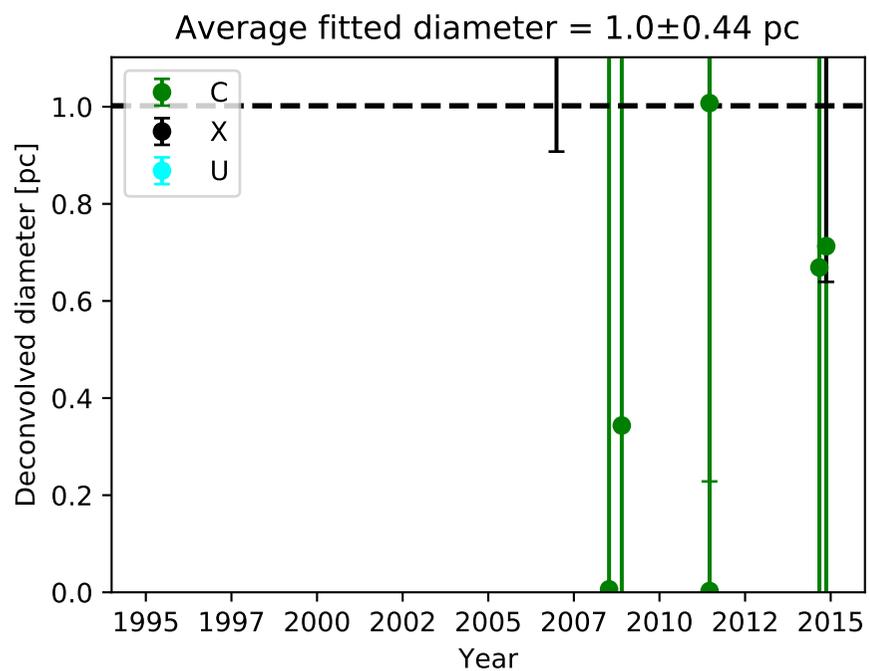

Classification: Slowly varying (S-var)

Figure notes:

Uncertainties calculated as described in Sect. 2.3.

Triangles represent $3\sigma$ upper limits for non-detections.

Sizes shown only for detections in bands C, X, and U.

Average diameters calculated according to Sect. 2.4 from the 3.6 cm and 6 cm measurements in experiments BB335A and BB335B.

# 0.2317+0.402 (W9)

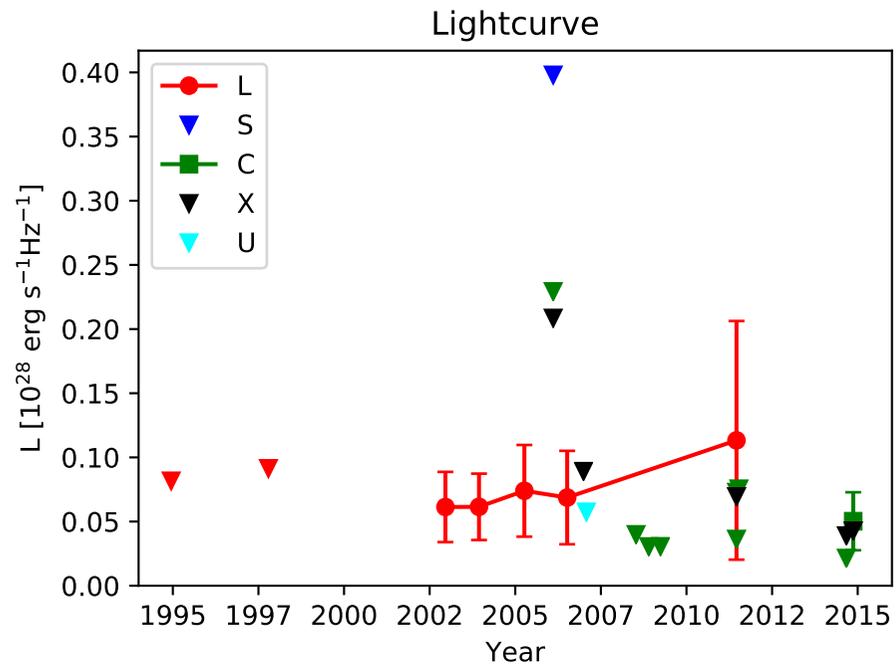
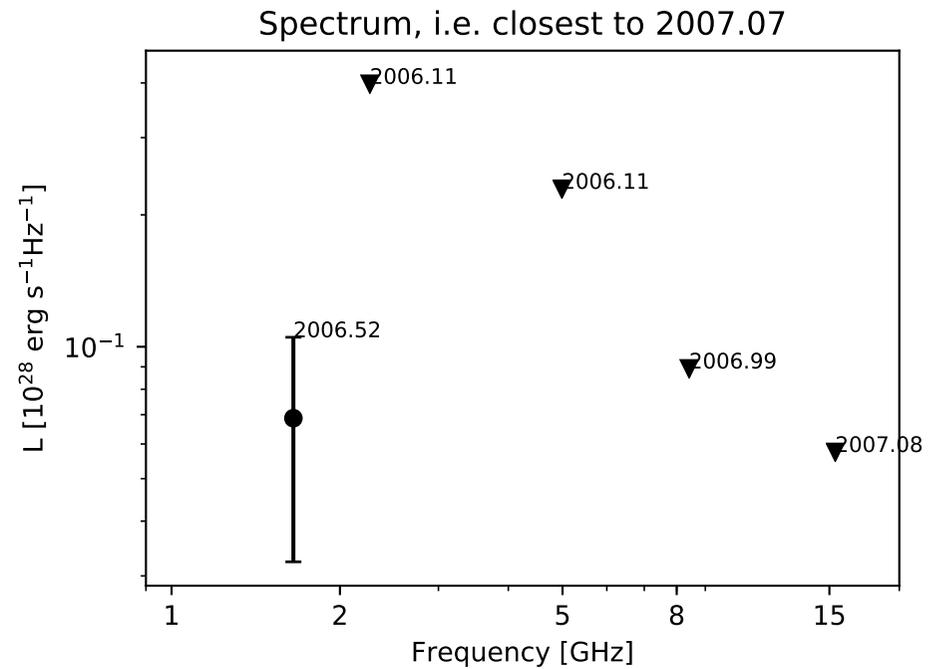
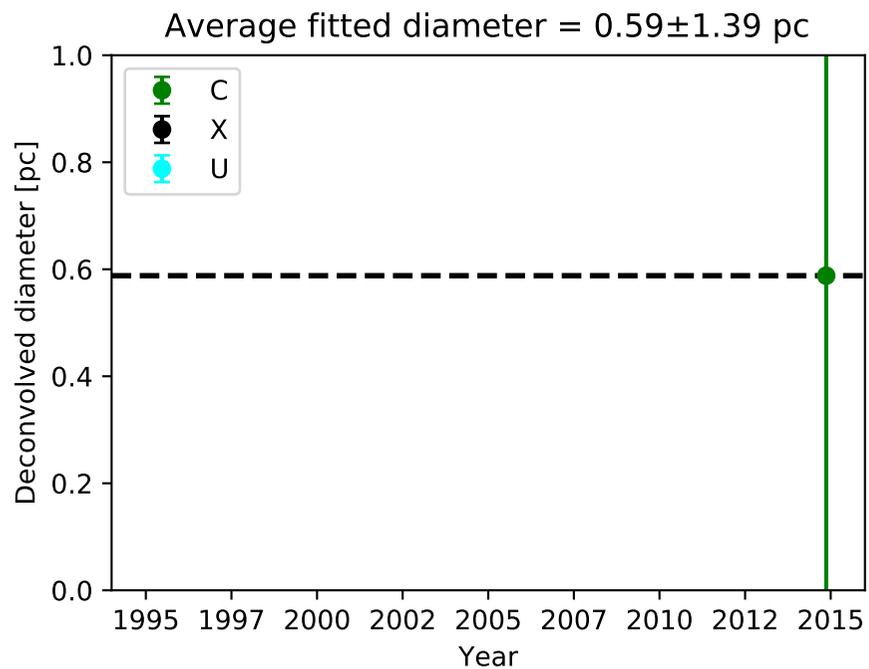

Classification: Slowly varying (S-var)

Figure notes:

Uncertainties calculated as described in Sect. 2.3.

Triangles represent 3σ upper limits for non-detections.

Sizes shown only for detections in bands C, X, and U.

Average diameters calculated according to Sect. 2.4 from the 3.6 cm and 6 cm measurements in experiments BB335A and BB335B.

# 0.2347+0.495

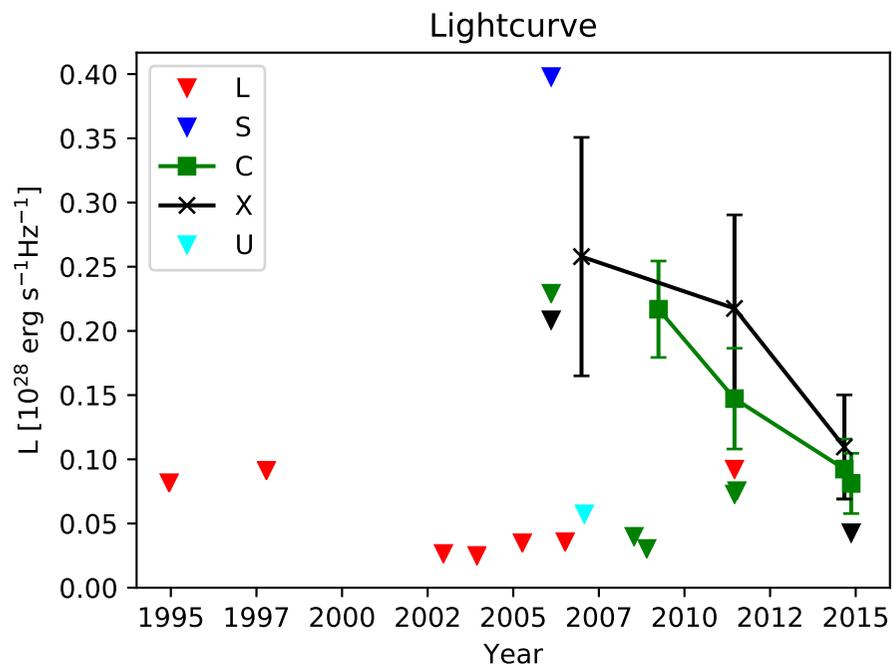
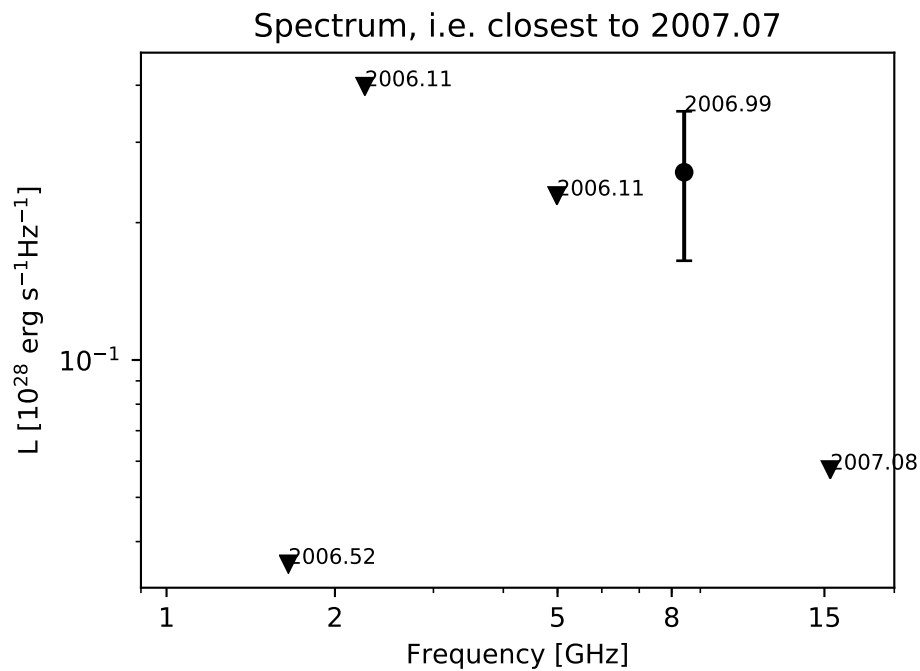
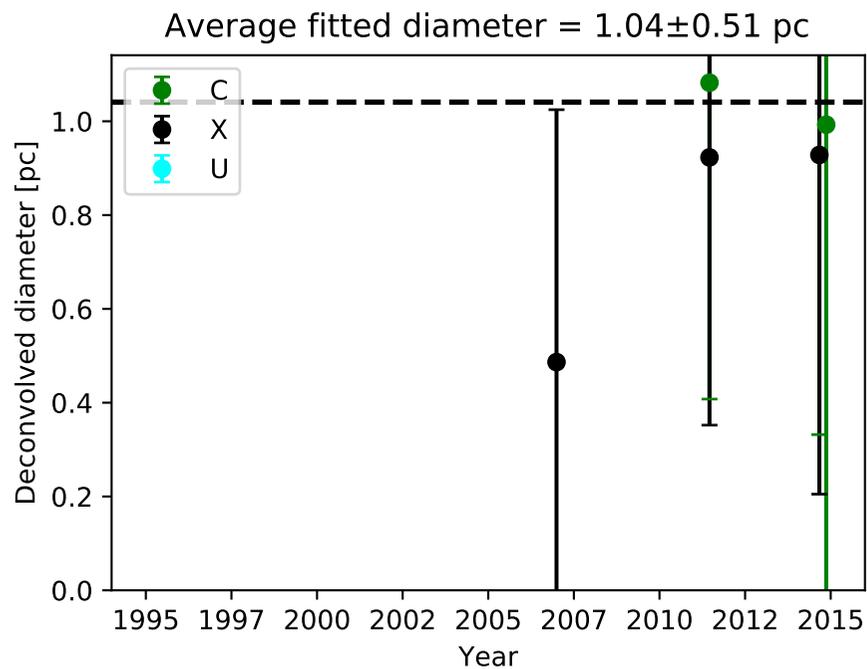

Classification: Slowly varying (S-var)

Figure notes:

Uncertainties calculated as described in Sect. 2.3.

Triangles represent $3\sigma$ upper limits for non-detections.

Sizes shown only for detections in bands C, X, and U.

Average diameters calculated according to Sect. 2.4 from the

3.6 cm and 6 cm measurements in experiments BB335A and BB335B.

# 0.2360+0.431 (W8)

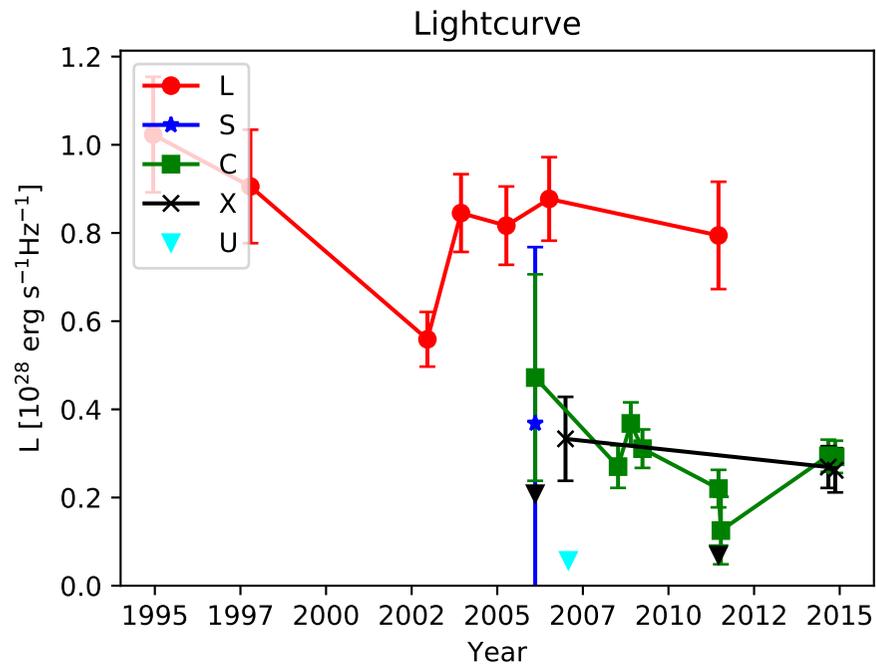
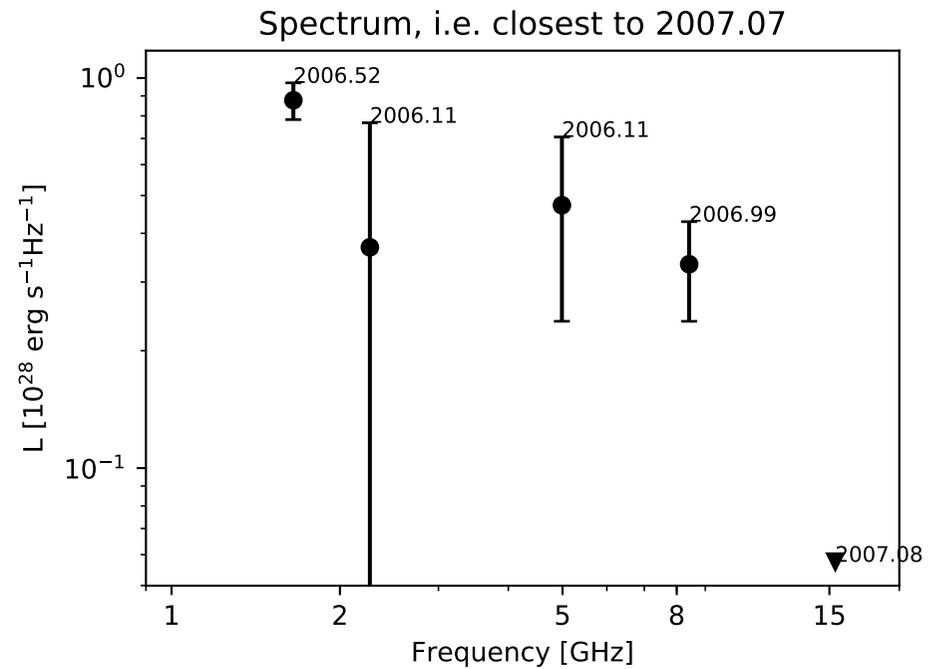
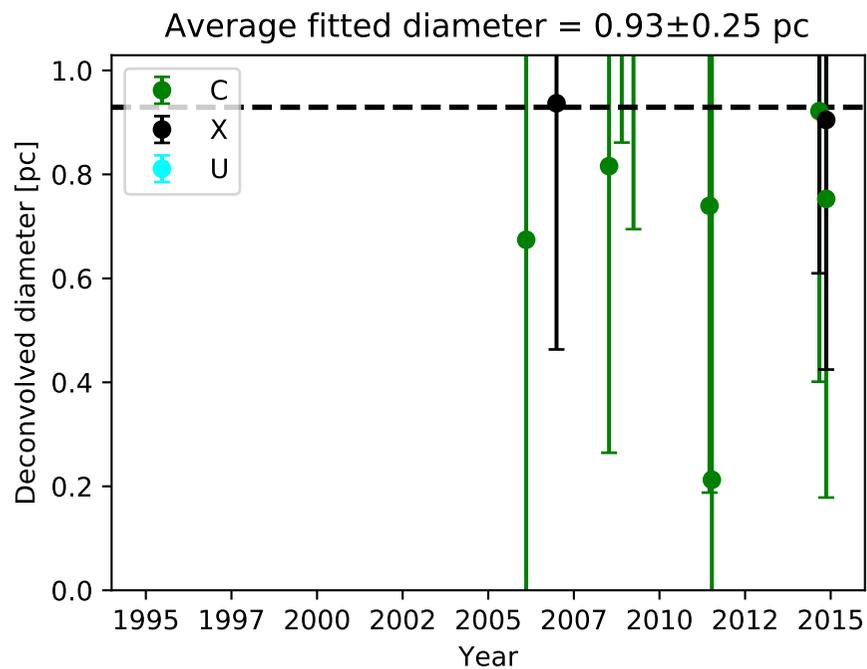

Classification: Slowly varying (S-var)

Figure notes:

Uncertainties calculated as described in Sect. 2.3.

Triangles represent 3σ upper limits for non-detections.

Sizes shown only for detections in bands C, X, and U.

Average diameters calculated according to Sect. 2.4 from the

3.6 cm and 6 cm measurements in experiments BB335A and BB335B.

# 0.2362+0.546 (W7)

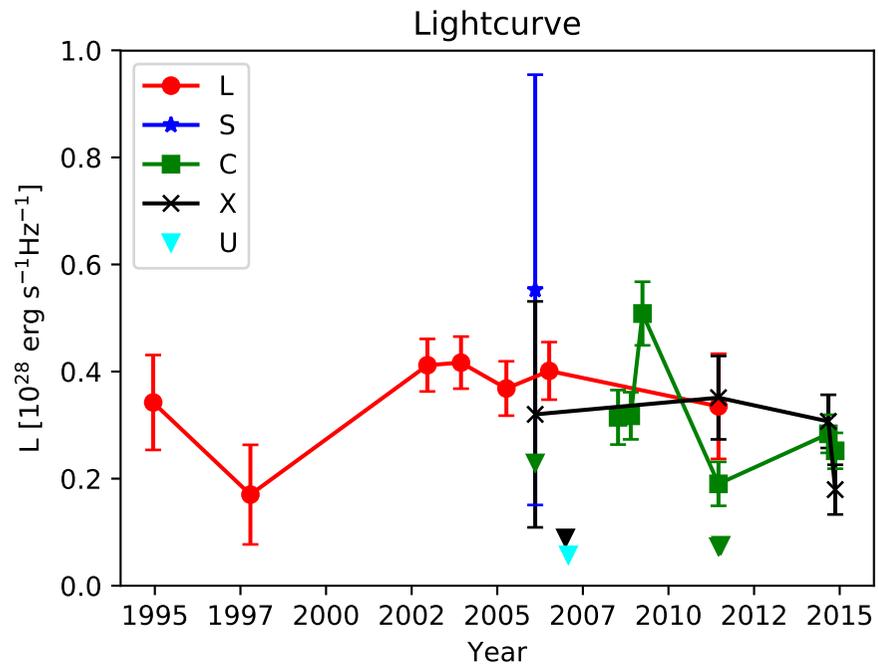

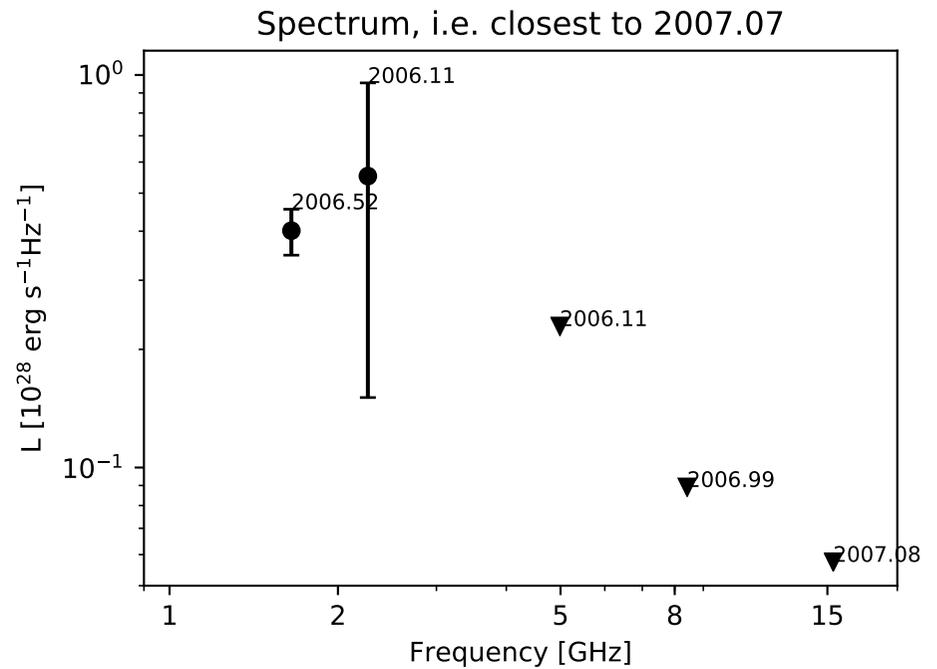

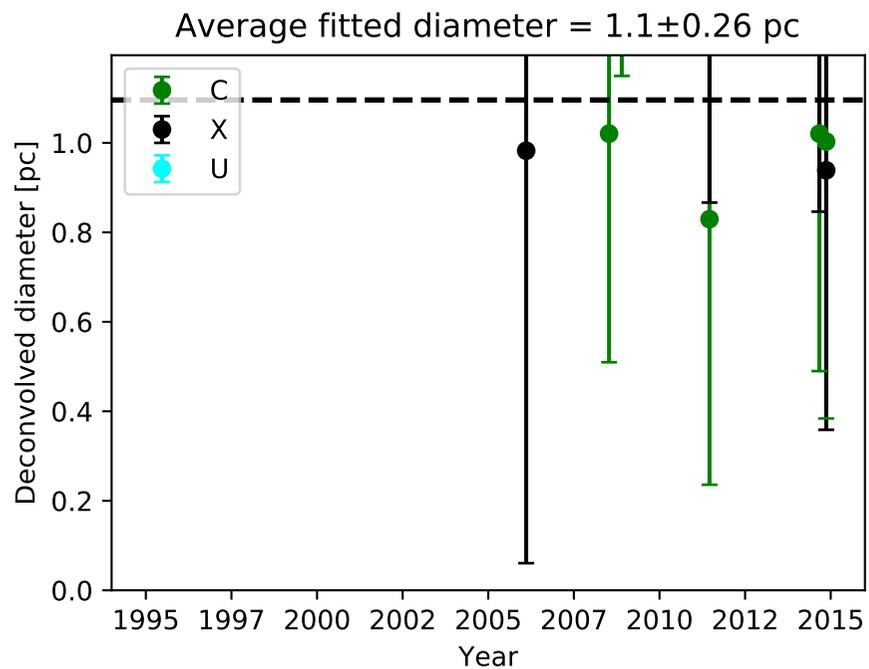

Classification: Fall

Figure notes:

Uncertainties calculated as described in Sect. 2.3.

Triangles represent $3\sigma$ upper limits for non-detections.

Sizes shown only for detections in bands C, X, and U.

Average diameters calculated according to Sect. 2.4 from the

3.6 cm and 6 cm measurements in experiments BB335A and BB335B.

# 0.2363+0.369

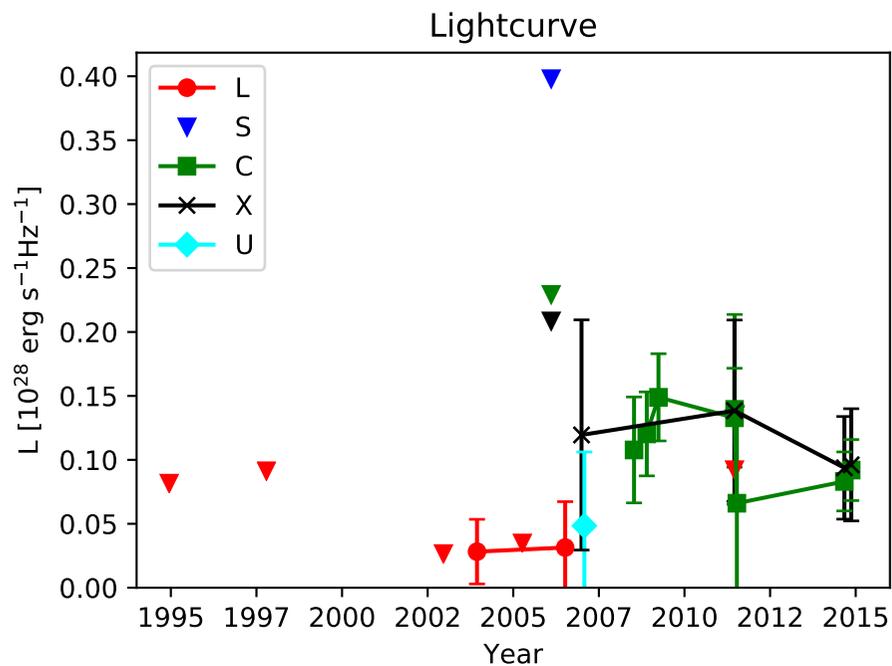
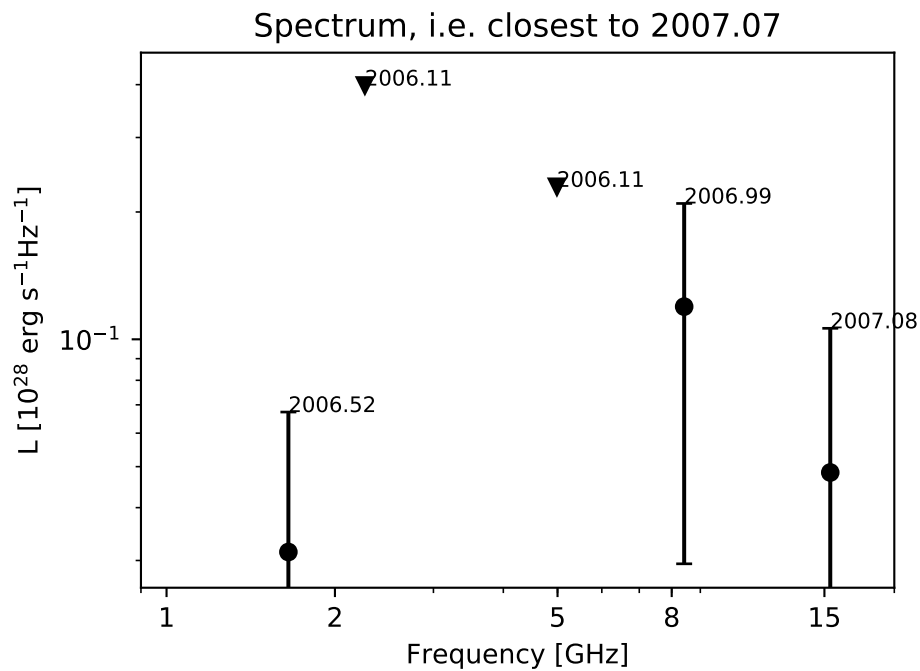
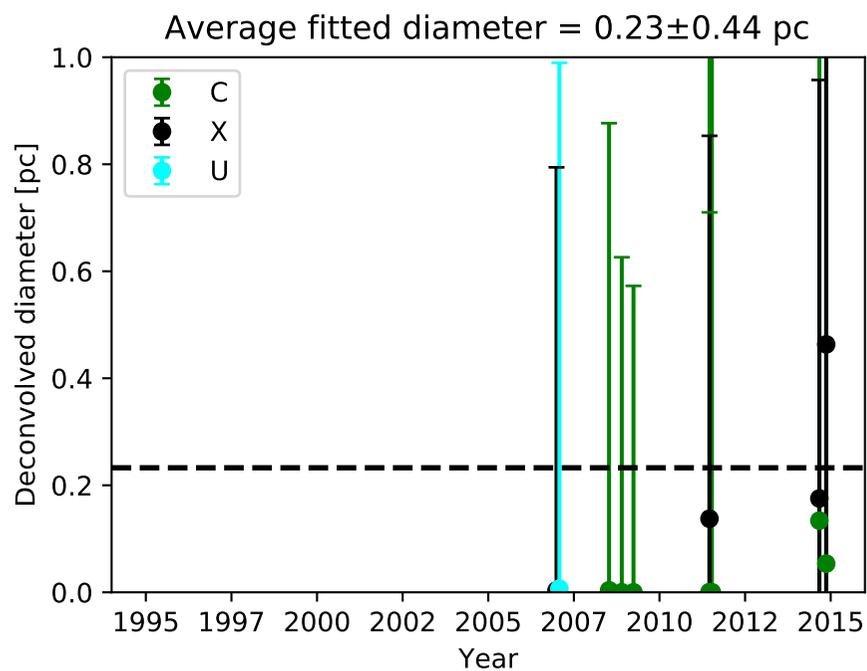

Classification: Slowly varying (S-var)

Figure notes:

Uncertainties calculated as described in Sect. 2.3.

Triangles represent $3\sigma$ upper limits for non-detections.

Sizes shown only for detections in bands C, X, and U.

Average diameters calculated according to Sect. 2.4 from the

3.6 cm and 6 cm measurements in experiments BB335A and BB335B.

# 0.2364+0.639

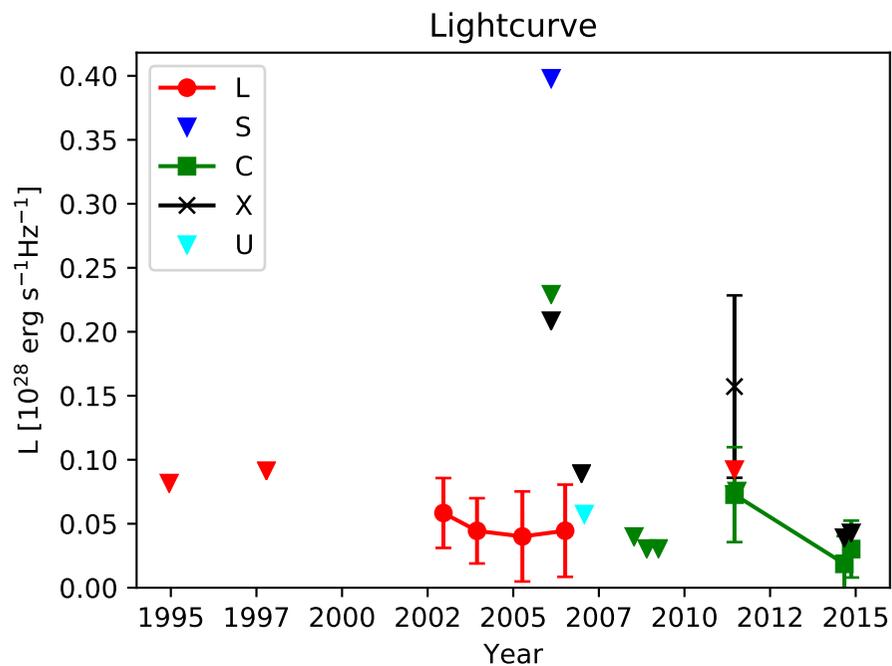
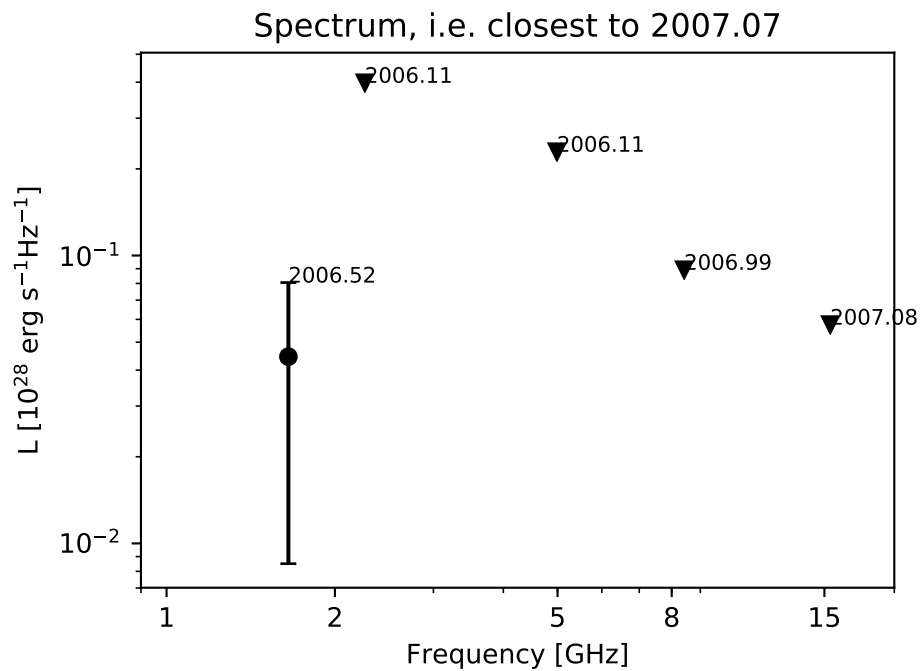
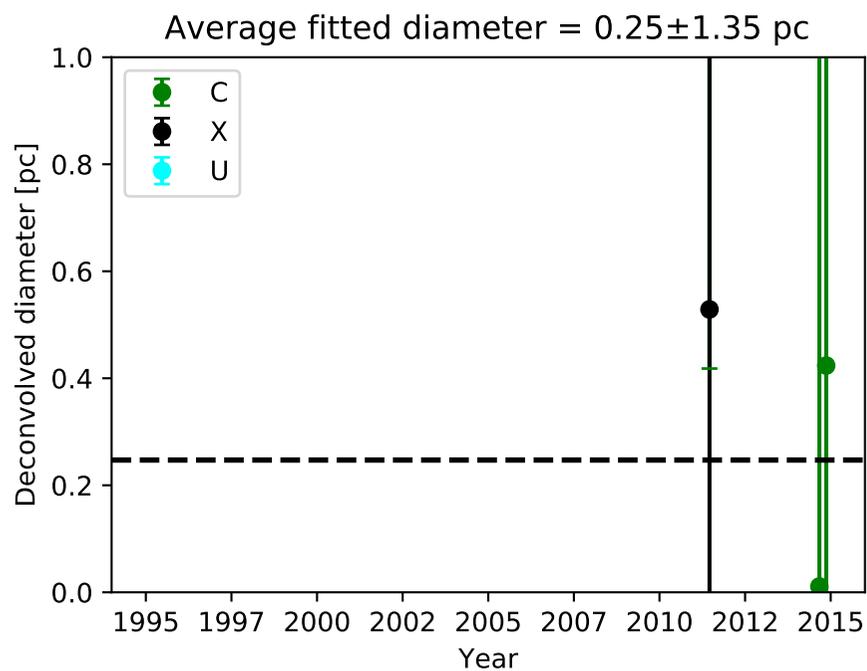

Classification: Slowly varying (S-var)

Figure notes:

Uncertainties calculated as described in Sect. 2.3.

Triangles represent 3σ upper limits for non-detections.

Sizes shown only for detections in bands C, X, and U.

Average diameters calculated according to Sect. 2.4 from the

3.6 cm and 6 cm measurements in experiments BB335A and BB335B.

# 0.2394+0.540 (W2)

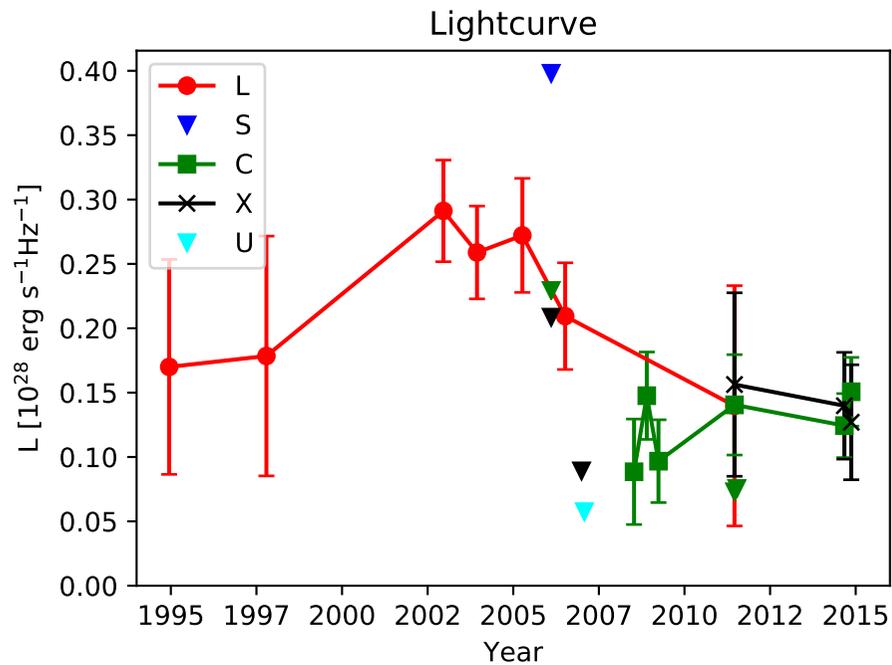
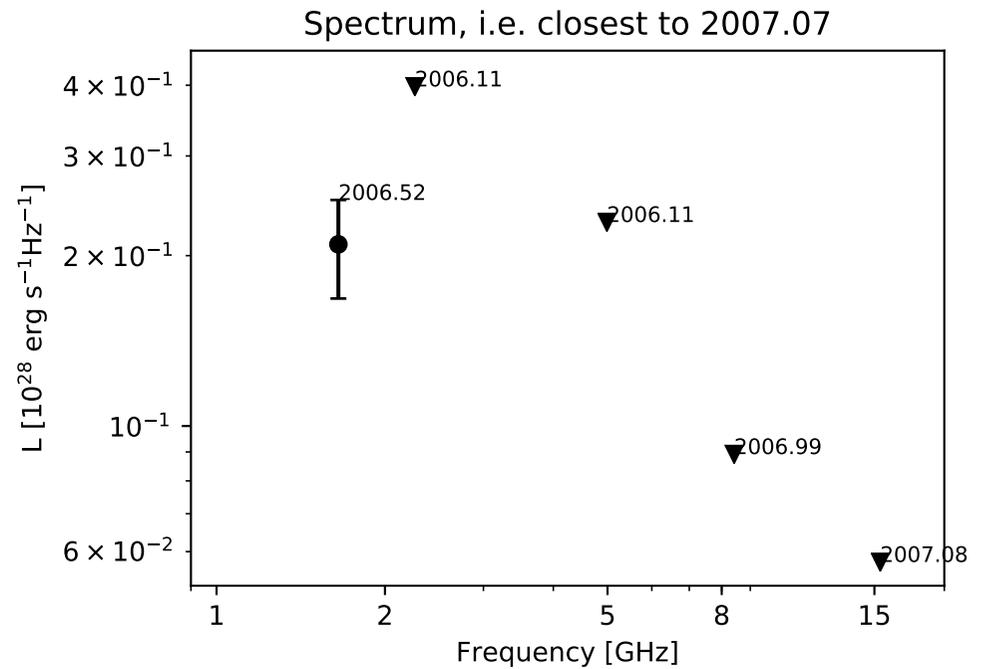
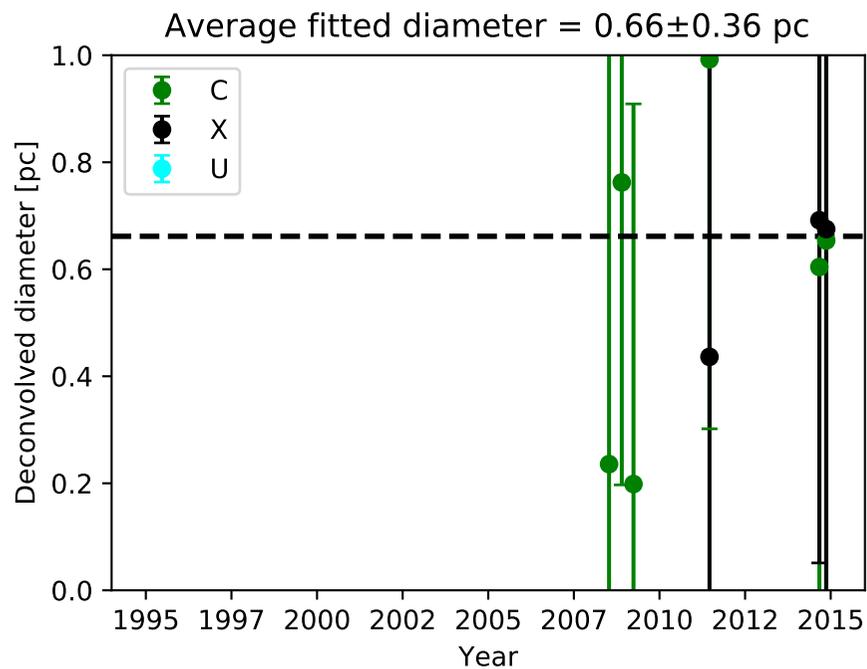

Classification: Slowly varying (S-var)

Figure notes:

Uncertainties calculated as described in Sect. 2.3.

Triangles represent $3\sigma$ upper limits for non-detections.

Sizes shown only for detections in bands C, X, and U.

Average diameters calculated according to Sect. 2.4 from the 3.6 cm and 6 cm measurements in experiments BB335A and BB335B.

# 0.2395+0.637

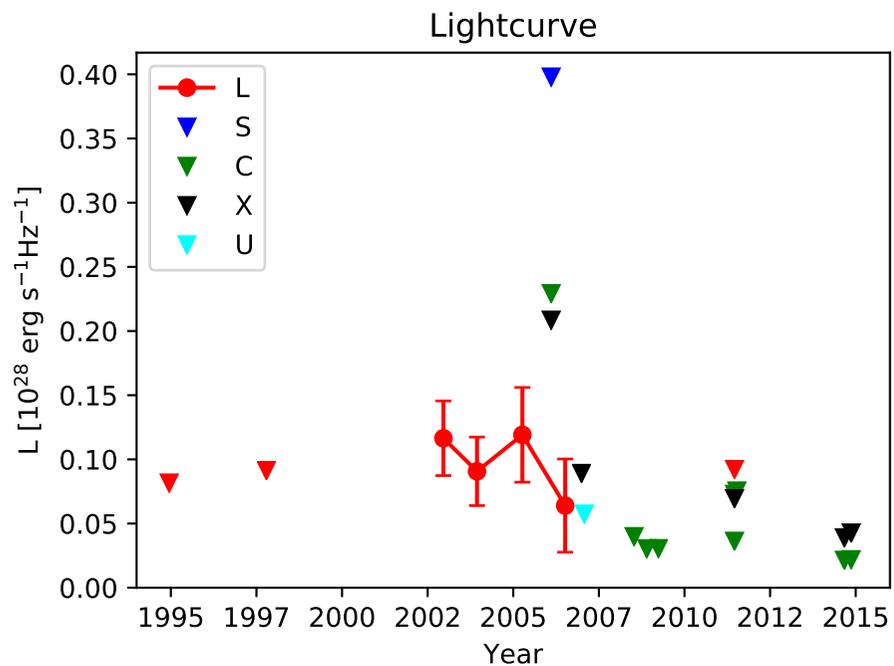
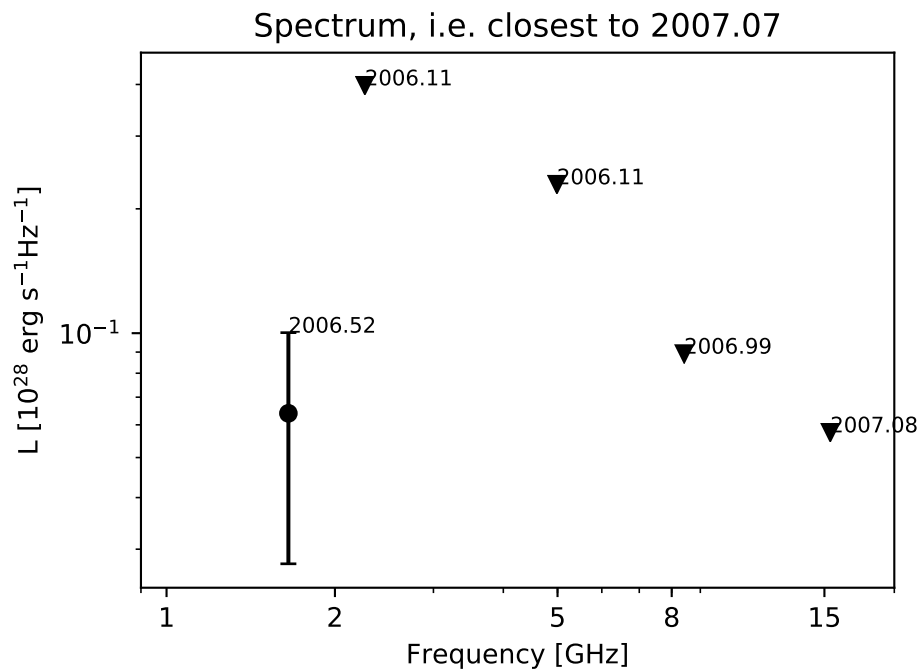
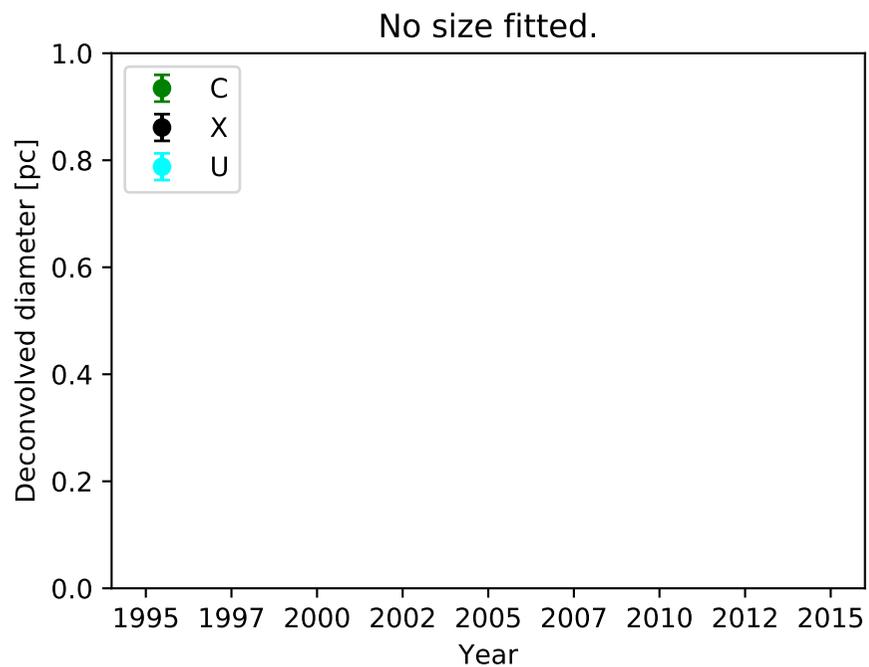

Classification: N/A

Figure notes:

Uncertainties calculated as described in Sect. 2.3.

Triangles represent 3σ upper limits for non-detections.

Sizes shown only for detections in bands C,X, and U.

Average diameters calculated according to Sect. 2.4 from the

3.6 cm and 6 cm measurements in experiments BB335A and BB335B.

# 0.2409+0.540

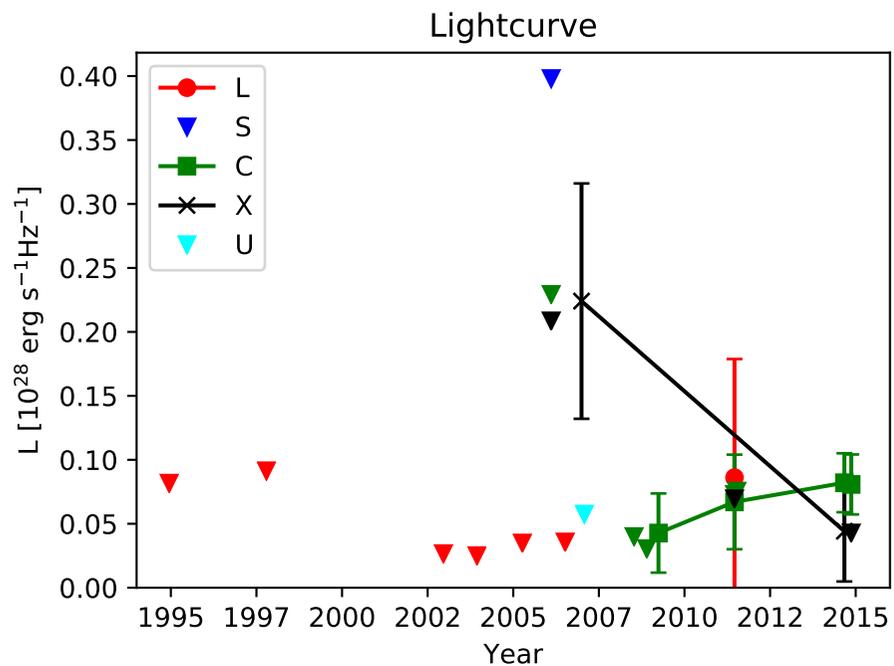
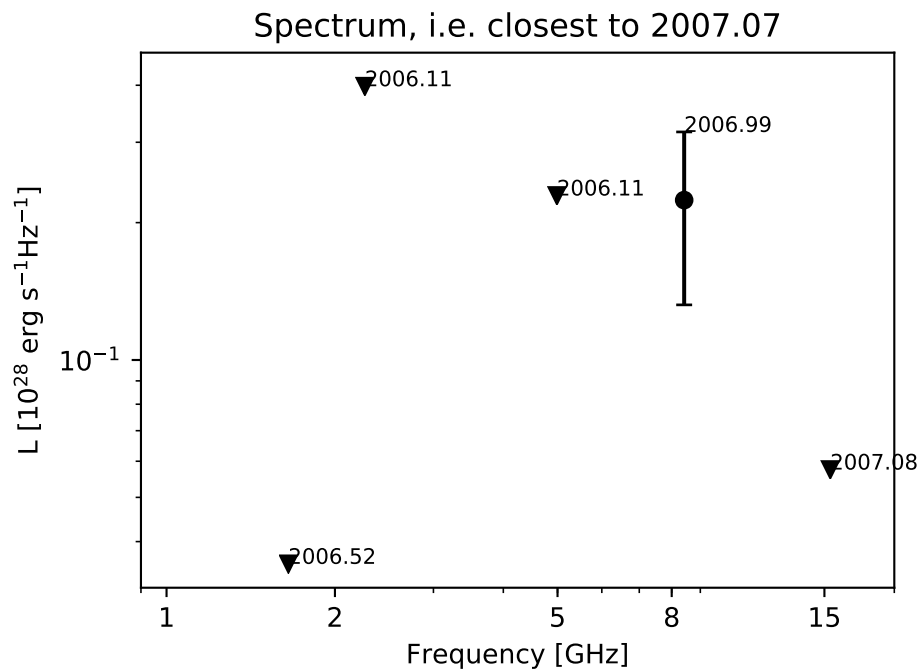
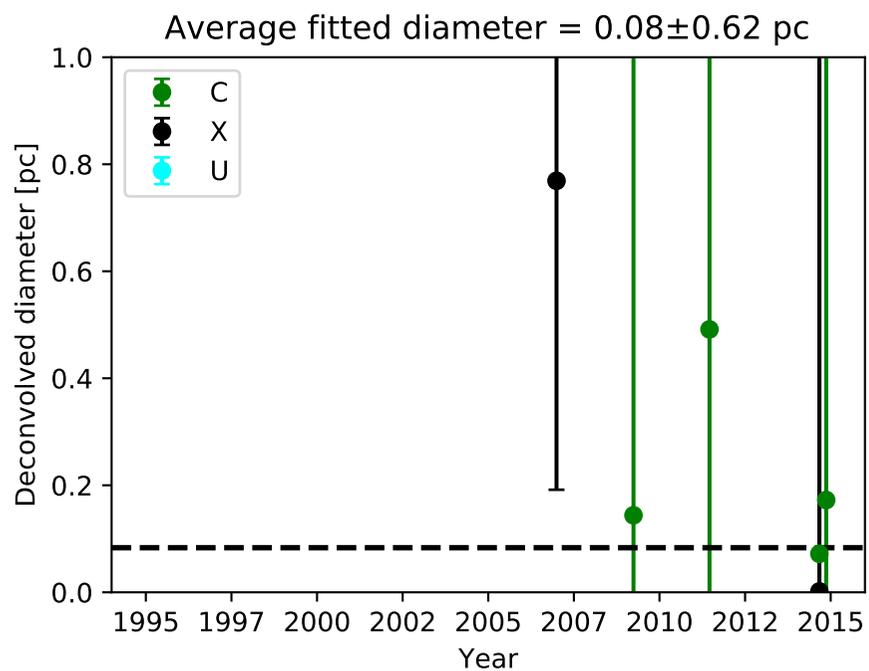

Classification: Rise

Figure notes:

Uncertainties calculated as described in Sect. 2.3.

Triangles represent 3σ upper limits for non-detections.

Sizes shown only for detections in bands C,X, and U.

Average diameters calculated according to Sect. 2.4 from the

3.6 cm and 6 cm measurements in experiments BB335A and BB335B.

# 0.2758+0.276 (E22)

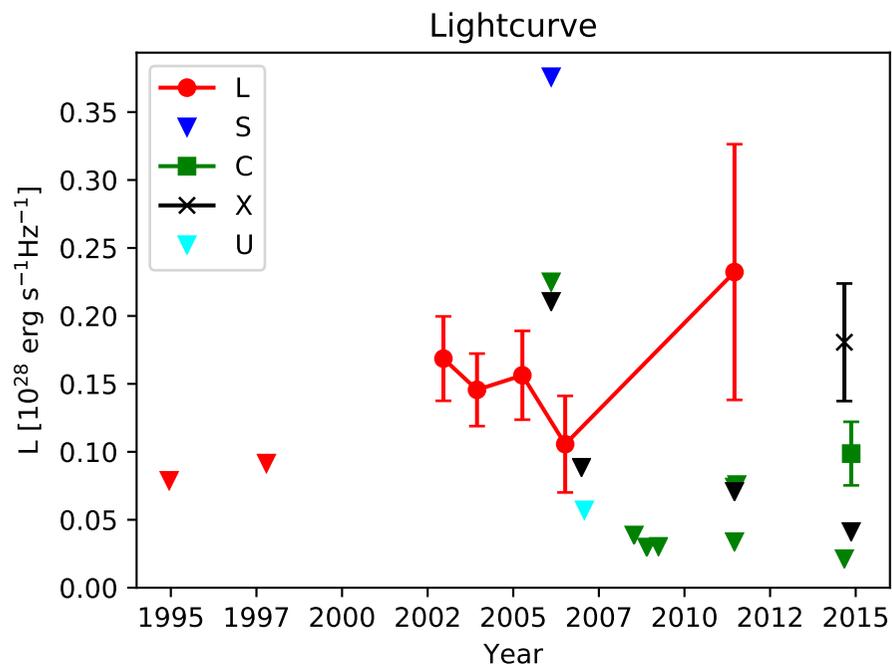
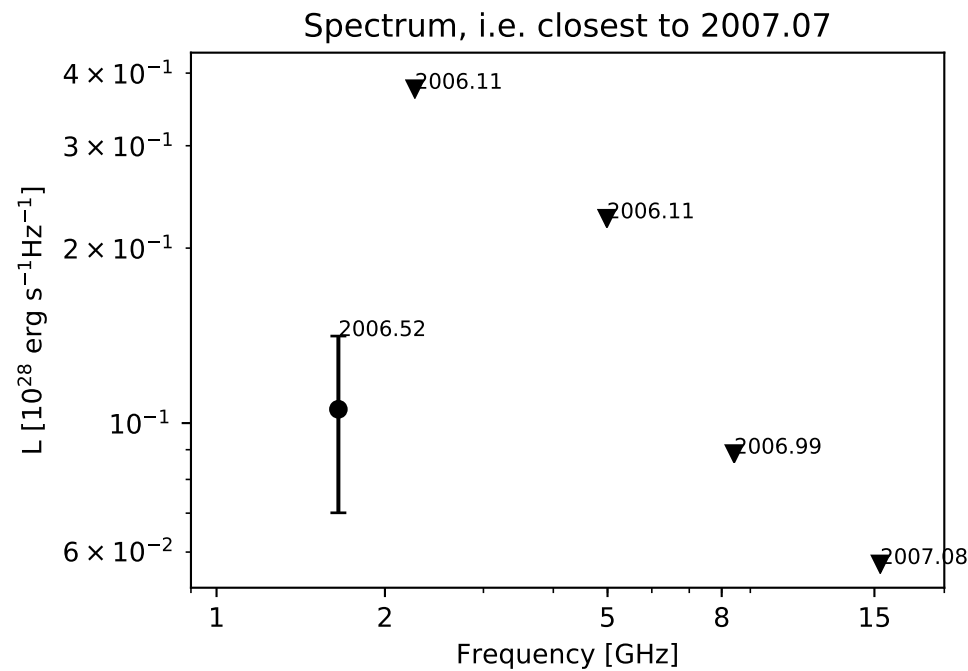
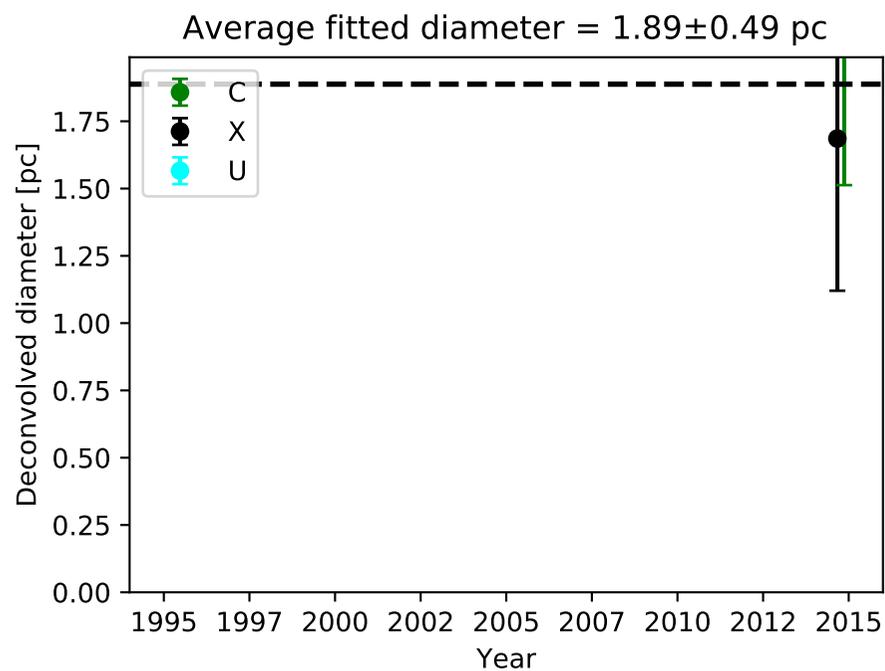

Classification: Rise

Figure notes:

Uncertainties calculated as described in Sect. 2.3.

Triangles represent 3σ upper limits for non-detections.

Sizes shown only for detections in bands C, X, and U.

Average diameters calculated according to Sect. 2.4 from the

3.6 cm and 6 cm measurements in experiments BB335A and BB335B.

# 0.2789+0.293 (E21)

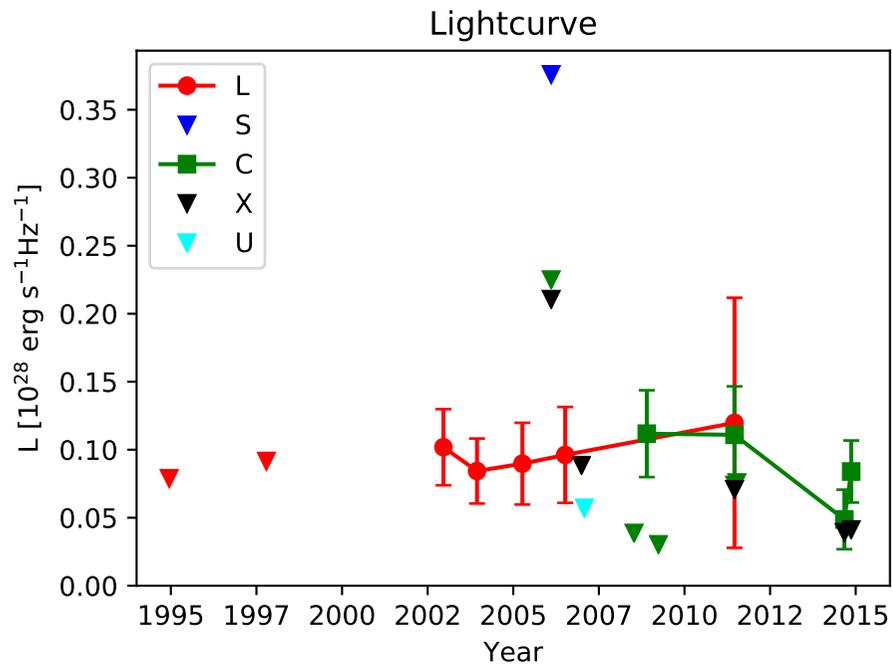
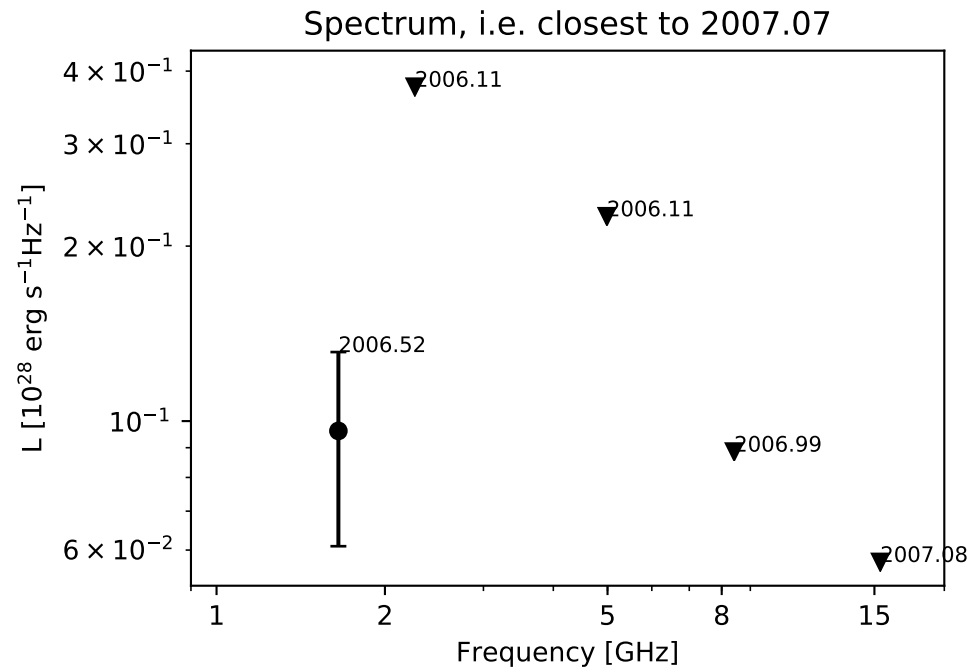
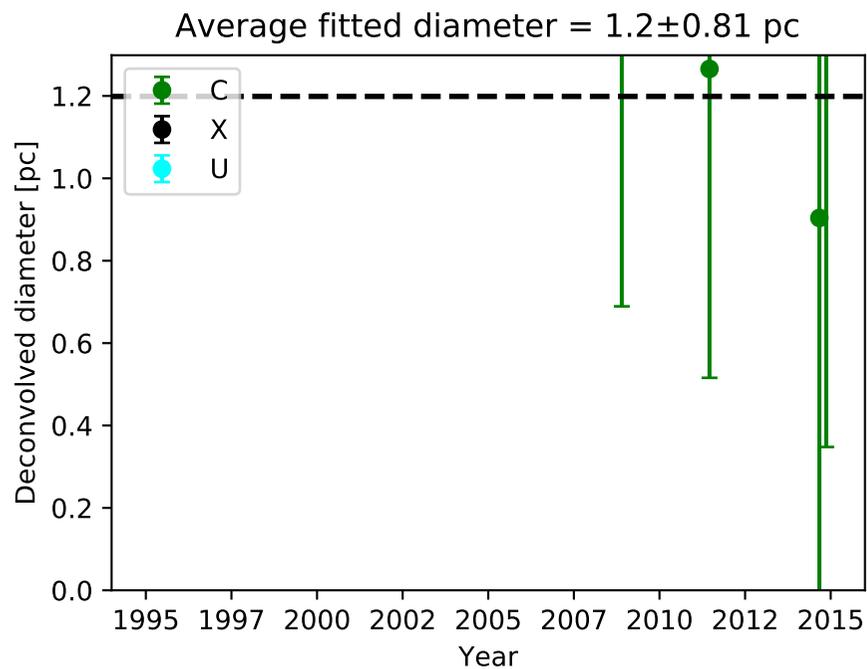

Classification: Slowly varying (S-var)

Figure notes:

Uncertainties calculated as described in Sect. 2.3.

Triangles represent 3σ upper limits for non-detections.

Sizes shown only for detections in bands C, X, and U.

Average diameters calculated according to Sect. 2.4 from the

3.6 cm and 6 cm measurements in experiments BB335A and BB335B.

# 0.2801+0.353

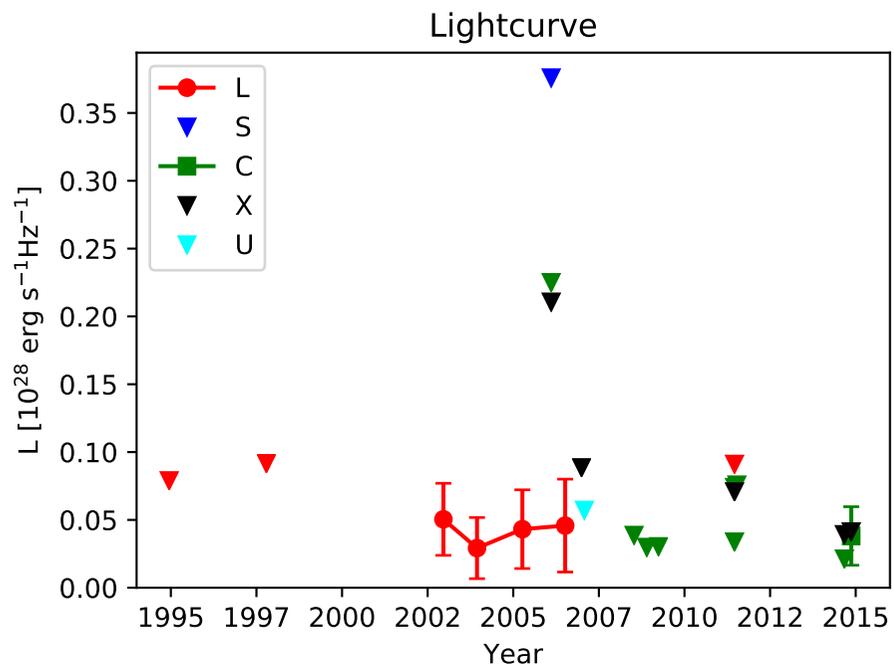
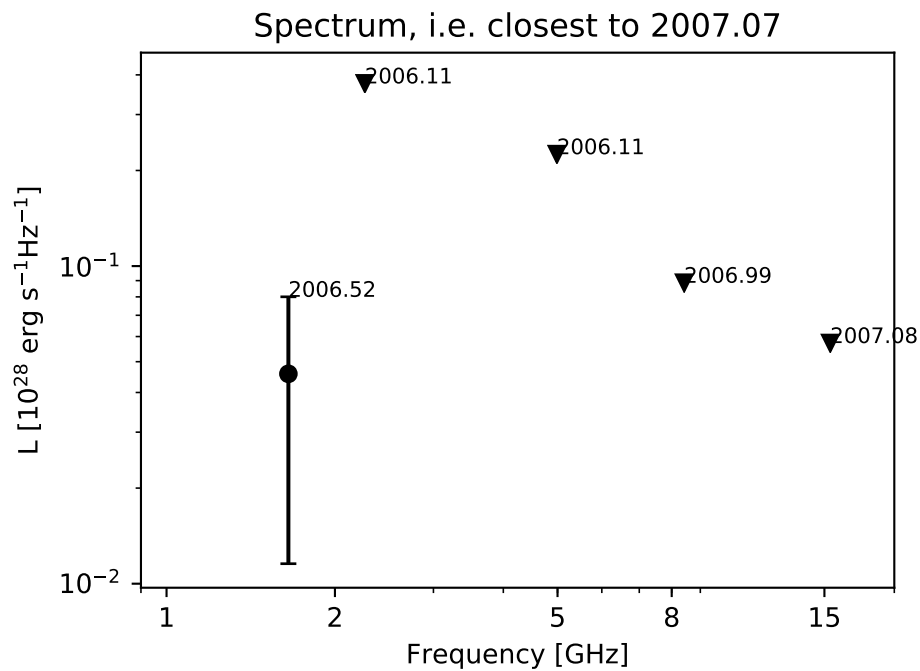
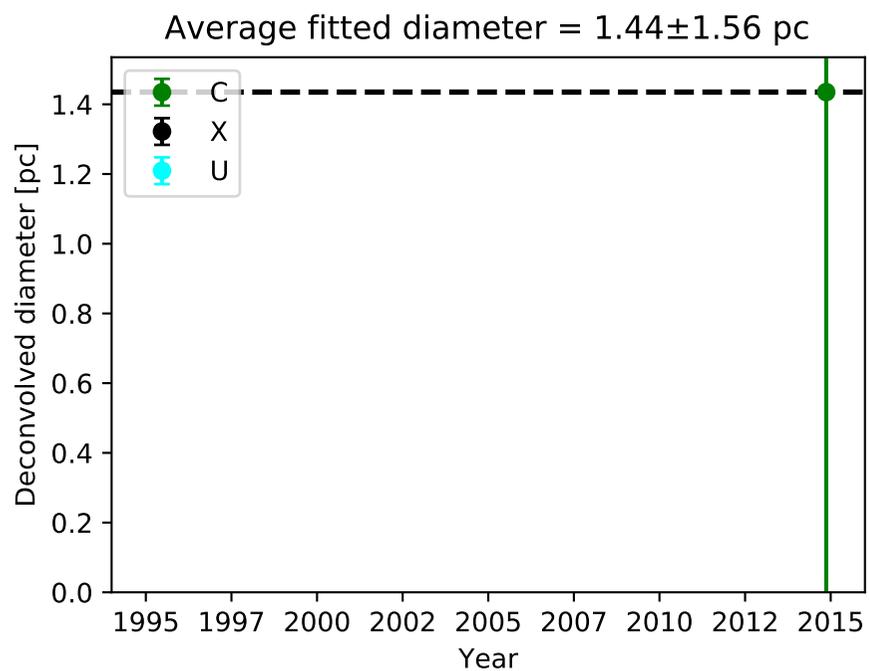

Classification: Slowly varying (S-var)

Figure notes:

Uncertainties calculated as described in Sect. 2.3.

Triangles represent 3σ upper limits for non-detections.

Sizes shown only for detections in bands C, X, and U.

Average diameters calculated according to Sect. 2.4 from the

3.6 cm and 6 cm measurements in experiments BB335A and BB335B.

# 0.2814+0.316

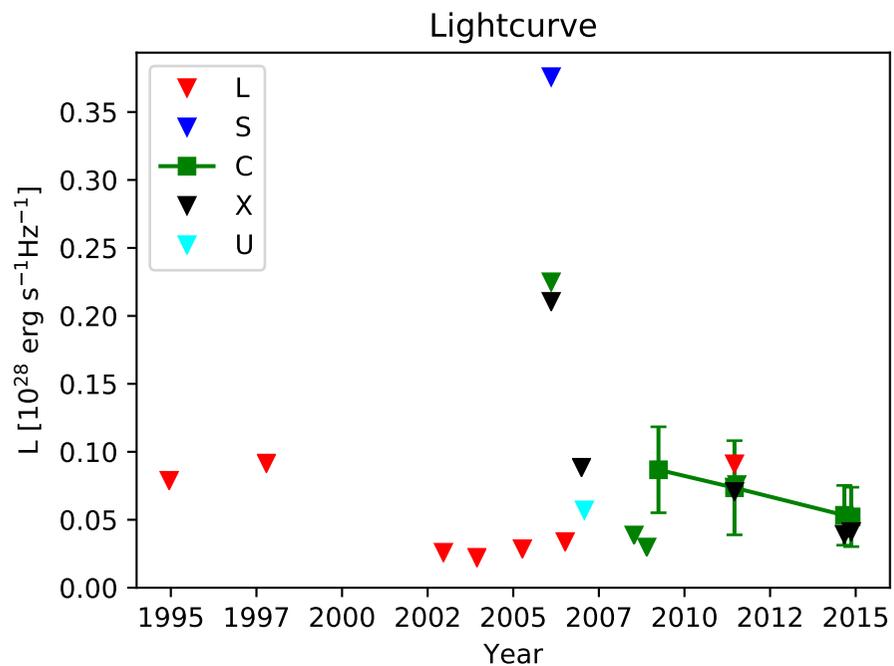

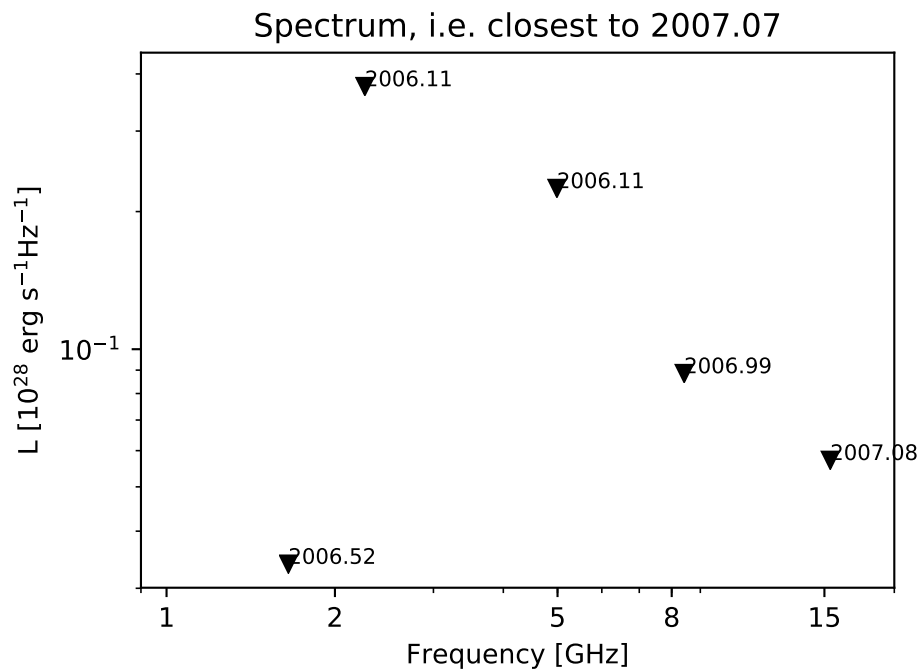

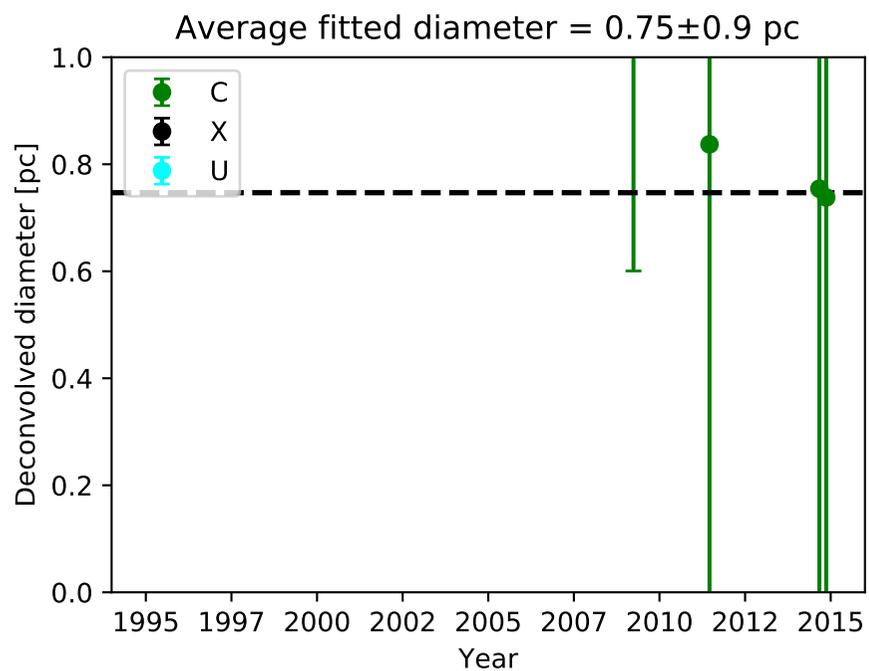

Classification: Slowly varying (S-var)

Figure notes:

Uncertainties calculated as described in Sect. 2.3.

Triangles represent 3σ upper limits for non-detections.

Sizes shown only for detections in bands C,X, and U.

Average diameters calculated according to Sect. 2.4 from the

3.6 cm and 6 cm measurements in experiments BB335A and BB335B.

# 0.2821+0.183 (E20)

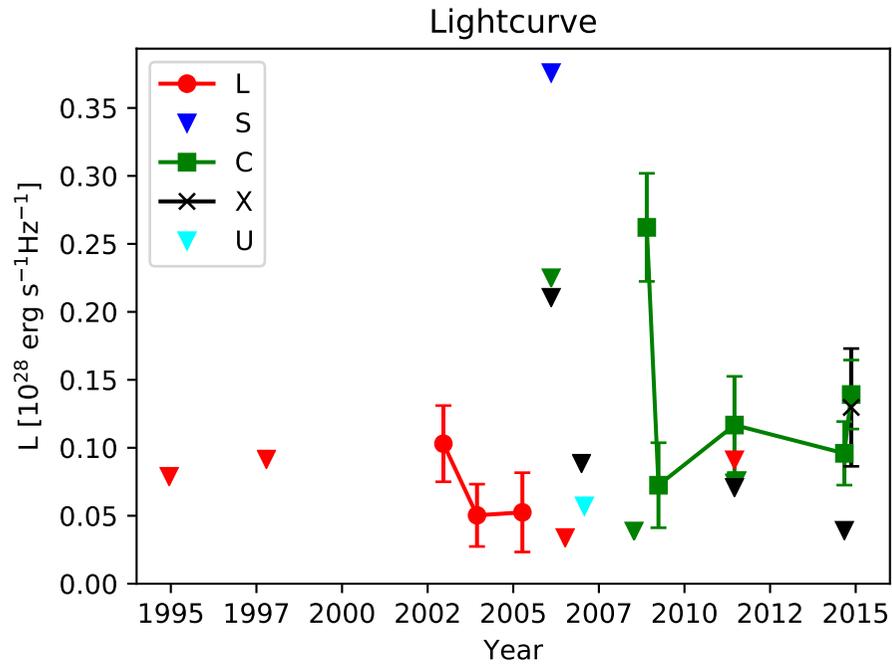

Lightcurve

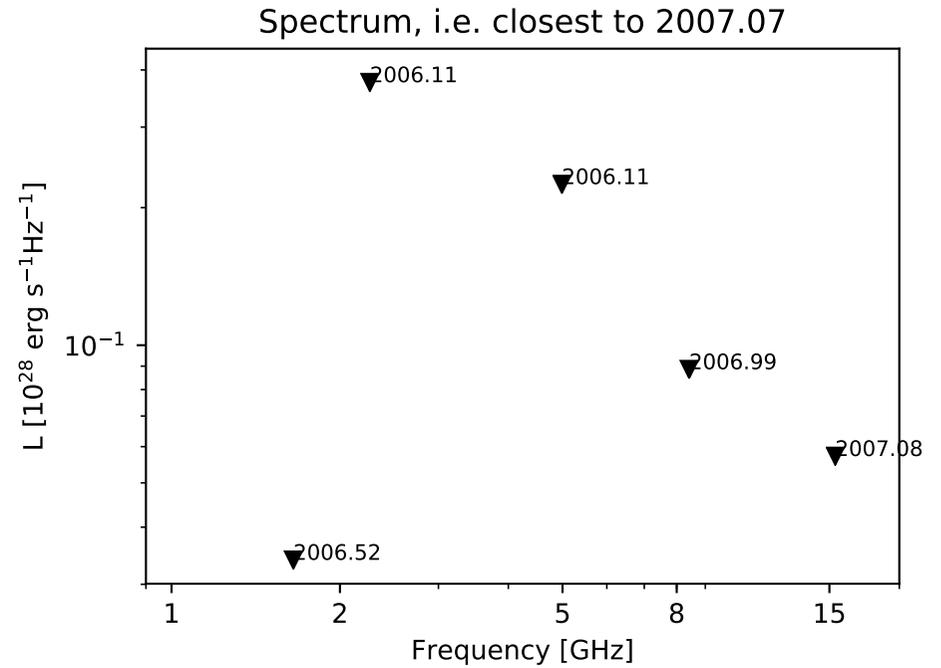

Spectrum, i.e. closest to 2007.07

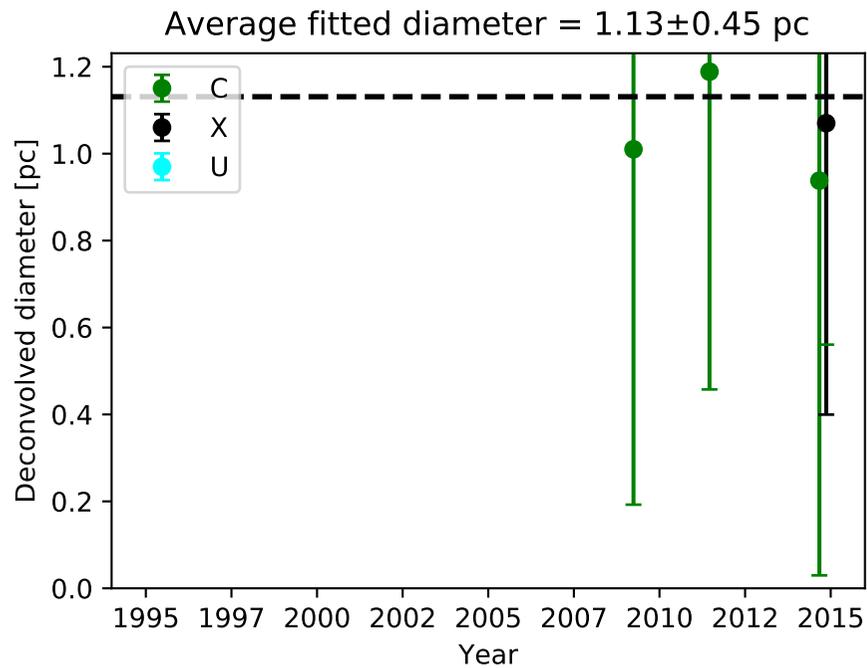

Average fitted diameter = 1.13±0.45 pc

Classification: Slowly varying (S-var)

Figure notes:

Uncertainties calculated as described in Sect. 2.3.

Triangles represent 3$\sigma$ upper limits for non-detections.

Sizes shown only for detections in bands C,X, and U.

Average diameters calculated according to Sect. 2.4 from the

3.6 cm and 6 cm measurements in experiments BB335A and BB335B.

# 0.2835+0.308

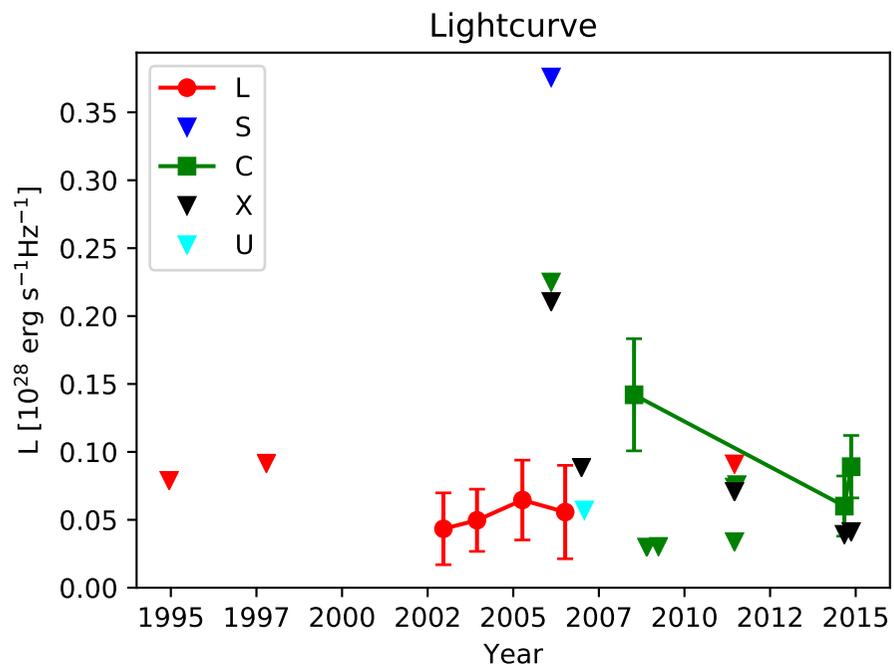

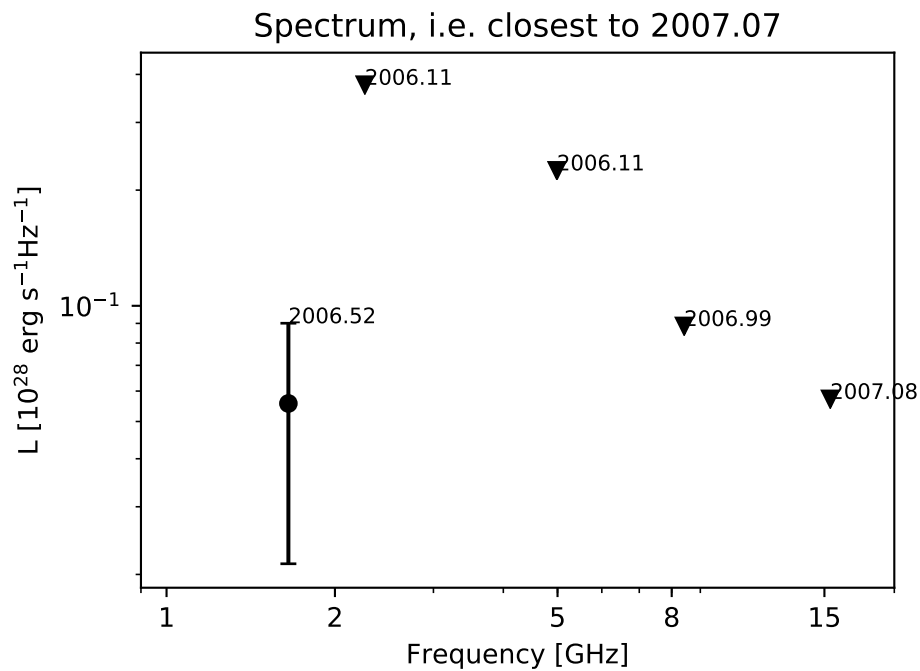

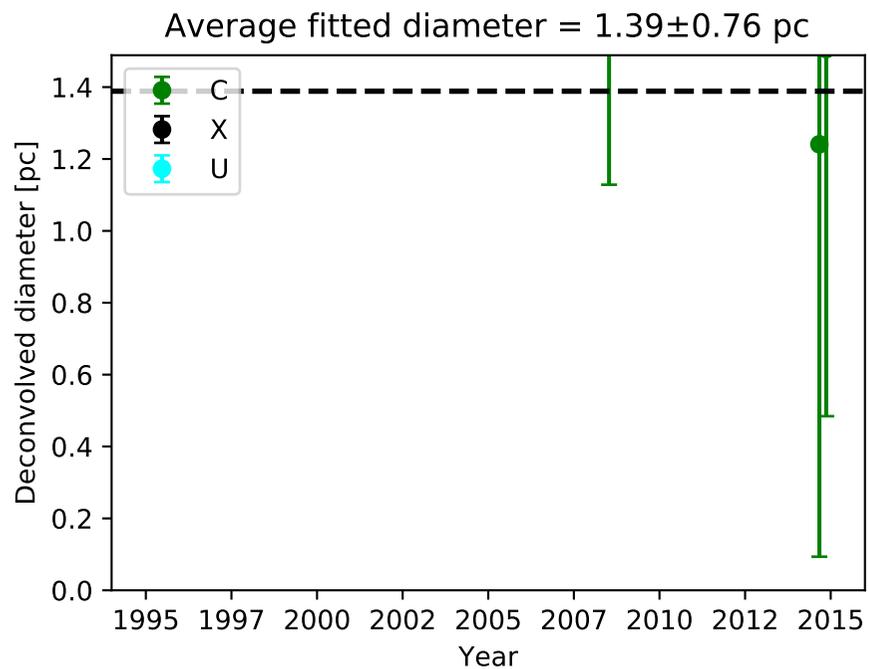

Classification: Rise

Figure notes:

Uncertainties calculated as described in Sect. 2.3.

Triangles represent 3σ upper limits for non-detections.

Sizes shown only for detections in bands C, X, and U.

Average diameters calculated according to Sect. 2.4 from the

3.6 cm and 6 cm measurements in experiments BB335A and BB335B.

# 0.2840+0.151 (E19)

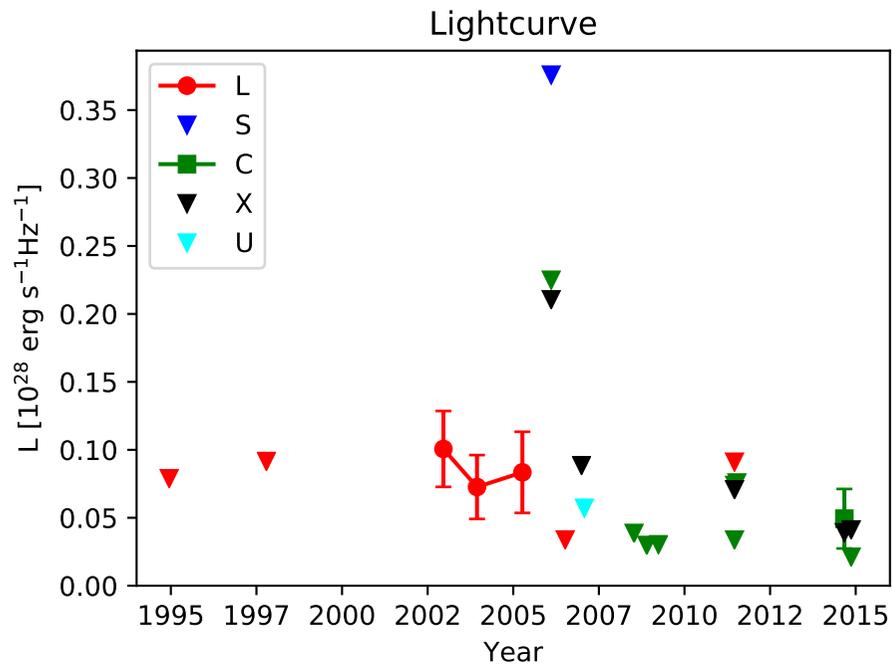
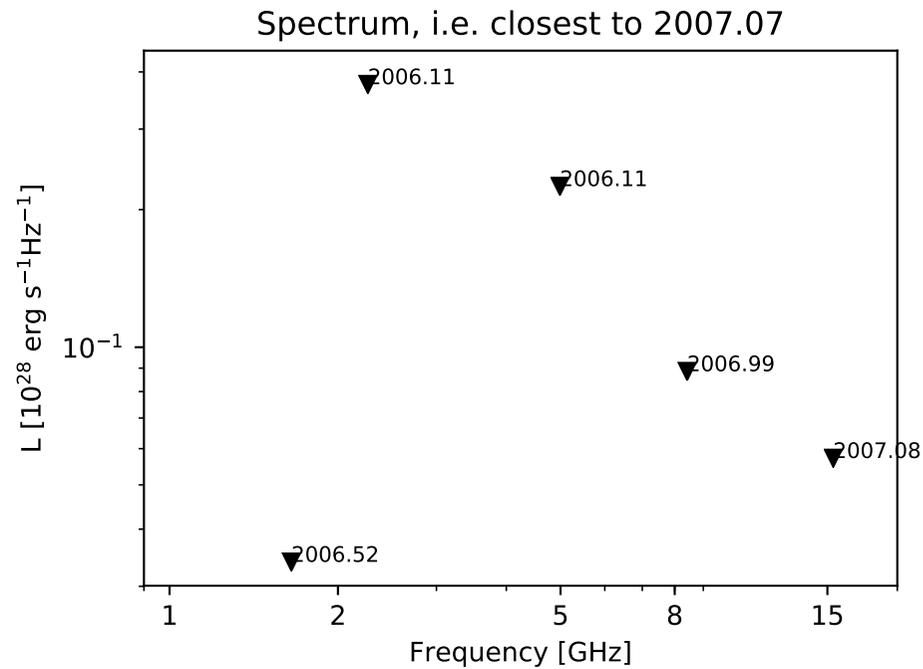
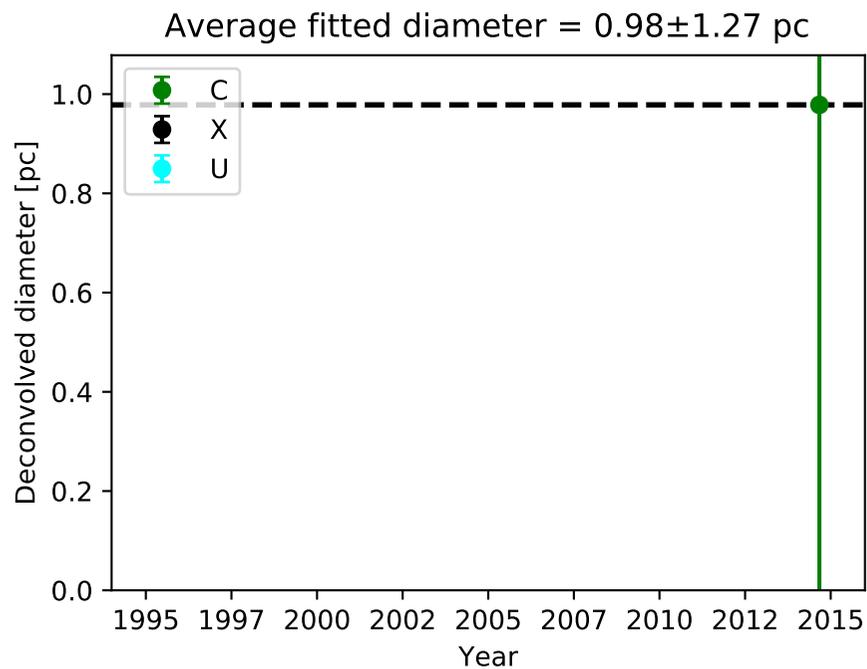

Classification: Slowly varying (S-var)

Figure notes:

Uncertainties calculated as described in Sect. 2.3.

Triangles represent 3σ upper limits for non-detections.

Sizes shown only for detections in bands C, X, and U.

Average diameters calculated according to Sect. 2.4 from the

3.6 cm and 6 cm measurements in experiments BB335A and BB335B.

# 0.2842+0.226 (E18)

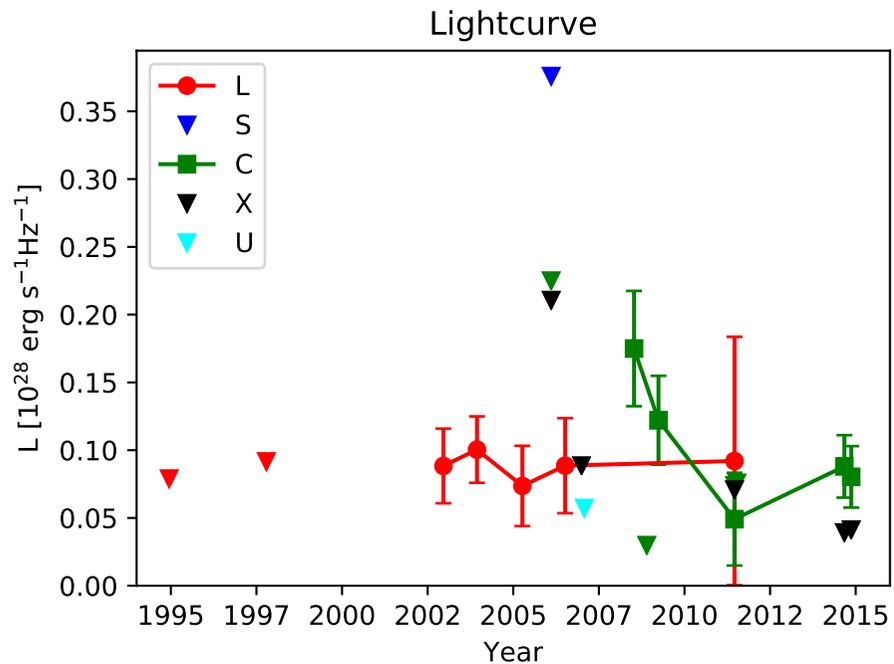
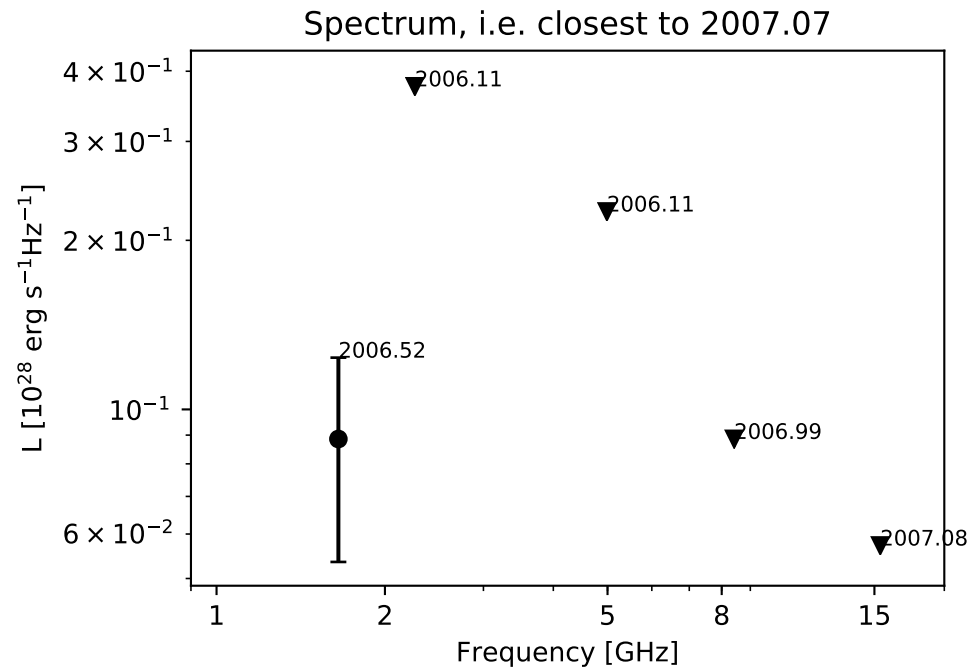
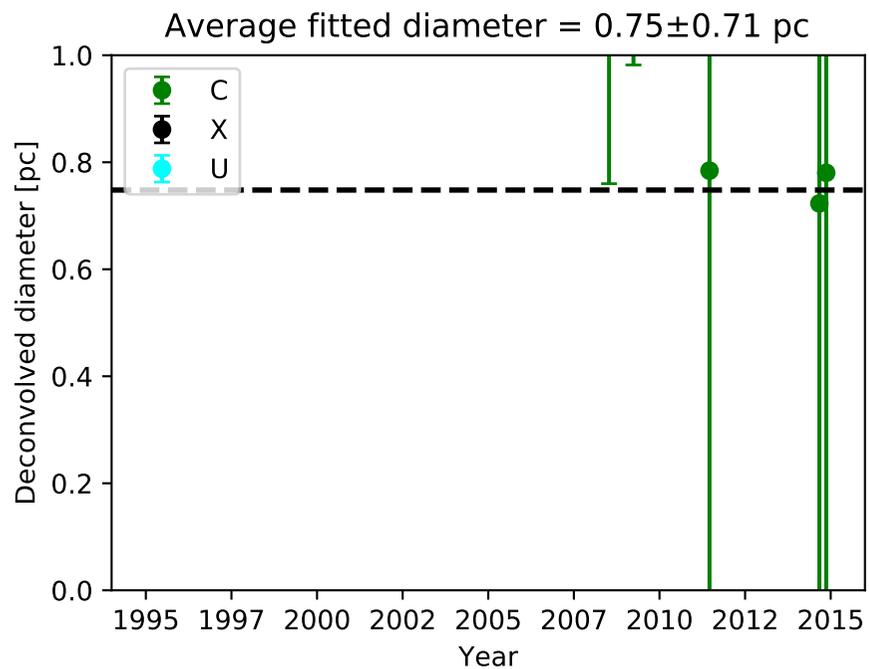

Classification: Slowly varying (S-var)

Figure notes:

Uncertainties calculated as described in Sect. 2.3.

Triangles represent 3σ upper limits for non-detections.

Sizes shown only for detections in bands C, X, and U.

Average diameters calculated according to Sect. 2.4 from the

3.6 cm and 6 cm measurements in experiments BB335A and BB335B.

# 0.2848+0.294

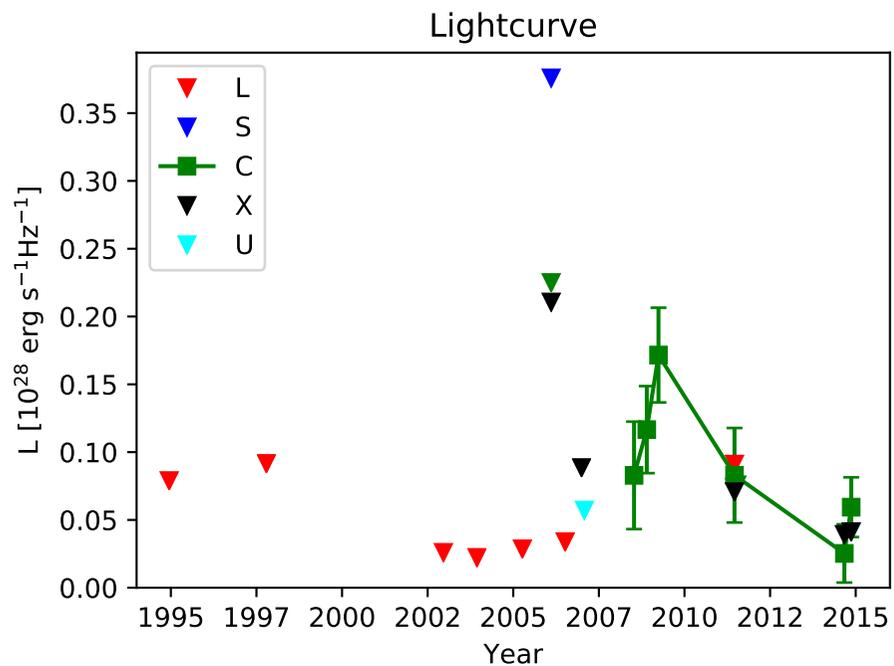

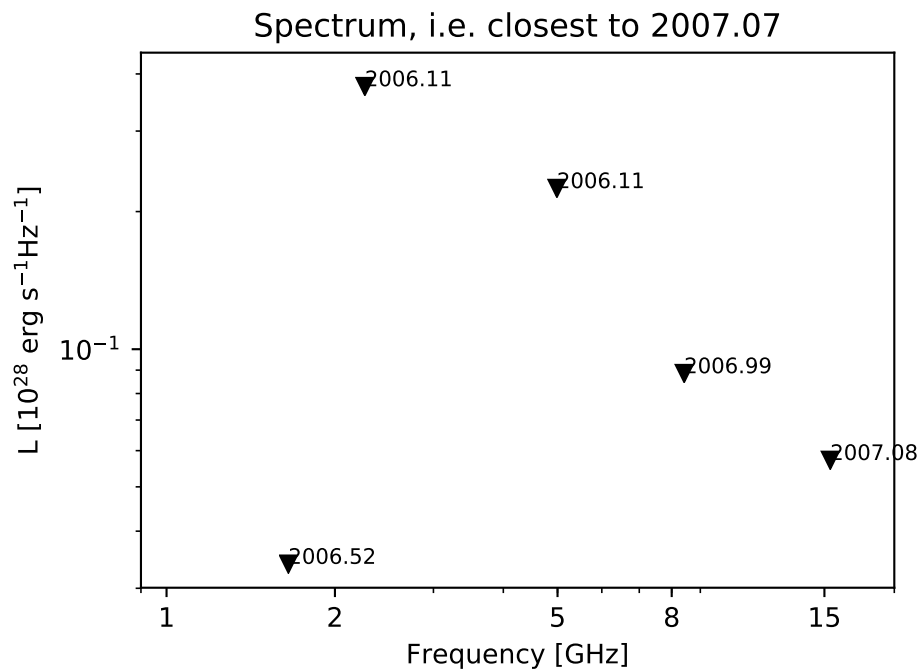

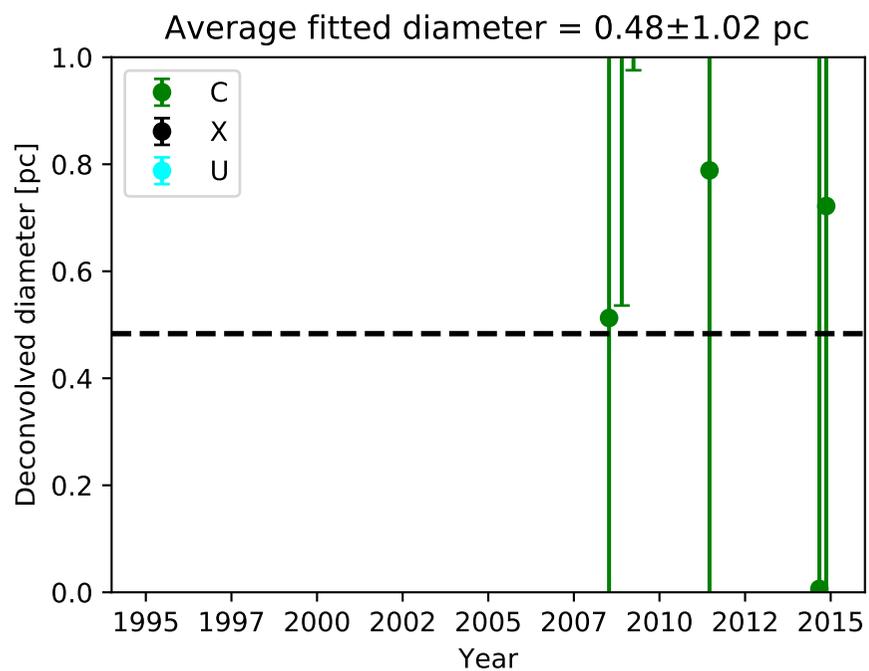

Classification: Fall

Figure notes:

Uncertainties calculated as described in Sect. 2.3.

Triangles represent 3σ upper limits for non-detections.

Sizes shown only for detections in bands C,X, and U.

Average diameters calculated according to Sect. 2.4 from the

3.6 cm and 6 cm measurements in experiments BB335A and BB335B.

# 0.2848+0.471 (E16)

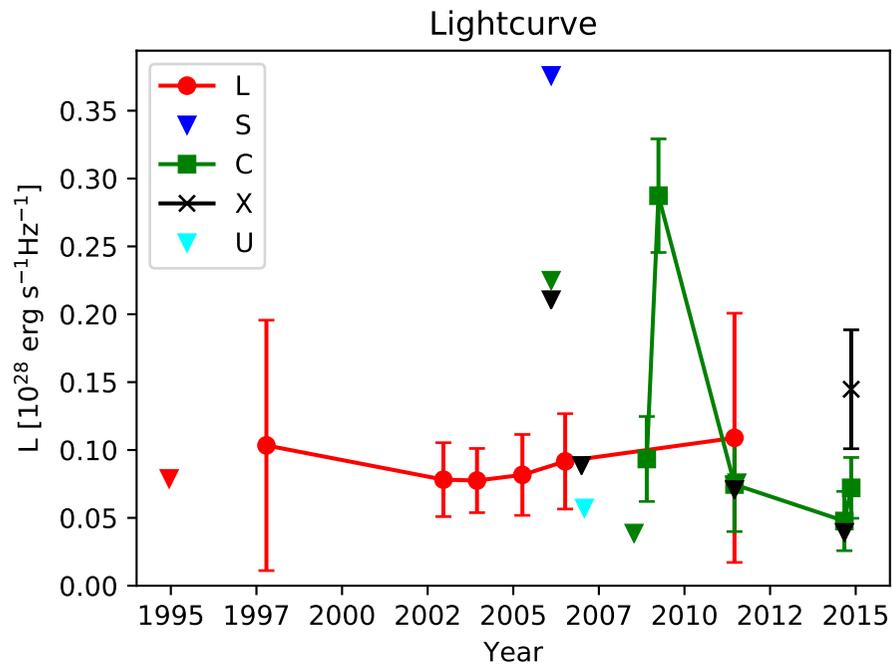

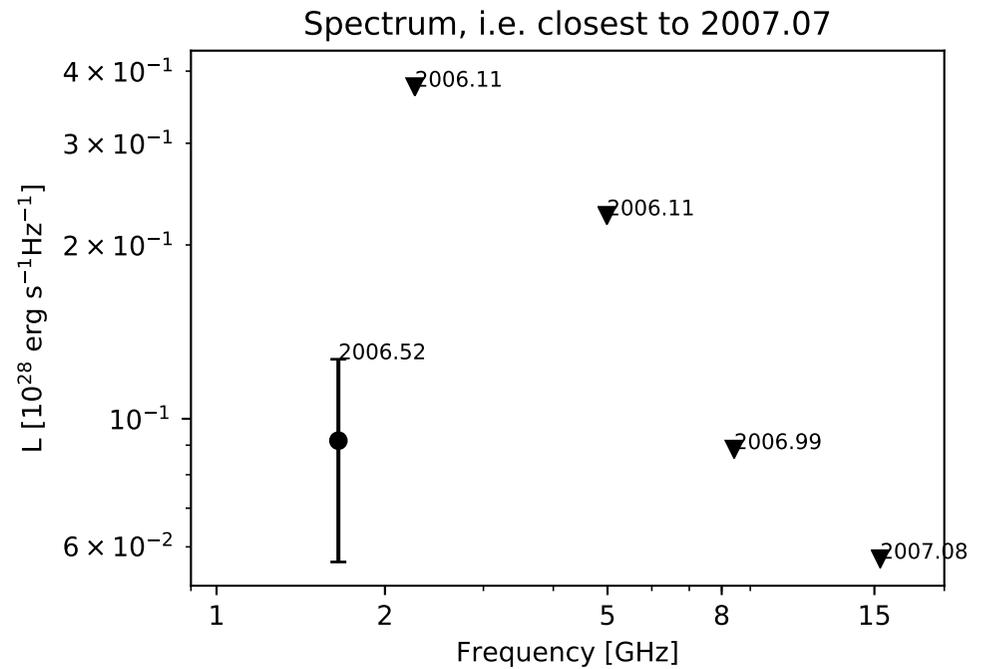

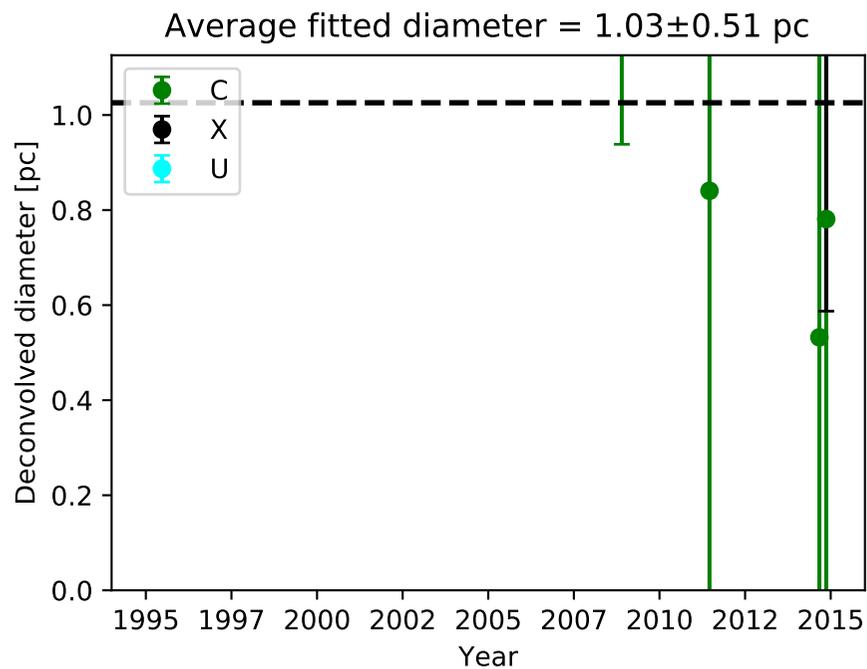

Classification: Fall

Figure notes:

Uncertainties calculated as described in Sect. 2.3.

Triangles represent 3σ upper limits for non-detections.

Sizes shown only for detections in bands C, X, and U.

Average diameters calculated according to Sect. 2.4 from the

3.6 cm and 6 cm measurements in experiments BB335A and BB335B.

# 0.2850+0.193 (E17)

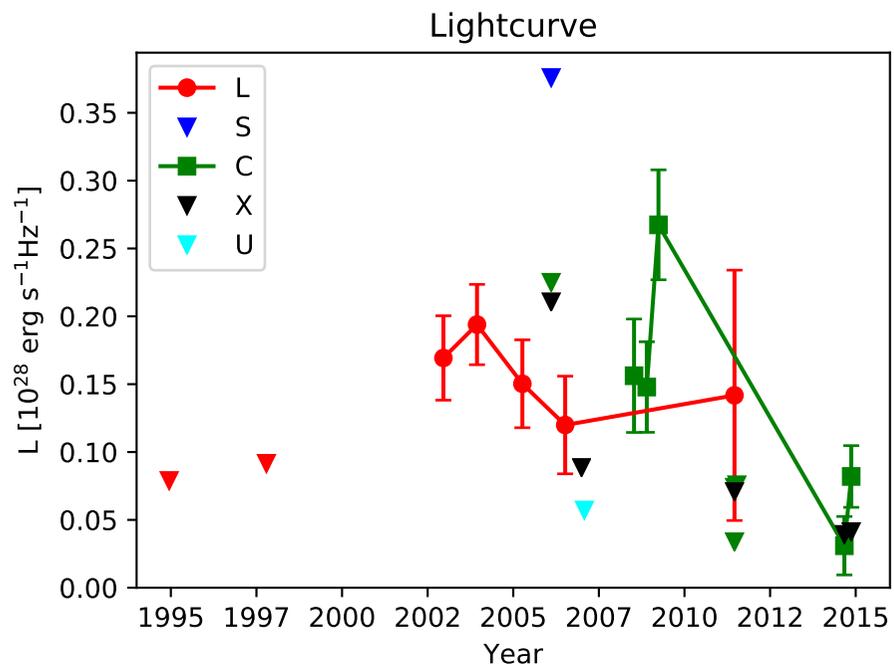
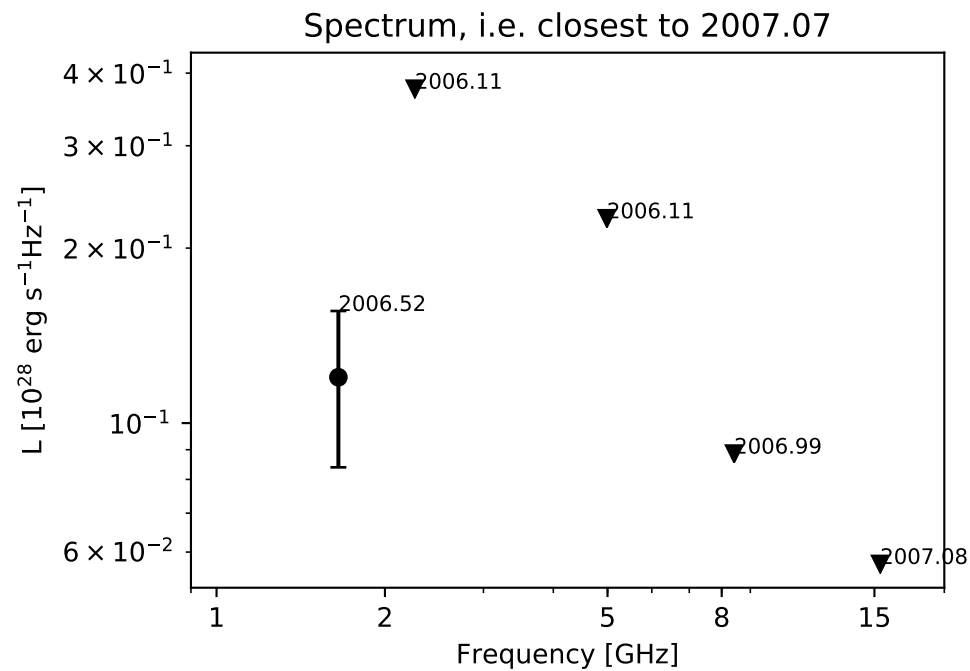
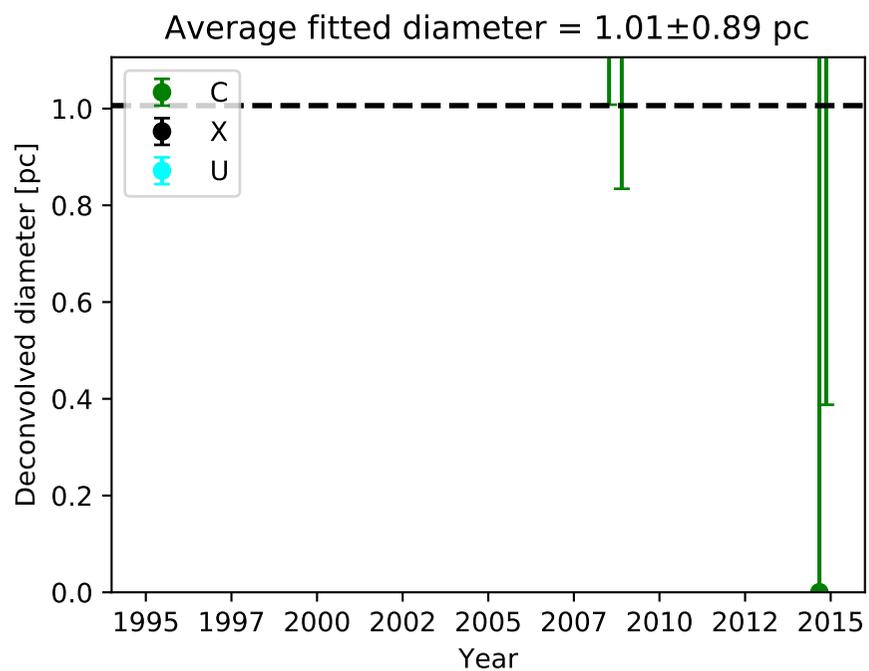

Classification: Fall

Figure notes:

Uncertainties calculated as described in Sect. 2.3.

Triangles represent 3σ upper limits for non-detections.

Sizes shown only for detections in bands C, X, and U.

Average diameters calculated according to Sect. 2.4 from the 3.6 cm and 6 cm measurements in experiments BB335A and BB335B.

# 0.2858+0.013

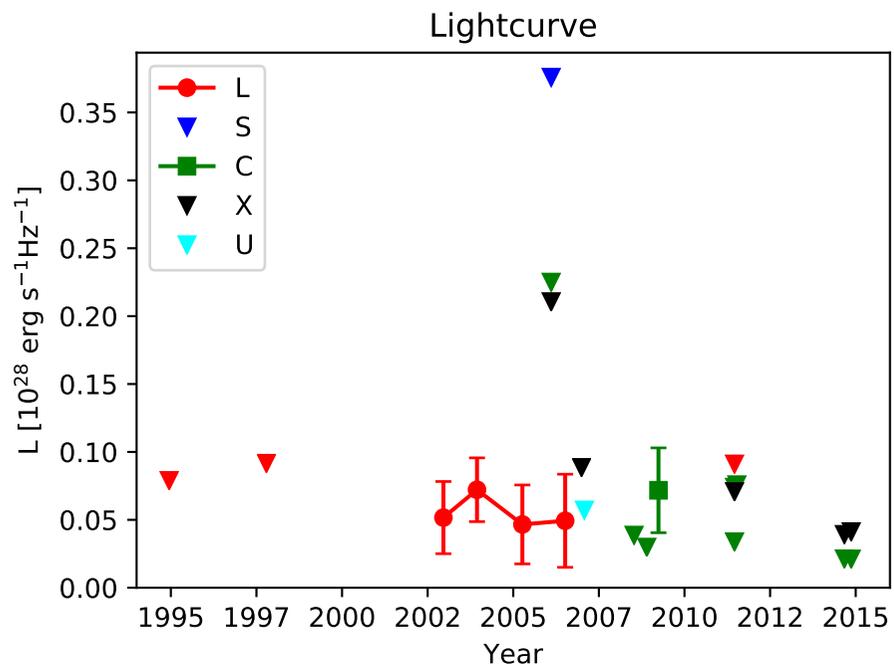
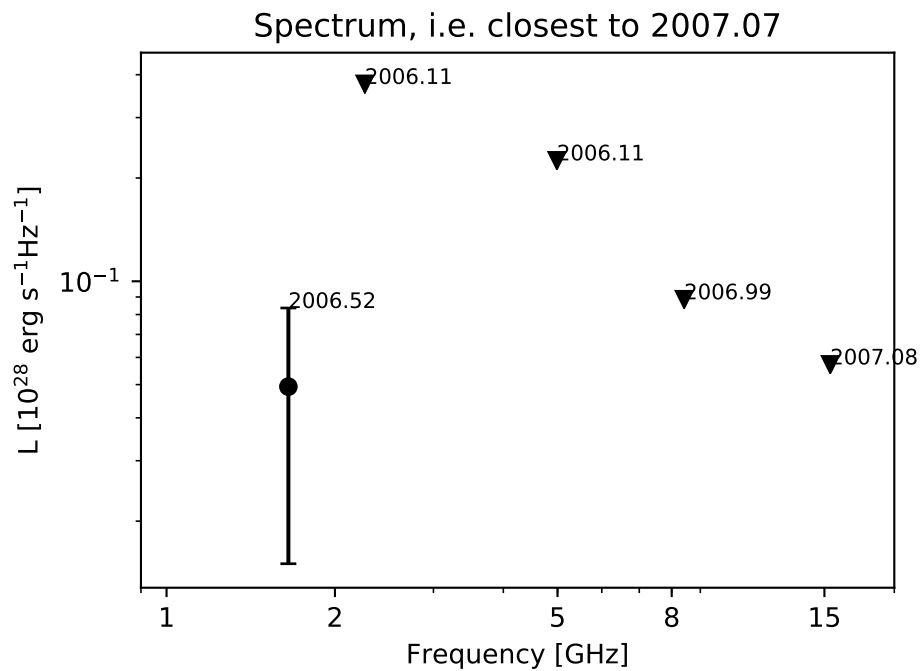
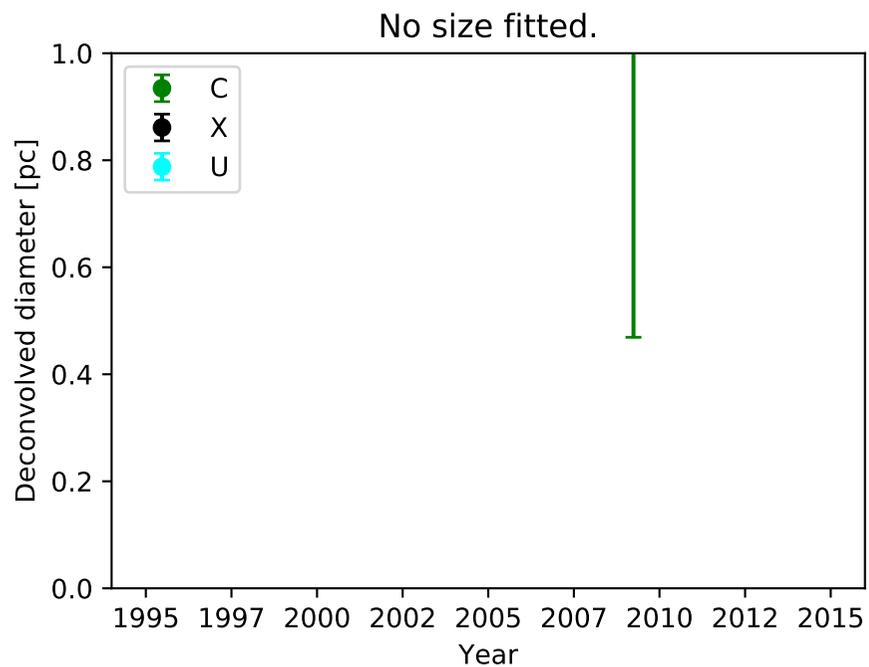

Classification: Slowly varying (S-var)

Figure notes:

Uncertainties calculated as described in Sect. 2.3.

Triangles represent 3σ upper limits for non-detections.

Sizes shown only for detections in bands C,X, and U.

Average diameters calculated according to Sect. 2.4 from the

3.6 cm and 6 cm measurements in experiments BB335A and BB335B.

# 0.2868+0.297 (E14)

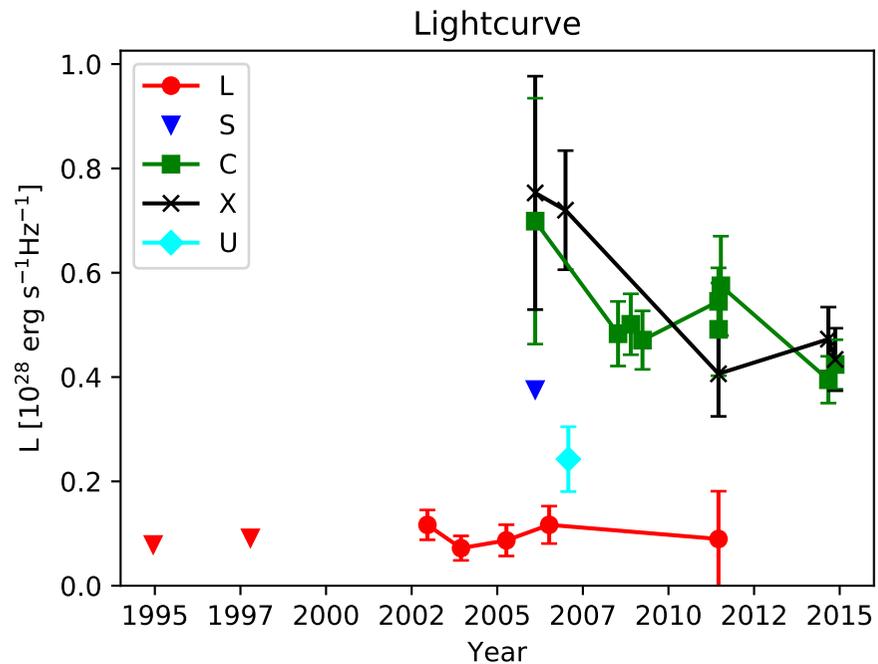
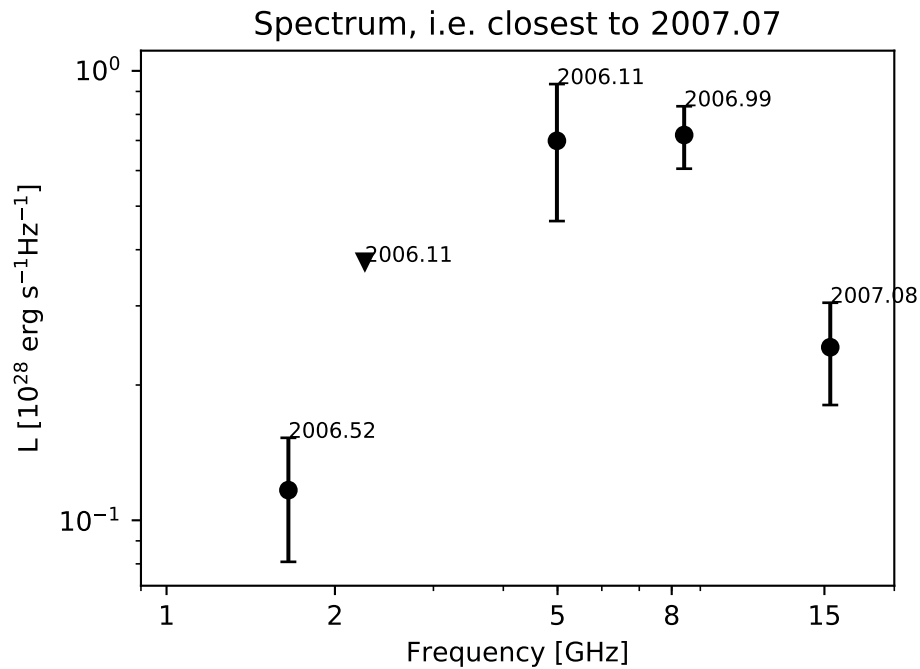
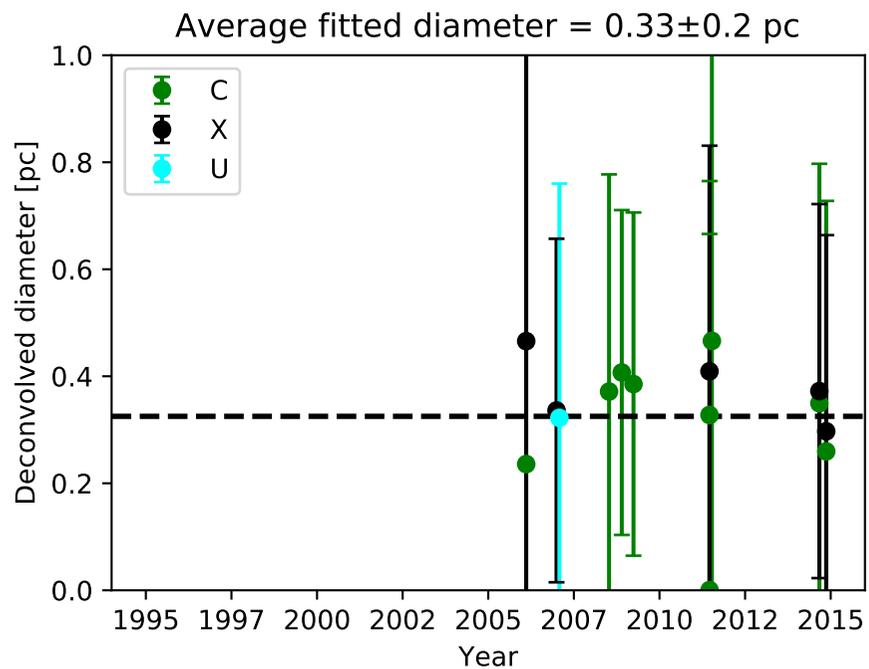

Classification: Slowly varying (S-var)

Figure notes:

Uncertainties calculated as described in Sect. 2.3.

Triangles represent 3σ upper limits for non-detections.

Sizes shown only for detections in bands C, X, and U.

Average diameters calculated according to Sect. 2.4 from the

3.6 cm and 6 cm measurements in experiments BB335A and BB335B.

# 0.2875+0.352

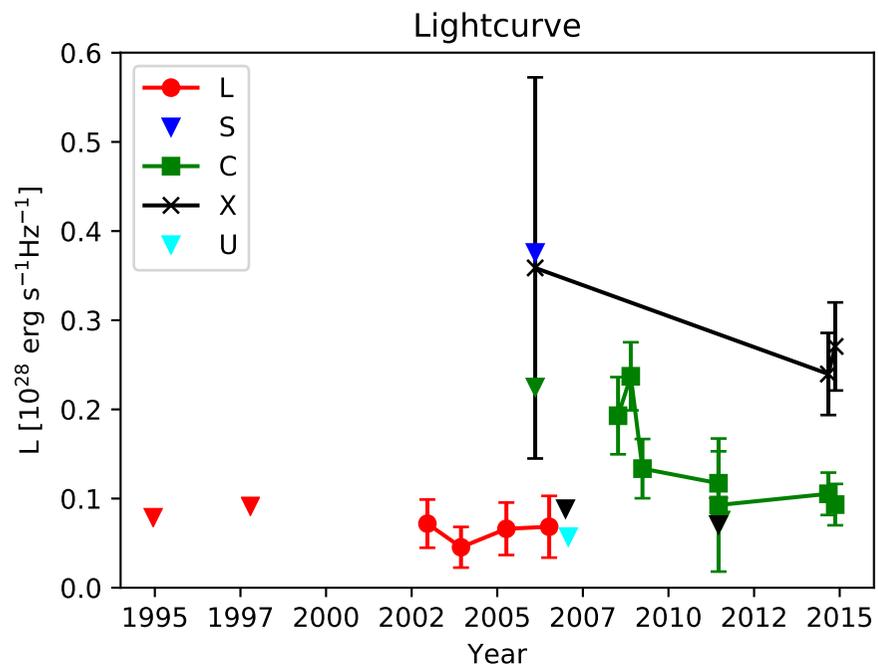

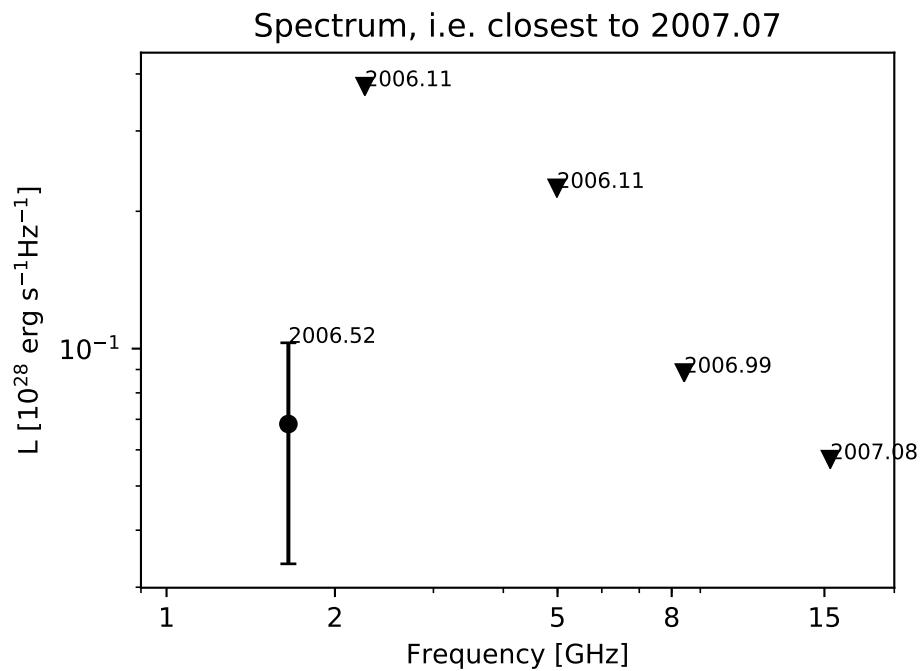

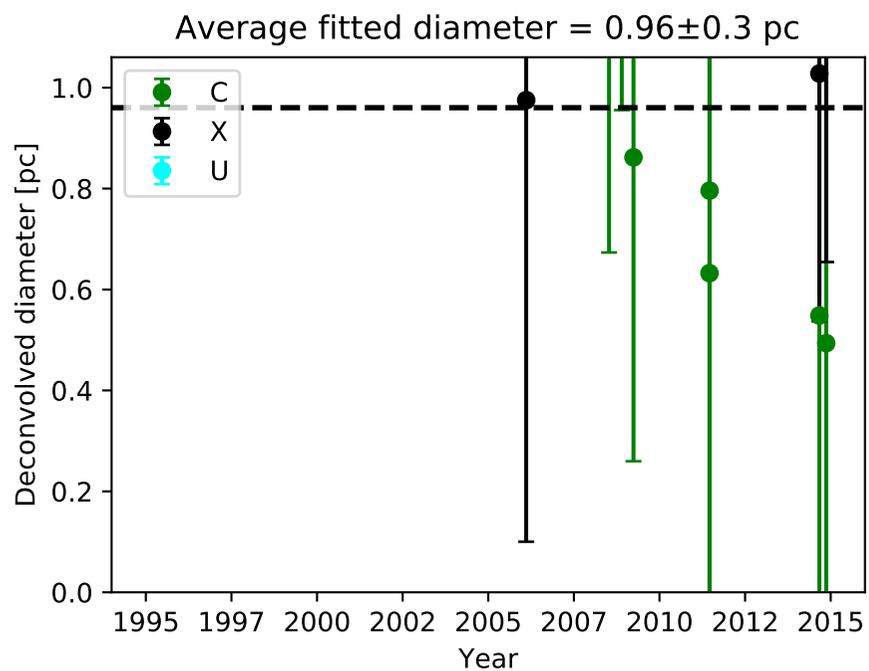

Classification: Fall

Figure notes:

Uncertainties calculated as described in Sect. 2.3.

Triangles represent 3σ upper limits for non-detections.

Sizes shown only for detections in bands C, X, and U.

Average diameters calculated according to Sect. 2.4 from the

3.6 cm and 6 cm measurements in experiments BB335A and BB335B.

# 0.2878+0.308 (E13)

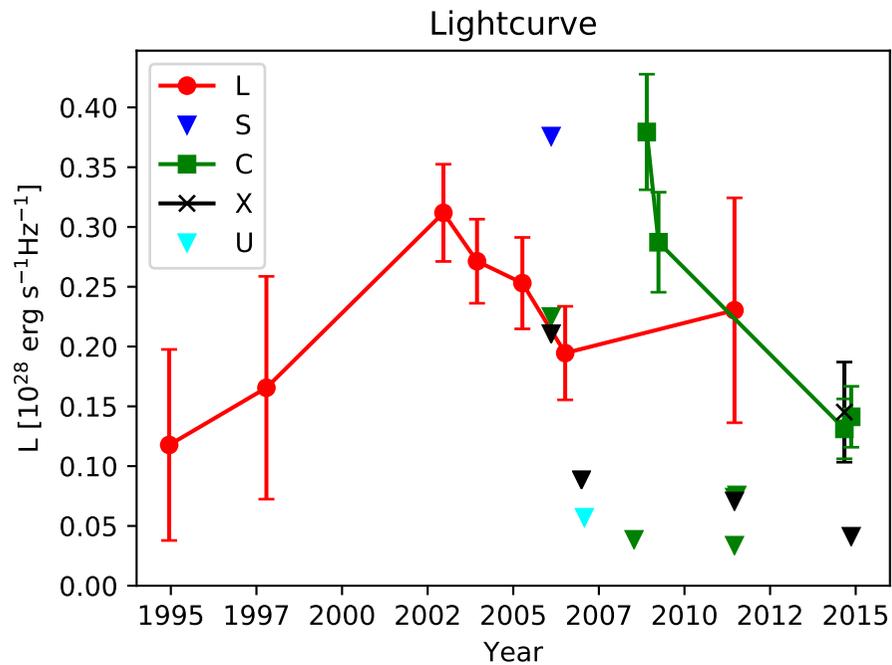
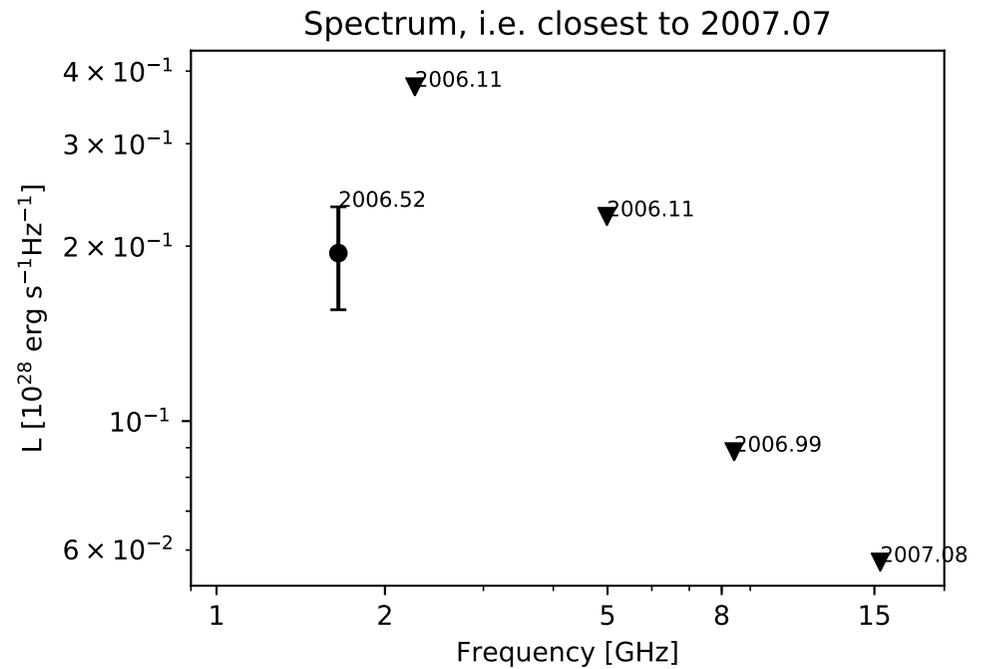
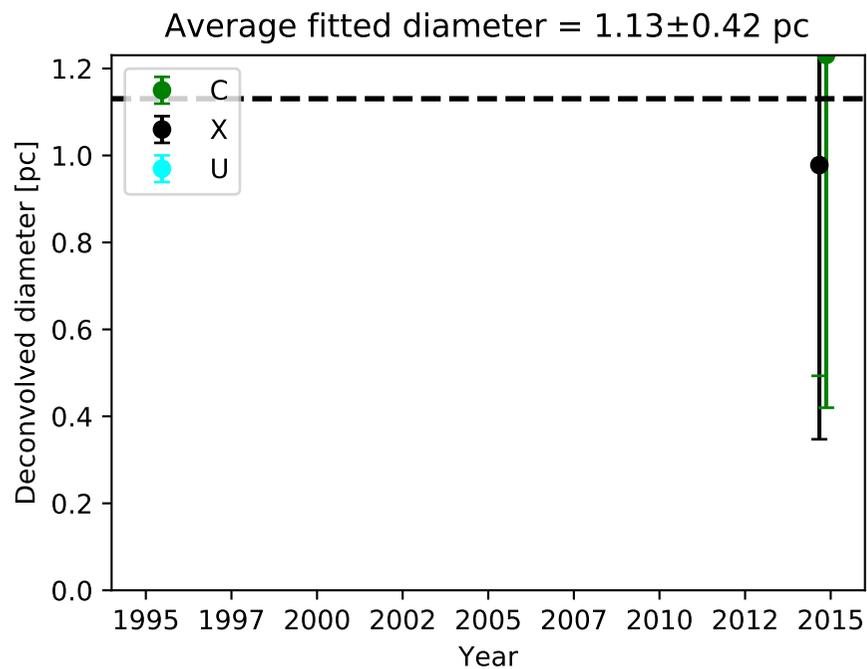

Classification: Fall

Figure notes:

Uncertainties calculated as described in Sect. 2.3.

Triangles represent 3$\sigma$ upper limits for non-detections.

Sizes shown only for detections in bands C,X, and U.

Average diameters calculated according to Sect. 2.4 from the

3.6 cm and 6 cm measurements in experiments BB335A and BB335B.

# 0.2881+0.369

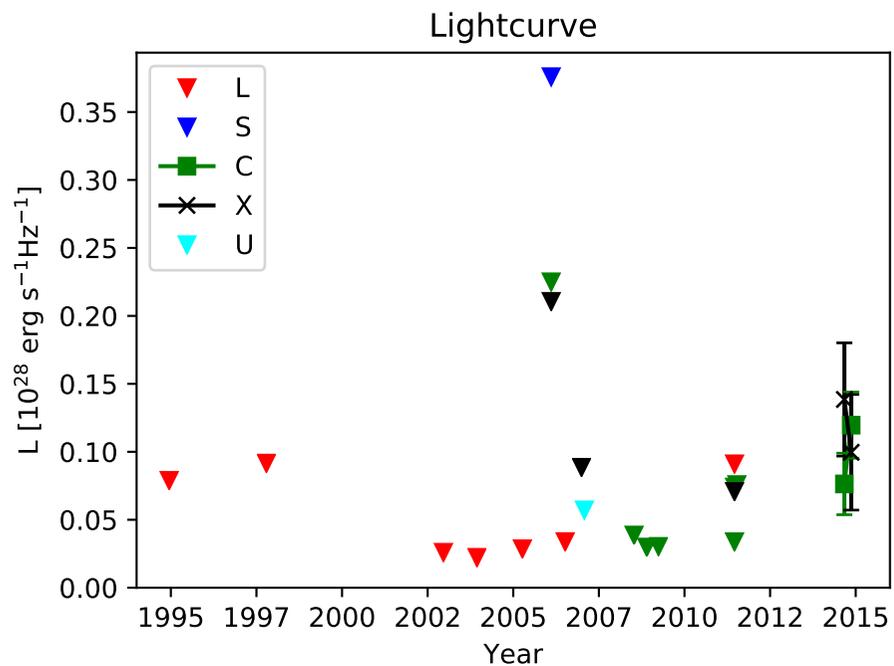
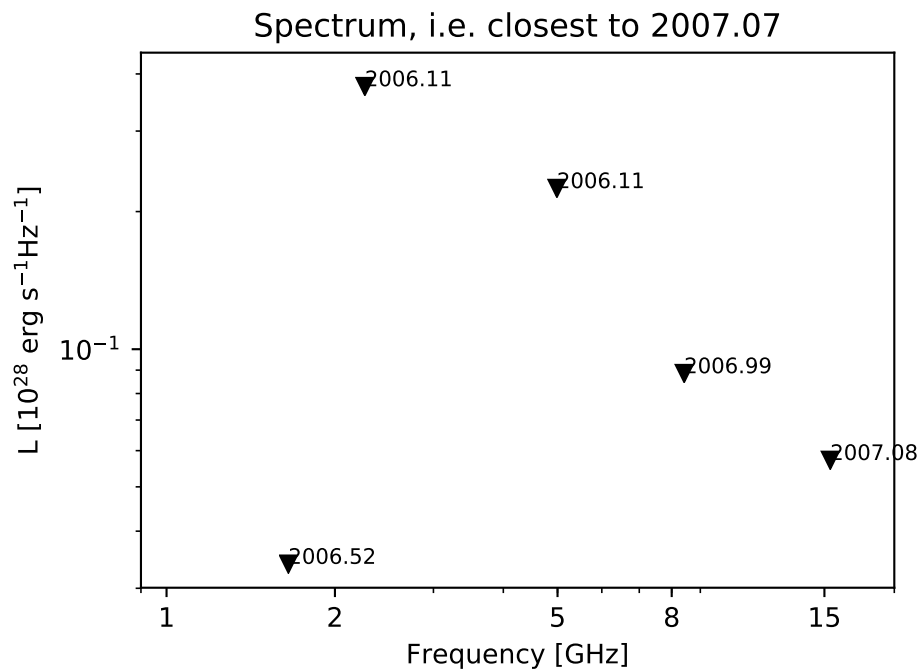
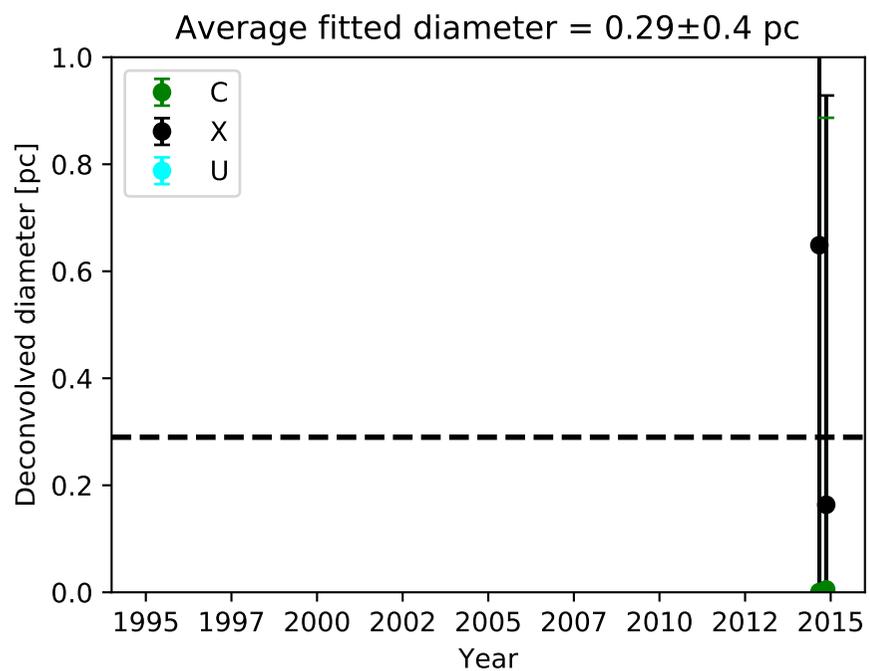

Classification: Rise

Figure notes:

Uncertainties calculated as described in Sect. 2.3.

Triangles represent $3\sigma$ upper limits for non-detections.

Sizes shown only for detections in bands C, X, and U.

Average diameters calculated according to Sect. 2.4 from the

3.6 cm and 6 cm measurements in experiments BB335A and BB335B.

# 0.2883+0.338

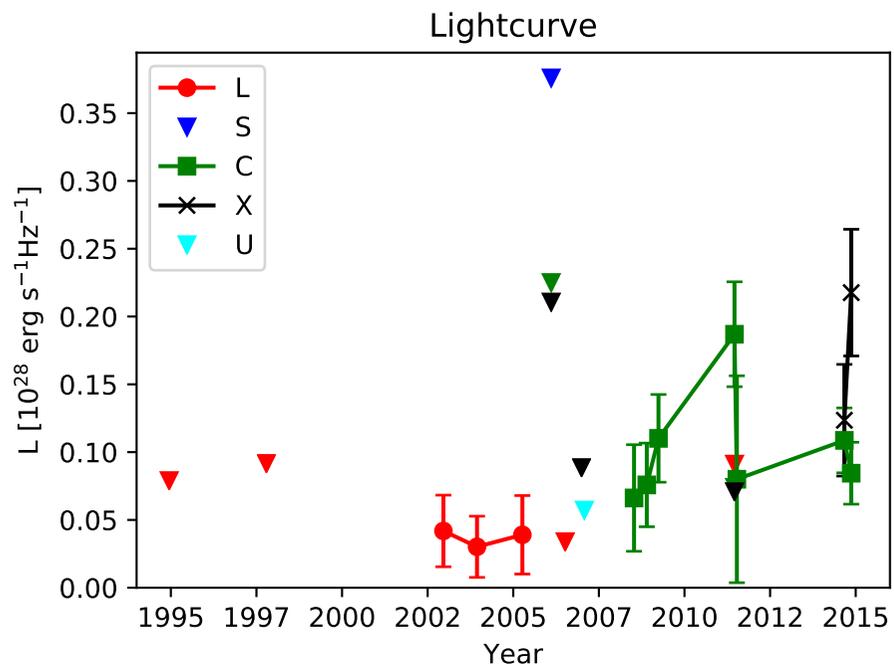

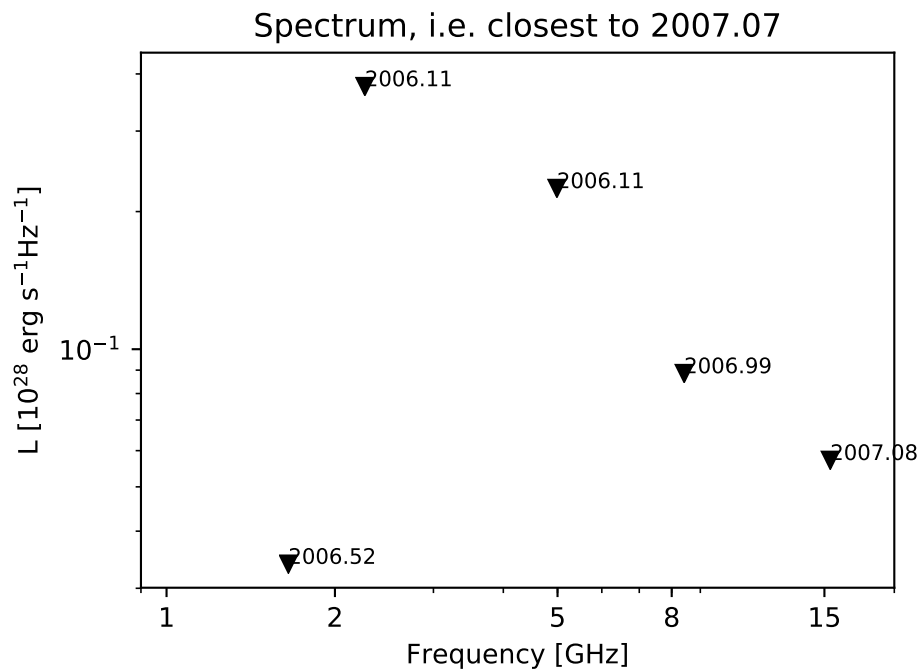

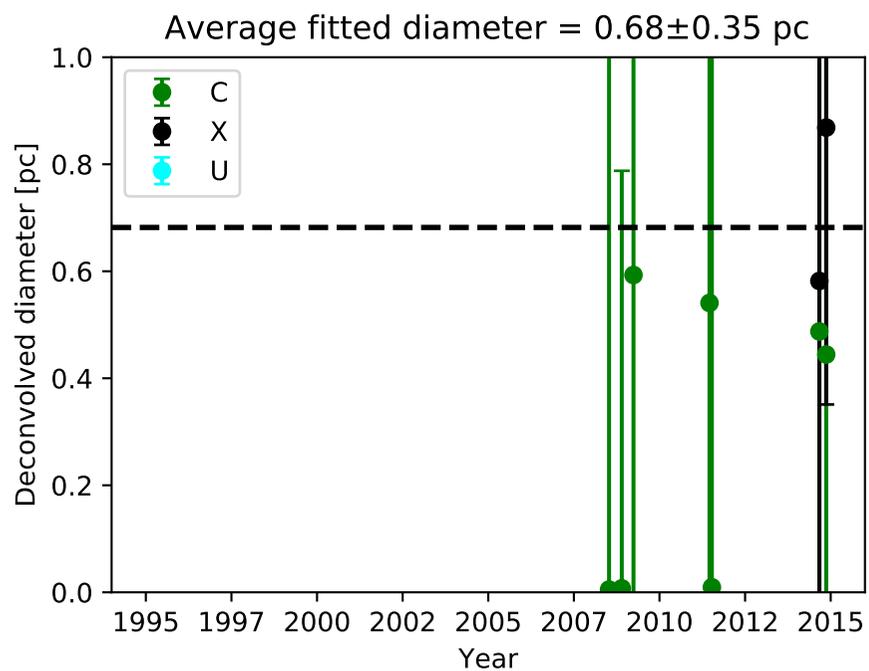

Classification: Slowly varying (S-var)

Figure notes:

Uncertainties calculated as described in Sect. 2.3.

Triangles represent 3σ upper limits for non-detections.

Sizes shown only for detections in bands C,X, and U.

Average diameters calculated according to Sect. 2.4 from the

3.6 cm and 6 cm measurements in experiments BB335A and BB335B.

# 0.2887+0.201

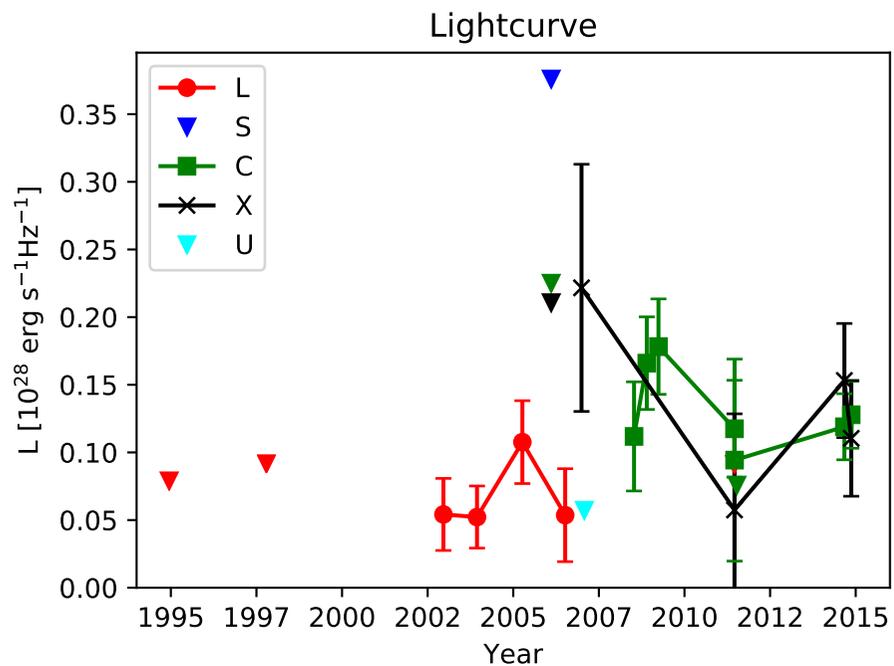

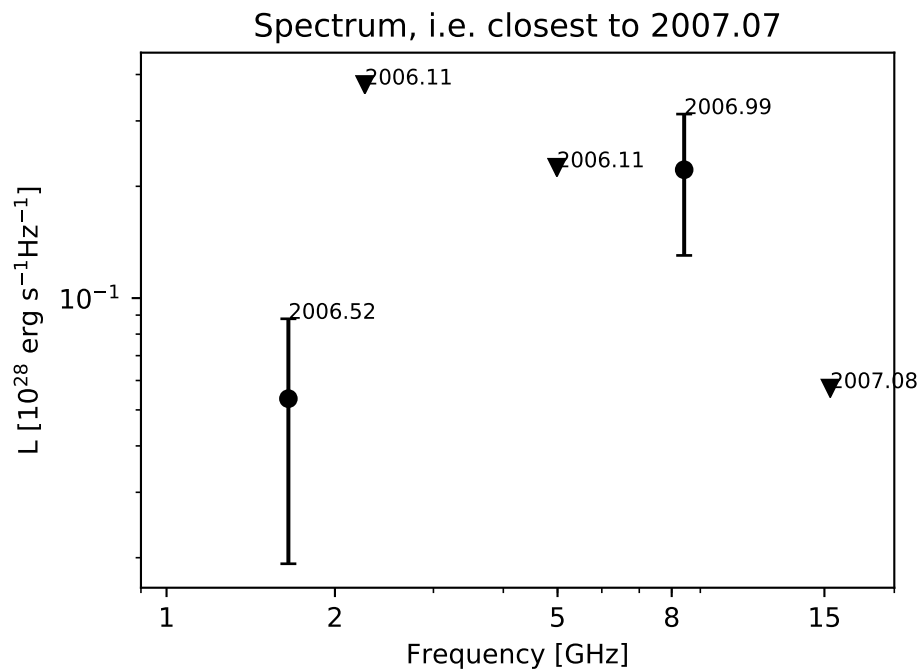

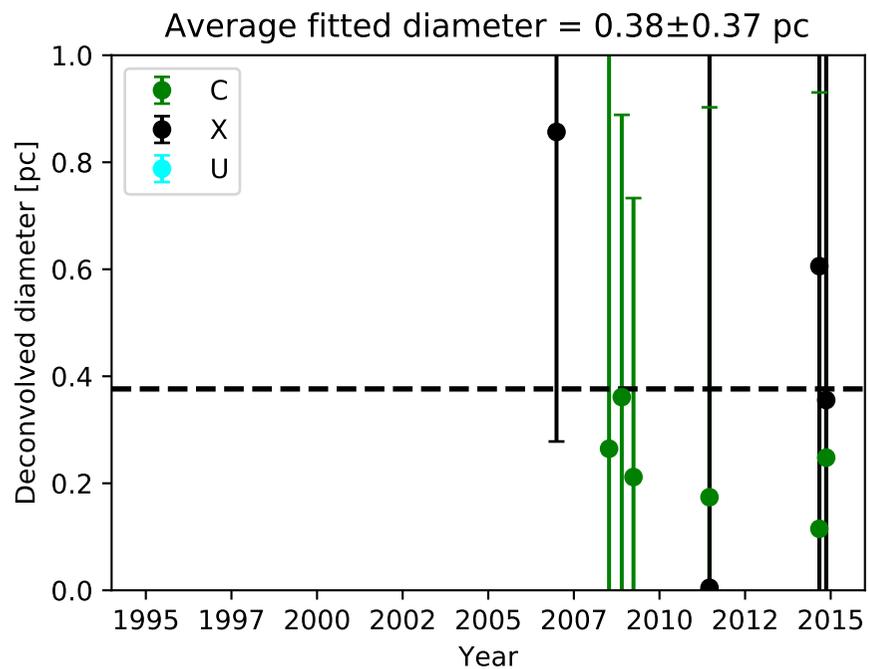

Classification: Slowly varying (S-var)

Figure notes:

Uncertainties calculated as described in Sect. 2.3.

Triangles represent 3σ upper limits for non-detections.

Sizes shown only for detections in bands C, X, and U.

Average diameters calculated according to Sect. 2.4 from the

3.6 cm and 6 cm measurements in experiments BB335A and BB335B.

# 0.2890+0.337

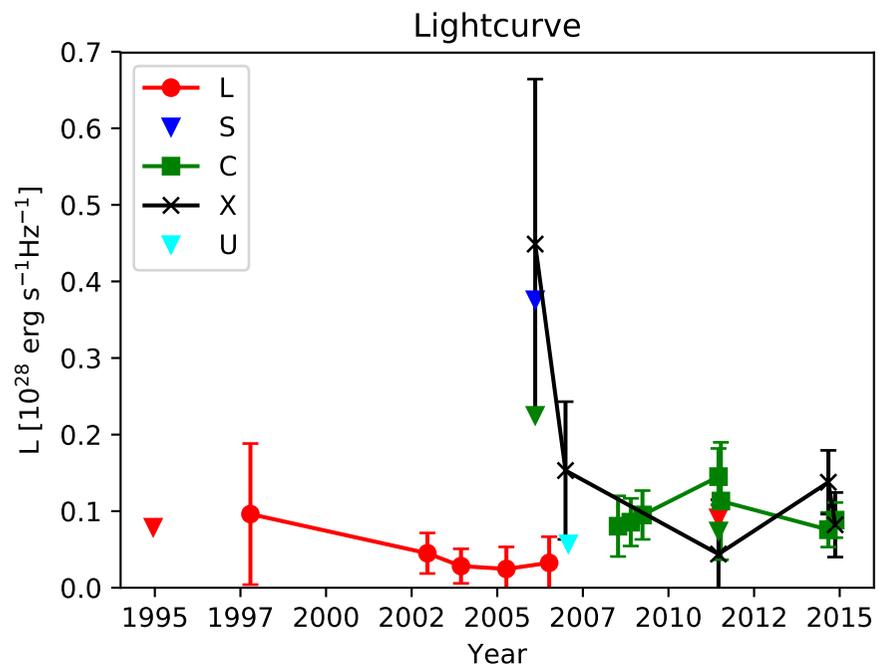
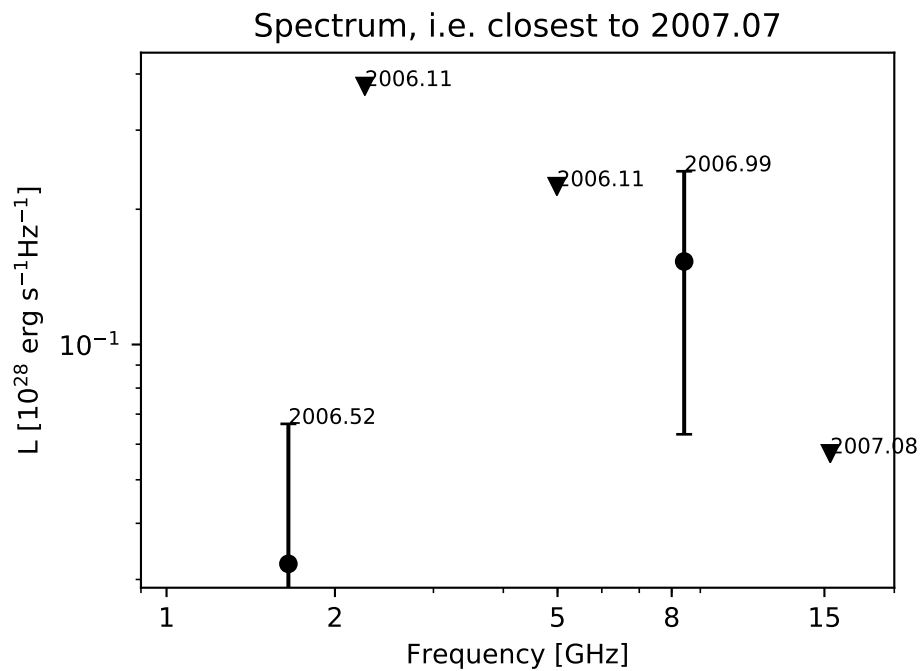
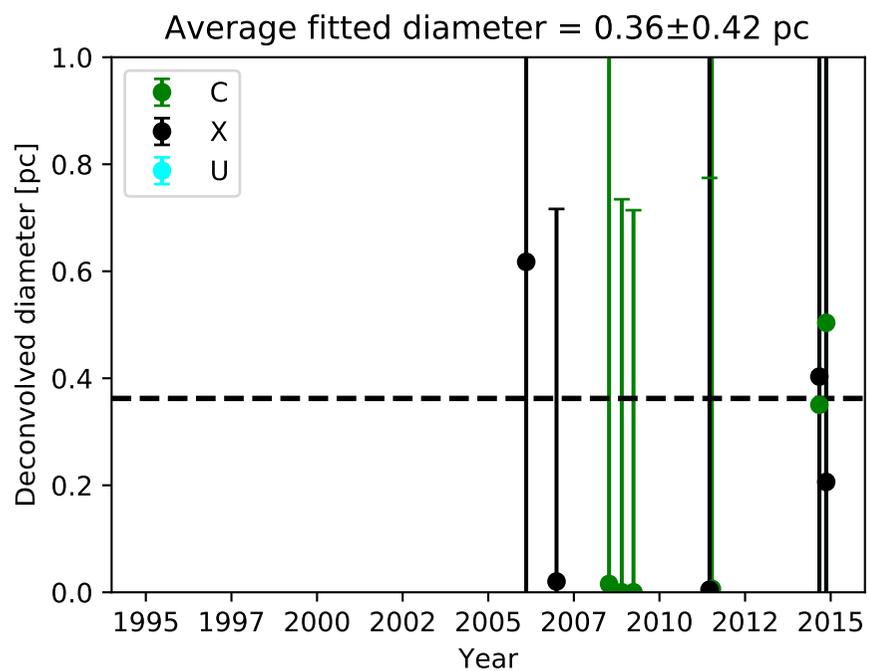

Classification: Slowly varying (S-var)

Figure notes:

Uncertainties calculated as described in Sect. 2.3.

Triangles represent 3σ upper limits for non-detections.

Sizes shown only for detections in bands C, X, and U.

Average diameters calculated according to Sect. 2.4 from the

3.6 cm and 6 cm measurements in experiments BB335A and BB335B.

# 0.2891+0.253

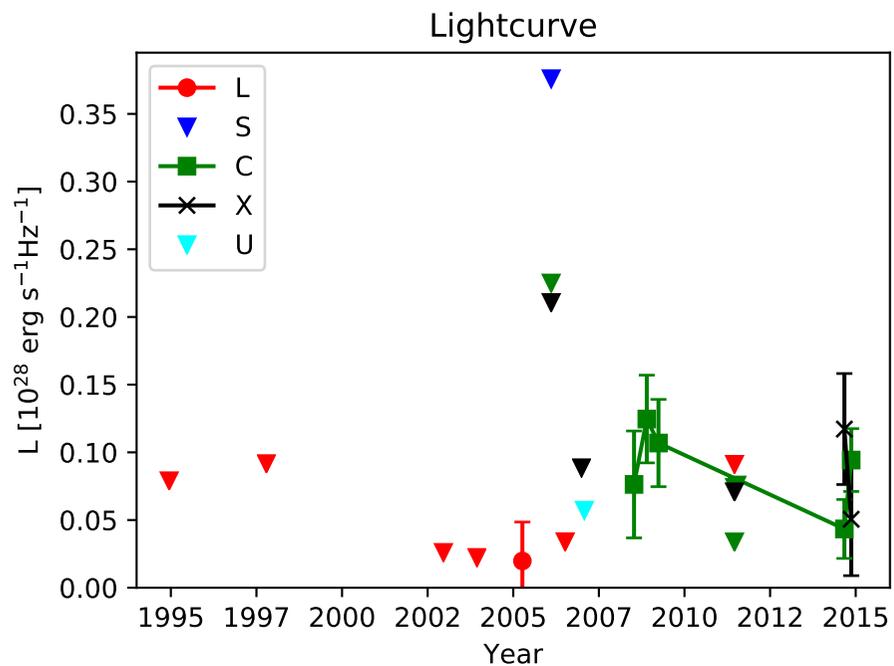

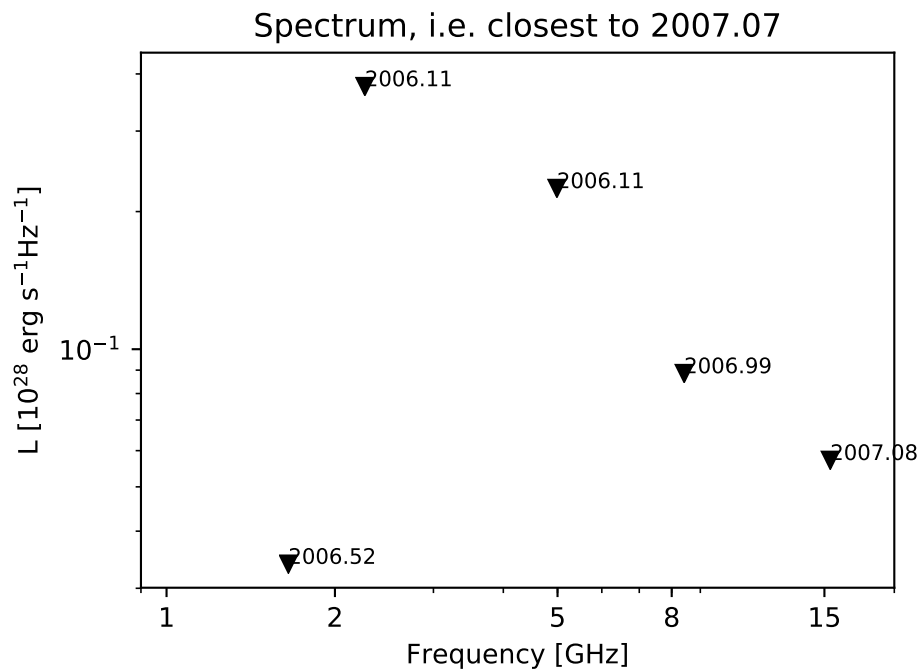

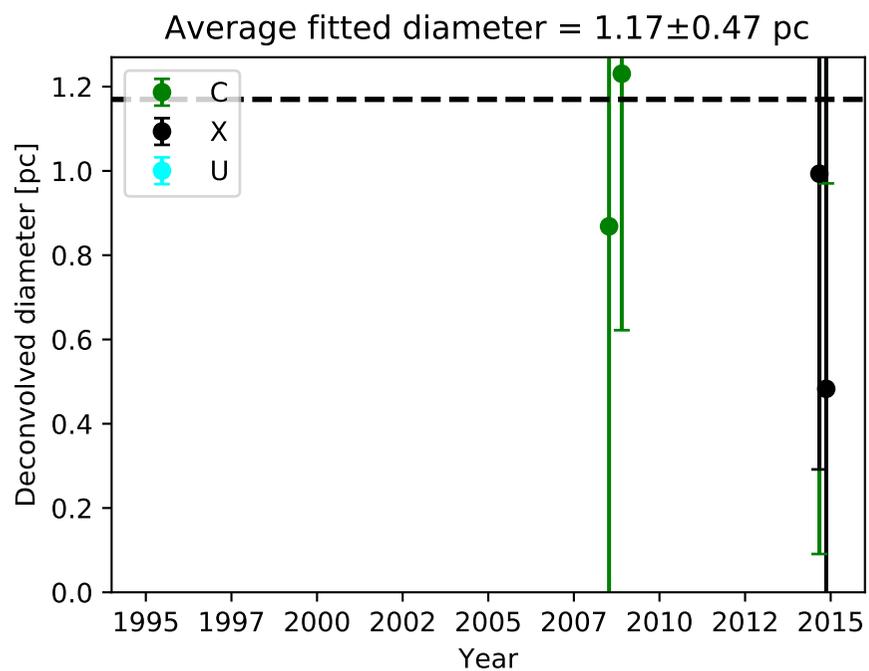

Classification: Slowly varying (S-var)

Figure notes:

Uncertainties calculated as described in Sect. 2.3.

Triangles represent 3σ upper limits for non-detections.

Sizes shown only for detections in bands C, X, and U.

Average diameters calculated according to Sect. 2.4 from the

3.6 cm and 6 cm measurements in experiments BB335A and BB335B.

# 0.2897+0.278

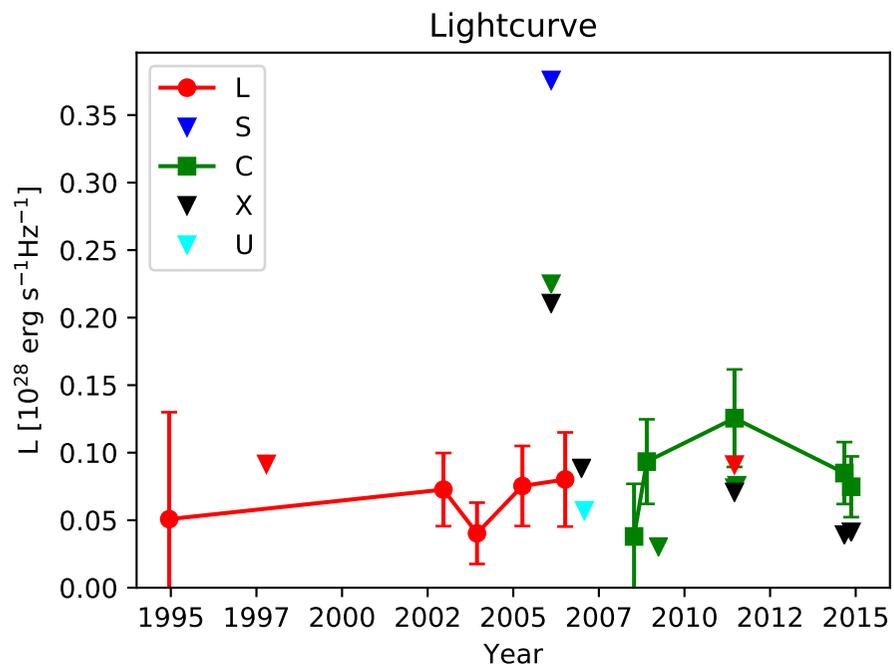
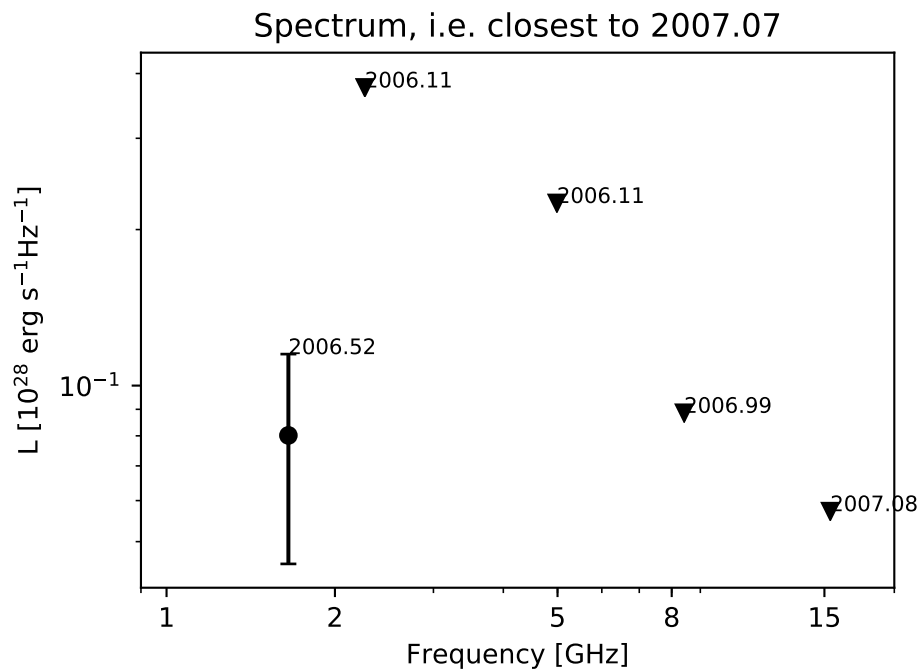
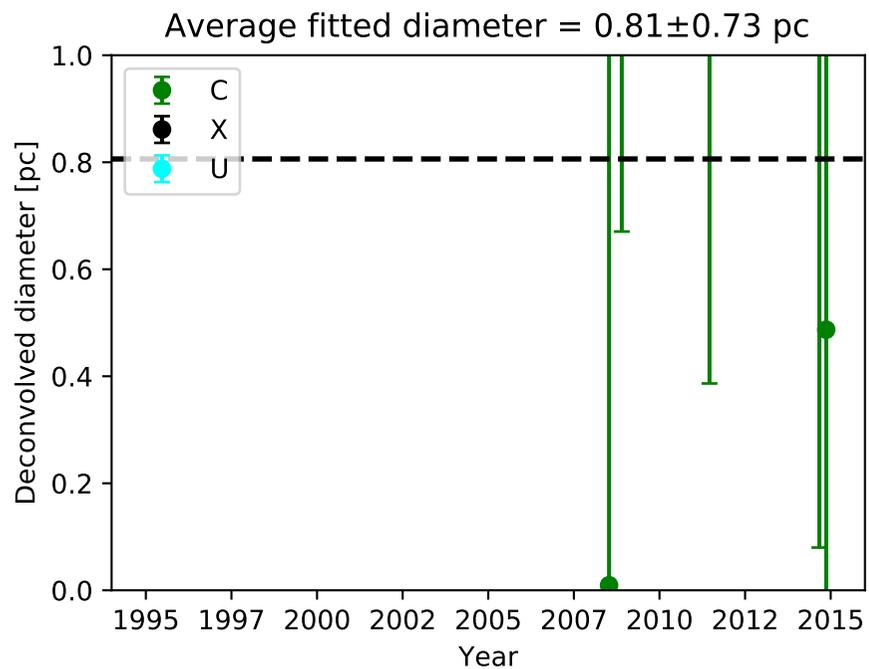

Classification: Slowly varying (S-var)

Figure notes:

Uncertainties calculated as described in Sect. 2.3.

Triangles represent 3σ upper limits for non-detections.

Sizes shown only for detections in bands C, X, and U.

Average diameters calculated according to Sect. 2.4 from the

3.6 cm and 6 cm measurements in experiments BB335A and BB335B.

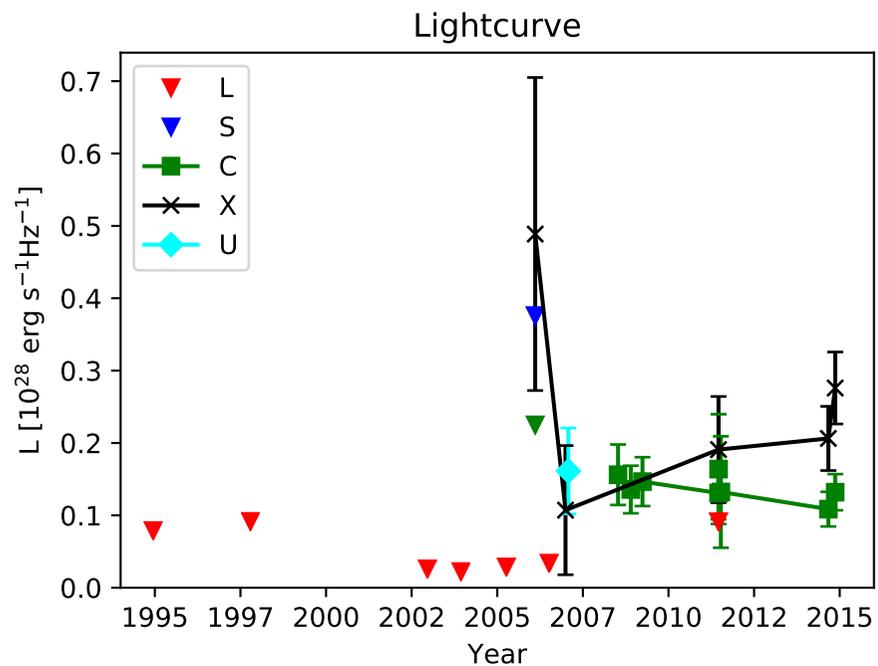

# 0.2910+0.325 (E11)

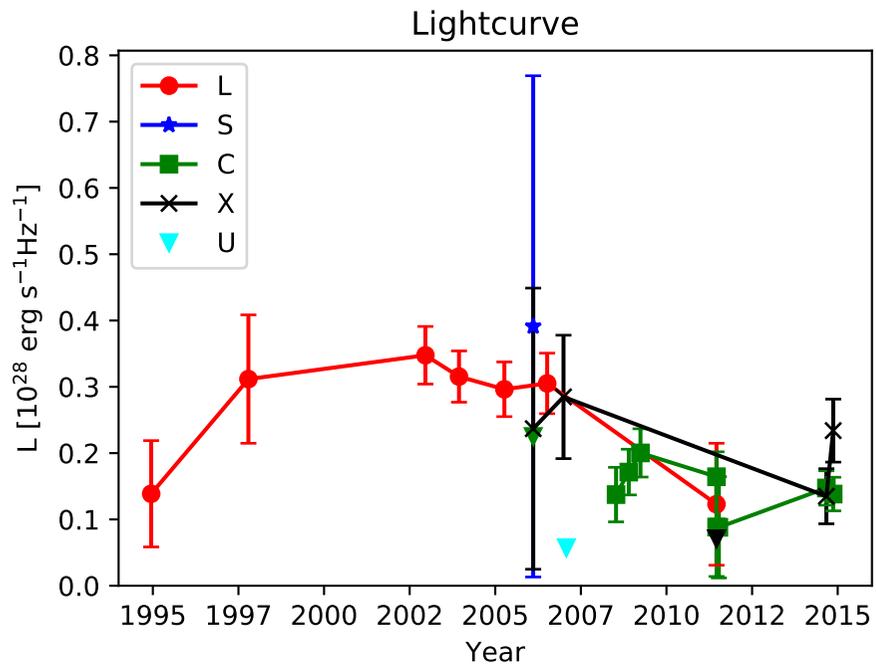
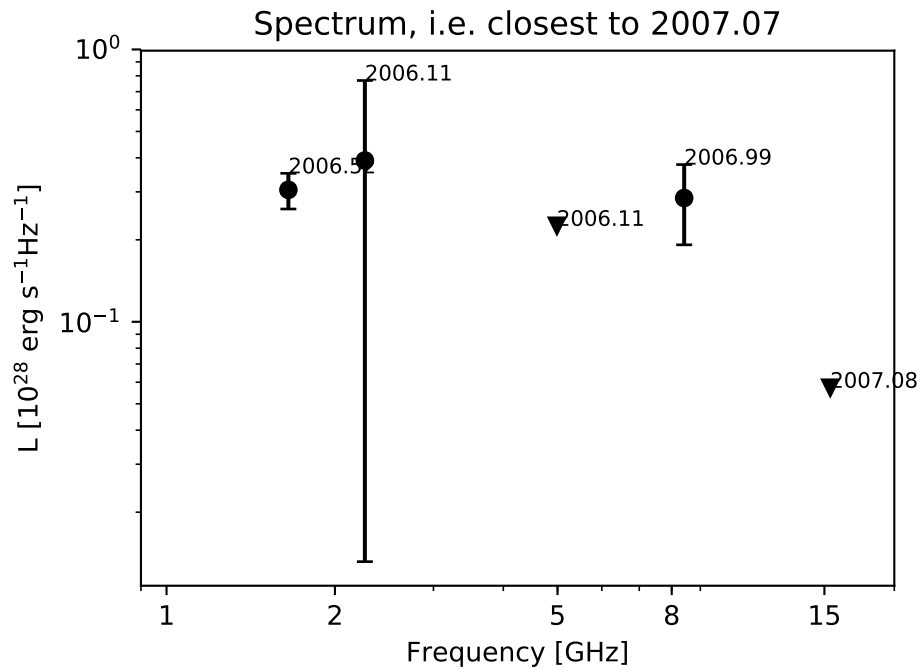
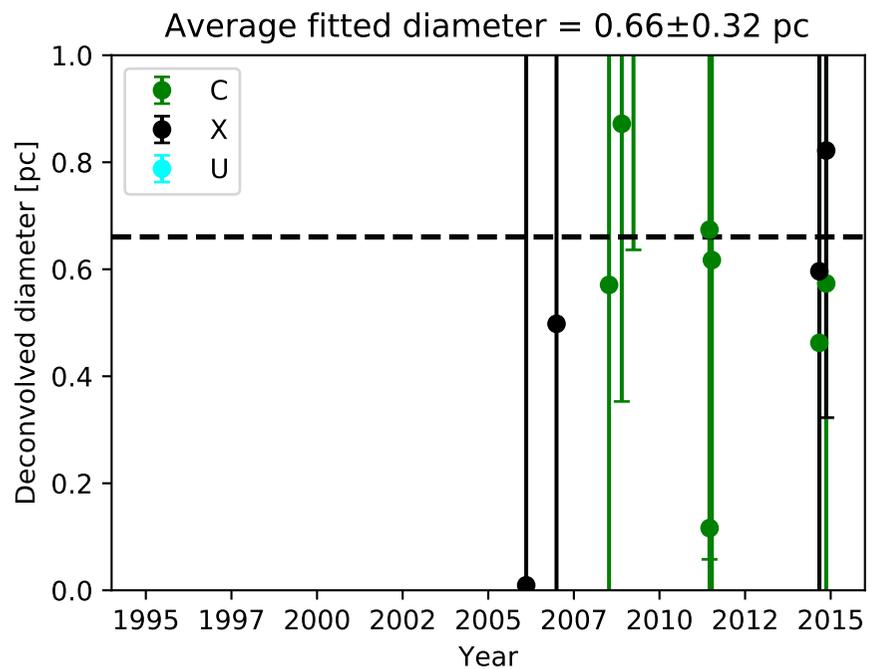

Classification: Slowly varying (S-var)

Figure notes:

Uncertainties calculated as described in Sect. 2.3.

Triangles represent 3σ upper limits for non-detections.

Sizes shown only for detections in bands C, X, and U.

Average diameters calculated according to Sect. 2.4 from the

3.6 cm and 6 cm measurements in experiments BB335A and BB335B.

# 0.2914+0.333

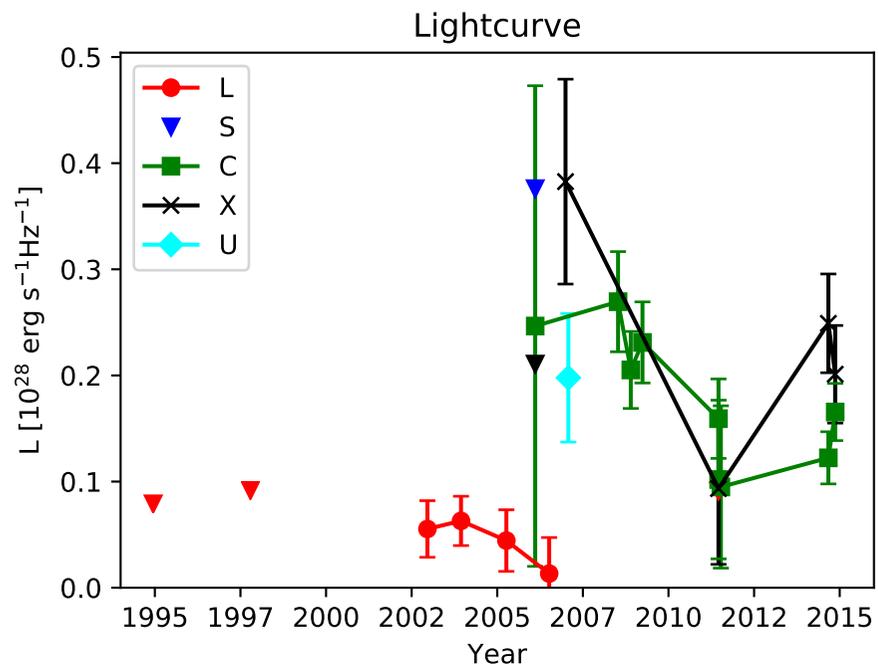
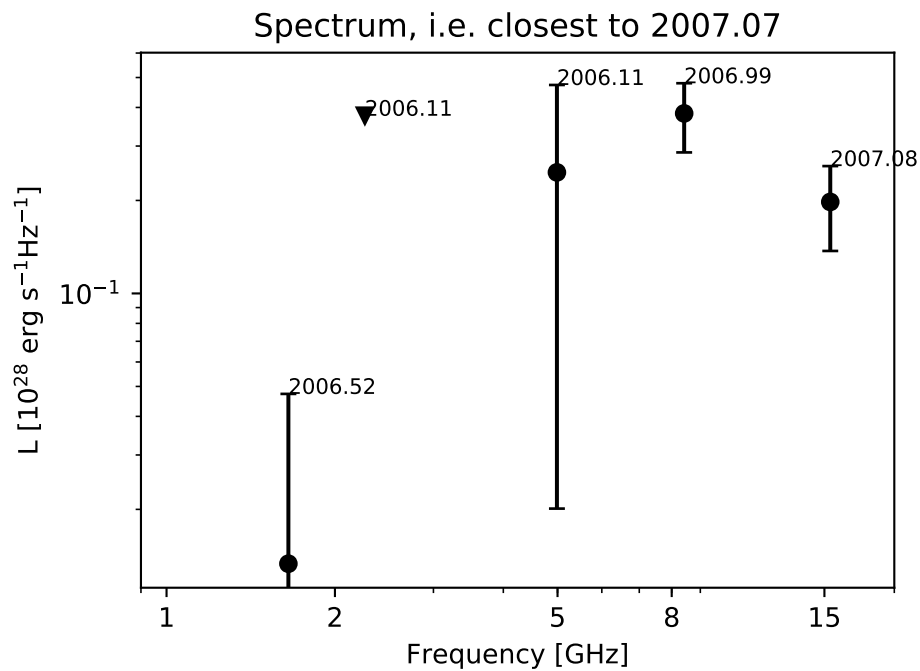
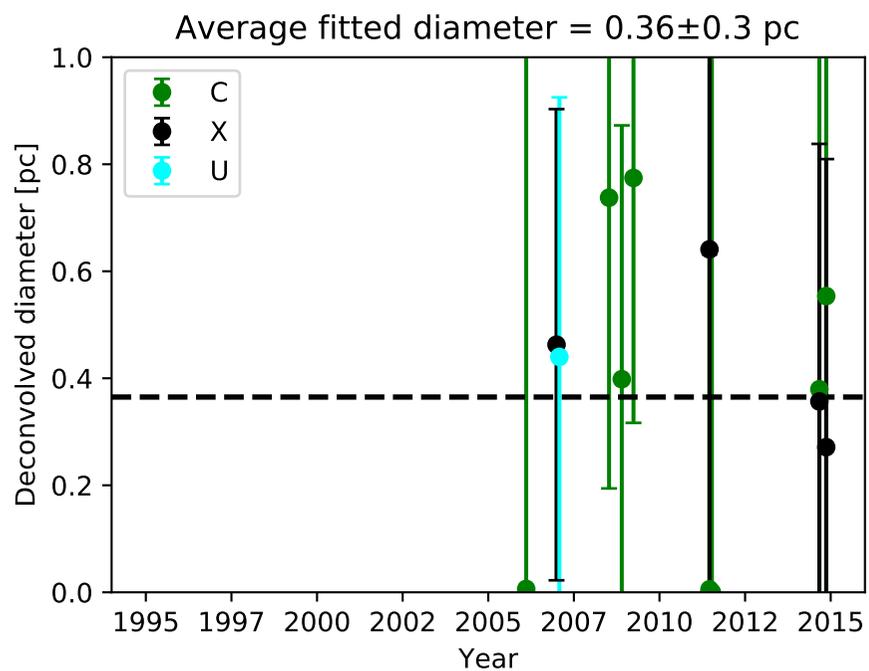

Classification: Fall

Figure notes:

Uncertainties calculated as described in Sect. 2.3.

Triangles represent 3σ upper limits for non-detections.

Sizes shown only for detections in bands C, X, and U.

Average diameters calculated according to Sect. 2.4 from the

3.6 cm and 6 cm measurements in experiments BB335A and BB335B.

# 0.2915+0.335 (E10)

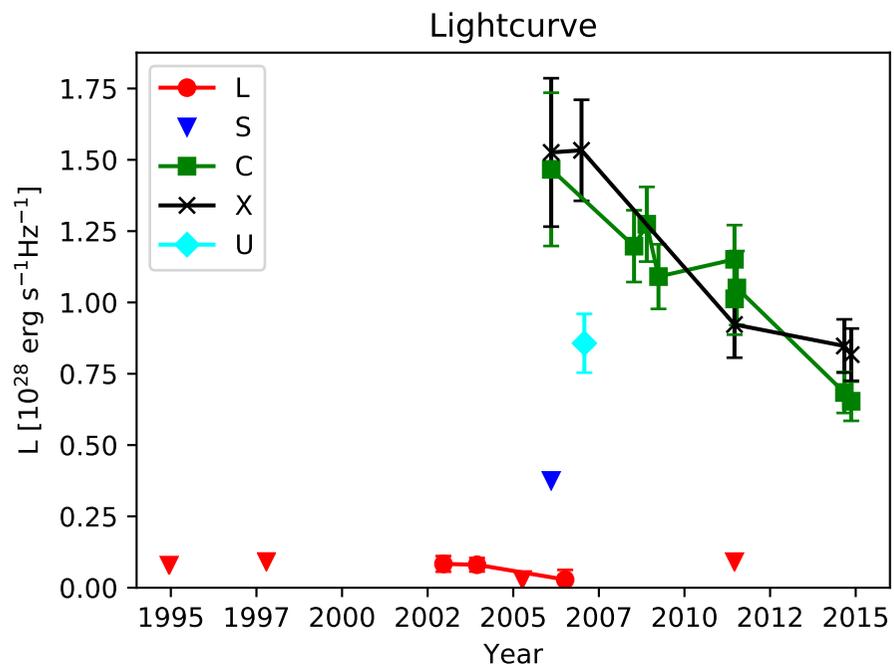
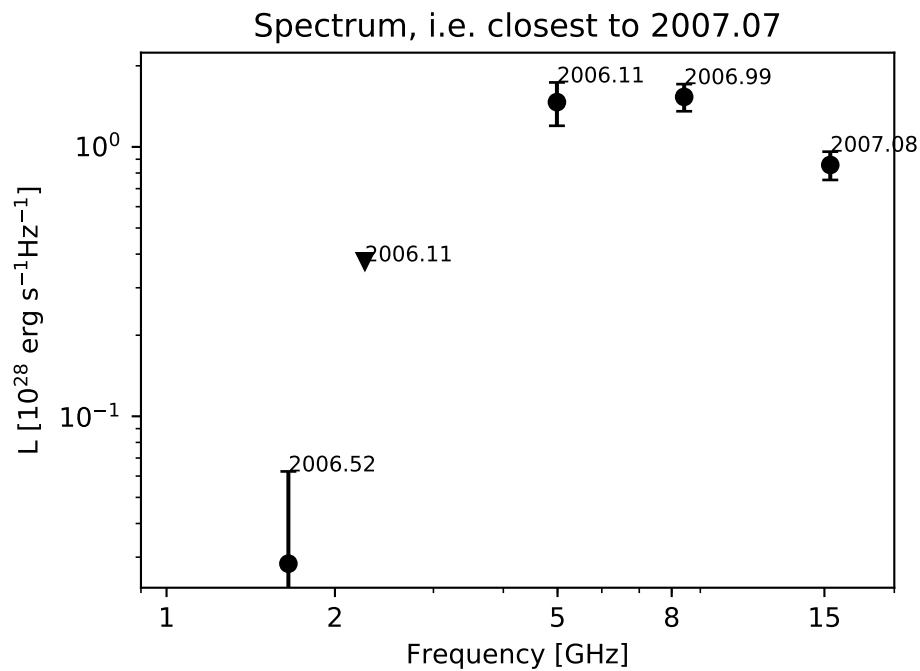
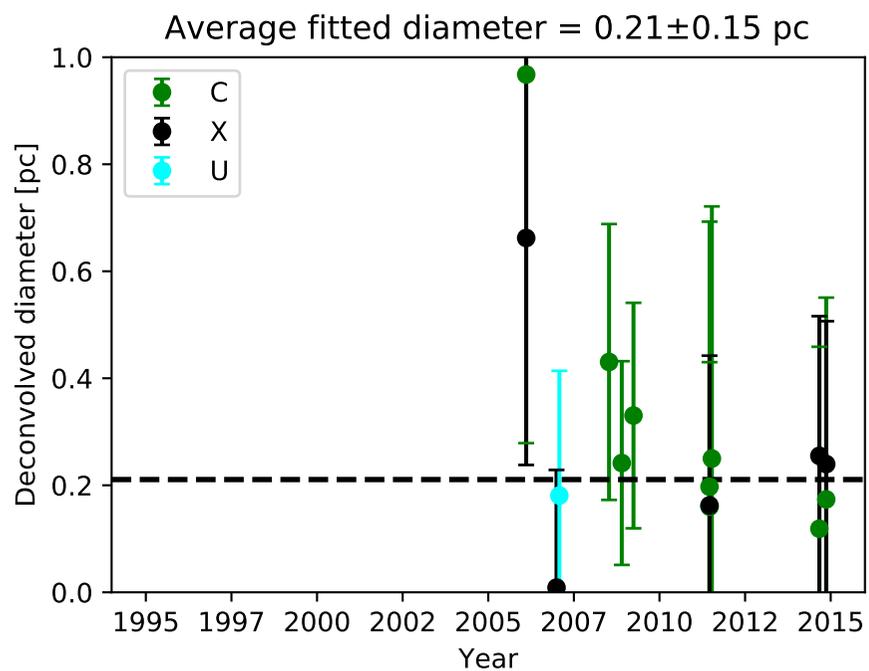

Classification: Fall

Figure notes:

Uncertainties calculated as described in Sect. 2.3.

Triangles represent 3σ upper limits for non-detections.

Sizes shown only for detections in bands C, X, and U.

Average diameters calculated according to Sect. 2.4 from the

3.6 cm and 6 cm measurements in experiments BB335A and BB335B.

# 0.2915+0.476 (E9)

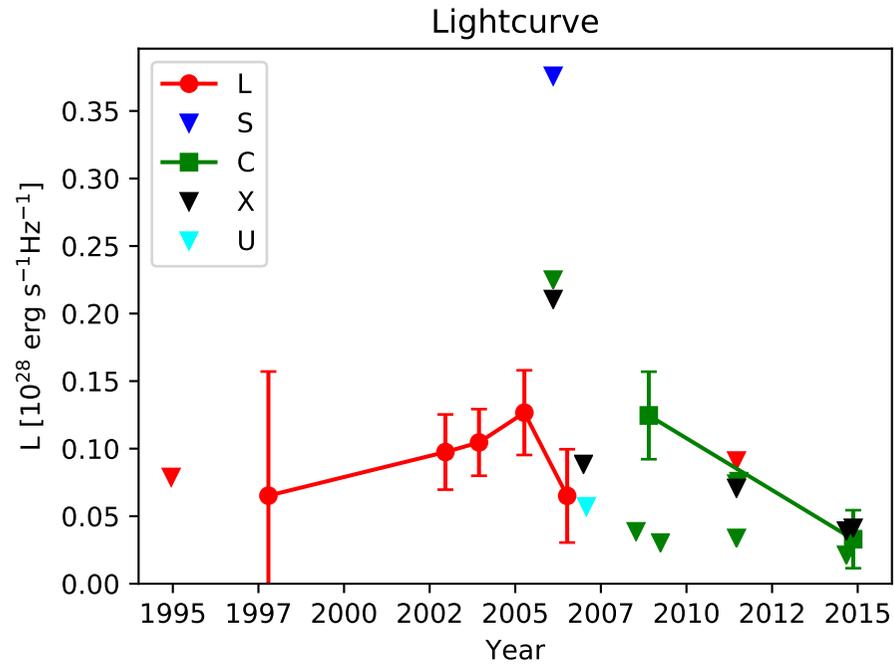

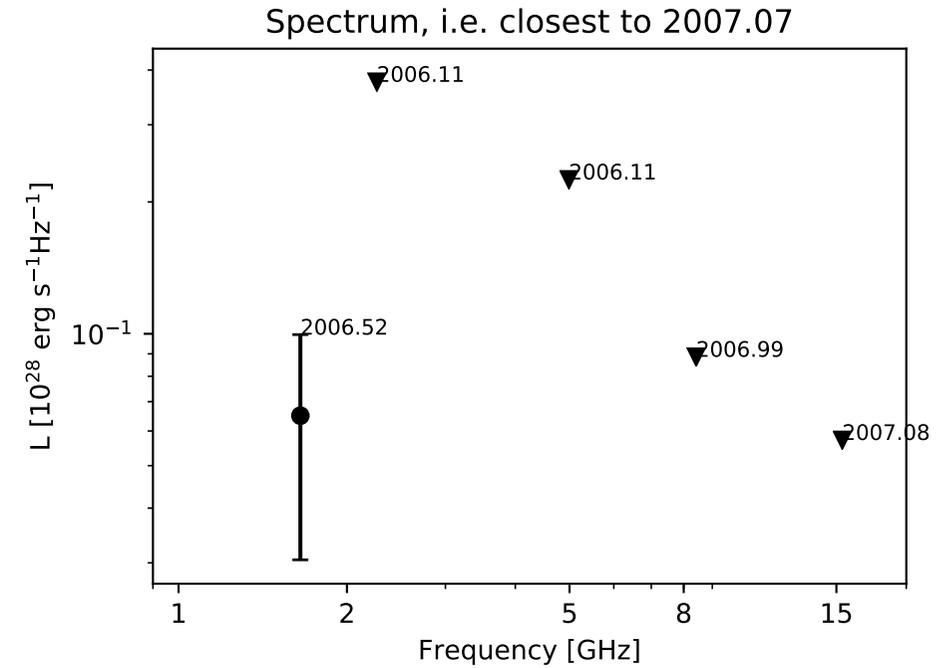

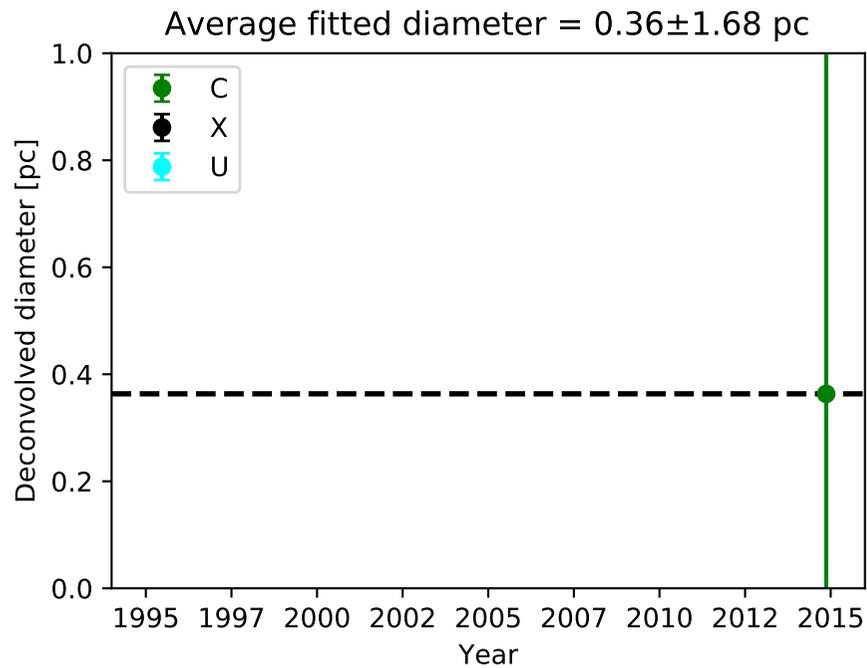

Classification: Slowly varying (S-var)

Figure notes:

Uncertainties calculated as described in Sect. 2.3.

Triangles represent $3\sigma$ upper limits for non-detections.

Sizes shown only for detections in bands C, X, and U.

Average diameters calculated according to Sect. 2.4 from the

3.6 cm and 6 cm measurements in experiments BB335A and BB335B.

# 0.2919+0.340

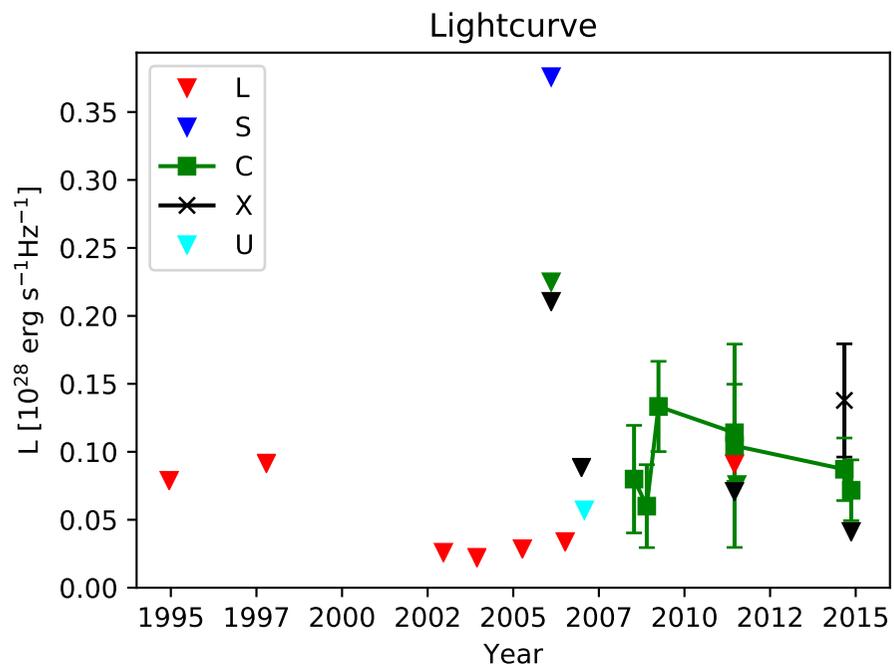

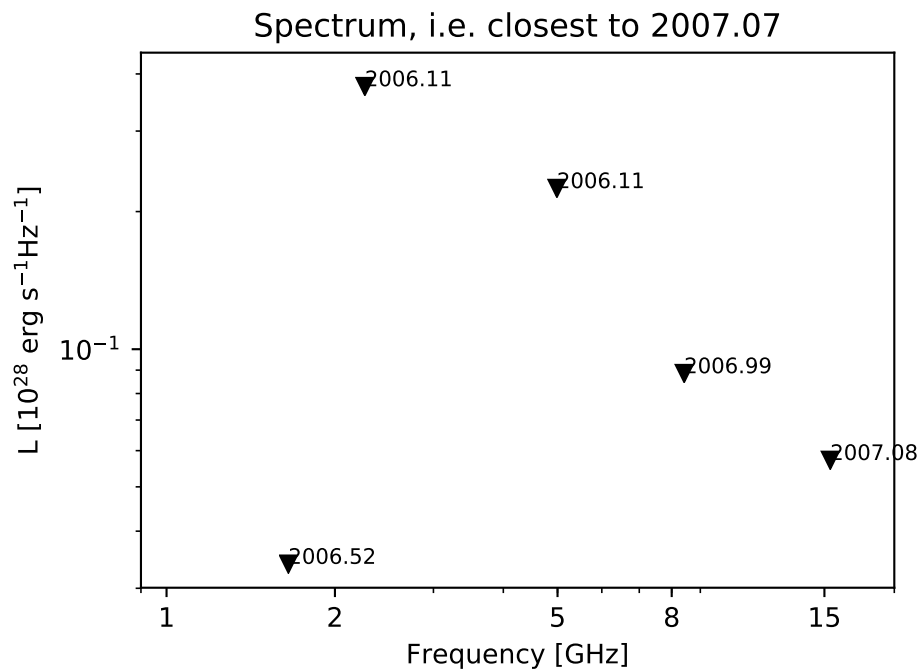

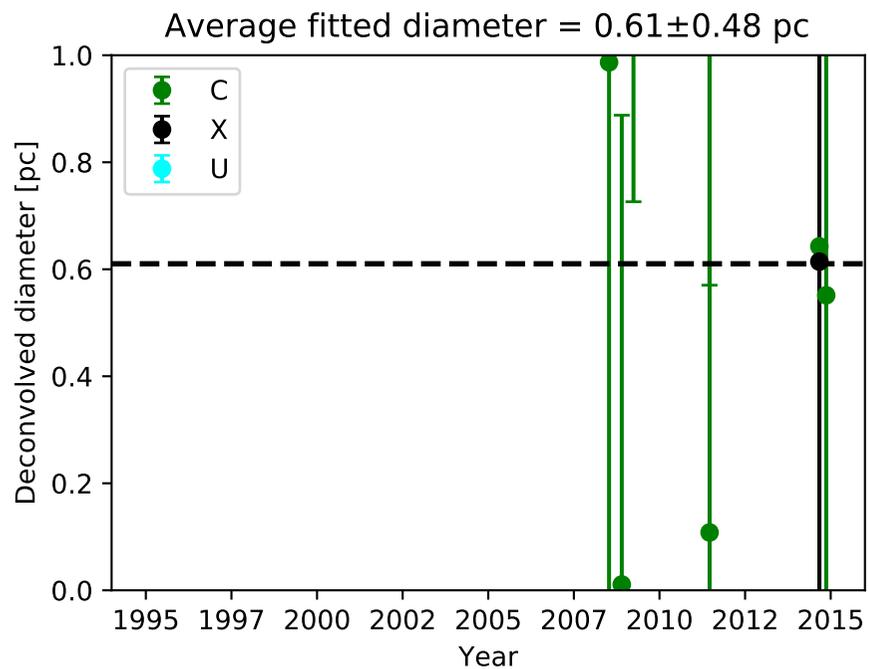

Classification: Slowly varying (S-var)

Figure notes:

Uncertainties calculated as described in Sect. 2.3.

Triangles represent 3σ upper limits for non-detections.

Sizes shown only for detections in bands C,X, and U.

Average diameters calculated according to Sect. 2.4 from the

3.6 cm and 6 cm measurements in experiments BB335A and BB335B.

# 0.2928+0.373 (E24)

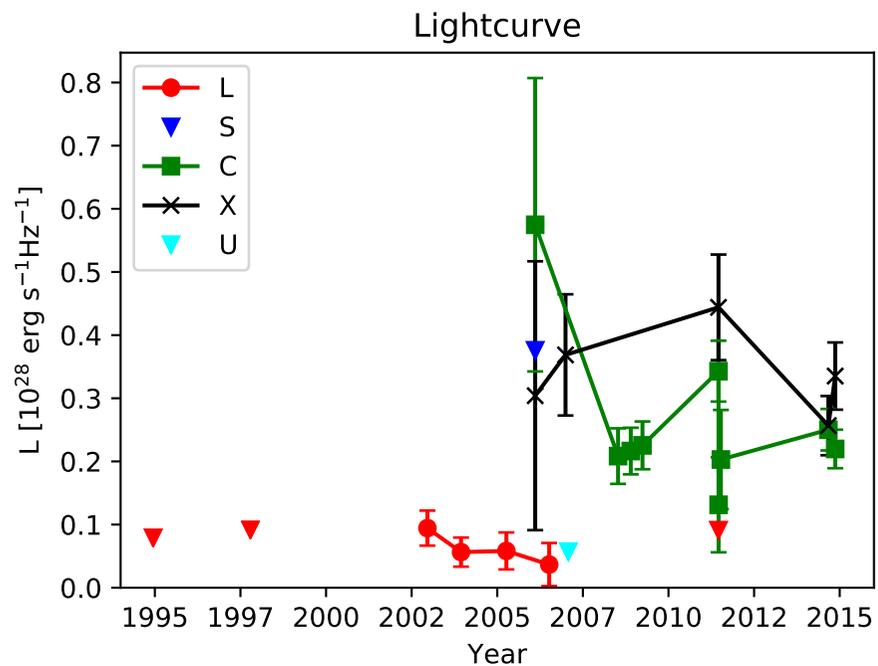

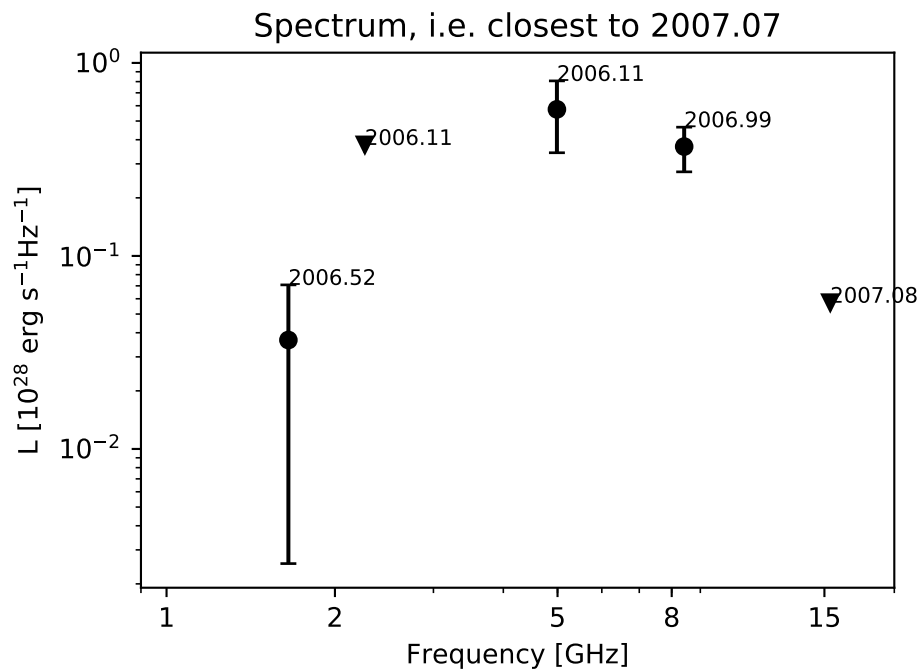

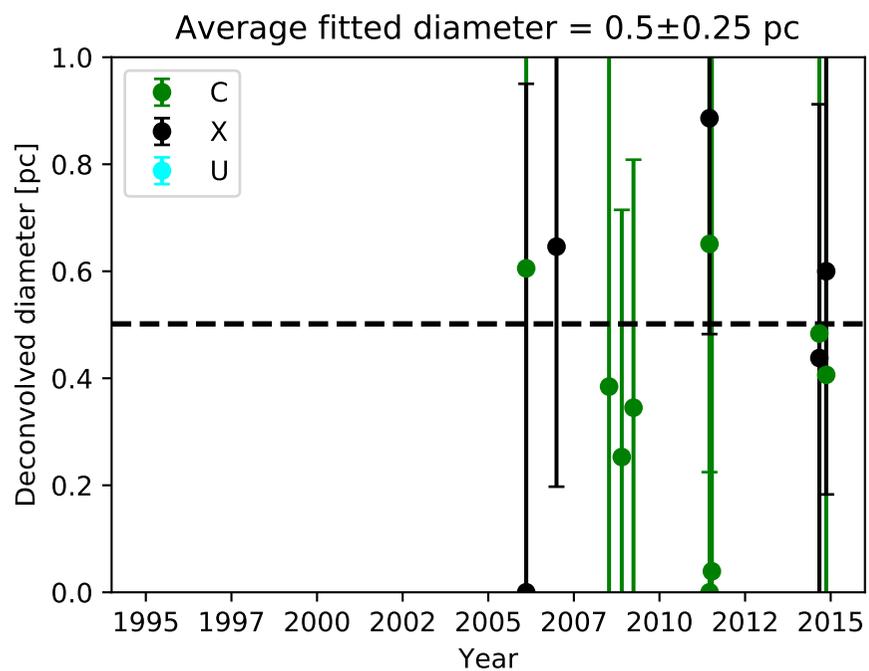

Classification: Slowly varying (S-var)

Figure notes:

Uncertainties calculated as described in Sect. 2.3.

Triangles represent $3\sigma$ upper limits for non-detections.

Sizes shown only for detections in bands C, X, and U.

Average diameters calculated according to Sect. 2.4 from the 3.6 cm and 6 cm measurements in experiments BB335A and BB335B.

# 0.2931+0.330 (E8)

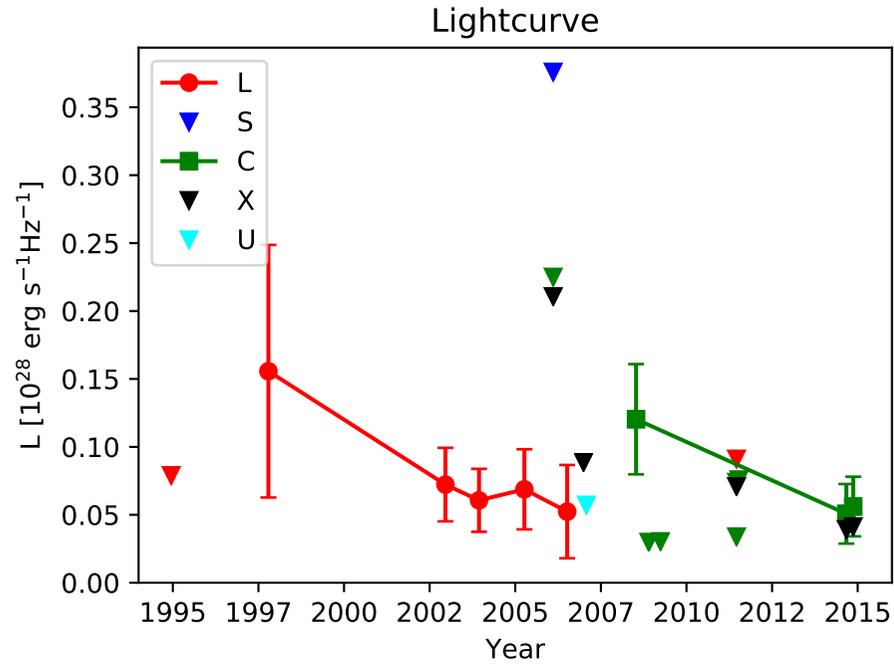
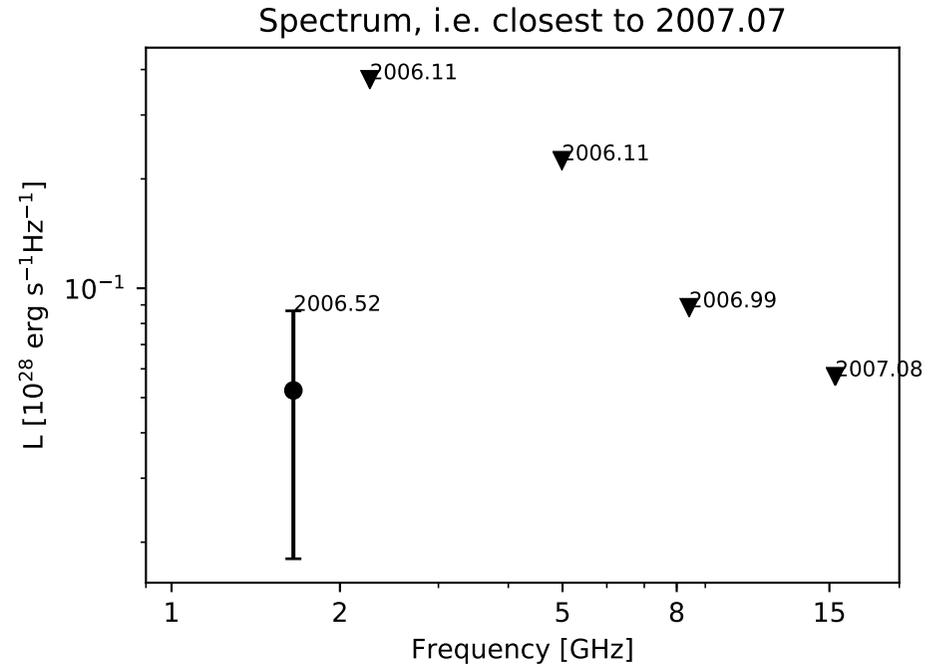
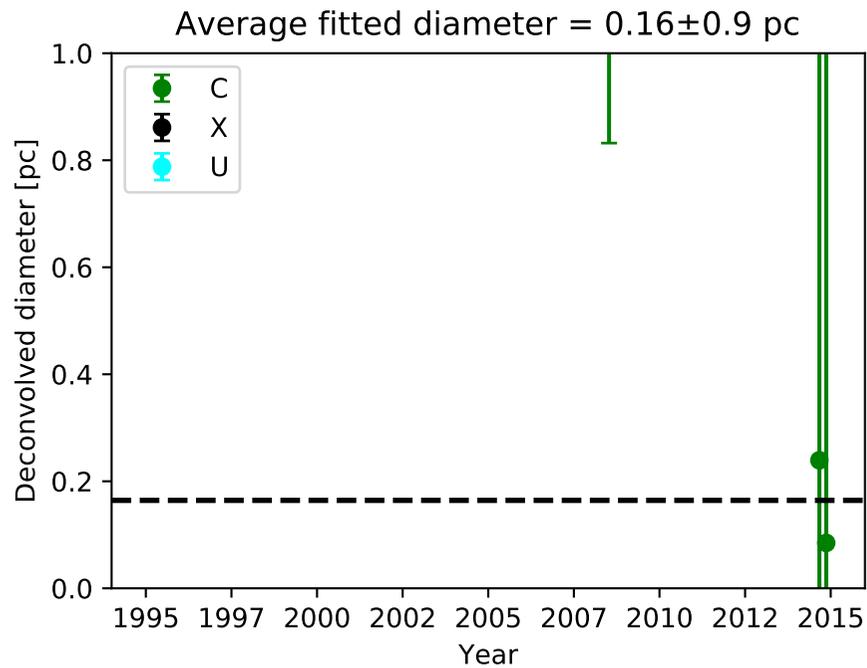

Classification: Slowly varying (S-var)

Figure notes:

Uncertainties calculated as described in Sect. 2.3.

Triangles represent 3σ upper limits for non-detections.

Sizes shown only for detections in bands C, X, and U.

Average diameters calculated according to Sect. 2.4 from the

3.6 cm and 6 cm measurements in experiments BB335A and BB335B.

# 0.2938+0.344

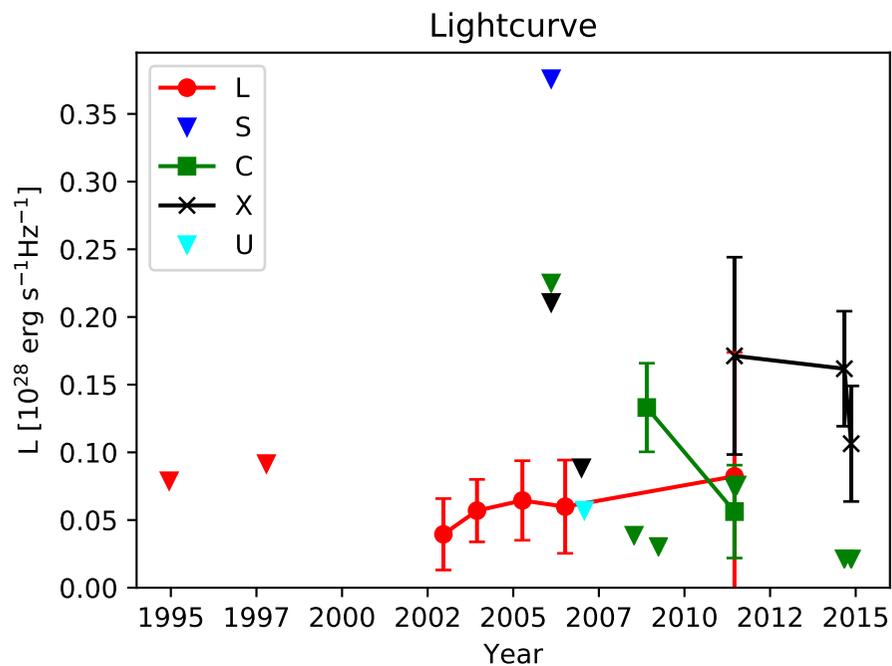

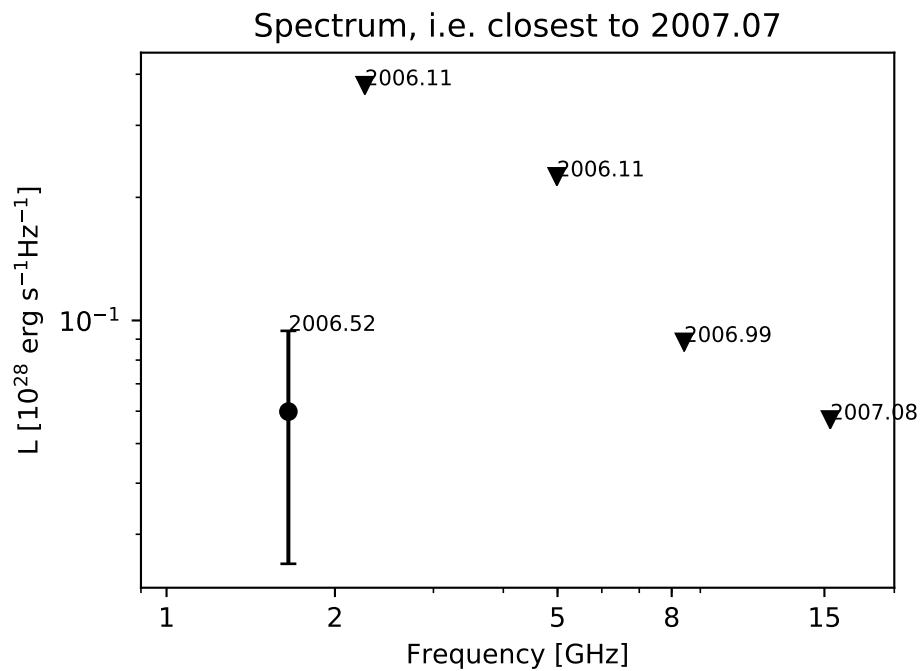

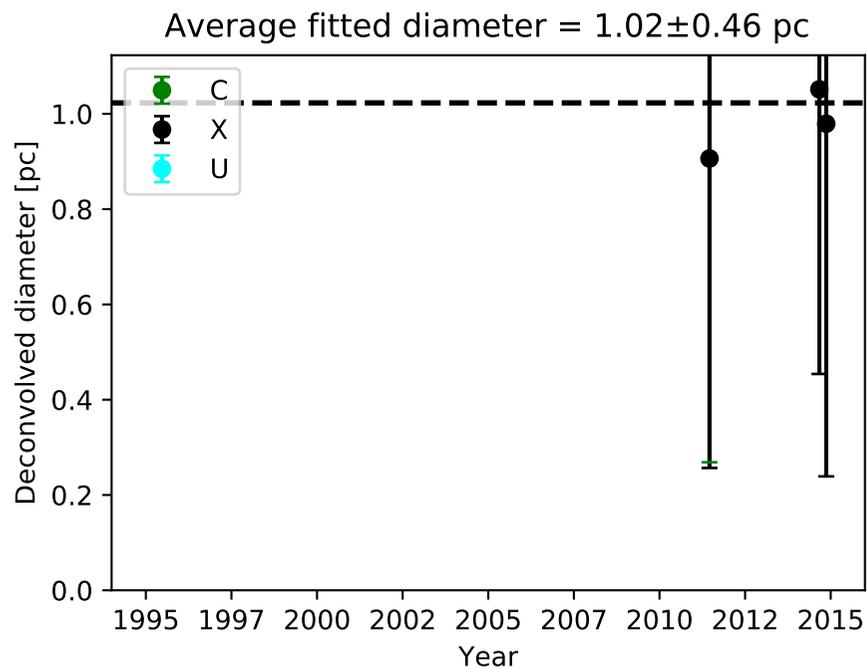

Classification: Fall

Figure notes:

Uncertainties calculated as described in Sect. 2.3.

Triangles represent 3σ upper limits for non-detections.

Sizes shown only for detections in bands C, X, and U.

Average diameters calculated according to Sect. 2.4 from the

3.6 cm and 6 cm measurements in experiments BB335A and BB335B.

# 0.2940+0.481

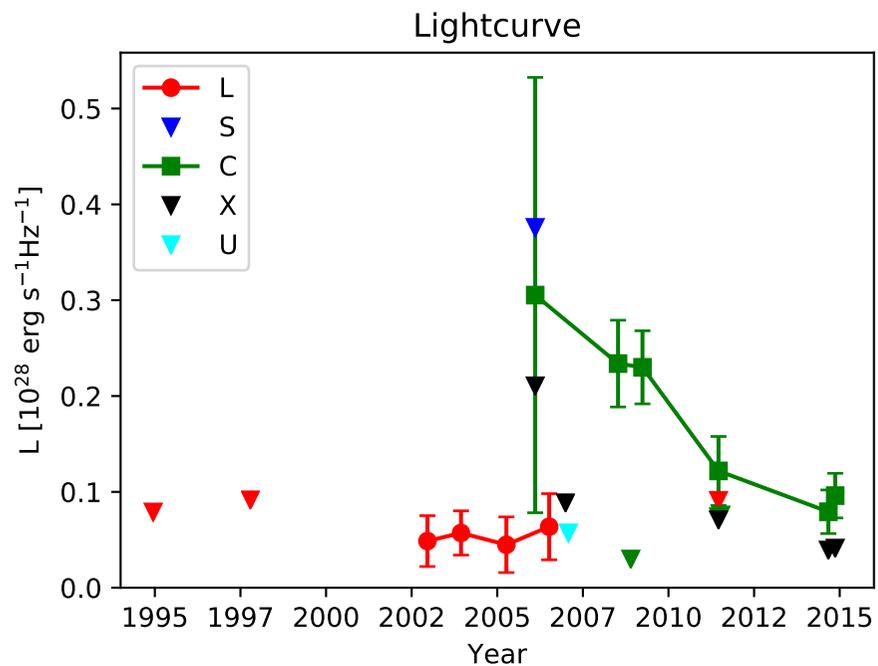
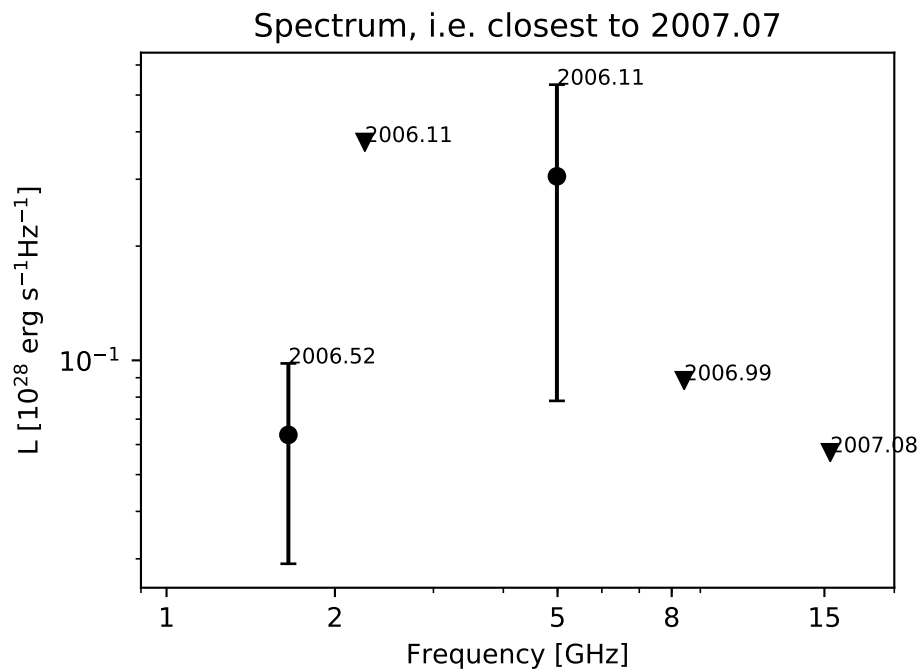
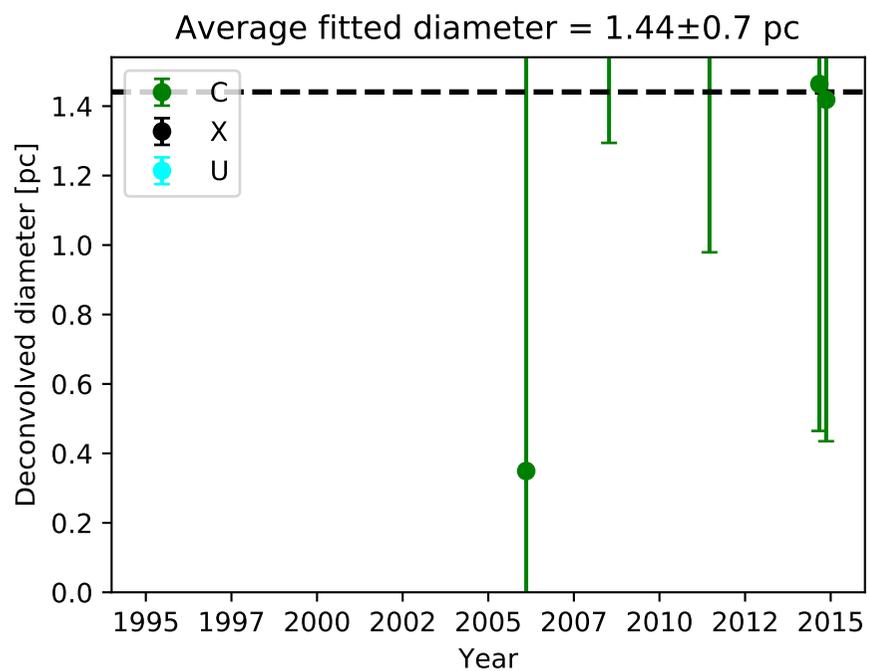

Classification: Slowly varying (S-var)

Figure notes:

Uncertainties calculated as described in Sect. 2.3.

Triangles represent 3σ upper limits for non-detections.

Sizes shown only for detections in bands C, X, and U.

Average diameters calculated according to Sect. 2.4 from the

3.6 cm and 6 cm measurements in experiments BB335A and BB335B.

# 0.2943+0.279 (E7)

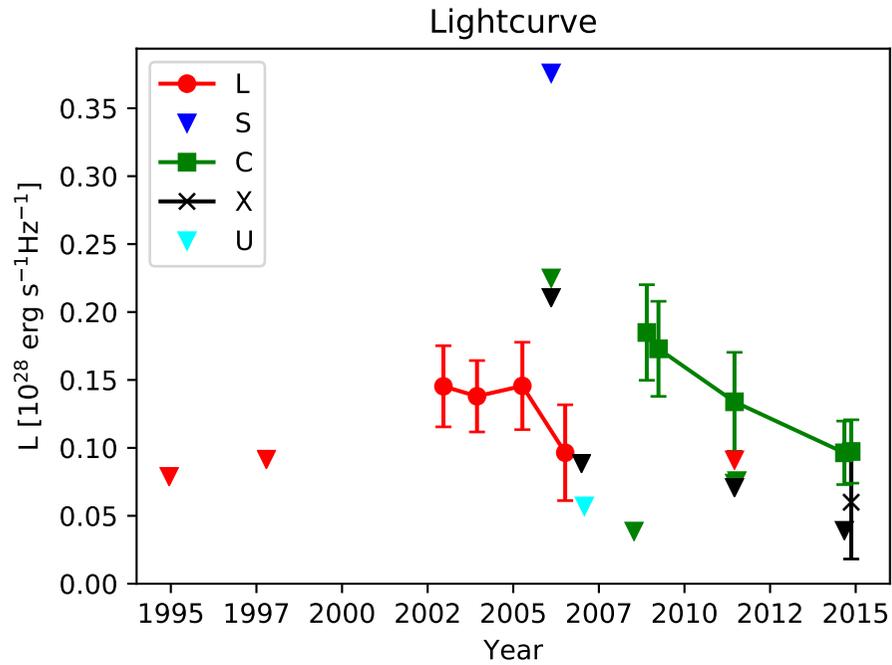

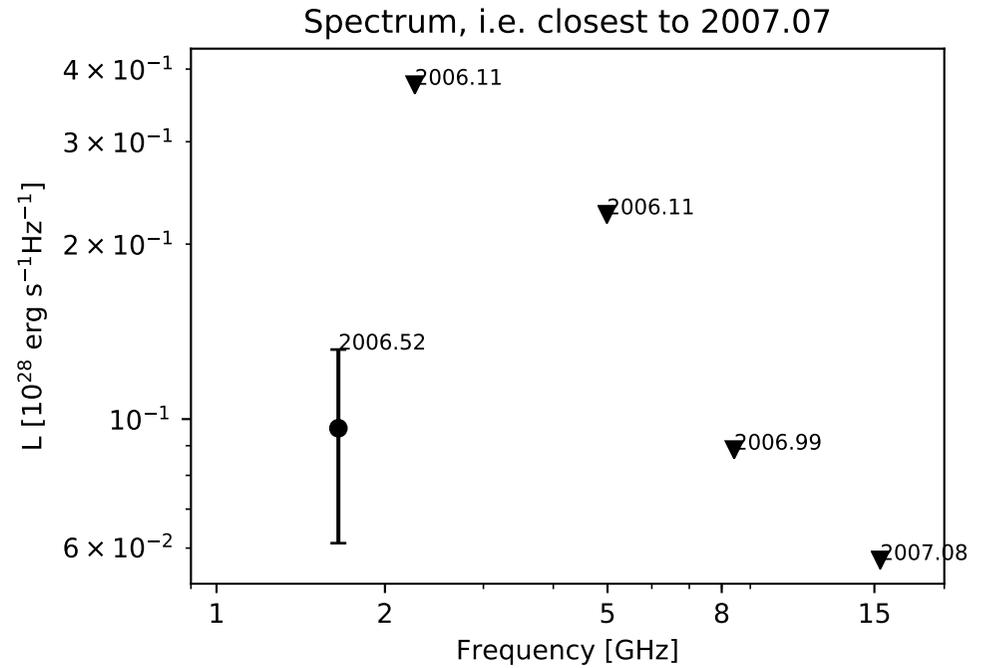

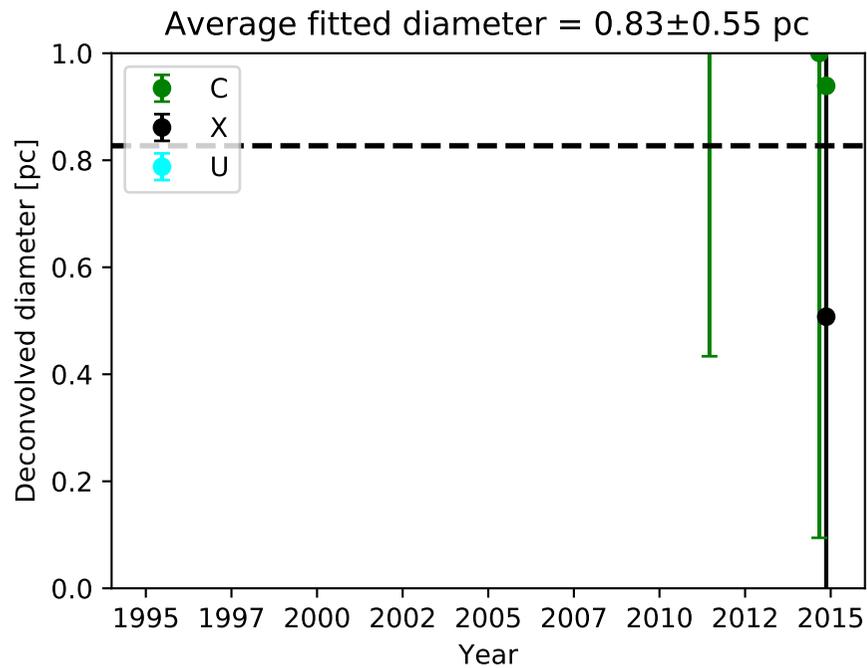

Classification: Slowly varying (S-var)

Figure notes:

Uncertainties calculated as described in Sect. 2.3.

Triangles represent 3σ upper limits for non-detections.

Sizes shown only for detections in bands C, X, and U.

Average diameters calculated according to Sect. 2.4 from the

3.6 cm and 6 cm measurements in experiments BB335A and BB335B.

# 0.2947+0.263 (E6)

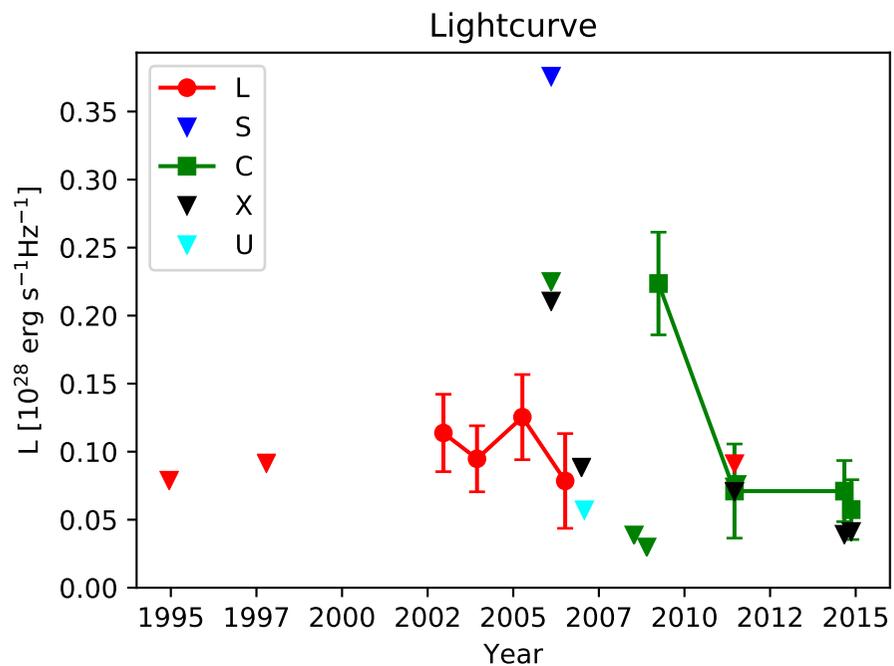

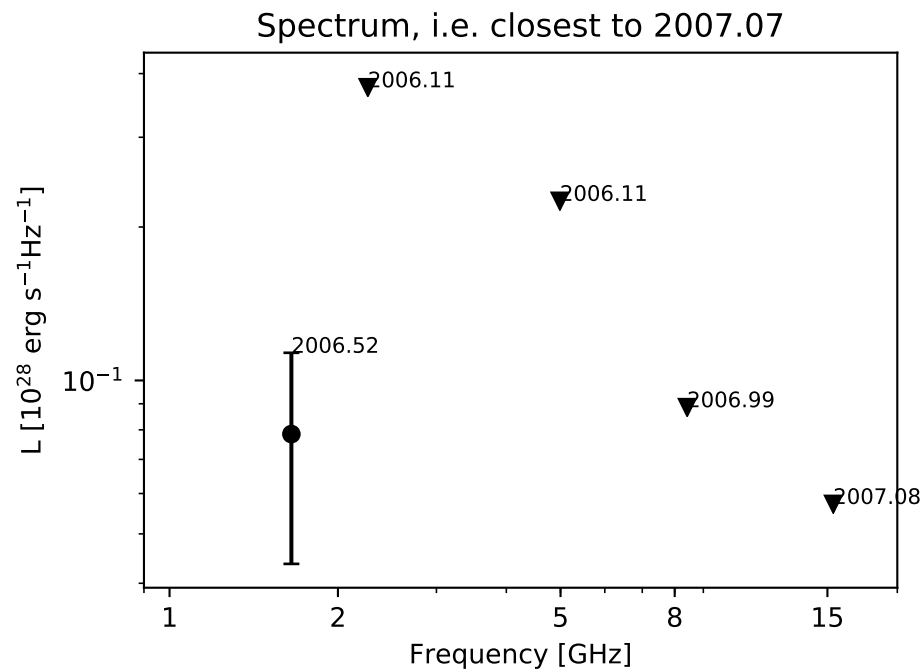

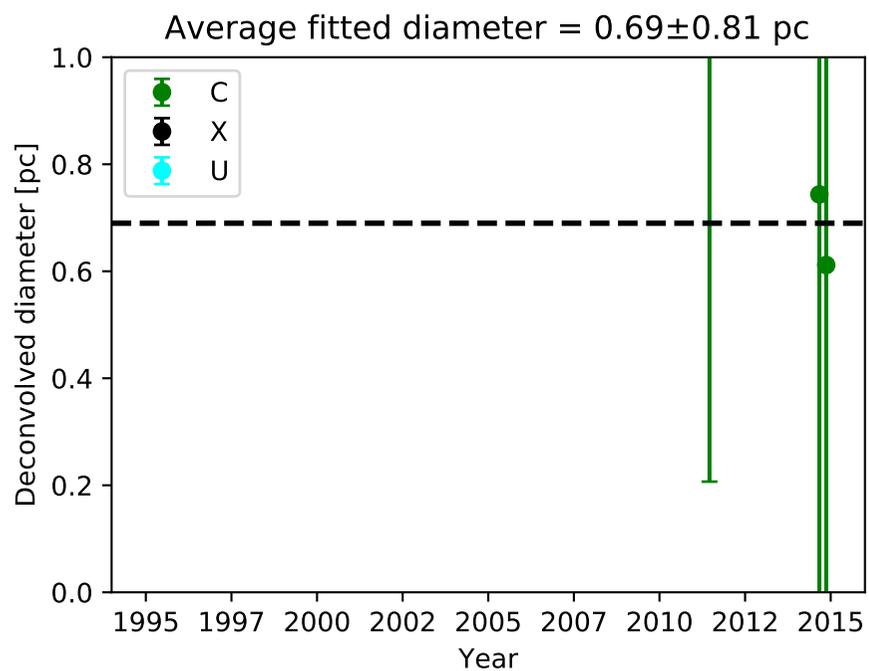

Classification: Slowly varying (S-var)

Figure notes:

Uncertainties calculated as described in Sect. 2.3.

Triangles represent 3σ upper limits for non-detections.

Sizes shown only for detections in bands C, X, and U.

Average diameters calculated according to Sect. 2.4 from the

3.6 cm and 6 cm measurements in experiments BB335A and BB335B.

# 0.2951+0.384

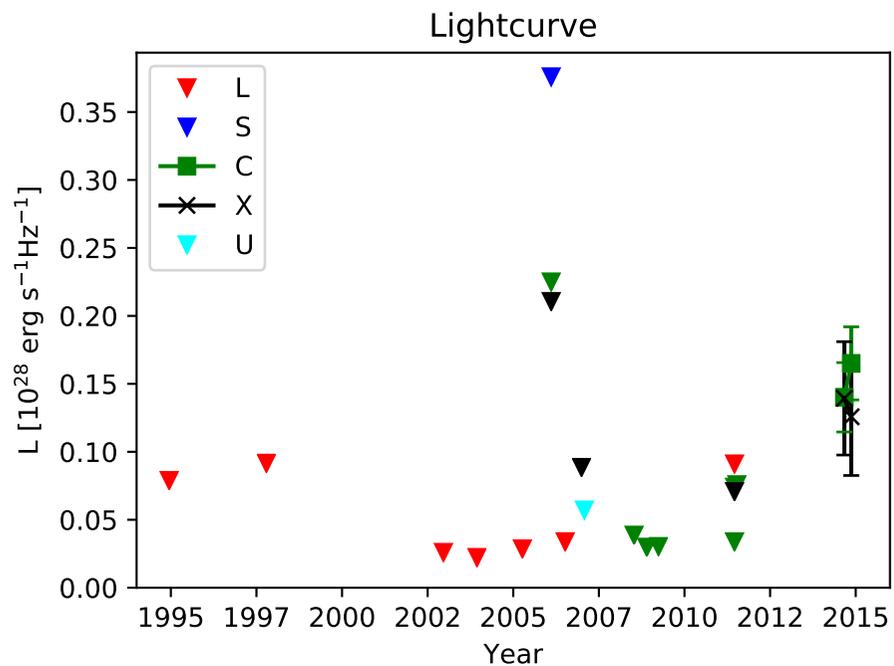
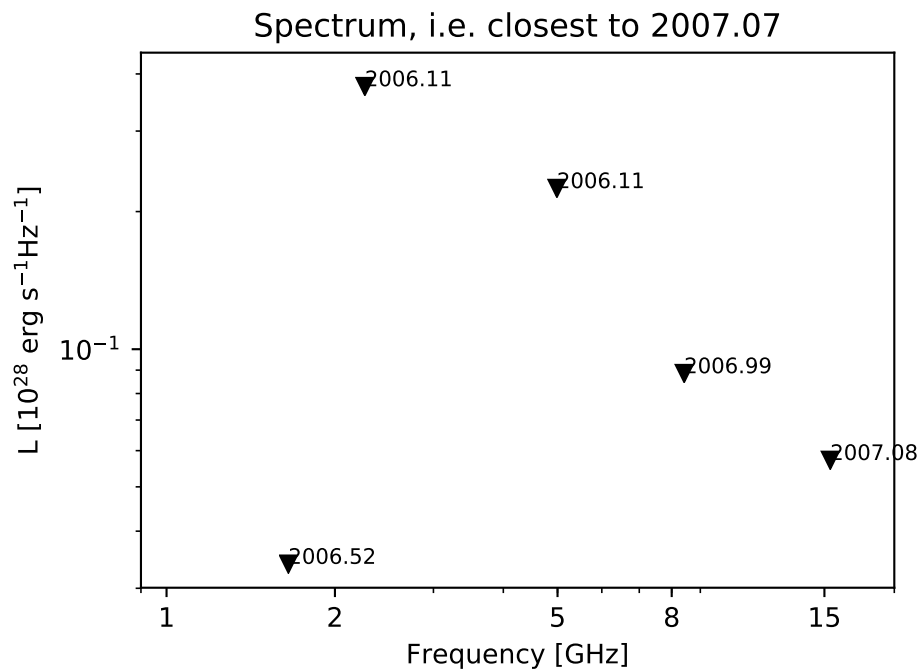
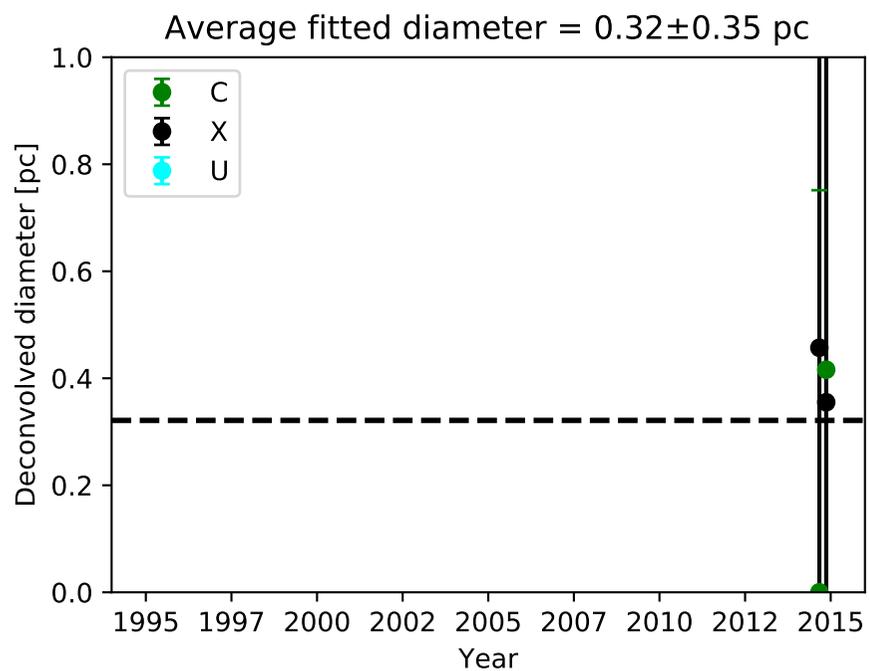

Classification: Rise

Figure notes:

Uncertainties calculated as described in Sect. 2.3.

Triangles represent $3\sigma$ upper limits for non-detections.

Sizes shown only for detections in bands C, X, and U.

Average diameters calculated according to Sect. 2.4 from the

3.6 cm and 6 cm measurements in experiments BB335A and BB335B.

# 0.2954+0.393

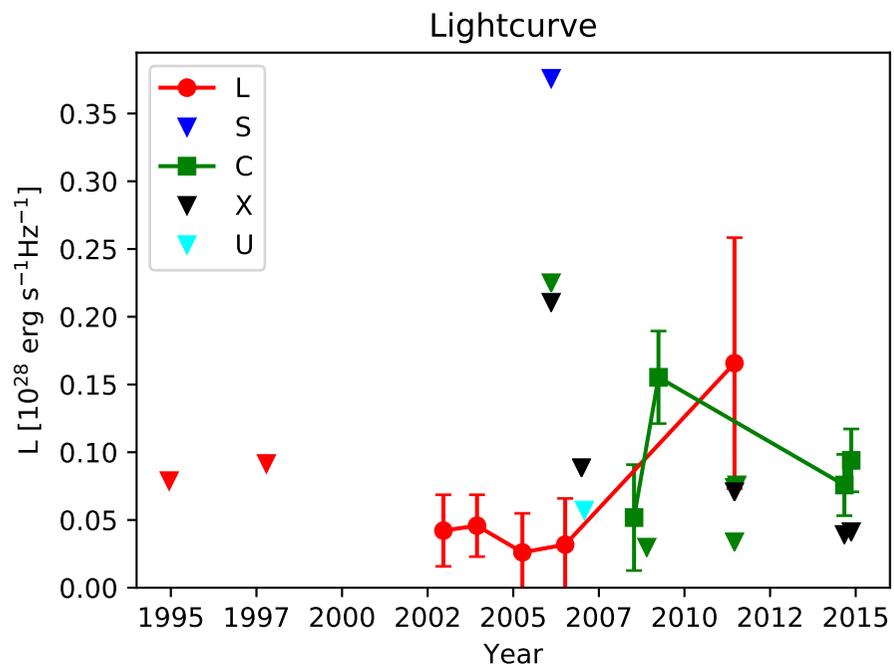
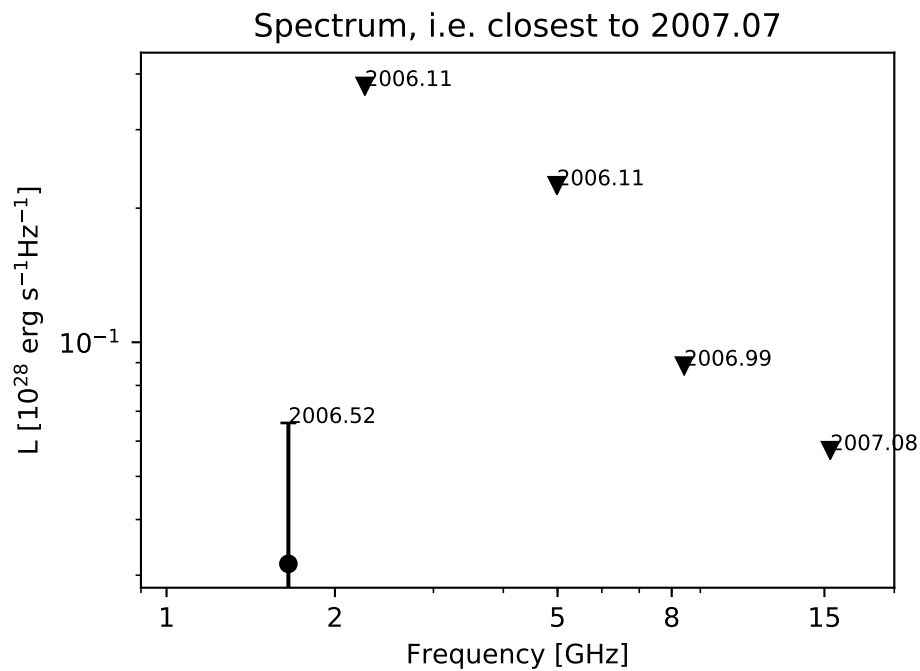
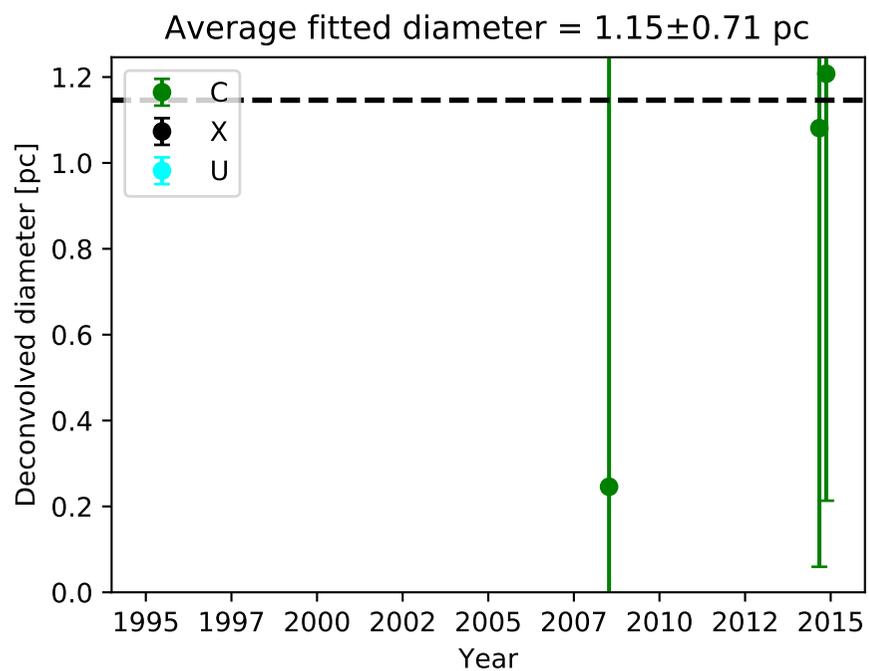

Classification: Slowly varying (S-var)

Figure notes:

Uncertainties calculated as described in Sect. 2.3.

Triangles represent 3σ upper limits for non-detections.

Sizes shown only for detections in bands C,X, and U.

Average diameters calculated according to Sect. 2.4 from the

3.6 cm and 6 cm measurements in experiments BB335A and BB335B.

# 0.2959+0.263

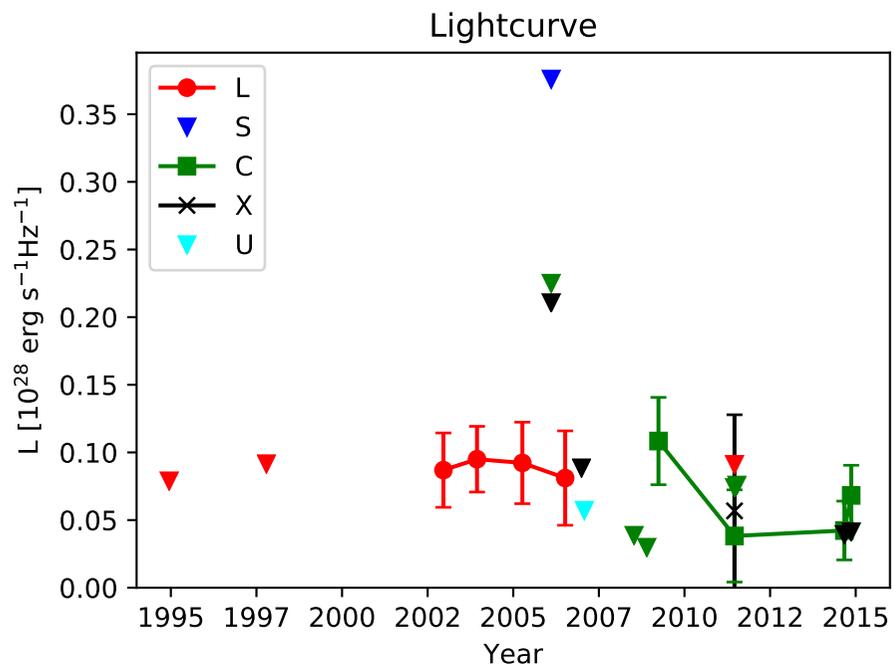

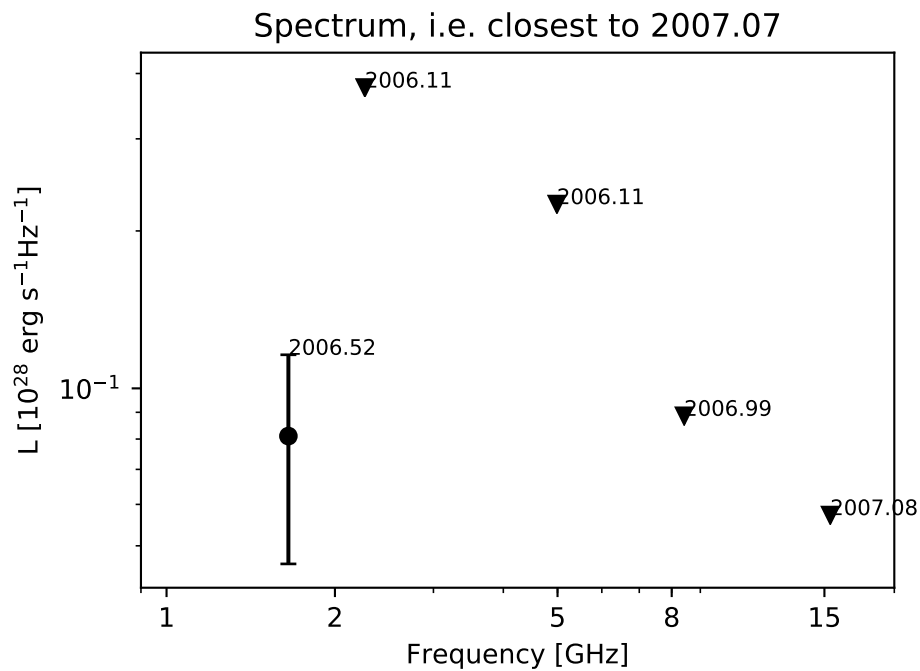

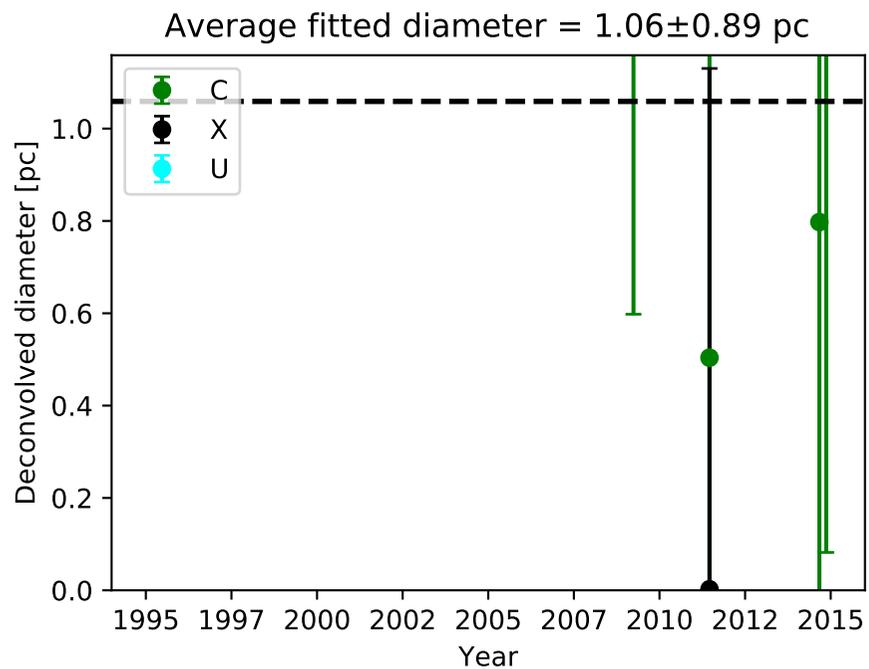

Classification: Slowly varying (S-var)

Figure notes:

Uncertainties calculated as described in Sect. 2.3.

Triangles represent 3σ upper limits for non-detections.

Sizes shown only for detections in bands C, X, and U.

Average diameters calculated according to Sect. 2.4 from the

3.6 cm and 6 cm measurements in experiments BB335A and BB335B.

# 0.2971+0.378

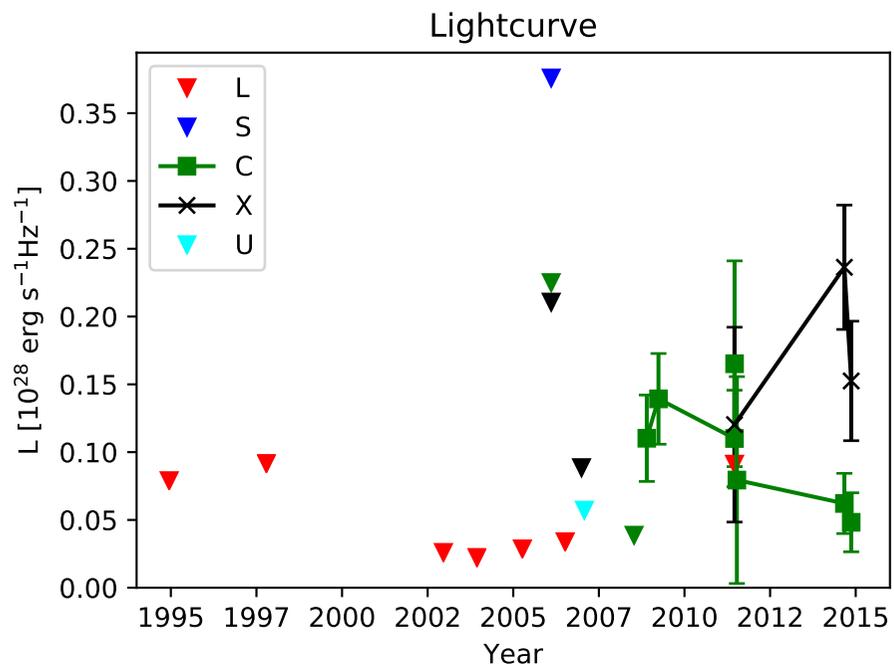

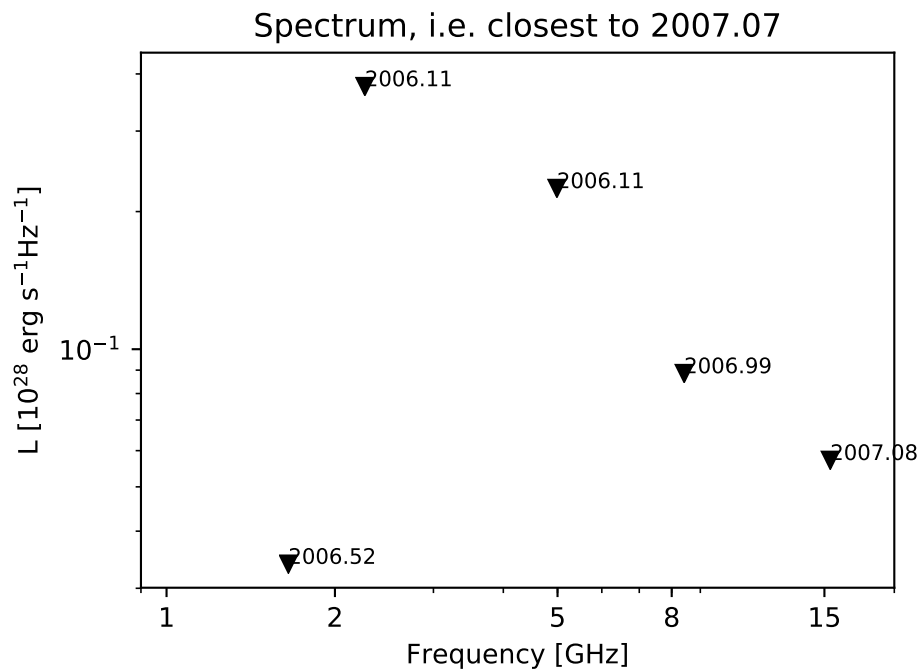

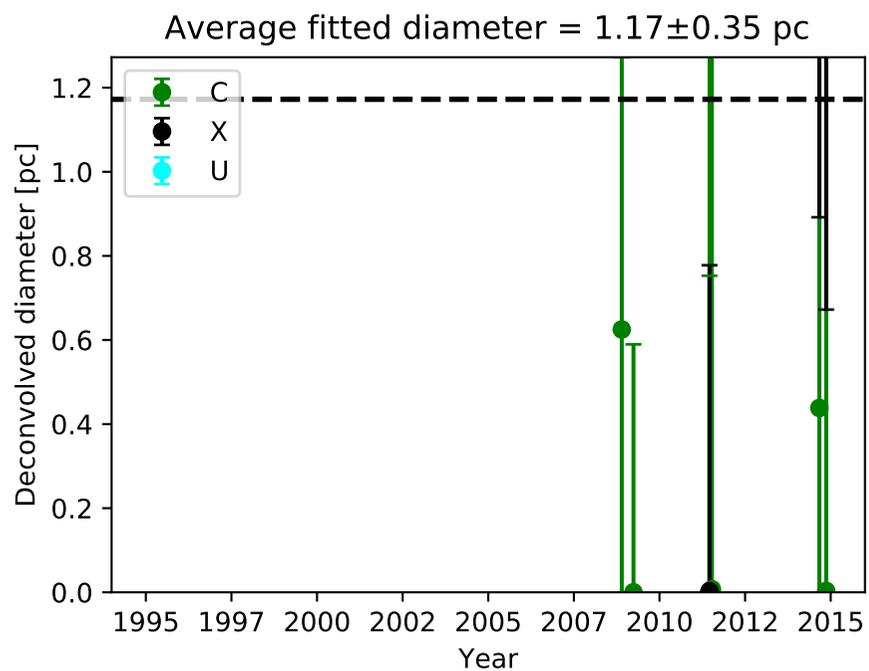

Classification: Slowly varying (S-var)

Figure notes:

Uncertainties calculated as described in Sect. 2.3.

Triangles represent 3σ upper limits for non-detections.

Sizes shown only for detections in bands C, X, and U.

Average diameters calculated according to Sect. 2.4 from the

3.6 cm and 6 cm measurements in experiments BB335A and BB335B.

# 0.2979+0.419 (E5)

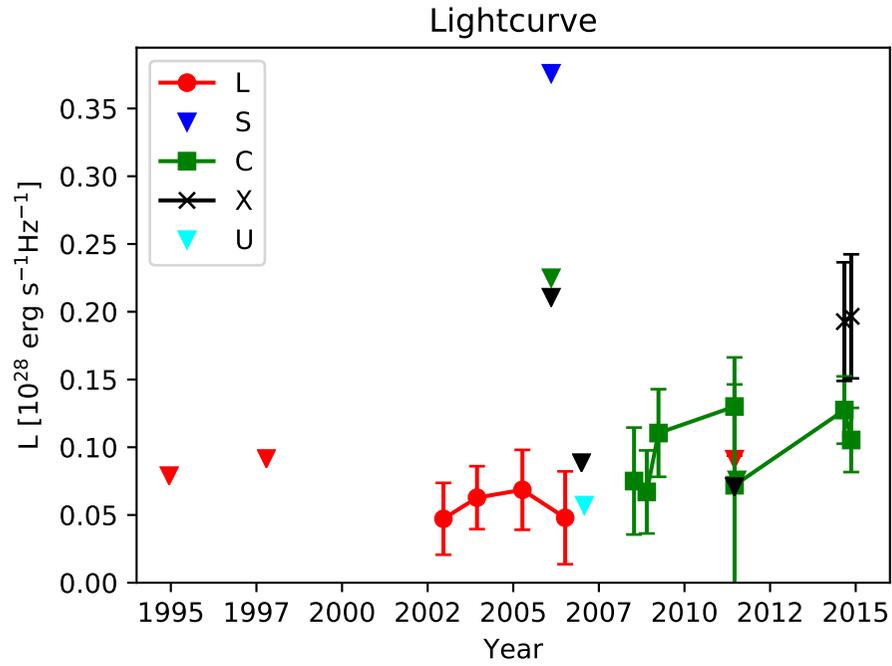
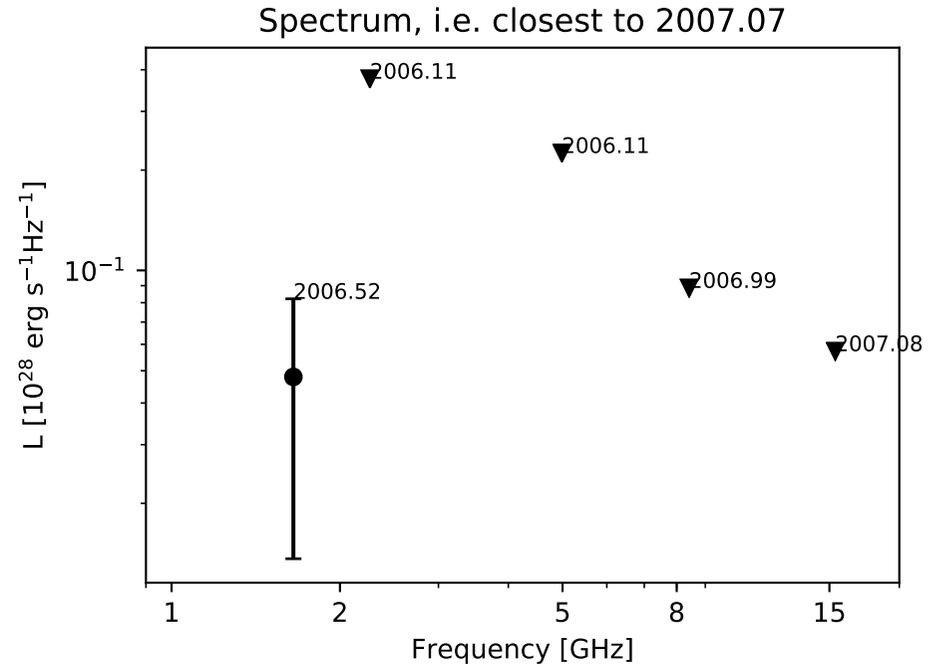
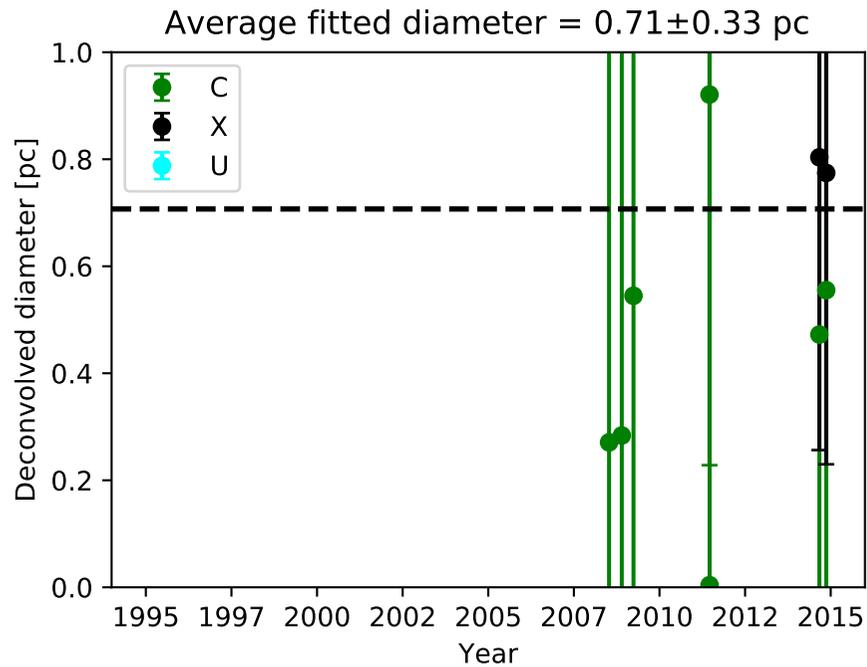

Classification: Slowly varying (S-var)

Figure notes:

Uncertainties calculated as described in Sect. 2.3.

Triangles represent 3σ upper limits for non-detections.

Sizes shown only for detections in bands C, X, and U.

Average diameters calculated according to Sect. 2.4 from the

3.6 cm and 6 cm measurements in experiments BB335A and BB335B.

# 0.2995+0.502

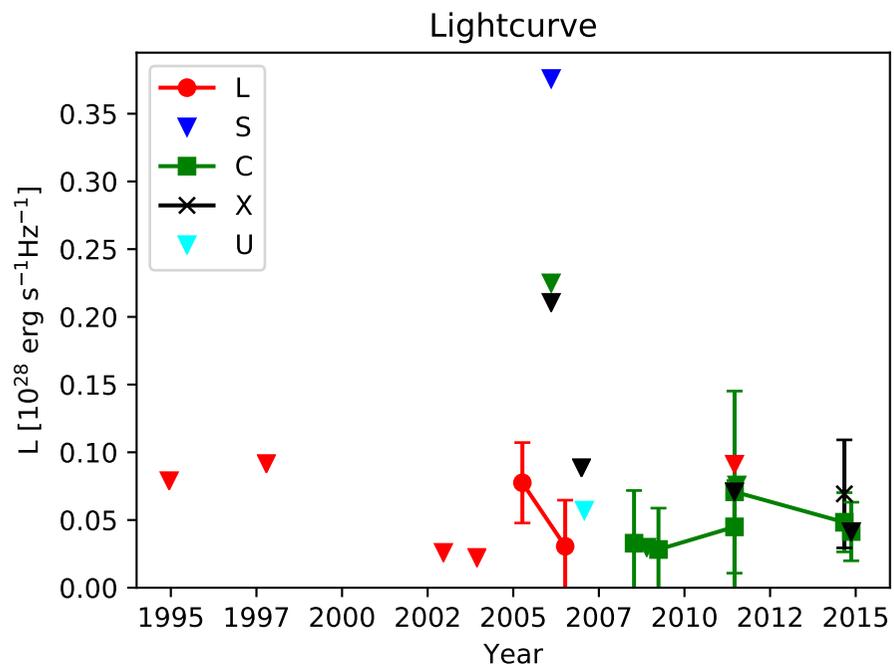

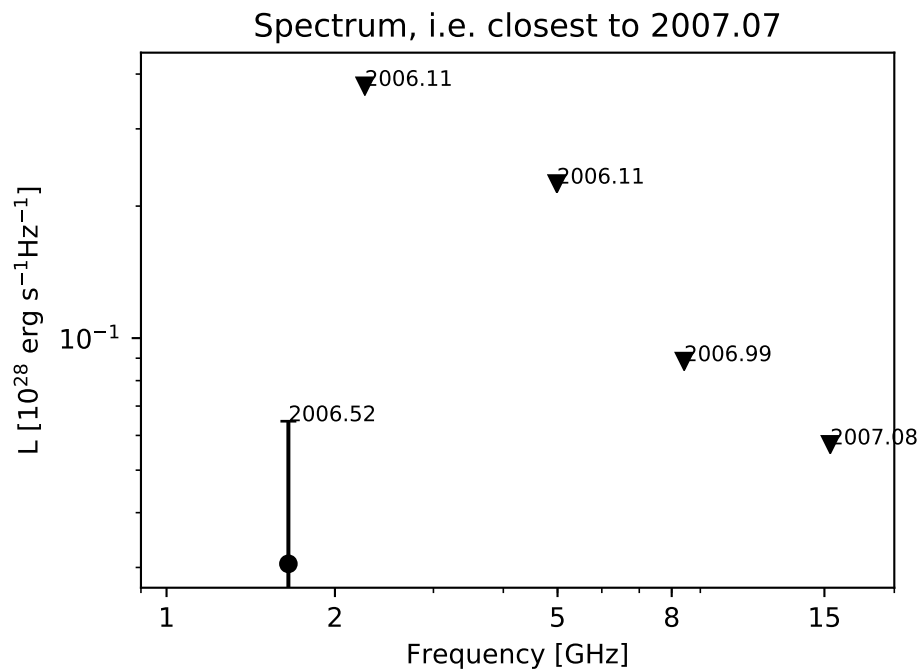

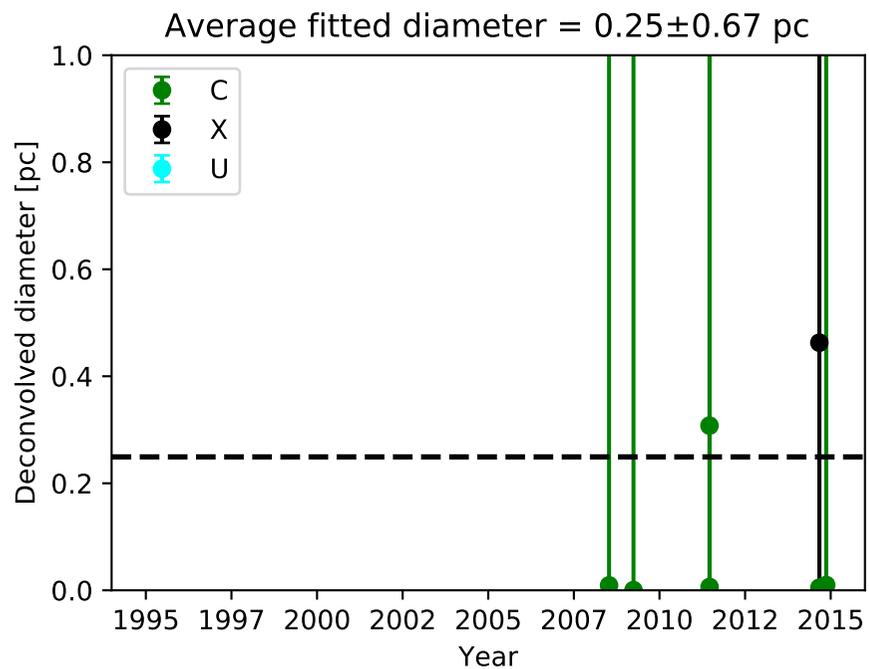

Classification: Slowly varying (S-var)

Figure notes:

Uncertainties calculated as described in Sect. 2.3.

Triangles represent $3\sigma$ upper limits for non-detections.

Sizes shown only for detections in bands C, X, and U.

Average diameters calculated according to Sect. 2.4 from the

3.6 cm and 6 cm measurements in experiments BB335A and BB335B.

# 0.3011+0.341

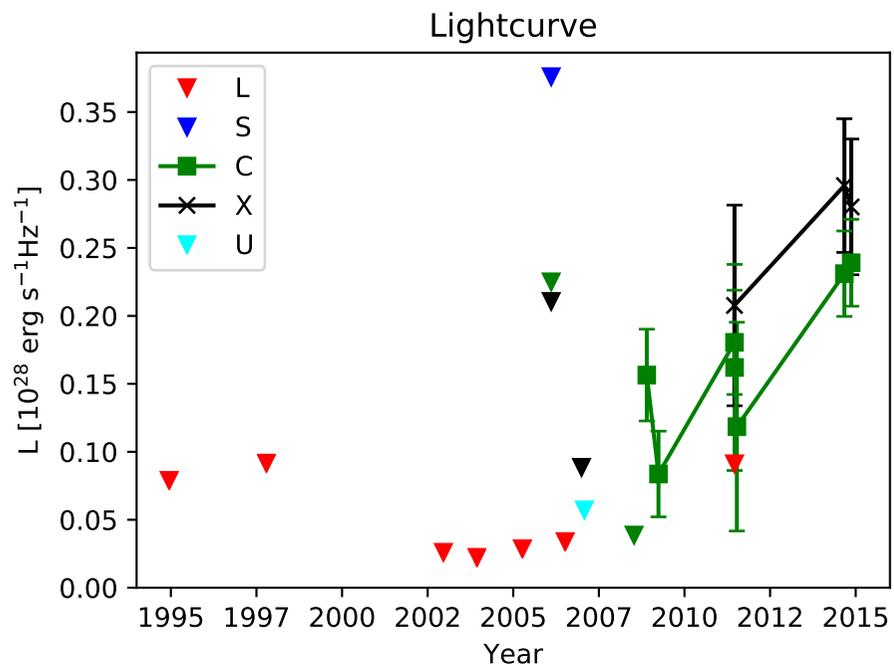
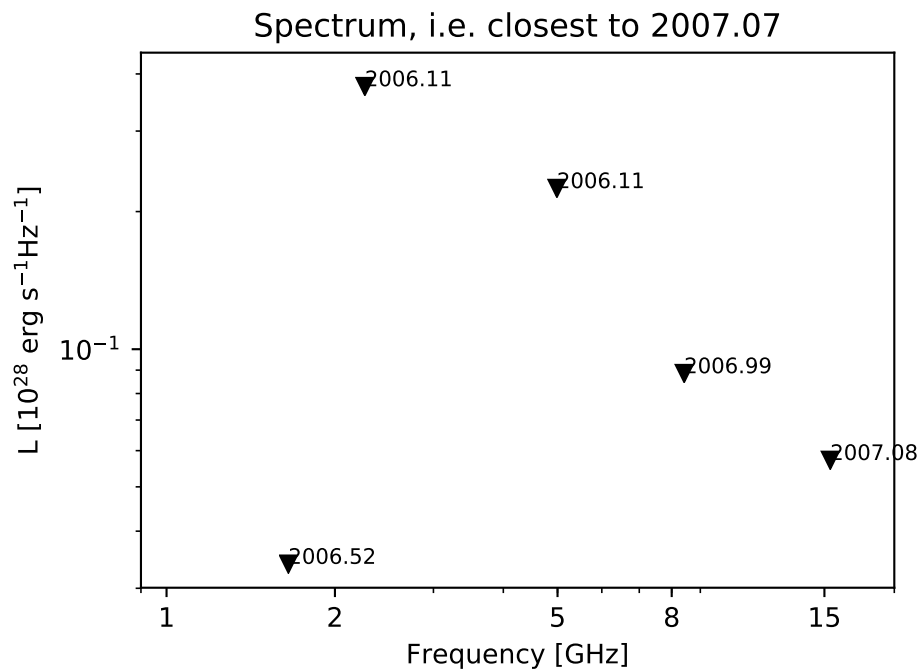
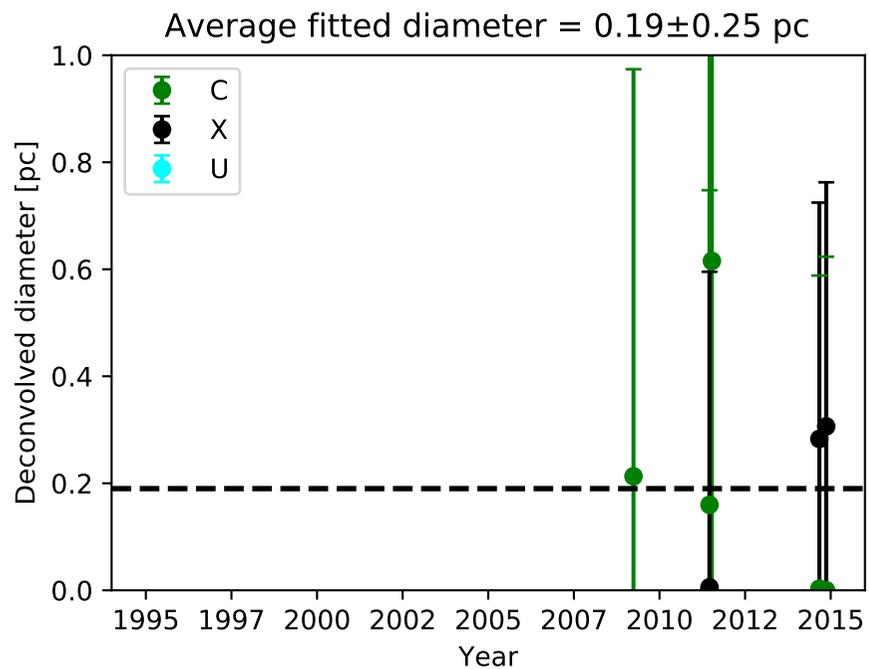

Classification: Rise

Figure notes:

Uncertainties calculated as described in Sect. 2.3.

Triangles represent $3\sigma$ upper limits for non-detections.

Sizes shown only for detections in bands C, X, and U.

Average diameters calculated according to Sect. 2.4 from the

3.6 cm and 6 cm measurements in experiments BB335A and BB335B.

# 0.3073+0.332 (E3)

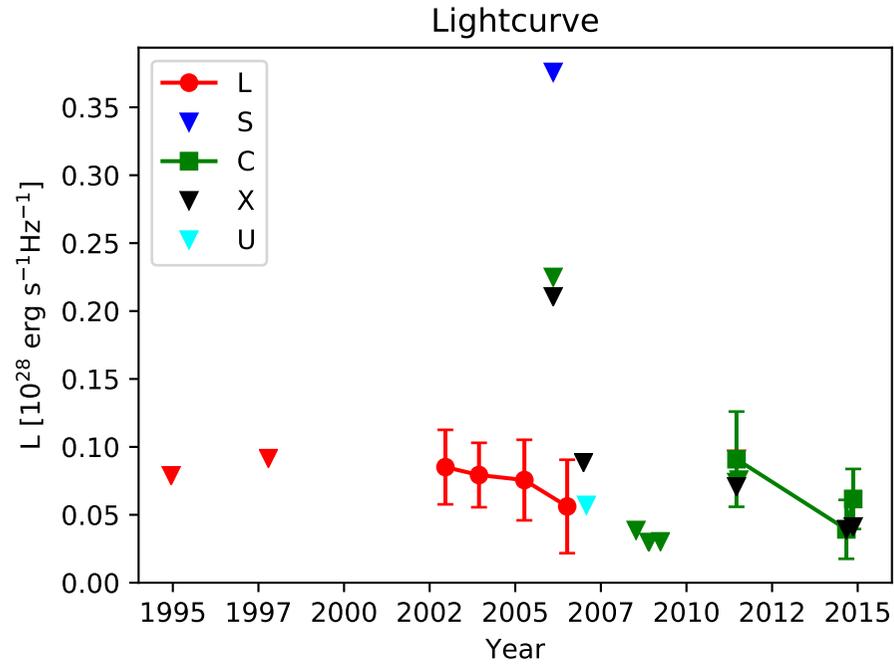

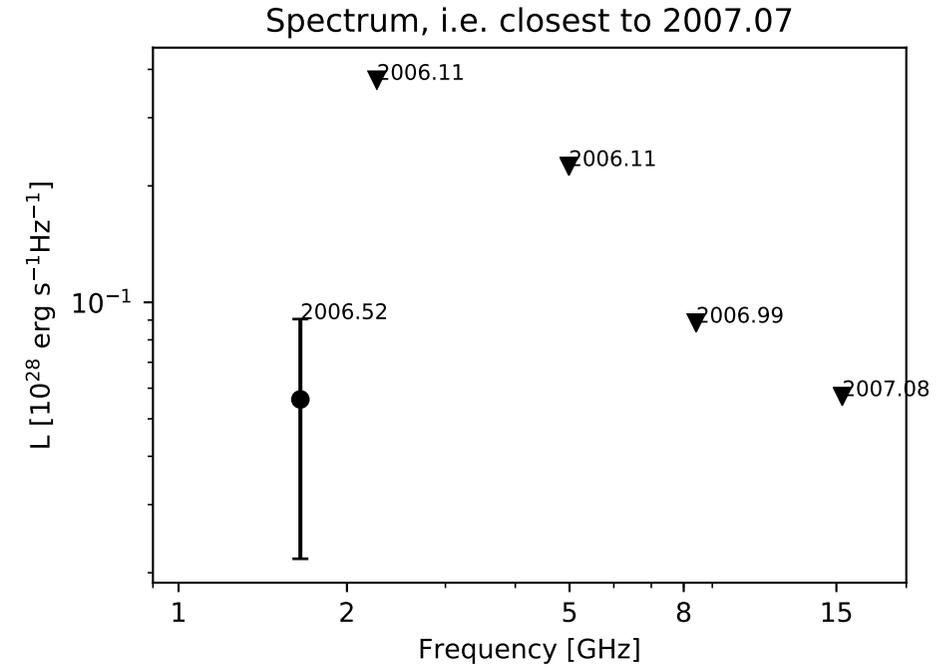

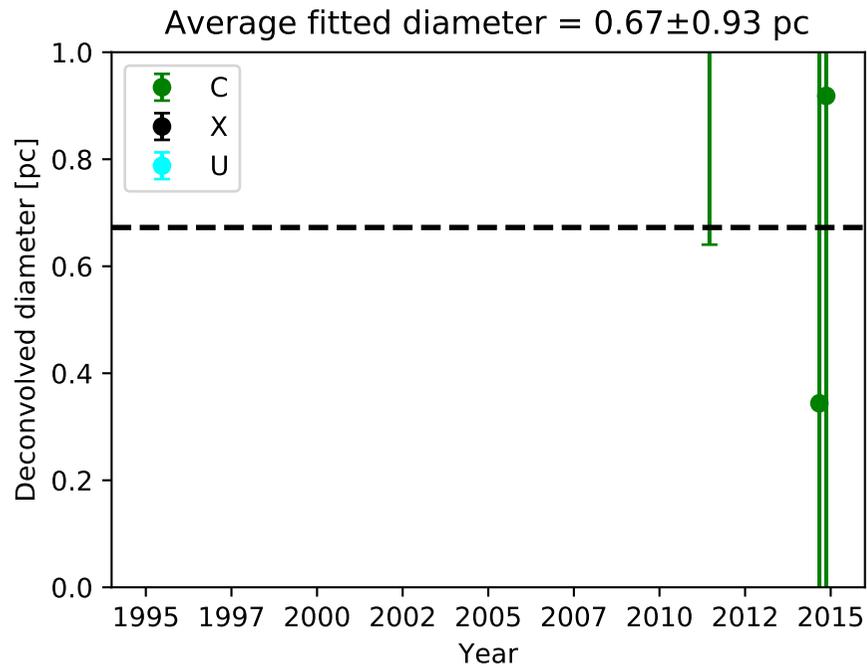

Classification: Rise

Figure notes:

Uncertainties calculated as described in Sect. 2.3.

Triangles represent 3σ upper limits for non-detections.

Sizes shown only for detections in bands C, X, and U.

Average diameters calculated according to Sect. 2.4 from the

3.6 cm and 6 cm measurements in experiments BB335A and BB335B.

# 0.3073+0.456

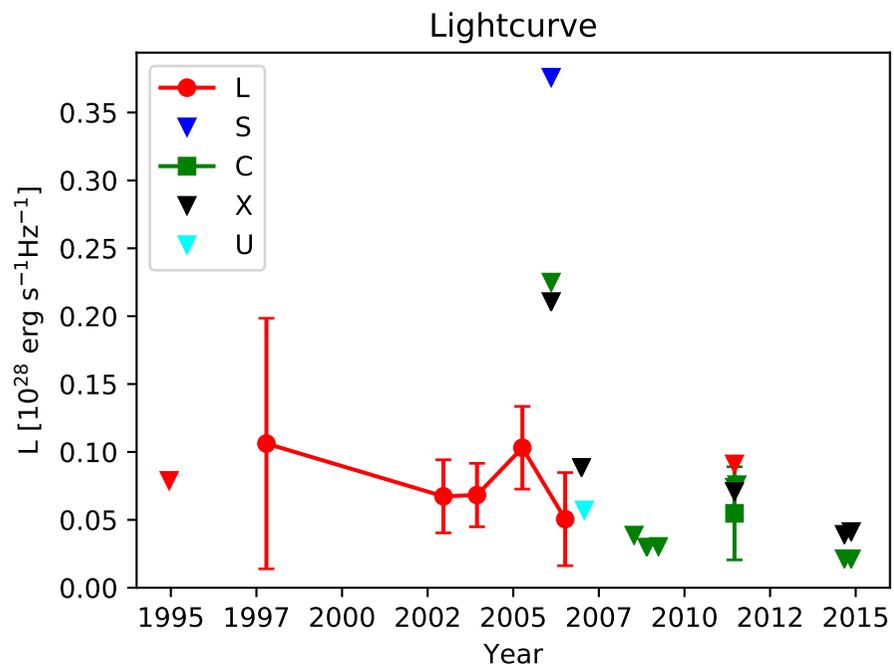

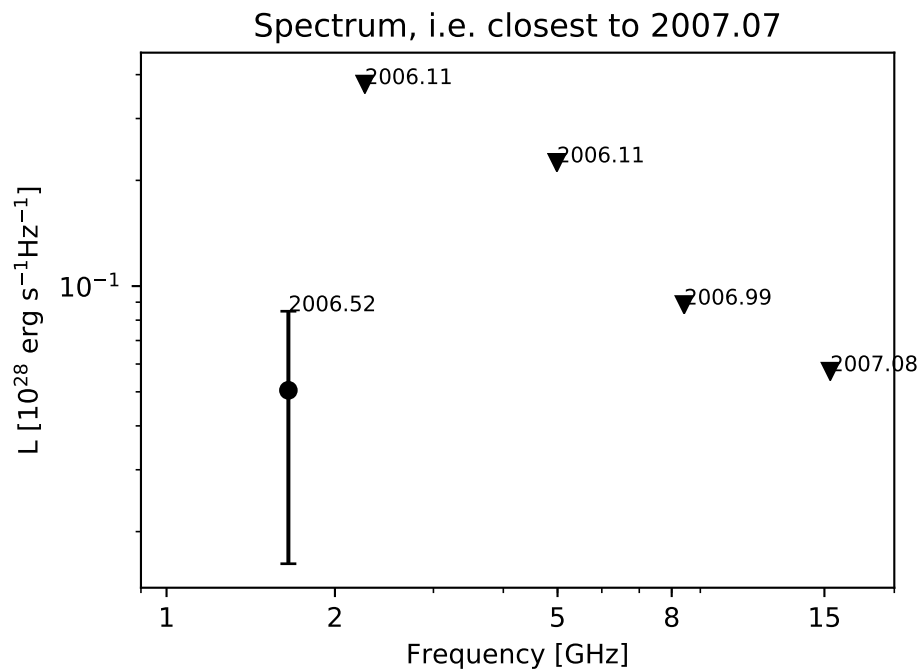

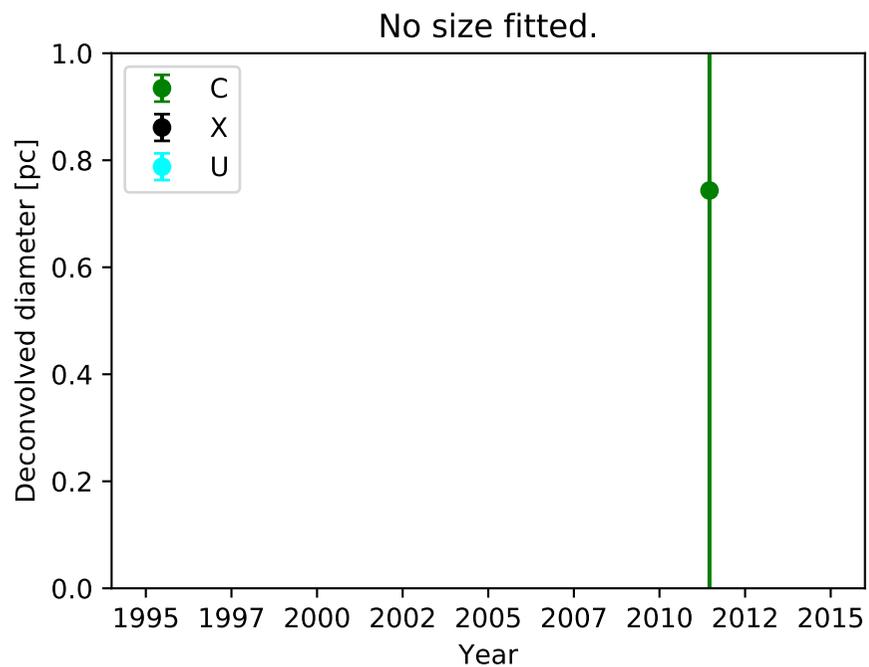

Classification: N/A

Figure notes:

Uncertainties calculated as described in Sect. 2.3.

Triangles represent $3\sigma$ upper limits for non-detections.

Sizes shown only for detections in bands C,X, and U.

Average diameters calculated according to Sect. 2.4 from the

3.6 cm and 6 cm measurements in experiments BB335A and BB335B.

# 0.3090+0.618

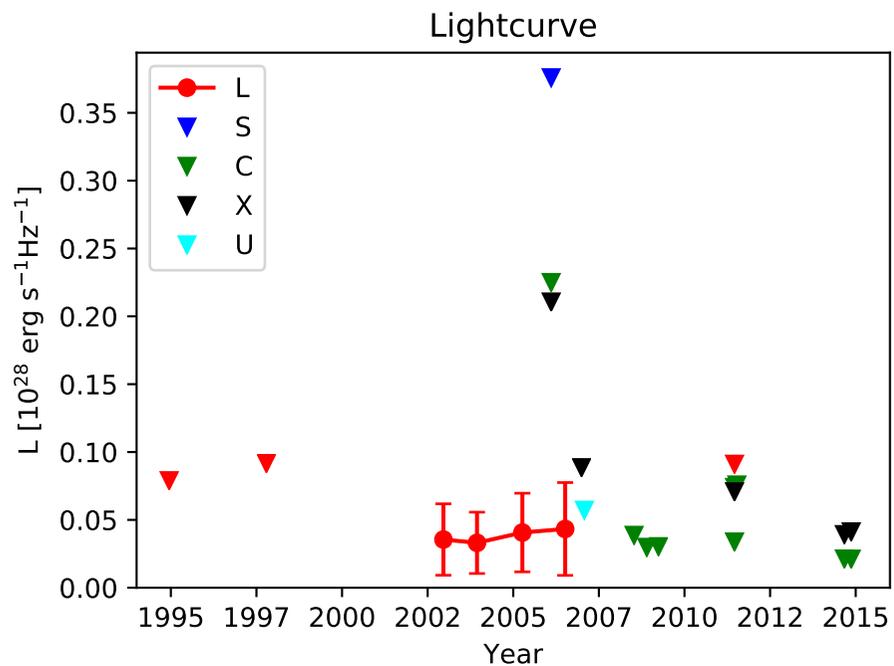
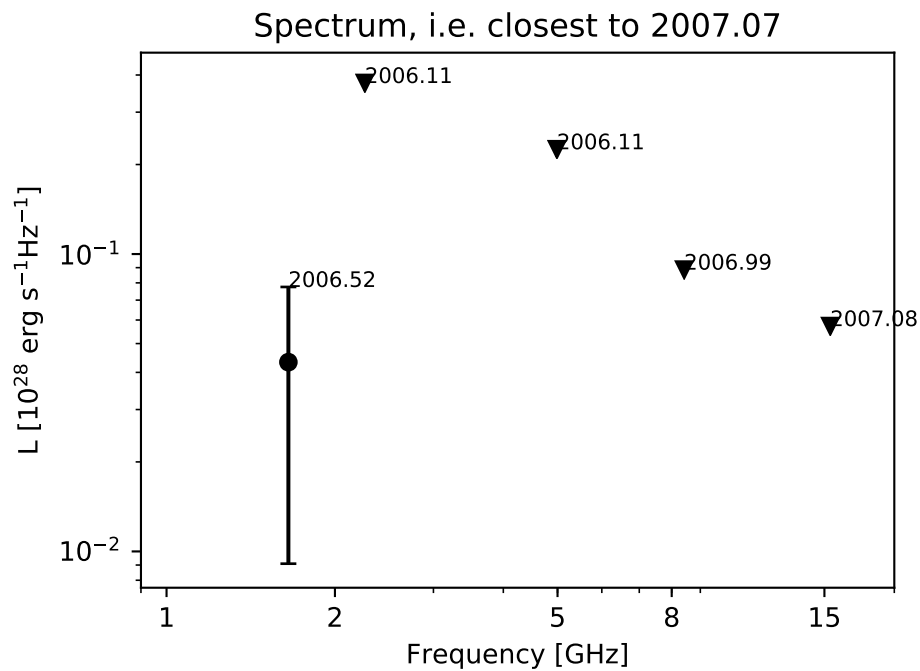
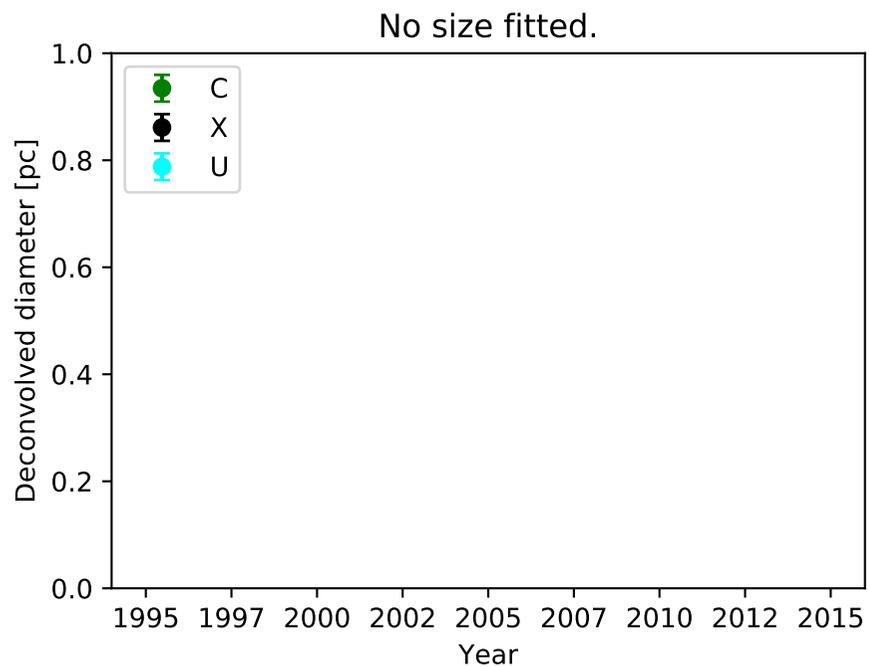

Classification: N/A

Figure notes:

Uncertainties calculated as described in Sect. 2.3.

Triangles represent 3$\sigma$ upper limits for non-detections.

Sizes shown only for detections in bands C,X, and U.

Average diameters calculated according to Sect. 2.4 from the

3.6 cm and 6 cm measurements in experiments BB335A and BB335B.

# 0.3103+0.575

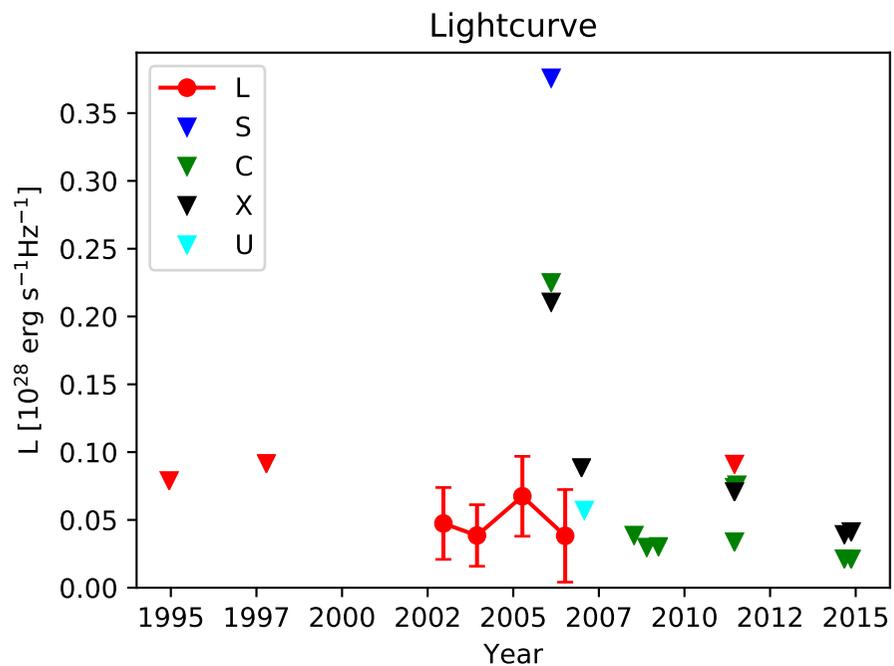

Lightcurve

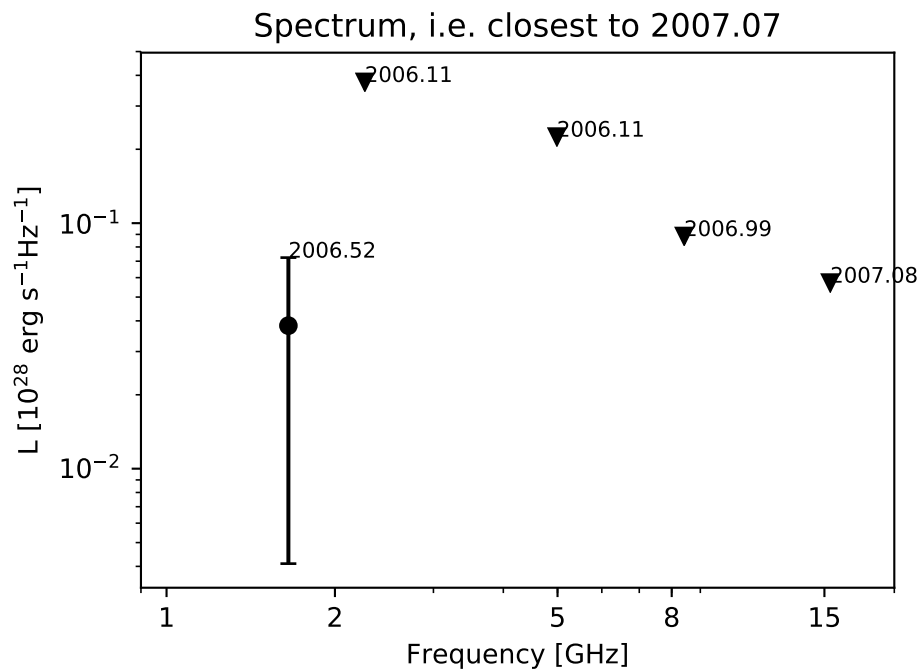

Spectrum, i.e. closest to 2007.07

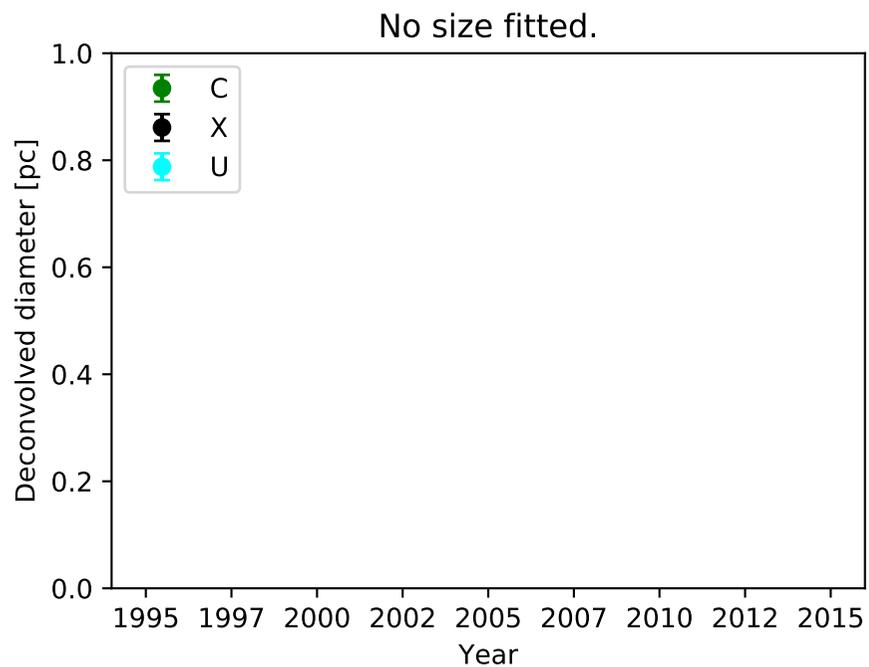

No size fitted.

Classification: N/A

Figure notes:

Uncertainties calculated as described in Sect. 2.3.

Triangles represent $3\sigma$ upper limits for non-detections.

Sizes shown only for detections in bands C,X, and U.

Average diameters calculated according to Sect. 2.4 from the

3.6 cm and 6 cm measurements in experiments BB335A and BB335B.

# 0.3125+0.494

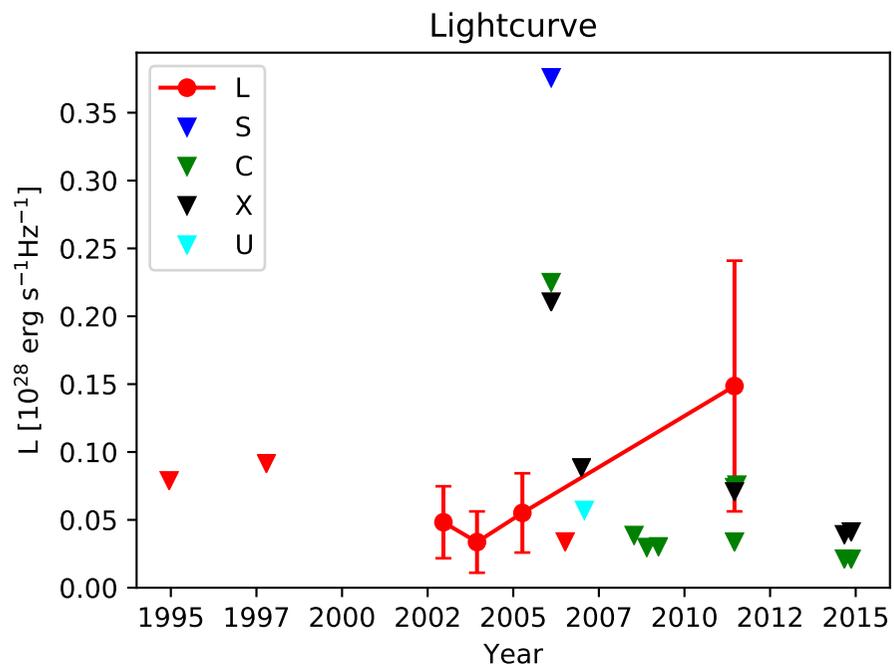

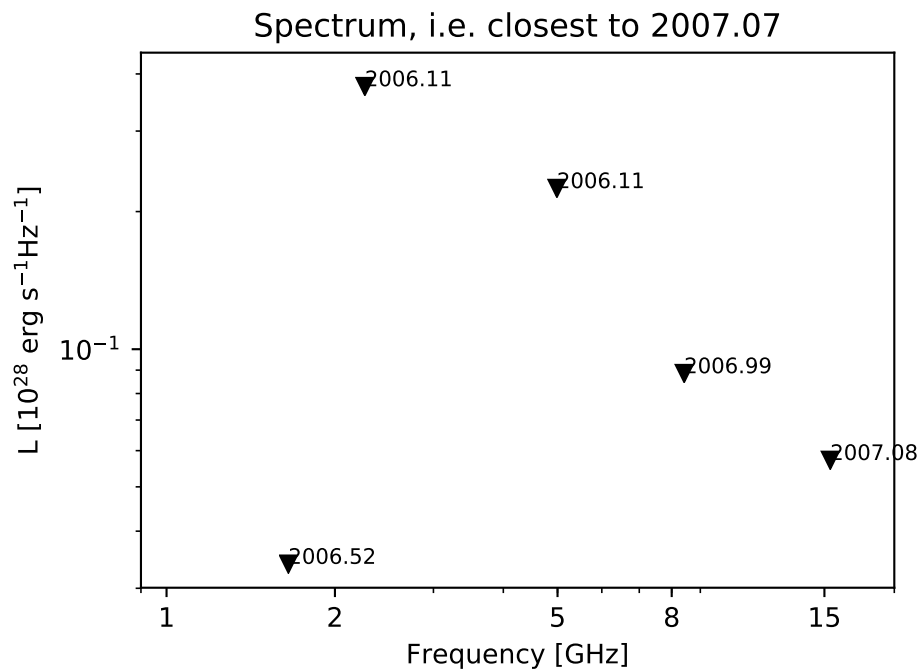

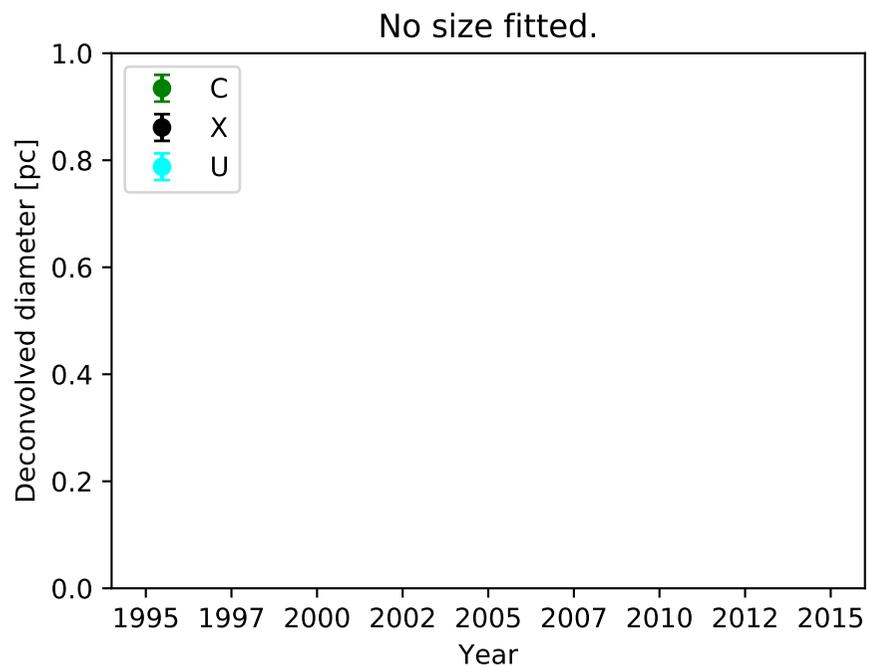

Classification: N/A

Figure notes:

Uncertainties calculated as described in Sect. 2.3.

Triangles represent $3\sigma$ upper limits for non-detections.

Sizes shown only for detections in bands C, X, and U.

Average diameters calculated according to Sect. 2.4 from the

3.6 cm and 6 cm measurements in experiments BB335A and BB335B.



\end{document}